\newcommand*{\ATLASLATEXPATH}{}
\documentclass[cernpreprint,texlive=2011,txfonts,UKenglish,maketitle=true]{\ATLASLATEXPATH atlasdoc}

\usepackage{subfigure}
\usepackage{bm}   
\usepackage{multirow}
\usepackage{float}
\usepackage[T1]{fontenc}

\usepackage{amssymb}
\usepackage{rotating}
\usepackage{slashed}

\usepackage[subfigure=true]{atlaspackage}
\usepackage{atlas-biblatex}

\usepackage{atlascontribute}

\usepackage[BSM=true]{atlasphysics}

\graphicspath{{logos/}{figures/}}

\usepackage{inclsqgl_summarypaper-defs}
\usepackage{arydshln}

\hypersetup{pdftitle={ATLAS draft},pdfauthor={The ATLAS Collaboration}}

\AtlasTitle{Summary of the searches for squarks and gluinos using $\sqrt{s} = 8 \TeV$ $pp$ collisions with the ATLAS experiment at the LHC}

\author{The ATLAS Collaboration}

\AtlasRefCode{SUSY-2014-06}

\PreprintIdNumber{CERN-PH-EP-2015-162}

\AtlasAbstract{A summary is presented of ATLAS searches for gluinos
and first- and second-generation squarks in final states containing jets and missing transverse momentum, with or without leptons or $b$-jets, in the $\sqrt{s}$ = 8 TeV data set collected at the Large Hadron Collider in 2012. This paper reports the results of new interpretations and statistical combinations of previously published analyses, as well as a new analysis. 
Since no significant excess of events over the Standard Model expectation is observed, the data are used to set limits in a variety of models.  
In all the considered simplified models that assume R-parity
conservation, the limit on the gluino mass exceeds 1150 GeV at 95\%
confidence level, for an LSP mass smaller than 100 GeV. Furthermore, exclusion limits are set for left-handed squarks in a phenomenological MSSM model, a minimal Supergravity/Constrained MSSM model, R-parity-violation scenarios, a minimal gauge-mediated supersymmetry breaking model, a natural gauge mediation model, a non-universal Higgs mass model with gaugino mediation and a minimal model of universal extra dimensions.}

\begin{document}

\tableofcontents

\newpage
\section{Introduction}
\label{sec:intro}

Supersymmetry (SUSY)
\cite{Miyazawa:1966,Ramond:1971gb,Golfand:1971iw,Neveu:1971rx,Neveu:1971iv,Gervais:1971ji,Volkov:1973ix,Wess:1973kz,Wess:1974tw} 
is a generalization of space-time symmetries that 
predicts new bosonic partners for the fermions and new fermionic partners for the bosons
of the Standard Model (SM). If R-parity is conserved~\cite{Fayet:1976et,Fayet:1977yc,Farrar:1978xj,Fayet:1979sa},
SUSY particles (called sparticles) are produced in pairs and the lightest supersymmetric particle (LSP) is stable. %
The scalar partners of  the left- and right-handed quarks, the squarks ($\squarkL$ and $\squarkR$ which mix to form two mass eigenstates $\tilde{q}_1$ and $\tilde{q}_2$, ordered by increasing mass), and the fermionic partners of the gluons, gluinos  ($\gluino$), could be produced in strong interaction processes at the 
Large Hadron Collider (LHC)  \cite{LHC:2008} and decay via cascades ending with a stable LSP. 
The rest of the cascade would yield final states with multiple jets and possibly leptons arising from the decay of sleptons ($\tilde\ell$), the superpartners of leptons, or
$W$, $Z$ and Higgs ($h$) bosons originating from the decays of charginos ($\tilde{\chi}^{\pm}$) or neutralinos ($\tilde{\chi}^{0}$), where the charginos and neutralinos are the mass eigenstates
formed from the linear superpositions of the superpartners of the charged and neutral electroweak and Higgs bosons. 
In the Minimal Supersymmetric extension of the Standard Model (MSSM) \cite{Fayet:1976et,Fayet:1977yc,Farrar:1978xj,Fayet:1979sa,Dimopoulos:1981zb}, there are four charginos, 
$\tilde{\chi}^{\pm}_1$ and $\tilde{\chi}^{\pm}_2$, and four neutralinos,
$\tilde{\chi}^{0}_i$ ($i=1$ to 4, ordered by increasing mass); unless stated otherwise, this is assumed in the following. 
In a large variety of models, the LSP is the lightest neutralino ($\ninoone$), which interacts weakly and is a possible candidate for dark matter \cite{Goldberg:1983nd}.
Undetected $\ninoone$ LSPs would result in substantial missing transverse momentum ($\bm{E}\mathrm{^{miss}_T}$, with magnitude $\met$). Significant $\met$ can also arise in R-parity-violating (RPV) scenarios in which the LSP decays to final states containing neutrinos or in scenarios where neutrinos are present in the cascade decay chains of the produced  SUSY particles.
Significant mass splitting between the top squark (stop) mass eigenstates $\tilde{t}_1$ and $\tilde{t}_2$ is possible due to the large top Yukawa coupling.\footnote{The masses of the $\tilde{t}_1$ and $\tilde{t}_2$ are the eigenvalues of the stop mass matrix. The stop mass matrix involves the top quark Yukawa coupling in the off-diagonal elements, which typically induces a large mass splitting. %
} 
 Because of the  SM weak isospin symmetry the mass of the left-handed bottom squark (sbottom, $\tilde{b}_{\rm L}$) is tied to the mass of the  left-handed stop ($\tilde{t}_{\rm L}$), and as a consequence the lightest sbottom ($\tilde{b}_1$) and stop ($\tilde{t}_1$) could be produced 
via the strong interaction with relatively large cross-sections at the LHC, either through direct pair production or in the decay of pair-produced gluinos.

The ATLAS experiment \cite{Aad:2008zzm} performed several searches for supersymmetric
particles in Run 1. No statistically significant excesses of events compared to the predictions 
of the Standard Model were observed. Therefore the results were  
expressed as model-independent limits on the production cross-sections of 
new particles and limits in the parameter space of supersymmetric or simplified models. 

The large cross-sections of squark and gluino production in strong interaction processes offer sensitivity to a broad range of SUSY models. 
This paper provides a summary of the results from inclusive searches for gluinos and first- and second-generation squarks performed by ATLAS, using data from proton--proton ($pp$) collisions at a centre-of-mass energy of 8~\TeV\ collected during Run 1 of the LHC. The results for direct production of third-generation squarks are reported elsewhere \cite{3rdGen-summarypaper}. 
In addition to summarizing already published searches for squarks and gluinos, this paper presents 
new signal regions, new interpretations and statistical combinations
of those searches, as well as an additional search using the Razor variable set \cite{RazorVariables}, thus improving the sensitivity to supersymmetric models.
In order to differentiate strongly produced SUSY events from the SM background, the searches typically require high $\met$ due to the presence of the LSP and possibly neutrinos, several high-$\pt$ jets and large deposited transverse energy. 
They are further classified according to the presence of leptons and $b$-jets. A first class of searches applies a veto on leptons  
 \cite{0-leptonPaper, MonojetPaper, multijetsPaper}, 
a second considers final states containing electrons and muons \cite{1lepPaper, dilepton-edgePaper, SS3LPaper}, and a third requires tau leptons in the final state \cite{TauStrongPaper}. 
A fourth class of searches concentrates on final states containing multiple $b$-jets \cite{3bjetsPaper}.

The paper is organized as follows. Section \ref{sec:susysignals} summarizes the SUSY signals in the strong production of gluinos and 
light-flavour squarks.  %
Section \ref{sec:detector} describes the  ATLAS experiment and the data sample used, and section \ref{sec:mcsamples} the Monte Carlo (MC) simulation samples used for background and signal modelling. The physics object reconstruction and identification are presented in section \ref{sec:objects}.
A description of the analysis strategy is given in section \ref{sec:strategy}, and the experimental signatures are presented in section  \ref{sec:signatures}. 
A summary of systematic uncertainties is presented in section \ref{sec:sysuncert}. 
Results obtained using the new signal regions with selections similar to those used in previous publications as well as the new analysis using the Razor variable set are reported in section \ref{sec:results}. The strategy used for the combination of the results from different analyses is discussed in section \ref{sec:combination}. 
Limits in phenomenological and simplified models are presented in section~\ref{sec:limits}.
Section \ref{sec:conclusion} is devoted to a summary and conclusions. 

\section{SUSY models}
\label{sec:susysignals}

Since no superpartners of any of the SM particles have been observed, SUSY, if realized in nature, must be a broken symmetry with a mechanism for breaking the symmetry taking place at a higher energy scale. 
It is difficult to construct a realistic model of spontaneously broken low-energy supersymmetry where the SUSY breaking arises solely as a consequence of the interactions of the particles of the MSSM \cite{Fayet:1974jb,Fayet:1974pd,O'Raifeartaigh:1975pr}. Therefore, it is often assumed that the SUSY breaking originates in a ``hidden'' sector, and its effects are transmitted to the MSSM by some unknown mechanism.  
Various such mechanisms have been proposed, such as gravity-mediated SUSY breaking (SUGRA) \cite{Chamseddine:1982jx,Barbieri:1982eh,Ibanez:1982ee,Hall:1983iz,Ohta:1982wn,Kane:1993td},  gauge-mediated SUSY breaking (GMSB)  \cite{Dine:1981gu,AlvarezGaume:1981wy,Nappi:1982hm,Dine:1993yw, Dine:1994vc,Dine:1995ag}  and anomaly-mediated SUSY breaking (AMSB) \cite{Giudice:1998xp,Randall:1998uk}. As a result, these models consider only a small part of the parameter space of the more general MSSM. 
In such SUSY models, the particle spectrum is typically specified by fixing parameters at the high scale. In order to translate this set of parameters into physically meaningful quantities that describe physics near the electroweak scale, it is necessary to evolve them using their renormalization group equations. %

Another approach to constraining SUSY at the electroweak scale is to use simplified models  
\cite{Alwall:2008ag,Alves:2011wf} which are based on an effective Lagrangian that only describes a small set of kinematically accessible particles, interactions, production cross-sections and branching ratios. The simplest case corresponds to considering one specific SUSY production process with a fixed decay chain. 

Several classes of phenomenological and simplified models, as well as a minimal Universal Extra Dimensions (mUED) scenario \cite{Cheng:2000,Cheng:2002}, covering different combinations of physics objects in the final state, are considered in this paper. Unless otherwise specified, R-parity is assumed to be conserved and the lightest neutralino, \ninoone, is taken to be the LSP. 
The phenomenological models include a scenario for the phenomenological MSSM (pMSSM) \cite{Djouadi:1998di,Berger:2008cq,CahillRowley:2012cb}, minimal Supergravity/Constrained MSSM (mSUGRA/CMSSM) \cite{Chamseddine:1982jx,Barbieri:1982eh,Ibanez:1982ee,Hall:1983iz,Ohta:1982wn,Kane:1993td}, bilinear R-parity violation (bRPV) \cite{brpv}, a minimal gauge-mediated supersymmetry breaking model (mGMSB)  \cite{Dine:1981gu,AlvarezGaume:1981wy,Nappi:1982hm,Dine:1993yw,Dine:1994vc,Dine:1995ag}, natural gauge mediation (nGM) \cite{ngm}, and a non-universal Higgs mass model with gaugino mediation (NUHMG) \cite{Covi:2007xj}.
The simplified models presented here include the pair production of gluinos or first- and second-generation squarks 
with various hypotheses for their decay chains (direct, one-step or two-step decay), as well as 
gluino decays via real or virtual third-generation squarks. 
Direct decays are those where the considered SUSY particles decay directly into SM particles and the LSP, e.g., $\squark \to q \ninoone$. One-step (two-step) decays refer to the cases where the decays occur via one (two) intermediate on-shell SUSY particle(s), e.g., 
$\squark \to q \chinoonepm \to q W \ninoone\; (\squark \to q \chinoonepm \to q W \ninotwo \to q W Z \ninoone)$.  In gluino decays via third-generation squarks, gluinos undergo a one-step decay to a stop or sbottom such as $\gluino \to t \stop \to t t \ninoone$, or decay directly to final states containing top or bottom quarks, for example $\gluino \to t t \ninoone$ if the stop is off-shell.   
In these simplified models, all supersymmetric particles which do not directly enter the production and decay chain are effectively decoupled, i.e. with masses set above a few TeV. %
The list of models considered is not comprehensive, and the searches presented here are sensitive to a larger class of decay patterns, mass combinations and hierarchies.

\subsection{Phenomenological models} \label{subsec:phenomodels}

\subsubsection{A phenomenological MSSM model} \label{subsubsec:pMSSM}

In the pMSSM scenario, no specific theoretical assumption is introduced at the scale of Grand Unification Theories (GUT), or associated with a SUSY breaking mechanism.
A short list of experimentally motivated considerations is used to reduce the 120 parameters of the MSSM to 19 real, weak-scale parameters: 
\begin{itemize}
\item R-parity is exactly conserved,
\item there are no new sources of CP violation beyond that already present in the quark mixing matrix,
\item Minimal Flavour Violation~\cite{D'Ambrosio:2002ex} is imposed at the electro weak scale,
\item the first two generations of squarks and sleptons with the same quantum numbers are mass-degenerate, and their Yukawa couplings are too small to affect sparticle production or precision observables.
\end{itemize}
The remaining 19 independent  parameters are: 10 squark and slepton masses, the gaugino masses ($M_1$, $M_2$, $M_3$, associated with the U(1)$_{\rm Y}$, SU(2)$_{\rm L}$, SU(3)$_{\rm C}$ gauge groups, respectively), the higgsino mass parameter ($\mu$), the ratio ($\tan\beta$) of the vacuum expectation values of the two Higgs fields, the mass of the pseudoscalar Higgs boson ($m_{A}$), and the trilinear couplings for the third generation ($A_{b}$, $A_{t}$ and $A_{\tau}$) \cite{Djouadi:1998di}. 

In the pMSSM model considered here only the left-handed squarks of the first two generations, the two lightest neutralinos $\ninotwo$ and $\ninoone$, and the lightest chargino $\chinoonepm$ are assumed to be within kinematic reach. Three gluino masses are considered, $m_{\gluino}$ = 1.6, 2.2 and 3.0 \TeV, while the masses of all other SUSY particles are kinematically decoupled with masses set to 5 \TeV.
The parameter $\tan\beta$ is set to 4. The model is further specified by four parameters: $m_{\squarkL}$, $\mu$, and $M_1$ and $M_2$, from which $m_{\ninoone}$, $m_{\ninotwo}$ and $m_{\chinoonepm}$  can be calculated. Either $M_1$ is fixed to 
60~\GeV\ and $M_2$ is varied independently, or $M_1$ is varied and $M_2$ is set to $(M_1 + m_{\squarkL})/2$.

\begin{figure}[h]
\centering
\includegraphics[width=0.25\textwidth]{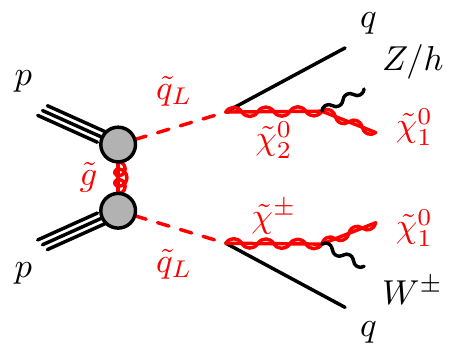}
\caption{Example of a one-step decay topology of the left-handed squark in the phenomenological MSSM.  
} \label{fig:feynman-pMSSM_qL}
\end{figure}

Left-handed squarks can be pair produced only via $t$-channel gluino exchange.  They can undergo a direct $\squarkL \to q \ninoone$ decay, or one-step decays: $\squarkL \to q + \ninotwo \to  q + Z/h + \ninoone$ or $\squarkL \to q + \chinoonepm \to q + W^{\pm} + \ninoone$. Here the lightest Higgs boson $h$ is assumed to have the SM decay branching fractions, and its mass is set to 125~\GeV. The $\chinoonepm$ always decays to $W^{\pm}$ and $\ninoone$ (figure \ref{fig:feynman-pMSSM_qL}). 
The branching fraction to a left-handed squark via the one-step decay with $\ninotwo$ ($\chinoonepm$) is $\sim$ 30\% (65\%). The branching fraction of the $\ninotwo \to h \ninoone$ decay is between 70\% and 90\%  depending on $m_{\squarkL}$. 

\subsubsection{Minimal Supergravity/Constrained MSSM and bilinear R-parity-violation models} \label{subsubsec:msugra}

The mSUGRA/CMSSM model is specified by five parameters: 
a universal scalar mass ($m_0$), a universal gaugino mass ($m_{1/2}$) , a universal trilinear scalar coupling ($A_0$), all defined at the grand unification scale, $\tan\beta$, and the sign of the higgsino mass parameter ($\mu$).
The dependence of the SUSY particle mass spectrum on these five parameters is such that all masses increase with increasing $m_{1/2}$, while squark and slepton masses also depend on $m_0$. 
In the mSUGRA/CMSSM model studied here the values $\tan\beta = 30$, $A_{0} = -2m_0$ and $\mu > 0$ are chosen, such that 
the lightest scalar Higgs boson mass is approximately 125~\GeV~in a large fraction of the $(m_0$, $m_{1/2})$ parameter space studied.

The bRPV scenario uses the same parameters as the mSUGRA/CMSSM model, but R-parity violation is allowed through the bilinear terms\footnote{In this notation, $L_i$ indicates a lepton SU(2)-doublet superfield, the Higgs SU(2)-doublet superfield $H_2$ contains the Higgs field that couples to up-type quarks, and the $\epsilon_i$ parameters have dimension of mass.} $\epsilon_iL_iH_2$, whose coupling parameters 
are determined by a fit to neutrino oscillation data \cite{Carvalho:2002bq} under the tree-level dominance scenario \cite{Grossman:2003gq}. In this scenario, the 
$\ninoone$ LSP decays promptly to $W\mu$, $W\tau$, $Z\nu_{\tau}$ or $h\nu_{\tau}$ (where
the $W/Z/h$ boson can either be on- or off-shell) with branching fractions which are weakly dependent on $m_0$ and $m_{1/2}$ and are typically on the order of 20--40\%, 20--40\%, 20--30\% and 0--20\%, respectively.

\subsubsection{Minimal gauge-mediated supersymmetry breaking model}  \label{subsubsec:gmsb}

In gauge-mediated SUSY breaking models, the LSP is a very light gravitino ($\gravino$). 
The mGMSB model is described by six parameters: the SUSY-breaking mass
scale felt by the low-energy sector ($\Lambda$), the mass of the SUSY breaking messengers 
(\Mmess), the number of SU(5) messenger fields (\Nfive), $\tan\beta$,  $\mu$ and the gravitino coupling scale factor (\Cgrav) which determines the lifetime of the next-to-lightest SUSY particle (NLSP). Four parameters are fixed to
the values previously used in refs.~\cite{Aad:2012rt,ATLAS:2012ag,Aad:2012ms}:
$\Mmess=250$~\TeV, $\Nfive=3$, $\mu>0$ and $\Cgrav=1$. With this choice of parameters  
the production of squark and/or gluino pairs is expected to dominate over other SUSY processes at the LHC. 
These SUSY particles decay into the NLSP, which subsequently decays to the LSP.
The experimental signatures are largely determined by the nature of the NLSP: this can be either the lightest stau ($\stau$), a
selectron or a smuon ($\slepton$), the lightest neutralino ($\ninoone$), or a sneutrino ($\snu$), leading to final states usually containing tau leptons, light leptons ($\ell=e,\mu$), photons, or neutrinos, respectively.

\subsubsection{Natural gauge mediation model}  \label{subsubsec:nGM}

\begin{figure}[h]
\centering
\includegraphics[width=0.25\textwidth]{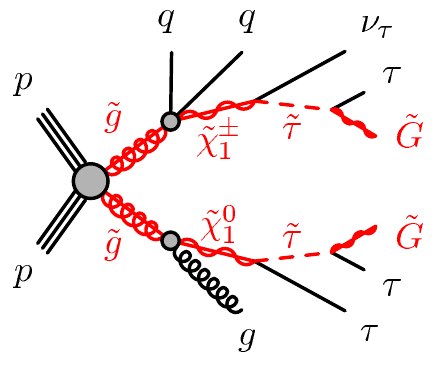}
\caption{Example of a gluino-pair production followed by the two possible decay chains within the nGM scenario. 
} \label{fig:feynman-nGM}
\end{figure}

In the nGM scenario, which assumes general gauge mediation \cite{ggm,ggm2}, the phenomenology depends on the nature of the NLSP~\cite{Barnard:2012au, Asano:2010ma}.  Various models assume that the mass hierarchies of squarks and sleptons are generated by the same physics responsible for breaking SUSY (for example refs.~\cite{Gabella:2007mg, Craig:2012ng}).  Typically in these models the third generation of squarks and sleptons is lighter than the other two, and together with the fact that sleptons only acquire small masses through hypercharge interactions in gauge mediation, this leads to a stau NLSP. 
In the model considered here, it is also assumed that the gluino is the only light coloured sparticle.
All squark and slepton mass parameters are set to 2.5 \TeV\ except the lightest stau mass, $m_{\tilde{\tau}}$, which is assumed to be smaller.  The parameters $M_1$ and $M_2$ are also set to 2.5 \TeV, while all trilinear 
coupling terms are set to zero. 
The value of $\mu$ is set to 400 \GeV\ to ensure that strong production dominates in the parameter space studied. 
This leaves the gluino mass $M_3$ and the stau mass $m_{\tilde{\tau}}$ as the only free parameters.  %
The chosen value of the $\mu$ parameter sets the masses of the $\chinoonepm$, $\ninoone$ and $\ninotwo$, which are almost mass-degenerate. 
The only light sparticles in the model are the stau, a light gluino, higgsino-dominated charginos and neutralinos, and a very light gravitino LSP. Therefore, the strong production process allowed in this model is gluino-pair production followed by the three possible decay chains: 
$ \gluino \rightarrow g \ninoonetwo \rightarrow g \stau \tau \rightarrow g \tau \tau \gravino$,  
$ \gluino \rightarrow q \bar{q} \ninoonetwo \rightarrow q \bar{q} \stau \tau \rightarrow q \bar{q} \tau \tau \gravino$ 
and  $\gluino \rightarrow q q \chinoonepm \rightarrow q q \nu_{\tau} \stau \rightarrow  q q \nu_{\tau} \tau \gravino$ (figure \ref{fig:feynman-nGM}), where the final-state quarks are almost exclusively top or bottom quarks.
A range of signals with varying gluino and stau masses is studied. The lightest Higgs boson mass is specifically set to 125~\GeV. 

\subsubsection{Non-universal Higgs mass models with gaugino mediation} \label{subsubsec:NUHMG}

The NUHMG model is specified with parameters $m_0=0$, $\tan \beta$ = 10, $\mu > 0$, $m_{H_{2}}^2$ = 0, and $A_{0}$ chosen to maximize the mass of the lightest Higgs boson. The ranges of the two remaining free parameters of the model, $m_{1/2}$ and $m_{H_{1}}^2$, are chosen such that the NSLP is a tau sneutrino with properties satisfying Big Bang nucleosynthesis constraints \cite{Covi:2007xj}. The squared mass terms of the two Higgs doublets, $m_{H_{1}}^2$ and $m_{H_{2}}^2$, are defined at the unification scale. This model is characterized by significant cross-sections for $\squark$ and $\gluino$ production. The gluino decays mainly to a light quark/squark pair $q\squark$ ($\approx50$\%), but also to $t\stop$~($\approx30$\%) or $b\sbottom$~($\approx20$\%), while the squark multi-step decays typically involve charginos, neutralinos and/or sleptons. 

\subsubsection{Minimal Universal Extra Dimensions model}  \label{subsubsec:mUED}

The mUED model is the minimal extension of the SM with one additional universal spatial dimension.
In this non-SUSY model, the Kaluza--Klein (KK) quark excitation's decay
chain to the lightest KK particle, the KK photon, gives a signature very similar to the supersymmetric
decay chain of a squark to the lightest neutralino.
The properties of the model depend on two parameters: the compactification radius $R_{\rm c}$ and the cut-off scale $\Lambda$. This cut-off is interpreted as the scale at which some new physics underlying the effective non-renormalizable UED framework becomes relevant.
The Higgs boson mass is fixed to 125~\GeV.

\subsection{Simplified models} \label{subsec:simplified}

The details of the simplified models considered are given below and summarized in tables \ref{tab:model1}--\ref{tab:model3}. 

\begin{table}[H]
\centering
\small
 \renewcommand{\arraystretch}{1.5}
  \resizebox{\columnwidth}{!}{%
\begin{tabular}{|l| c | c|l | l | l |}  \hline 
Diagram & Production  &  Parameters & Mass relation  &Branching ratio & Result \\  \hline
fig. \ref{fig:feynman-direct}(a)& $\squark\squark$& $m_{\squark}, m_{\ninoone} $ & $m_{\squark}>m_{\ninoone}$& BR$(\squark\rightarrow q\ninoone )=1$&     fig. \ref{fig:limit-sqsq_direct}\\[1mm] \hline  %
 
fig. \ref{fig:feynman-direct}(b)& $\gluino\squark$& $ m_{\gluino}, m_{\ninoone}$ & $m_{\squark}= 0.96\ m_{\gluino}>m_{\ninoone}$& BR$(\squark\rightarrow q\ninoone )=1$&    
fig. \ref{fig:limit-sqgl_direct}(a)\\ %
&&& &BR$(\gluino \rightarrow \squark q )=1$& \\  \cdashline{3-6}%
& & $m_{\squark}, m_{\gluino}$ & $m_{\ninoone}=(0, 395, 695)\GeV$& If
$m(\gluino) > m(\squark)$: &    fig. \ref{fig:limit-sqgl_direct}(b)\\[1mm] %
 & & & & BR$(\squark\rightarrow q\ninoone )=1$, BR$(\gluino\rightarrow \squark q )=1$ & \\
 & & & & If $m(\squark) > m(\gluino)$: & \\
 & & & & BR$(\gluino\rightarrow
q q \ninoone )=1$, BR$(\squark\rightarrow \gluino q )=1$ & \\ \hline
 
fig. \ref{fig:feynman-direct}(c)& $\gluino\gluino$& $m_{\gluino}, m_{\ninoone}$ & $m_{\gluino}>m_{\ninoone}$& BR$(\gluino\rightarrow qq\ninoone )=1$&  fig. \ref{fig:limit-glgl_direct}\\[1mm]  \hline %
 
 fig. \ref{fig:feynman-direct}(d)&$\gluino\gluino$& $m_{\gluino}$ & $m_{\ninoone}=0$& BR$(\gluino\rightarrow g\ninoone )=1$&  fig. \ref{fig:limit-gluino_gluon}(a) \\[1mm]  \cdashline{3-4}\cdashline{6-6} %
  && $m_{\ninoone}$ & $m_{\gluino}=850\GeV$& &  fig. \ref{fig:limit-gluino_gluon}(b)\\[1mm]  \hline %

\end{tabular}
}
\caption{Simplified models of squark and gluino production with direct
  decays to $\ninoone$. For each model the diagram of the decay
  topology, the model parameters and assumptions about mass relations
  and branching ratios are listed. The last column refers to the
  experimental results presented in section \ref{subsec:simplifiedlimits}. Horizontal dashed lines separate different mass or branching ratio assumptions within a model.}
\label{tab:model1}
\end{table}

\begin{table}[H]
\centering
\small
 \renewcommand{\arraystretch}{1.5}
 \resizebox{\columnwidth}{!}{%
\begin{tabular}{|l| c |c| l | l | l |}  \hline  
Diagram&Production    &  Parameters & Mass relation  &Branching ratio & Result \\  \hline
fig. \ref{fig:feynman-onestep}(a) & $\squark\squark$ &$m_{\squark}, m_{\ninoone}$ & $m_{\chinoonepm}=(m_{\squark}+m_{\ninoone})/2$ & 
BR$(\squark\rightarrow qW\ninoone)=1$ & fig. \ref{fig:limit-sqsq_onestep}(a) \\[2mm] \cdashline{3-4}\cdashline{6-6}%
 & &$m_{\squark}$ & $m_{\ninoone}=60\GeV$ & & fig. \ref{fig:limit-sqsq_onestep}(b)  \\ %
 & &$x=\Delta m(\chinoonepm,\ninoone)/\Delta m(\squark,\ninoone)$ & & &  \\ \hline
 
fig. \ref{fig:feynman-onestep}(b) & $\gluino\gluino$ &$m_{\gluino}, m_{\ninoone}$ & $m_{\chinoonepm}=(m_{\gluino}+m_{\ninoone})/2$ & 
BR$(\gluino\rightarrow qqW\ninoone)=1$ & fig.  \ref{fig:limit-glgl_onestep}(a)  \\[2mm] \cdashline{3-4}\cdashline{6-6}%
 & &$m_{\gluino}$ & $m_{\ninoone}=60\GeV$ & & fig.  \ref{fig:limit-glgl_onestep}(b) \\ %
 & &$x=\Delta m(\chinoonepm,\ninoone)/\Delta m(\gluino,\ninoone)$ & & &  \\ \hline

fig. \ref{fig:feynman-twostep}(a) & $\squark\squark$ &$m_{\squark}, m_{\ninoone}$ & $m_{\chinoonepm,\ninotwo}=(m_{\squark}+m_{\ninoone})/2$ &
BR$(\squark\rightarrow q(\ell\nu/\ell\ell)\ninoone)=1$ & fig. \ref{fig:limit-sqsq_twostepSlepton} \\ %
&&&$m_{\slepton,\snu}=(m_{\chinoonepm,\ninotwo}+m_{\ninoone})/2$&$\ell\equiv(e,\mu)$& \\[1mm] \cdashline{5-6}
 &  & & & BR$(\squark\rightarrow q(\tau\nu/\tau\tau/\nu\nu)\ninoone)=1$ & fig. \ref{fig:limit-sqsq_twostepTau} \\ %
&&&&$\ell\equiv\tau$& \\[1mm] \hline

fig. \ref{fig:feynman-twostep}(b) & $\squark\squark$ &$m_{\squark}, m_{\ninoone}$ & $m_{\chinoonepm}=(m_{\squark}+m_{\ninoone})/2$ &
BR$(\squark\rightarrow qWZ\ninoone)=1$ & fig. \ref{fig:limit-sqsq_twostepWWZZ} \\ %
&&&$m_{\ninotwo}=(m_{\chinoonepm}+m_{\ninoone})/2$&& \\[2mm] \hline

fig. \ref{fig:feynman-twostep}(c) & $\gluino\gluino$ &$m_{\gluino}, m_{\ninoone}$ & $m_{\chinoonepm,\ninotwo}=(m_{\gluino}+m_{\ninoone})/2$ &
BR$(\gluino\rightarrow qq(\ell\nu/\ell\ell)\ninoone)=1$ & fig. \ref{fig:limit-glgl_twostepSlepton} \\ %
&&&$m_{\slepton,\snu}=(m_{\chinoonepm,\ninotwo}+m_{\ninoone})/2$&$\ell\equiv(e,\mu)$& \\[1mm]\cdashline{5-6}
 &  & & & BR$(\gluino\rightarrow qq(\tau\nu/\tau\tau/\nu\nu)\ninoone)=1$ & fig.  \ref{fig:limit-glgl_twostepTau} \\ %
&&&&$\ell\equiv\tau$& \\[1mm] \hline

fig. \ref{fig:feynman-twostep}(d) & $\gluino\gluino$ &$m_{\gluino}, m_{\ninoone}$ & $m_{\chinoonepm}=(m_{\gluino}+m_{\ninoone})/2$ &
BR$(\gluino\rightarrow qqWZ\ninoone)=1$ & fig. \ref{fig:limit-glgl_twostepWWZZ} \\ %
&&&$m_{\ninotwo}=(m_{\chinoonepm}+m_{\ninoone})/2$&& \\[2mm] \hline

\end{tabular}
}
\caption{Simplified models of squark and gluino production with one- and two-step decays to $\ninoone$. For each model the diagram of the decay topology, the model parameters and assumptions about mass relations and branching ratios are listed. The last column refers to the experimental results presented in section \ref{subsec:simplifiedlimits}. Horizontal dashed lines separate different mass or branching ratio assumptions within a model.}
\label{tab:model2}
\end{table}

\begin{table}[H]

\centering
\small
 \renewcommand{\arraystretch}{1.5}
  \resizebox{\columnwidth}{!}{%
\begin{tabular}{|l| c |c| l | l | l |}  \hline 
Diagram  & Parameters &  Mass relation  &Branching ratio & Result  \\  \hline
fig. \ref{fig:feynman-Gluino-top-onshell}(a)& $m_{\gluino}, m_{\stopone}$ & $m_{\gluino}>m_{\stopone}+m_t$& BR$(\gluino\rightarrow \stopone  t)=1$& fig. \ref{fig:limit-Gtt-onshell}\\ %
 &&$m_{\ninoone}=60\GeV$&BR$(\stopone\rightarrow t \ninoone )=1$&\\[1mm] \hline

fig. \ref{fig:feynman-Gluino-top-onshell}(b)&$m_{\gluino}, m_{\stopone}$ &$m_{\gluino}>m_{\stopone}+m_t$  &BR$(\gluino\rightarrow \stopone  t)=1$ &    fig. \ref{fig:limit-Gtt-bChi}\\ %
&  & $m_{\chinoonepm} = 2 m_{\ninoone}$  & BR$(\stopone\rightarrow b \chinoonepm)=1$  &   \\
&  & $m_{\ninoone}=60\GeV$ & BR$(\chinoonepm\rightarrow W^*\ninoone)=1$  &   \\[1mm] \hline
  
 fig. \ref{fig:feynman-Gluino-top-onshell}(c)&$m_{\gluino}, m_{\stopone}$ &$m_{\gluino}>m_{\stopone}+m_t$   & BR$(\gluino\rightarrow \stopone  t)=1$&    fig. \ref{fig:limit-Gtt-charm}\\ %
      &      & $m_{\ninoone}=m_{\stopone}-20\GeV$   &   BR$(\stopone\rightarrow c \ninoone)=1$&     \\[1mm]   \hline
                
fig. \ref{fig:feynman-Gluino-top-onshell}(d)&$m_{\gluino}, m_{\stopone}$ & $m_{\gluino}>m_{\stopone}+m_t$  & BR$(\gluino\rightarrow \stopone  t)=1$ &    fig. \ref{fig:limit-Gtt-RPV}\\ %
&  &  & BR$(\stopone\rightarrow sb)=1$  &   \\ \hline
  
fig. \ref{fig:feynman-Gluino-bottom-onshell}&$m_{\gluino}, m_{\sbottomone}$ &  $m_{\gluino}>m_{\sbottomone}+m_b$& BR$(\gluino\rightarrow \sbottomone  b)=1$&  fig. \ref{fig:limit-Gbbonshell}\\ %
&  &$m_{\ninoone}=60\GeV$  & BR$(\sbottomone\rightarrow b \ninoone )=1$ &   \\[1mm] \hline  

fig. \ref{fig:feynman-Gluino-topbottom-offshell}(a)&$m_{\gluino},
m_{\ninoone}$ &  $m_{\gluino}\ll m_{\stopone}$& If
$m_{\gluino} > 2m_t+m_{\ninoone}$: BR$(\gluino\rightarrow \ttbar \ninoone)=1$ &    
fig.  \ref{fig:limit-Gtt-offshell} \\ %
 & & &   If $m_{\gluino}<2m_t+m_{\ninoone}$: & \\
 & & &   BR$(\gluino\rightarrow tWb\ninoone)+$BR$(\gluino\rightarrow WbWb\ninoone)=1$ & \\ [1mm]\hline 
  
fig. \ref{fig:feynman-Gluino-topbottom-offshell}(b)&$m_{\gluino}, m_{\ninoone}$ &  $2m_b+m_{\ninoone} < m_{\gluino}\ll m_{\sbottomone}$& BR$(\gluino\rightarrow \bbbar \ninoone)=1$&    
fig. \ref{fig:limit-Gbb} \\[1mm] \hline %
  
fig. \ref{fig:feynman-Gluino-topbottom-offshell}(c)&$m_{\gluino}, m_{\ninoone}$ & $m_b+m_t+m_{\chinoonepm}<m_{\gluino}\ll m_{\stopone},m_{\sbottomone} $& BR$(\gluino\rightarrow tb \chinoonepm)=1$&    fig. \ref{fig:limit-Gtb}\\    %
&  &$m_{\chinoonepm}=m_{\ninoone}+2\GeV$ & BR$(\chinoonepm\rightarrow \ninoone ff^\prime)=1$&    \\[1mm]    \hline
   
 \hline

\end{tabular}
}
\caption{Simplified models of gluino pair production with decays via third-generation squarks. For each model the diagram of the decay topology, the model parameters and assumptions about  mass relations and branching ratios are listed. The last column refers to the experimental results presented in section \ref{subsec:simplifiedlimits}. Horizontal dashed lines separate different mass or branching ratio assumptions within a model.}

\label{tab:model3}
\end{table}

\subsubsection{Direct decays of squarks and gluinos} \label{subsubsec:direct}

\begin{figure}[h]
\centering
\subfigure[]{\includegraphics[width=0.25\textwidth]{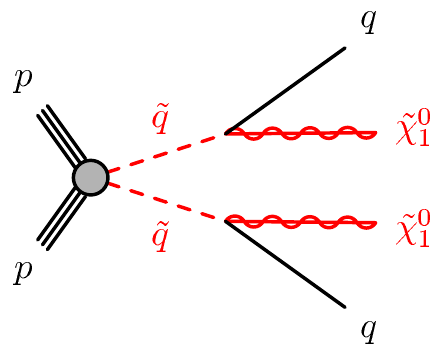}}\hspace{0.05\textwidth}
\subfigure[]{\includegraphics[width=0.25\textwidth]{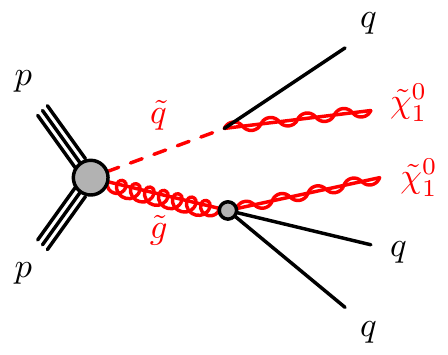}}\\
\subfigure[]{\includegraphics[width=0.25\textwidth]{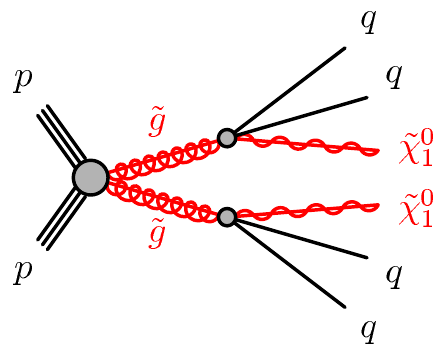}}\hspace{0.05\textwidth}
\subfigure[]{\includegraphics[width=0.25\textwidth]{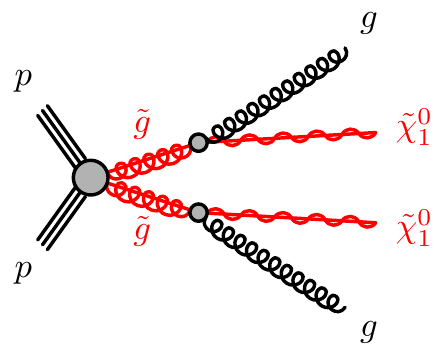}}
\caption{The decay topologies of (a) squark-pair production,  (b) squark--gluino production or (c,d) gluino-pair production, in the simplified models with direct decays. 
} \label{fig:feynman-direct}
\end{figure}

Simplified models with direct decay of the pair-produced strongly interacting supersymmetric particles assume the production of gluino pairs with decoupled squarks, light-flavour squark pairs with decoupled gluinos, or light-flavour squarks and gluinos; all other superpartners except the lightest neutralino are decoupled. This assumption forces squarks or gluinos to decay directly to quarks or gluons and the lightest neutralino, as shown in figure \ref{fig:feynman-direct}. In the case of squark--gluino production, the masses of the light-flavour squarks are set to 0.96 times the mass of the gluino as suggested in refs.~\cite{TomSteve1,TomSteve2}, and gluinos can decay via on-shell squarks as $\gluino \to \squark q  \to q q \ninoone $.   %
For models with decoupled gluinos two scenarios have been considered: a scenario with eight mass-degenerate light-flavour squarks ($\squark_{\rm L}$ and $\squark_{\rm R}$, with $\squark = \tilde{u}, \tilde{d}, \tilde{s}, \tilde{c}$), or a scenario with only one accessible light-flavour squark \cite{Mahbubani:2012qq}.  Changing the number of light-flavoured squarks affects only the cross-section but not the kinematics of the events. The free parameters in these models are $m_{\squark}$ or $m_{\gluino}$, and $m_{\ninoone}$. 

An additional set of simplified models with direct decay of pair-produced gluinos assumes that all squarks and sleptons are much heavier than the gluino, which remains relatively light and decays promptly into a gluon and a neutralino \cite{Toharia:2005gm}, as shown in figure \ref{fig:feynman-direct}(d). The free parameters in these models are $m_{\gluino}$ and $m_{\ninoone}$.

\subsubsection{One-step decays of squarks and gluinos} \label{subsubsec:onestep}

\begin{figure}[H]
\centering
\subfigure[]{\includegraphics[width=0.25\textwidth]{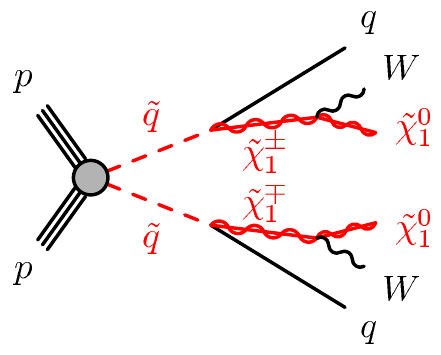}}\hspace{0.05\textwidth}
\subfigure[]{\includegraphics[width=0.25\textwidth]{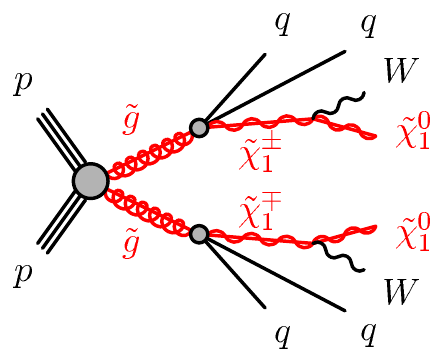}}
\caption{The decay topologies of (a) squark-  or (b) gluino-pair production, in the simplified models with one-step decays. 
} \label{fig:feynman-onestep}
\end{figure}

Simplified models with one-step decays of the pair-produced squarks or gluinos assume that these particles
decay via the $\chinoonepm$ into a $W$ boson and the $\ninoone$, as shown in figure \ref{fig:feynman-onestep}.  
The free parameters in these models are $m_{\squark}$ or $m_{\gluino}$, and either $m_{\chinoonepm}$ with a fixed $m_{\ninoone}$ = 60~\GeV\ or  $m_{\ninoone}$ with $m_{\chinoonepm}=(m_{\gluino/\squark}+m_{\ninoone})/2$.

\subsubsection{Two-step decays of squarks and gluinos} \label{subsubsec:twostep}

Two categories of simplified models with two-step decays of squarks and gluinos are considered: models with and without sleptons.

In the two-step models with sleptons the pair-produced squarks or gluinos decay with equal probability to
either the lightest chargino or the next-to-lightest neutralino ($\ninotwo$). These subsequently decay via left-handed sleptons (or sneutrinos)
which then further decay into a lepton (or neutrino) and the lightest neutralino. In these models, the free parameters
are the mass of the initially produced particle and the mass of the lightest neutralino. The masses of the intermediate charginos or neutralinos are equal
and set to be $m_{\chinoonepm,\ninotwo}=(m_{\gluino/\squark}+m_{\ninoone})/2$, 
while the slepton and sneutrino masses are set to be $m_{\sleptonL,\snu}=(m_{\chinoonepm/\ninotwo}+m_{\ninoone})/2$.  All three slepton flavours are mass-degenerate in this model. A separate model in which the slepton is exclusively a $\stau$ is also considered. 

In the second category, two-step models without sleptons, the initial supersymmetric particle decays via the lightest chargino, which itself decays into a $W$ boson and the next-to-lightest neutralino. The latter finally decays into a $Z$ boson and the lightest neutralino. The lightest chargino mass is fixed at $m_{\chinoonepm}=(m_{\gluino/\squark}+m_{\ninoone})/2$ and the next-to-lightest neutralino mass is set to be $m_{\ninotwo}=(m_{\chinoonepm}+m_{\ninoone})/2$. 

\begin{figure*}[h]
\centering
\subfigure[]{\includegraphics[width=0.25\textwidth]{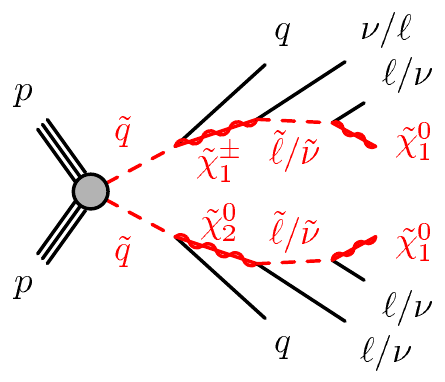}}\hspace{0.05\textwidth}
\subfigure[]{\includegraphics[width=0.25\textwidth]{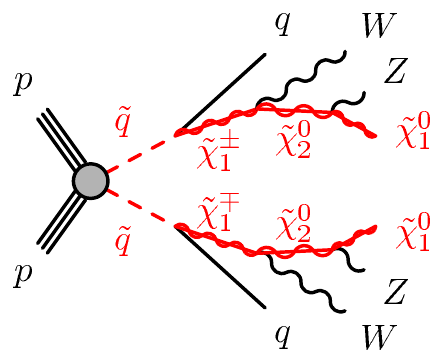}}\\
\subfigure[]{\includegraphics[width=0.25\textwidth]{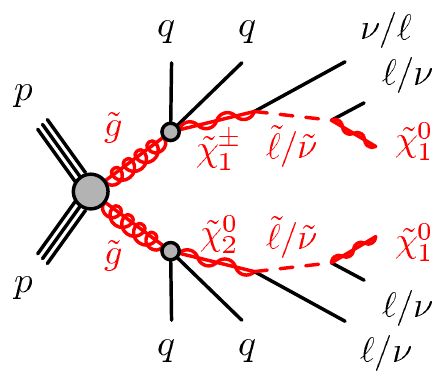}}\hspace{0.05\textwidth}
\subfigure[]{\includegraphics[width=0.25\textwidth]{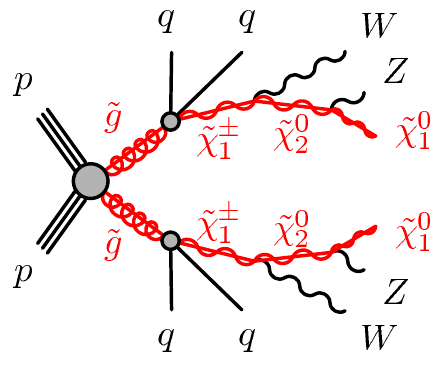}}\\
\caption{Examples of decay topologies of (a, b) squark- or (c, d) gluino-pair production, in the simplified models with two-step decays with (left) or without (right) sleptons.
} \label{fig:feynman-twostep}
\end{figure*}

These two categories of simplified models with two-step decays of squarks and gluinos are illustrated in figure ~\ref{fig:feynman-twostep}. 

\subsubsection{Gluino decays via third-generation squarks} \label{subsubsec:gttgbb}

\begin{figure*}[h]
\centering
\subfigure[]{\includegraphics[width=0.25\textwidth]{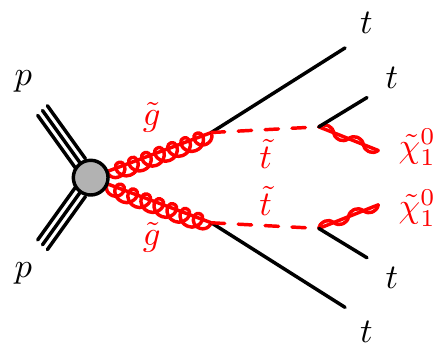}}\hspace{0.05\textwidth}
\subfigure[]{\includegraphics[width=0.25\textwidth]{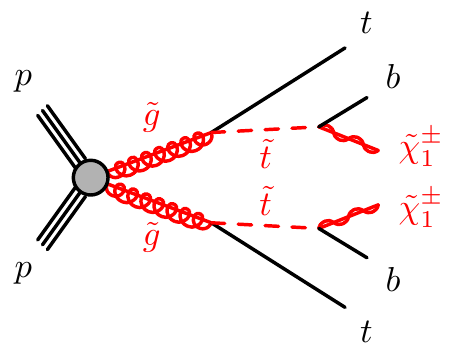}}\hspace{0.05\textwidth}\\
\subfigure[]{\includegraphics[width=0.25\textwidth]{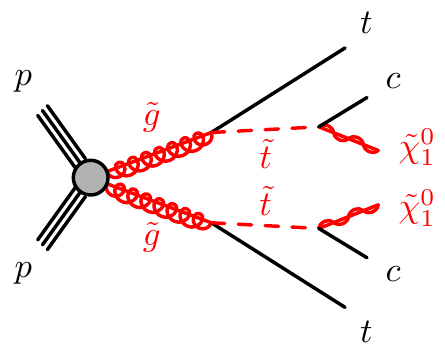}}\hspace{0.05\textwidth}
\subfigure[]{\includegraphics[width=0.25\textwidth]{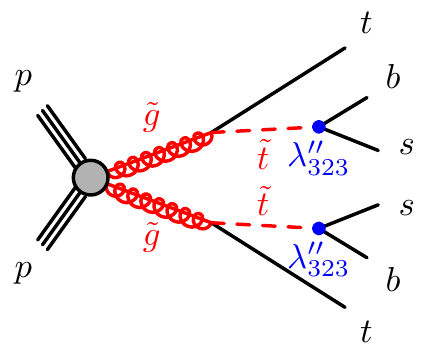}}
\caption{Decay topologies in the gluino--stop simplified models with the top squark decays: (a) $\stopone \to t \ninoone$, (b) $\stopone \to b \chinoonepm$, (c) $\stopone \to c \ninoone$ and (d) $\stopone \to s b$  with R-parity and baryon number violation, with a strength determined by the parameter $\lambda^{\prime\prime}_{323}$. 
} \label{fig:feynman-Gluino-top-onshell}
\end{figure*}

\begin{figure}[h]
\centering
\includegraphics[width=0.25\textwidth]{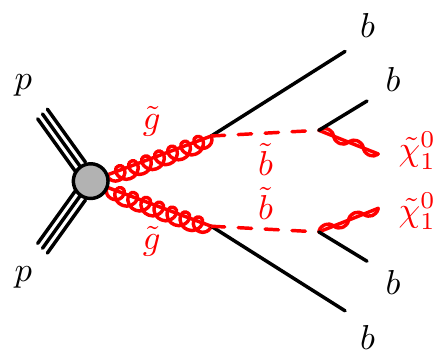}
\caption{The decay topology in the gluino--sbottom simplified models, with the bottom squark decay $\sbottomone \to b\ninoone$. 
} \label{fig:feynman-Gluino-bottom-onshell}
\end{figure}

Two classes of simplified models with gluino decays via third-generation squarks are considered.
In the first, the lightest stop or sbottom is lighter than the gluino, such that $\stopone$ or $\sbottomone$ are produced via gluino-pair production followed by $\gluino \to \stopone t$ or  $\gluino \to \sbottomone b$ decays. 
Gluino--stop models within this class assume that the $\stopone$ is the lightest squark while all other squarks are heavier than the gluino, and 
$m^{}_{\gluino}>m^{}_{\stopone}+m_t$ such that the branching ratio for $\gluino \to \stopone t$ decays is 100\%. 
Top squarks are assumed to decay via either $\stopone \to t \ninoone$, $\stopone \to b \chinoonepm$, $\stopone \to c \ninoone$, or via  $\stopone \to s b$ with R-parity and baryon number violation, as illustrated in figure \ref{fig:feynman-Gluino-top-onshell}. 
For the model with the $\stopone \to b \chinoonepm$ decay, the chargino mass is assumed to be twice the mass of the neutralino, and the chargino decays into a neutralino and a $W$ boson. 
In the model with the $\stopone \to c \ninoone$ decay, which proceeds via a loop and is most relevant when the $\stopone \to bW\ninoone$ decay is kinematically forbidden, the mass gap between the $\stopone$ and the lightest neutralino is fixed to 20~\GeV. Using gluino-pair production to probe this decay is particularly interesting because it is complementary to the direct pair production of  $\stopone$, which is more difficult to extract from the background for this specific decay mode  ~\cite{MonojetPaper}.
Gluino--sbottom models within this class assume that the $\sbottomone$ is the lightest squark, all other squarks are heavier than the gluino, and $m^{}_{\gluino}>m^{}_{\sbottomone}+m_b$ such that the branching ratio for $\gluino \to \sbottomone b$ decays is 100\%. The bottom squarks are assumed to decay exclusively via $\sbottomone \to b\ninoone$ (figure \ref{fig:feynman-Gluino-bottom-onshell}). 

\begin{figure*}[h]
\centering
\subfigure[]{\includegraphics[width=0.25\textwidth]{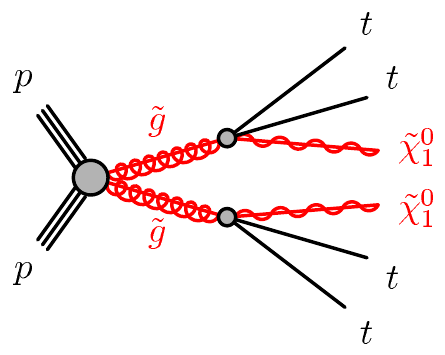}}\hspace{0.05\textwidth}
\subfigure[]{\includegraphics[width=0.25\textwidth]{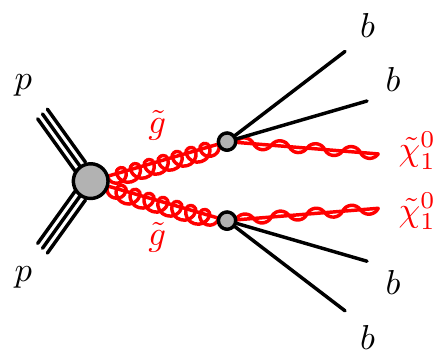}}\hspace{0.05\textwidth}
\subfigure[]{\includegraphics[width=0.25\textwidth]{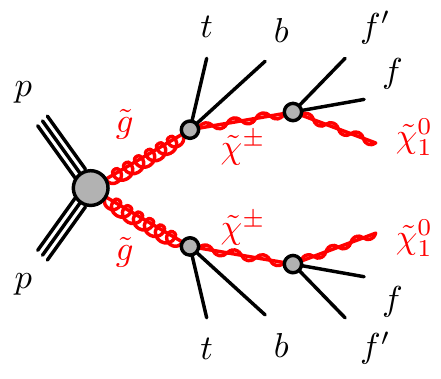}}
\caption{The decay topologies in the (a) gluino--off-shell--stop, (b) gluino--off-shell--sbottom and (c) gluino--off-shell--stop/sbottom simplified models.
}\label{fig:feynman-Gluino-topbottom-offshell}
\end{figure*}

In the second class of simplified models with gluino decays via top or bottom squarks, all sparticles apart from the 
gluino and the neutralino have masses well above the \TeV\ scale such that 
the $\stopone$ or the $\sbottomone$ are only produced off-shell via prompt decay of the gluinos and have little impact on the kinematics of the final state. 
For the gluino--off-shell--stop model illustrated in figure \ref{fig:feynman-Gluino-topbottom-offshell}(a), the $\stopone$ is assumed to be the lightest squark, but $m^{}_{\gluino}<m^{}_{\stopone}$. 
A three-body decay $\gluino \to t\bar{t}\ninoone$ via an off-shell stop is assumed for the gluino with a branching 
ratio of 100\%. 
For the configuration $m_{\gluino} < 2m_t + m_{\ninoone} $, decays of the gluino involve an off-shell top quark, e.g. the four-body decay $\gluino \to t W b \ninoone$. Only four- and five-body decays of this type are considered, because for higher multiplicities the gluinos do not decay promptly. 
For the gluino--off-shell--sbottom model shown in figure \ref{fig:feynman-Gluino-topbottom-offshell}(b), the $\sbottomone$ is assumed to be the lightest squark but with $m^{}_{\gluino}<m^{}_{\sbottomone}$. 
A three-body decay $\gluino \to b\bar{b}\ninoone$ via an off-shell sbottom is assumed for the gluino with a branching 
ratio of 100\%. In the gluino--off-shell--stop/sbottom model illustrated in figure \ref{fig:feynman-Gluino-topbottom-offshell}(c), the $\sbottomone$ and $\stopone$ are the lightest squarks, with 
$m^{}_{\gluino}<m^{}_{\sbottomone, \stopone}$. Pair production of gluinos is the only process taken 
into account, with gluinos decaying via off-shell stops or sbottoms, and a branching ratio 
of 100\% assumed for  $\stopone \to b+\chinoonepm$ and $\sbottomone \to t+\chinoonepm$ decays. 
The mass difference between charginos and neutralinos is set to 2~\GeV, 
such that the fermions produced in $\chinoonepm \to \ninoone + ff'$ decays do not contribute  to 
the event selection, and gluino decays result in effective three-body decays  
$bt\ninoone$.  %

\newpage
\section{The ATLAS detector and data sample}
\label{sec:detector}

The ATLAS detector~\cite{Aad:2008zzm} is a multi-purpose particle
physics detector with a forward-backward symmetric cylindrical
geometry and nearly 4$\pi$ coverage in solid angle.\footnote{
ATLAS uses a right-handed coordinate system with its origin at the nominal
interaction point in the centre of the detector. The positive $x$-axis is defined by the direction from the interaction point to the centre
of the LHC ring, with the positive $y$-axis pointing upwards, while the beam direction defines the $z$-axis. Cylindrical coordinates $(r,\phi)$ are used in the transverse
plane, $\phi$ being the azimuthal angle around the $z$-axis. The pseudorapidity $\eta$ is
defined in terms of the polar angle $\theta$ by $\eta=-\ln\tan(\theta/2)$.
}   
The inner tracking detector (ID) consists of pixel and silicon microstrip detectors 
covering the pseudorapidity region $|\eta|<2.5$, surrounded by a transition radiation tracker (TRT) 
which enhances electron identification in the region $|\eta|<2.0$.  
The ID is surrounded by a thin superconducting solenoid providing an axial 2 T magnetic field and by
a fine-granularity lead/liquid-argon (LAr) electromagnetic calorimeter covering $|\eta|<3.2$.
A steel/scintillator-tile calorimeter provides hadronic coverage in
the central pseudorapidity range ($|\eta|<1.7$). 
The endcap and forward regions ($1.5<|\eta|<4.9$) of the hadronic calorimeter are made of LAr active layers with either copper or tungsten as the absorber material. 
An extensive muon spectrometer
with an air-core toroid magnet system surrounds the calorimeters.
Three layers of high-precision tracking chambers
provide coverage in the range $|\eta|<2.7$, while dedicated fast chambers allow triggering in the region $|\eta|<2.4$.
The ATLAS trigger system \cite{atlastrigger} consists of three levels; the first level (L1) is a hardware-based system, while the second and
third levels are software-based systems and are together called the High Level Trigger (HLT).

The data used in these searches were collected from March to December 2012 with the LHC operating at a centre-of-mass energy of 8~\TeV.
After the application of beam, detector and data quality requirements,
the total integrated luminosity ranges from 20.1 to 20.3~\ifb, depending on the triggers used for the event selection,  with a relative uncertainty of $\pm 2.8$\%. The uncertainty is derived following the methodology detailed in ref.~\cite{Aad:2013ucp}.
During the data-taking period, the peak instantaneous luminosity per LHC fill was typically $7\times 10^{33}$ cm$^{-2}$ s$^{-1}$, while the average number of $pp$ interactions per LHC bunch crossing ranged from approximately 6 to 40, with a mean value of 21. 
In order to maximize the efficiency of selecting the various final states used by the analyses included in this paper, different triggers or combinations of triggers were used: 
\met~ triggers, multi-jet triggers, combined \met+jet, lepton+\met~ or lepton+jet+\met triggers, single-lepton or dilepton triggers.  
Details of the trigger selections used in the published ATLAS searches included in this paper are not discussed here and can be found in the corresponding publications \cite{0-leptonPaper, MonojetPaper, multijetsPaper,1lepPaper, dilepton-edgePaper, SS3LPaper,TauStrongPaper,3bjetsPaper}.

\section{Monte Carlo simulated samples}
\label{sec:mcsamples}

The simulated event samples for the SM backgrounds are summarized in table~\ref{tab:MCsamples}, together with the choices of Monte Carlo generator, cross-section calculation, set of tunable parameters (tune) used for the underlying event and parton distribution functions (PDFs). The \powheg+\pythia{} $\ttbar$ sample is used for all analyses except for the analysis that requires high jet multiplicities (at least seven to at least ten jets) and large missing transverse momentum \cite{multijetsPaper}, which uses the \sherpa\ $\ttbar$ sample. The \sherpa\ Drell--Yan samples have a lepton filter requiring  $\pt^{\ell_1(\ell_2)}>9~(5)$~\GeV~and 
$|\eta^{\ell_1(\ell_2)}|<2.8$. This filter prevents its use in analyses requiring the presence of soft leptons in the final state. Such analyses instead use \alpgen\ samples with a lepton \pT\ threshold at 5~\GeV. 
When using the baseline \powheg+\pythia{} top quark pair production sample, in some of the analyses events are reweighted in bins of $\pt(t\bar{t})$ to match the top quark pair differential cross-section measured in ATLAS data  \cite{Aad:2012hg,atlas-ttjets}. The exact usage of MC simulated samples together with the additional samples used to assess modelling uncertainties are detailed in  the corresponding publication of each analysis.

\begin{table}[h]
\centering
\footnotesize
\renewcommand\arraystretch{1.1}
\begin{tabular}{| ccccc |}
\hline
\multirow{2}{*}{Process} &  Generator &Cross-section  & \multirow{2}{*}{Tune}   & \multirow{2}{*}{PDF set}  \\
 &   & order in $\alpha_{\rm s}$ & & \\
\hline \hline

$W(\rightarrow \ell \nu)$+jets  & \sherpa~1.4.1 ~\cite{Gleisberg:2008ta} &   NNLO~\cite{Catani:2009sm} &\sherpa{} default  & CT10 ~\cite{CT10pdf} \\
\hline

 $Z/\gamma^{*}(\rightarrow \ell \ell)$+jets & \sherpa~1.4.1  &   NNLO~\cite{Catani:2009sm} &\sherpa{} default  & CT10 \\
Drell--Yan  & \sherpa~1.4.1  & NNLO \cite{Melnikov:2006kv} & \sherpa\ default  & CT10 \\
($8<m_{\ell\ell}<40$~\GeV) &&&&\\

\multirow{2}{*}{$Z/\gamma^{*}(\rightarrow \ell \ell)$ + jets}  &\alpgen~2.14~\cite{Mangano:2002ea} & \multirow{2}{*}{NNLO \cite{Melnikov:2006kv}} & \multirow{2}{*}{AUET2 \cite{AUET2}}  & \multirow{2}{*}{CTEQ6L1 \cite{Pumplin:2002vw}}\\
  & + \herwig~6.520~ \cite{Corcella:2000bw,herwig65long} & & &\\
($10<m_{\ell\ell}<60$~\GeV) & + \jimmy~\cite{Butterworth:1996zw} & & &\\

\hline

$\gamma$+jets & \sherpa~1.4.1  &   LO &\sherpa{} default  & CT10 \\
\hline

\multirow{2}{*}\ttbar &\powheg~1.0~\cite{Nason:2004rx,Frixione:2007vw,Alioli:2010xd} & \multirow{2}{*}{NNLO+NNLL~\cite{Czakon:2013goa,Czakon:2011xx}}& \Perugia & \multirow{2}{*}{CT10}  \\ 
& + \pythia~6.426 \cite{Sjostrand:2006za} &&\cite{Cooper:2011gk,Skands:2010ak}&\\
\ttbar &\sherpa~1.4.1 & NNLO+NNLL & \sherpa{} default & CT10  \\ 

\hline

Single top & & & & \\
        \multirow{2}{*}{$t$-channel} & \AcerMC~3.8~\cite{Kersevan:2004yg}  & \multirow{2}{*}{NNLO+NNLL~\cite{Kidonakis:2011wy}}  & \multirow{2}{*}{AUET2B  \cite{ATL-PHYS-PUB-2011-014}} & \multirow{2}{*}{CTEQ6L1 } \\
        &  + \pythia~6.426 &&&\\
        \multirow{2}{*}{$s$-channel, $Wt$} &  \mcatnlo~4.03 \cite{Frixione:2002ik,Frixione:2003ei} & \multirow{2}{*}{NNLO+NNLL~\cite{Kidonakis:2010tc, Kidonakis:2010ux}} & \multirow{2}{*}{AUET2B} & \multirow{2}{*}{CT10} \\
        &  + \herwig~6.520 &&&\\
\hline

 \multirow{2}{*}{ \ttbar+W/Z boson} &\Madgraph~5 1.3.28~\cite{Alwall:2011uj}  &   \multirow{2}{*}{NLO~\cite{Lazopoulos:2008de,Garzelli:2012bn,Campbell:2012dh}} &  \multirow{2}{*}{AUET2B} &  \multirow{2}{*}{CTEQ6L1} \\ 
 &+ \pythia~6.426&&&\\
\hline

Dibosons & & & & \\
 $WW$, $WZ$, $ZZ$,  &  \multirow{2}{*}{\sherpa~1.4.1} & \multirow{2}{*}{NLO~\cite{Campbell:1999ah,Campbell:2011bn}} &\multirow{2}{*}{\sherpa{} default}  & \multirow{2}{*}{CT10} \\
   $W\gamma$ and $Z\gamma$&&&&\\
\hline

\hline
\end{tabular}      
\caption{The Standard Model background Monte Carlo simulation samples used in this paper. The generators, the order in $\alpha_{\rm s}$ of cross-section calculations used for yield normalization (leading order (LO), next-to-leading order (NLO), next-to-next-to-leading order (NNLO), next-to-next-to-leading logarithm (NNLL)), tunes used for the underlying event and PDF sets are shown. 
For the $\gamma$+jets process the LO cross-section is taken directly from the MC generator. 
\label{tab:MCsamples} }
\end{table}

Signal samples for the pMSSM, mSUGRA, mGMSB, nGM and mUED models, as well as the samples for the simplified models of  gluino-mediated top squark production (for $m_{\tilde g}-m_{\ninoone} > 2 m_t$) are generated with \herwig++~2.5.2~\cite{Bahr:2008pv}. 
Samples for all the other simplified models are generated with up to one extra parton in the matrix element using  \Madgraph~5 1.3.33 interfaced to \pythia~6.426. The MLM matching scheme \cite{Mangano:2006rw} is applied with a scale parameter that is set to a quarter of the mass of the lightest sparticle in the hard-scattering matrix element, with a maximum value of 500 GeV. The signal samples used for the bRPV and NUHMG  models are generated with \pythia~6.426. 

For the gluino--off-shell--stop model in the region $m_{\tilde g}-m_{\ninoone}<2m_t$, the production of gluino pairs is generated with \Madgraph ~5 1.3.33. The events are subsequently combined with separately generated gluino decays $\gluino\to f\bar f'f''\bar f''' b\bar b\ninoone$ based on the full matrix element amplitude (also using  \Madgraph{}), preserving spin-dependent distributions. A summary of the studies related to event generation in this model can be found in appendix \ref{AppGttExt}.  
Potential effects of the gluino lifetime (displaced decays, hadronization), which are strongly model dependent, have been neglected.%

The ATLAS underlying-event tune AUET2B \cite{AUET2} is used for  \Madgraph~5 and \pythia~6 samples while the UE-EE-3C tune \cite{UEEE3} is used for  \herwig++ samples. The parton distribution functions from CTEQ6L1 \cite{Pumplin:2002vw} are used for all signal samples.

For all except the mUED sample, the signal cross-sections are calculated to next-to-leading order in the strong coupling constant, including the resummation of soft gluon emission at next-to-leading-logarithmic accuracy (NLO+NLL) \cite{Beenakker:1996ch,Kulesza:2008jb,Kulesza:2009kq,Beenakker:2009ha,Beenakker:2011fu}. In each case the nominal cross-section and its uncertainty are taken from an ensemble of cross-section predictions using different PDF sets and factorization and renormalization scales, as described in ref.~\cite{Kramer:2012bx}. 
For the mUED model, the cross-section is taken at leading order from \herwig++.
For the \mSUGRA{} and NUHMG samples,  \textsc{Susy-Hit} \cite{Djouadi:2006bz} and \textsc{Sdecay}~1.3b \cite{Muhlleitner:2003vg}, 
 interfaced to the  \textsc{Softsusy}~3.1.6  spectrum generator \cite{Allanach:2001kg}, are used to calculate the sparticle mass spectra and decay tables, and to ensure consistent electroweak symmetry breaking.

The decays of tau leptons are simulated directly in the generators in the case of event samples produced with \sherpa, \herwig++~2.5.2 and \pythia~8.165, while in all other cases \Tauola~2.4~\cite{Jadach:1993tau,Golonka:2006} is used. 

Standard Model background samples are passed through either the full ATLAS detector simulation \cite{:2010wqa} based on \textsc{Geant4} \cite{Agostinelli:2002hh}, or through a fast simulation using a parameterization of the performance of the ATLAS electromagnetic and hadronic calorimeters \cite{ATLAS:2010bfa} and \textsc{Geant4} elsewhere; the latter applies to $W/Z/\gamma$+jets samples with boson $\pt<280$ GeV and \powheg{}+\pythia{} $t\bar{t}$ samples. 
All SUSY signal samples are passed through the fast simulation, with the exception of the \mSUGRA{} model samples which are produced with the \textsc{Geant4} simulation. The fast simulation of SUSY signal events was validated against full \textsc{Geant4} simulation for several signal models. 
Differing pile-up (multiple $pp$ interactions in the same or neighbouring bunch-crossings) conditions as a function of the instantaneous luminosity are taken into account by overlaying simulated minimum-bias events (simulated using \pythia~8 with the MSTW2008LO PDF set~\cite{Sherstnev:2007nd} and the A2 tune \cite{ATL-PHYS-PUB-2011-014})  onto the hard-scattering process and reweighting events according to the distribution of the mean number of interactions observed in data.

\section{Object reconstruction and identification}
\label{sec:objects}

This paper summarizes different analyses which are combined to improve the sensitivity to a variety of possible topologies originating from the production and decay of squarks and gluinos.  Although different event selections are used among these analyses, they share common definitions of the  reconstructed objects. 
Analysis-specific exceptions to these definitions are detailed in the corresponding publication of each analysis. 

The reconstructed primary vertex of the event is required to be consistent with the beamspot envelope and to have at least five associated tracks with $\pt > 400$~\MeV. When more than one such vertex is found, the vertex with the largest  $\sum \pt^2$ of the associated tracks is chosen.

Jet candidates are reconstructed using the anti-$k_{t}$ jet clustering algorithm~\cite{Cacciari:2008gp,Cacciari:2005hq} with a
radius parameter of $0.4$. The inputs to this algorithm are topological clusters \cite{Lampl:2008,Aad:2011he} of calorimeter cells seeded by those with energy significantly above the measured noise (topoclusters).  
The local cluster weighting (LCW)  calibration method \cite{Issever:2004qh,Aad:2011he} is used to classify topoclusters as being either of electromagnetic or hadronic origin, and based on this classification it applies energy corrections derived from MC simulations and measurements in data. 
The jets are corrected for energy from pile-up using the method suggested in ref. \cite{Cacciari:2007fd}:
a contribution equal to the product of the jet area and the median energy density of the event is subtracted
from the jet energy \cite{ATLAS-CONF-2013-083}.
Further corrections, referred to as the jet energy scale (JES) corrections, are derived from MC simulation and data and used to calibrate on average the energies of jets to the scale of their constituent particles \cite{ATLAS-CONF-2013-004,Aad:2011he}. 
Only jet candidates with $\pt > 20$~\GeV{} and $|\eta|<4.5$ after all corrections are retained. 
To remove events with jets from detector noise and non-collision backgrounds, events
are rejected if they include jets failing to satisfy the ``loose'' quality criteria described in ref. \cite{Aad:2011he}.

A neural-network-based algorithm \cite{ATLAS-CONF-2011-102} is used to identify jets  containing a $b$-hadron %
 ($b$-jets). It uses as inputs the output weights of several algorithms exploiting the impact parameter of the inner detector tracks,  secondary vertex reconstruction and the topology of $b$- and $c$-hadron decays inside the jet. The algorithm used has an efficiency of 70\% for tagging $b$-jets, determined with simulated $t\bar{t}$ events \cite{ATLAS-CONF-2014-004}.  
For this efficiency, the algorithm provides a rejection factor of approximately 140 for light-quark and gluon jets, and of approximately 5 for charm jets \cite{ATLAS-CONF-2014-046}.
Candidate $b$-jets are required to have $\pt>40$~GeV and $|\eta|<2.5$.

Electrons are reconstructed from energy clusters in the electromagnetic
calorimeter matched to tracks in the inner detector \cite{Aad:2014fxa} and are required to have $\pt>10$~\GeV\ and   $|\eta|<2.47$. 
The preselected electron candidates are required to pass a variant of the ``medium'' selection  \cite{Aad:2014fxa}, which was modified in 2012 to reduce the impact of pile-up. 

Photon candidates, which in the analyses presented are used only for the measurement of the missing transverse momentum, are required to have $\pt>10$~\GeV\ and  $|\eta| < 1.37$ or $1.52<|\eta| < 2.47$, to satisfy photon shower shape and electron rejection criteria \cite{ATLAS-CONF-2012-123}, and to be isolated.%

Muon candidates are formed by combining information from the muon spectrometer and inner tracking detectors \cite{Aad:2014zya}. The preselected muon candidates are required to have $\pt > 10$~\GeV\ and $|\eta| < $ 2.4 or 2.5, depending on the analysis.

Reconstruction of hadronically decaying tau leptons starts from jets with $\pt > 10$~\GeV\  \cite{ATLAS:TauID2013}, and an $\eta$- and $\pt$-dependent energy calibration to the tau energy scale for hadronic decays is applied \cite{ATLAS:TES2013}.
Tau lepton candidates must have one or three associated track(s) with a charge sum of $\pm1$,  and satisfy $\pt > 20$~\GeV\ and $|\eta|<2.5$.
The ``loose'' and ``medium'' working points  \cite{ATLAS:TauID2013} are used and correspond to efficiencies of approximately 70\% and 60\%, independent of \pt, with rejection factors of 10 and 20 against jets misidentified as tau candidates, respectively. 

After these selections, ambiguities between candidate jets with $|\eta|<2.8$ and leptons (electrons and muons) are resolved as follows.
First, any such jet candidate lying within a distance 
$\Delta R=\sqrt{(\Delta\eta)^2+(\Delta\phi)^2} = 0.2$ of a preselected electron is discarded;  then any lepton candidate within a distance
$\Delta R=0.4$ of any surviving jet candidate is discarded.
In analyses requiring the presence of one lepton (electron or muon) in the final state,  electrons are also required to be well separated from muon candidates with $\Delta R(e,\mu)>0.01$. If two preselected electrons are found within an angular distance 
$\Delta R(e,e)=0.05$ of each other, only the electron with the higher \pt is kept. 
Finally, in the analyses that require the presence of at least one or two opposite-sign leptons in the final state, any event containing a preselected electron in the transition region between the barrel and endcap electromagnetic calorimeters, $1.37 < |\eta| < 1.52$, is rejected. 

The measurement of the missing transverse momentum vector is based on the transverse momenta of all electron, photon, jet and muon candidates, and all calorimeter energy clusters not associated with such objects \cite{Aad:2012re}. 
Fully calibrated electrons and photons with $\pt >$10~\GeV\ and jets with $\pt >$ 20~\GeV\ are used.
Energy deposits not associated with these objects are also taken into account in the $\bm{E}\mathrm{^{miss}_T}$ calculation using an energy-flow algorithm that considers calorimeter energy deposits as well as ID tracks \cite{ATLAS-CONF-2013-082}. 
In the $\met$ measurement tau leptons are not distinguished from jets and it has been checked that this does not introduce a bias in any
kinematic variables used in the analyses.

Corrections derived from data control samples are applied to account for differences between data and simulation for the lepton trigger and reconstruction efficiencies, momentum/energy scale and resolution, and for the efficiency and mis-tag rate for tagging jets originating from $b$-quarks.

\section{Analysis strategy}
\label{sec:strategy}

A search for squarks and gluinos under various decay mode assumptions necessitates many different event selections targeting the wide range of experimental signatures. 
This section summarizes the common analysis strategy and statistical techniques that are employed in all searches included in this paper. 
Signal regions (SRs) are defined using the Monte Carlo simulation of the signal processes and the SM backgrounds, and are optimized to maximize the expected significance for each model considered. 
To estimate the SM backgrounds in a consistent and robust fashion, corresponding control regions (CRs) are defined for each of the signal regions. 
They are chosen to be non-overlapping with the SR selections in order to provide independent data samples enriched 
in particular background sources. The CR selections are optimized to have negligible SUSY signal contamination for the models under investigation, while minimizing as much as possible the systematic uncertainties arising from the extrapolation of the CR event yields to the expectations in the SR. 
Cross-checks of the background estimates are performed using several validation regions (VRs) selected with requirements such that these regions do not overlap with the CR and SR selections, again with a low probability of signal contamination. 

Several  classes of profile likelihood fits that utilize the observed numbers of events in the various regions are employed in the analyses \cite{HFpaper}. In some analyses, the shape of a final discriminating variable in the SRs is also used. 
A background-only fit is used to determine the compatibility of the observed event yield in each SR with the corresponding SM background expectation. This fit uses as constraints only the observed event yields or the shape of the discriminating variable distributions from the CRs associated with the SR, but not the SR itself. It is assumed that signal events from physics beyond the Standard Model (BSM) do not contribute to these yields. 
The numbers of observed and predicted events in each of these CRs are described using Poisson probability density functions. 
The systematic uncertainties and the MC statistical uncertainties on the expected values are included in the fit as nuisance parameters which are constrained by Gaussian distributions with  widths corresponding to the sizes of the uncertainties considered and Poisson distributions, respectively. Correlations of a given nuisance parameter across the various regions, between the various backgrounds, and possibly the signal, are taken into account. The product of the various probability density functions forms the likelihood, which the fit maximizes by adjusting the inputs to the fit and the nuisance parameters. The inputs to the fit for each of the SRs are the number of events observed in each of the CRs, and the corresponding number of events expected from simulation, the extrapolation factors obtained from the simulation which relate the number of predicted SM background events in their associated CR to that predicted in the SR, and the number of events predicted by the simulation in each region for the other background processes. 
The background fit results are cross-checked in validation regions. The data in the validation regions are not used to constrain the fits; they are only used to compare the results of the fits to statistically independent observations. 

A model-independent fit is used to set upper limits on the number of BSM signal events in each SR. This fit proceeds in the same way as the background-only fit, except that the number of events observed in the SR is added as an input to the fit, and the BSM signal strength, constrained to be non-negative, is added as a free parameter. The observed and expected upper limits at 95\% confidence level (CL) on the number of events from BSM phenomena for each signal region ($S_{\rm obs}^{95}$ and $S_{\rm exp}^{95}$) are derived using the $CL_{\rm S}$ prescription \cite{Read:2002hq}, neglecting any
possible signal contamination in the control regions; an uncertainty on $S_{\rm exp}^{95}$ is also computed from the $\pm 1\sigma$
uncertainty on the expectation. These limits, when normalized by the integrated luminosity of the data sample, may be interpreted as upper limits on the visible cross-section of BSM physics ($\langle\epsilon\sigma\rangle_{\rm obs}^{95}$), where the visible cross-section is defined as the product of production cross-section, acceptance and efficiency. 
The model-independent fit is also used to compute the one-sided $p$-value ($p_0$) of the background-only hypothesis which quantifies the statistical significance of an excess.

Model-dependent fits are used to set exclusion limits on the signal cross-sections for specific SUSY models. Such a fit proceeds in the same way as the model-independent fit, except that signal contamination in the CRs is taken into account as well as the yield in the signal region and, in some analyses, the model shape information. Correlations between signal and background systematic uncertainties are taken into account where appropriate. 
The systematic uncertainties on the signal expectations originating from detector effects and the theoretical uncertainties on the signal acceptance are included in the fit. 
The impact of the theoretical uncertainties on the signal cross-section is shown on the limit plots obtained (section~\ref{sec:limits}).
Numbers quoted in the text are evaluated from the observed exclusion limit based on the nominal signal cross-section minus its $1\sigma$ theoretical uncertainty. 

Background-only and model-independent fit results are presented in this paper only for new analyses or signal regions which are not available in earlier ATLAS publications.
In the context of this publication, model-dependent exclusion fits for various simplified and phenomenological models are combined to include results from different searches for each model individually, in order to maximize the expected exclusion reach for each model. Where possible a full statistical combination of non-overlapping searches is applied, as explained in section \ref{sec:combination}.

\section{Experimental signatures}
\label{sec:signatures}
 
This paper summarizes and combines the results of several individual inclusive squark and gluino analyses previously published by the ATLAS experiment. 
Each of these searches uses one or more sets of signal regions targeting specific experimental signatures which originate from different squark or gluino decay modes and mass hierarchies. 
Several extensions to the previously published searches in the form of additional signal regions are also included, along with one new analysis channel.
The full list of searches and their signal regions used in this paper is presented in table \ref{tab:signal_regions}, together with the corresponding references. 
The details of the signal region selections for all searches listed in table \ref{tab:signal_regions} can be found in appendix \ref{AppSRdefs}. 
The details of the control and validation region selections, together with the strategies used for the estimation of the background processes, can be found in the corresponding publications. 
The new analysis and extended signal regions, which are also presented in table \ref{tab:signal_regions}, are discussed in more detail in the subsequent subsections. 
Each signal region is referred to with an acronym, listed in table \ref{tab:signal_regions}, indicating the analysis origin, so for example the `2jl' region from the 0-lepton + 2--6 jets + \met analysis is referred to as `0L\_2jl'.  
The correspondence between the searches and the various models probed is provided in table \ref{tab:AnalysesvsModels} and a summary of the limits in simplified models presented in the respective papers is given in table~\ref{tab:old_limits}.
The 0-lepton + 2--6 jets + \met\ and 1-lepton (soft+hard) + jets + \met\ statistical combination, referred to as (0+1)-lepton combination, is used to probe the models for which both analyses have comparable sensitivity. 

\begin{table}[H]
\centering
\small
\begin{tabular}{| l | c | l |} \hline 
Short analysis name and corresponding reference                  &   Acronym      &  Signal region name \\  \hline \hline
Monojet  \cite{MonojetPaper}                                                    &   MONOJ    &  M1, M2, M3  \\  
0-lepton + 2--6 jets + \met  \cite{0-leptonPaper}                        &      0L         &  2jl, 2jm, 2jt, 2jW, 3j, 4jW, 4jl-, 4jl, 4jm, 4jt, \\
								                               &                   &  5j, 6jl, 6jm, 6jt, 6jt+ \\      
0-lepton + 4--5 jets + \met ($\star$)                                              &   0L             &  4jt+, 5jt \\   
0-lepton + 7--10 jets + \met \cite{multijetsPaper}                        &   MULTJ     &  8j50, 9j50, 10j50 (multi-jet+flavour stream), \\
                                                                                                  &                    & 7j80, 8j80,  (multi-jet+flavour stream), \\ 
                                                                                                  &                    & 8j50, 9j50, 10j50 (multi-jet+\MJ{} stream)  \\  
                                                                                                  
0-lepton Razor  (\textbullet)                                                      &  0LRaz       & SR$_{\rm loose}$, SR$_{\rm tight}$  \\   \hline                                                                                                 
1-lepton (soft+hard) + jets + \met  \cite{1lepPaper}                  &  1L(S,H)    & 3-jet/5-jet/3-jet inclusive (soft lepton), \\
                                                                                                  &                    & 3-jet/5-jet/6-jet (hard lepton) \\ 
1-lepton (hard) + 7 jets + \met ($\star$)                                    &  1L(H)        &  7-jet  \\   
2-leptons (soft) + jets + \met  \cite{1lepPaper}                          &  2L(S)        & 2-jet (soft dimuon) \\
2-leptons (hard) + jets + \met  \cite{1lepPaper}                         &  2LRaz      & $\leq$ 2-jet/3-jet  \\
                                                                                                                                                                                                      
2-leptons off-Z \cite{dilepton-edgePaper}                                  & 2L-offZ       & SR-2j-bveto, SR-2j-btag,  \\ 
                                                                                                   &                     & SR-4j-bveto, SR-4j-btag, SR-loose \\ 
Same-sign dileptons or 3-leptons + jets + \met \cite{SS3LPaper}                                        &   SS/3L       & SR3b, SR0b, SR1b, SR3Llow, SR3Lhigh  \\ \hline 
Taus + jets + \met \cite{TauStrongPaper}                                  &   TAU          &  1$\tau$ (Loose, Tight), \\
                                                                                                  &                    & 2$\tau$ (Inclusive, GMSB, nGM, bRPV), \\
                                                                                                  &                    & $\tau+l$ (GMSB, nGM, bRPV, mSUGRA)  \\ \hline 
0/1-lepton + 3$b$-jets + \met \cite{3bjetsPaper}                        &   0/1L3B     &  SR-0l-4j-A, SR-0l-4j-B, SR-0l-4j-C, \\
                                                                                                   &                   & SR-0l-7j-A , SR-0l-7j-B, SR-0l-7j-C, \\
                                                                                                   &                   & SR-1l-6j-A, SR-1l-6j-B, SR-1l-6j-C \\  \hline 
\end{tabular}
\caption{ List of analysis names referring to the experimental signatures addressed, with references to the appropriate publications; their acronyms; and all signal region names. The new analysis is denoted with (\textbullet), while the extended signal regions are denoted with ($\star$). The details of the signal region selections for all searches listed in the table can be found in appendix \ref{AppSRdefs}. 
}
\label{tab:signal_regions}
\end{table}

\begin{sidewaystable}[htbp]
\renewcommand\arraystretch{1.4}
\begin{center}
\tiny
\hspace*{-0.05\textwidth}\begin{tabular}{| l | c | c | c | c | c | c | c | c | c | c | c | c | c |}
\hline

           & (0+1)-lepton     & MONOJ & 0L        & MULTJ         &  0LRaz  & 1L(S,H)    & 1L(H) & 2L(S) & 2LRaz  & 2L-offZ & SS/3L    & TAU      & 0/1L3B \\ 
Model &  combination     &              &             &                     &               &                  &          &            &             &             &               &              &              \\ \hline \hline

pMSSM &   &  & \checkmark &  &  &  &  &  &  &  & & & \\ \hline
mSUGRA/CMSSM & \checkmark &  &  & \checkmark &  &  & \checkmark & &  &  & \checkmark  & \checkmark & \checkmark \\ \hline
mSUGRA/CMSSM with bRPV &  &  & \checkmark & \checkmark &  & \checkmark & \checkmark & & &  & \checkmark & \checkmark &  \\ \hline
mGMSB &  &  &  &  &  &  &  & &  &  & \checkmark & \checkmark &  \\ \hline
nGM &  &  & \checkmark &  &  & \checkmark &  & & &  &  & \checkmark & \\ \hline
NUHMG & \checkmark &  &  &  &  &  &  & & &  &  & &  \\ \hline
mUED &  &  & \checkmark &  &  & \checkmark &  & \checkmark & \checkmark &  & \checkmark &  & \\ \hline \hline
$\squark\squark$ production, $\squark\rightarrow q\ninoone$ &  & \checkmark & \checkmark &  & \checkmark &  &  & &  &  &  & &  \\ \hline
$\gluino\gluino$ production, $\gluino\rightarrow qq\ninoone$ &  &  & \checkmark &  &  &  &  & &  &  &  &  & \\ \hline
$\squark\gluino$ production, $\squark\rightarrow q\ninoone$, $\gluino\rightarrow qq\ninoone$ &  &  & \checkmark &  &  &  &  & & &  &  &  & \\ \hline
$\gluino\gluino$ production, $\gluino\rightarrow g\ninoone$  &  &  & \checkmark &  &  &  &  & & &  &  & &  \\ \hline
$\squark\squark$ production, $\squark\rightarrow qW\ninoone$ & \checkmark &  &  &  &  &  &  & & &  &  & &  \\ \hline
$\gluino\gluino$ production, $\gluino\rightarrow qqW\ninoone$ & \checkmark &  &  & \checkmark &  &  &  & & &  & \checkmark &   & \\ \hline
$\squark\squark$ production, $\squark\rightarrow q(\ell\ell/\ell\nu/\nu\nu)\ninoone$ & \checkmark &  &  &  &  &  &  & & \checkmark & \checkmark & \checkmark &   & \\ \hline
$\gluino\gluino$ production, $\gluino\rightarrow qq(\ell\ell/\ell\nu/\nu\nu)\ninoone$ & \checkmark &  &  & \checkmark &  & \checkmark &  & & \checkmark & \checkmark  & \checkmark &  & \\ \hline
$\squark\squark$ production, $\squark\rightarrow q(\tau\tau/\tau\nu/\nu\nu)\ninoone$ &  &  &  &  &  &  &  & & &  &  & \checkmark & \\ \hline
$\gluino\gluino$ production, $\gluino\rightarrow qq(\tau\tau/\tau\nu/\nu\nu)\ninoone$ &  &  &  &  &  &  &  & & &  & &  \checkmark & \\ \hline
$\squark\squark$ production, $\squark\rightarrow qWZ\ninoone$ &  &  & \checkmark &  &  &  &  &  & &  & \checkmark &  & \\ \hline
$\gluino\gluino$ production, $\gluino\rightarrow qqWZ\ninoone$ & \checkmark &  &  & \checkmark &  &  &  & & &  & \checkmark &  & \\ \hline
$\gluino\gluino$ production, $\gluino\rightarrow \ttbar\ninoone$ (off-shell stop) &  &  &  & \checkmark &  & \checkmark &  & & &  & \checkmark &  & \checkmark  \\ \hline
$\gluino\gluino$ production, $\gluino\rightarrow \stopone t$, $\stopone\rightarrow t\ninoone$  &  &  &  & \checkmark &  &  &  & & &   &  &  & \checkmark \\ \hline
$\gluino\gluino$ production, $\gluino\rightarrow \stopone t$, $\stopone\rightarrow b \chinoonepm$  &  &  &  &  &  &  &  & & &  & \checkmark &  & \checkmark \\ \hline
$\gluino\gluino$ production, $\gluino\rightarrow \stopone t$, $\stopone\rightarrow c\ninoone$  & \checkmark &  &  &  &  &  &  & & &  & \checkmark &  & \\ \hline
$\gluino\gluino$ production, $\gluino\rightarrow \stopone t$, $\stopone\rightarrow bs$ &  &  &  & \checkmark &  &  &  & & &  & \checkmark & &  \\ \hline
$\gluino\gluino$ production, $\gluino\rightarrow tb\ninoone$ &  &  &  &  &  &  &  & &  &  &  & &  \checkmark \\ \hline
$\gluino\gluino$ production, $\gluino\rightarrow b\bar{b}\ninoone$ (off-shell sbottom)  &  &  & \checkmark &  &  &  &  & & &  &  & &  \checkmark \\ \hline
$\gluino\gluino$ production, $\gluino\rightarrow \sbottomone b$, $\sbottomone \rightarrow b\ninoone$ &  &  &  &  &  &  &  & &  &  &  & & \checkmark  \\ \hline
\end{tabular}
\end{center}
\caption{\label{tab:AnalysesvsModels}
Searches used to probe each of the phenomenological  models described in section \ref{subsec:phenomodels} and simplified models described in section \ref{subsec:simplified}. 
}
\end{sidewaystable}

\begin{table}[H]
\centering
\scriptsize
\begin{tabular}{| l | l | l | l |}  \hline 
Analysis acronym                &   Process                               	   							& 95\% CL limit                                    & Assumptions \\  \hline \hline
0L \cite{0-leptonPaper}   &  $\gluino\gluino$, $\gluino\rightarrow g\ninoone$, $\gluino\rightarrow qq\ninoone$  &   $m_{\gluino}> 1330\GeV$  & $m_{\ninoone}=0\GeV$\\ 
				 &  $\squark\squark$, $\squark\rightarrow q\ninoone$ &    $m_{\squark}> 850\GeV$  &$m_{\ninoone}=0\GeV$, mass degenerate \squark  \\ 
 				  &  $\squark\squark$, $\squark\rightarrow q\ninoone$ &   $m_{\squark}> 440\GeV$   & $m_{\ninoone}=0\GeV$, single flavour \squark \\ 
				  &  $\gluino\gluino$, $\gluino\rightarrow qqW\ninoone$  &   $m_{\gluino}> 1100\GeV$ & $m_{\ninoone}=0\GeV$\\ 
				 &  $\squark\squark$, $\squark\rightarrow qW\ninoone$ &   $m_{\squark}> 700\GeV$    & $m_{\ninoone}=0\GeV$\\
				   &  $\gluino\gluino$, $\gluino\rightarrow \stopone t$, $\stopone\rightarrow c\ninoone$  &   $m_{\gluino}> 1100\GeV$ & $m_{\stopone}=400\GeV$, $m_{\ninoone}=m_{\stopone}-20\GeV$ \\ \hline
MULTJ   \cite{multijetsPaper} & $\gluino\gluino$, $\gluino\rightarrow \ttbar\ninoone$   &   $m_{\gluino}> 1100\GeV$  & $m_{\ninoone}< 350\GeV$ \\
				  &  $\gluino\gluino$, $\gluino\rightarrow qqW\ninoone$  &   $m_{\gluino}> 1000\GeV$ & $m_{\ninoone}<200\GeV$\\ 
				 &  $\gluino\gluino$, $\gluino\rightarrow qqWZ\ninoone$  &   $m_{\gluino}> 1100\GeV$  & $m_{\ninoone}<300\GeV$\\ \hline
1L(S,H),  &    $\gluino\gluino$, $\gluino\rightarrow qqW\ninoone$ &    $m_{\gluino}> 1200\GeV$ &  $x=\Delta m(\chi^{\pm}_1, \ninoone)/ \Delta m ({\tilde{g}}, \ninoone) = 1/2$, $m_{\ninoone}= 60\GeV$ \\
2L(S),				&  $\squark\squark$, $\squark\rightarrow qW\ninoone$   &   $m_{\squark}> 700\GeV$    & $x=\Delta m(\chi^{\pm}_1, \ninoone)/ \Delta m ({\tilde{q}}, \ninoone) = 1/2$, $m_{\ninoone}< 200\GeV$\\
2LRaz \cite{1lepPaper} 				&  $\gluino\gluino$, $\gluino\rightarrow qq\ell\nu\ninoone$  &   $m_{\gluino}> 1320\GeV$  & $m_{\ninoone}= 100\GeV$\\ 
				&  $\squark\squark$, $\squark\rightarrow q\ell\nu\ninoone$  &   $m_{\squark}> 840\GeV$  & $m_{\ninoone}= 40\GeV$  \\  
				&  $\gluino\gluino$, $\gluino\rightarrow \stopone t$, $\stopone\rightarrow c\ninoone$  &   $m_{\gluino}> 1200\GeV$ & 
				 $m_{\stopone}=200\GeV$, $m_{\ninoone}=m_{\stopone}-20\GeV$  \\ 
       	                   &  $\gluino\gluino$, $\gluino\rightarrow qqWZ\ninoone$  &   $m_{\gluino}> 1140\GeV$  & $m_{\ninoone}< 200\GeV$ \\ \hline
2L-offZ  \cite{dilepton-edgePaper} &$\gluino\gluino$, $\gluino\rightarrow qq(\ell\ell/\ell\nu/\nu\nu)\ninoone$ &$m_{\gluino}> 1170\GeV$&$m_{\ninoone}= 50\GeV$\\ 
				& $\squark\squark$, $\squark\rightarrow q(\ell\ell/\ell\nu/\nu\nu)\ninoone$ &$m_{\gluino}> 780\GeV$&$m_{\ninoone}= 50\GeV$\\ \hline
 SS/3L  \cite{SS3LPaper} & $\gluino\gluino$, $\gluino\rightarrow \ttbar\ninoone$   &   $m_{\gluino}> 950\GeV$  & \\
 				& $\gluino\gluino$, $\gluino\rightarrow \stopone t$, $\stopone\rightarrow s b$   &   $m_{\gluino}> 850\GeV$  & \\
				&  $\gluino\gluino$, $\gluino\rightarrow qqW\ninoone$  &   $m_{\gluino}> 860\GeV$ & $m_{\ninoone}<400\GeV$\\ 
				&  $\gluino\gluino$, $\gluino\rightarrow qqWZ\ninoone$  &   $m_{\gluino}> 1040\GeV$  & $m_{\ninoone}<520\GeV$\\
				&  $\squark\squark$, $\squark\rightarrow qWZ\ninoone$   &   $m_{\squark}> 670\GeV$    & $m_{\ninoone}<300\GeV$\\
				&$\gluino\gluino$, $\gluino\rightarrow qq(\ell\ell/\ell\nu/\nu\nu)\ninoone$ &$m_{\gluino}> 1200\GeV$&$m_{\ninoone}< 660\GeV$ \\
				& $\squark\squark$, $\squark\rightarrow q(\ell\ell/\ell\nu/\nu\nu)\ninoone$ &$m_{\gluino}> 780\GeV$&$m_{\ninoone}< 460\GeV$ \\ \hline 
TAU\cite{TauStrongPaper} & 	$\gluino\gluino$, $\gluino\rightarrow qq(\tau\tau/\tau\nu/\nu\nu)\ninoone$ & $m_{\gluino}> 1090\GeV$&	nGM model, $\stau$ is NLSP\\	\hline
 0/1L3B  \cite{3bjetsPaper} & $\gluino\gluino$, $\gluino\rightarrow \ttbar\ninoone$   &   $m_{\gluino}> 1340\GeV$  & $m_{\ninoone}< 400\GeV$ \\ \hline

\end{tabular}
\caption{The 95\% CL exclusion limits obtained in published ATLAS searches listed in table~\ref{tab:signal_regions} for the indicated processes and related assumptions. 
A dedicated search for $\tilde{c}\tilde{c}$ pair production \cite{scharmPaper} excludes charm squark masses up to 490 GeV for $m_{\ninoone}< 200\GeV$ (95\% CL). }
\label{tab:old_limits}
\end{table}

\newpage
\subsection{Final states with high-$\pt$ jets, missing transverse momentum and no electrons or muons}\label{no_lepton_states}

Several searches to address final states without electrons or muons, containing high-$\pt$ jets and missing transverse momentum, have been performed in ATLAS. 
These searches are split according to the jet multiplicity into three categories: searches with at least one, two to six and seven to ten jets. They are presented in table \ref{tab:signal_regions} as Monojet, 0-lepton + 2--6 jets + $\met$ (extended with two additional signal regions) and 0-lepton + 7--10 jets + $\met$, respectively. Events with reconstructed electrons or muons are vetoed in these searches. A new search using kinematic variables, known as Razor variables \cite{RazorVariables}, which provide longitudinal and transverse information about each event (listed as 0-lepton Razor in table \ref{tab:signal_regions}), has also been performed and is included in the results presented in this paper.  

The monojet (MONOJ) analysis, originally designed to search for direct production of top squarks ($\stop$), each decaying into a charm quark and a neutralino  ($\ninoone$) \cite{MonojetPaper}, targets final states characterized by at least one high-$\pt$ jet (with $\pt > $ 150 GeV and $|\eta| < $ 2.8) and large missing transverse momentum. 
Signal regions have been specifically optimized for models with a very small mass difference ($\leq$ 20~\GeV) between the top squark and the neutralino. The event selection makes use of the presence of initial-state radiation (ISR) jets to identify signal events, and the squark-pair system is boosted, leading to large \met. 
Three signal regions which are based only on different selection criteria related to the jet \pt\ and \met\  have been used to bring additional sensitivity to models with very small mass differences between SUSY particles. These signal regions do not impose any criteria to specifically select events originating from the top squarks and as such they can be used to select events in which squarks are produced in pairs and decay directly via $\squark \to q \ninoone$ with a small $\squark$--$\ninoone$ mass difference. 

The 0-lepton + 2--6 jets + $\met$ (0L) search \cite{0-leptonPaper} targets final states where each initial squark yields one jet and $E_{\rm T}^{\rm miss}$ and each initial gluino yields two jets and $\met$. Additional decay modes can include the production of charginos via $\tilde{q} \rightarrow q \tilde{\chi}^{\pm}_{1}$ and $\tilde{g} \rightarrow q\bar{q} \tilde{\chi}^{\pm}_{1}$, where the subsequent decay of these charginos to a $W$ boson and $\tilde{\chi}^{0}_{1}$ can lead to final states with larger jet multiplicity. 
The search strategy is optimized for various squark and gluino masses, for a range of models. 
Fifteen inclusive signal regions are characterized by increasing the minimum jet-multiplicity from two to six (for jets with $\pt > $ 40 GeV and $|\eta| < $ 2.8), and are based on different selection criteria on the effective mass $\meff^{\rm incl}$, defined as the scalar sum of  $\met$ and the $\pt$ of the jets; the ratio of $\met/\meff^{N_{\rm{j}}}$, where $\meff^{N_{\rm{j}}}$ is $\meff$ constructed from only the leading $N_{\rm j}$ jets; and the minimum azimuthal angle between jets and $\met$. 
Two of the signal regions are designed to improve sensitivity to models with the cascade $\tilde{q}$ or $\tilde{g}$ decay via $\tilde{\chi}^{\pm}_{1}$ to $W$ and $\ninoone$, in cases where the $\tilde{\chi}^{\pm}_{1}$ is nearly degenerate in mass with the $\tilde{q}$ or $\tilde{g}$. 
These signal regions place additional requirements on the invariant masses $m(W_{\rm cand})$ of the candidate $W$ bosons reconstructed from a single high-mass jet, or from a pair of jets.

Following  the same analysis strategy, two additional signal regions are included in this paper, which are optimized to increase the sensitivity of the 0L search for left-handed squarks within the pMSSM model described in section \ref{sec:susysignals}. These two signal regions target the two one-step decays of $\squarkL$,  $\squarkL \to q \chinopm \to q W^{\pm} \ninoone$ and $\squarkL \to q \ninotwo \to q (Z/h) \ninoone$ and are obtained by optimizing on two variables, $\met / \meff^{N_{\rm{j}}}$ and $\meff^{\rm incl}$, in the channels with at least four or at least five jets.  All other selection criteria are exactly the same as for the corresponding channels described in the original publication.  
The two new signal regions, named 4jt+ and 5jt following the naming convention from ref.~\cite{0-leptonPaper}, are  summarized in table \ref{tab:pMSSM_qL_SR}.

\begin{table}[H]
\centering
\begin{tabular}{| l | c | c |}
\hline 
Signal region name  & {\bf 0L\_4jt+} & {\bf 0L\_5jt}  \\
\hline \hline
Number of jets $\geq$ & 4 & 5  \\
\hline
$\met/\meff^{N_{\rm{j}}} \geq $ & 0.30 & 0.15  \\
\hline
$\meff^{\rm incl}$ [\GeV] $\geq$ & 2200 & 1900 \\ 
\hline
\end{tabular}
\caption{\label{tab:pMSSM_qL_SR} Additional 0L signal regions optimized to increase the sensitivity of the search for left-handed squarks within the pMSSM.  }
\end{table}

A high jet multiplicity is expected from the decays of gluino pairs via a top squark, or via squarks involving the production of  $\tilde{\chi}^{\pm}$ and $\tilde{\chi}^{0}_{2}$ in their decay chain, and is the main topology targeted by the 0-lepton + 7--10 jets + $\met$ (MULTJ) analysis \cite{multijetsPaper}. 
The sensitivity of the search is enhanced by the subdivision into two categories.
First, in the multi-jet+flavour stream, an event classification based on the number of jets ($\pt > $ 50 GeV and $|\eta| < $ 2) and number of $b$-jets ($\pt > $ 40 GeV and $|\eta| < $ 2.5) gives enhanced sensitivity to models which predict either more or fewer $b$-jets than the SM background. 
In the second category (multi-jet+$\MJ{}$ stream), which targets models with large numbers of objects in the final state, the jets reconstructed with the jet radius parameter $R$ = 0.4 are reclustered into large composite jets using the anti-$k_{t}$ algorithm with $R$ = 1.0. The event variable $M_{J}^{\Sigma}$ is computed %
 as the sum 
of the masses of the composite jets: $\MJ \equiv \sum_{j} m_j^{R=1.0}$, where the composite jets satisfy $p_{\rm T}^{R=1.0} > 100 \GeV$ and  $|\eta^{R=1.0}|<1.5$.
In total, nineteen signal regions are defined, based on different selection criteria on the total number of jets, number of $b$-jets, $\met/\sqrt{H_{\rm T}}$ (where $H_{\rm T}$ is the scalar sum of the $\pt$ of all jets) and on the event variable $M_{J}^{\Sigma}$.

The Razor variable set is designed to group together visible final-state particles associated with heavy produced sparticles, and in doing so contains information about the mass scale of those sparticles. 
The events are selected using a combination of \met\ triggers which 
are fully efficient for the event selections considered in this search. 
The new 0-lepton Razor (0LRaz) analysis presented here selects events with at least two high-$\pt$ jets and $\met$.  The baseline object selection and event cleaning, as well as the choice of MC generators for SM background processes and the approach for calculating systematic uncertainties exactly follow those of the 0L search \cite{0-leptonPaper}. Two signal regions are identified by optimizing criteria on the Razor variables to give the best expected sensitivity in the model with squark pair production followed by the direct decay of the squarks. One signal region, SR$_{\rm loose}$, targets models with small mass splittings which typically have softer visible objects, while the other signal region, SR$_{\rm tight}$, is designed to target models with high squark masses which typically contain harder visible objects. 
Appendix \ref{AppRazor} describes in detail the construction of the event variables, optimization strategy for these signal regions and corresponding control and validation regions, explicitly showing the distributions of the variables used for the selection, and the impact of the selection on the expected SM background and signal yields. An overview of the selection criteria for the two signal regions used in this search is given in table  \ref{tab:Razor_SR}. %

\begin{table}[H]
\small
\centering                                  
\begin{tabular}{|r|ccccc|ccccc|}
\hline
  & \multicolumn{5}{|c|}{\bf 0LRaz\_SR$_{\rm loose}$} & \multicolumn{5}{|c|}{\bf 0LRaz\_SR$_{\rm tight}$}  \\ \hline
\hline
\met [GeV] $>$ &\multicolumn{10}{|c|}{160} \\ \hline   
$\pt^{\rm jet_{1,2}}$ [GeV] $>$ & \multicolumn{5}{|c|}{ 150 }  & \multicolumn{5}{|c|}{ 200 } \\ \hline
$\Delta\phi({\rm jet_{1,2}}, \met)$ $>$  & \multicolumn{5}{|c|}{0.4} & \multicolumn{5}{|c|}{1.4} \\  \hline
$R$           $>$  & \multicolumn{5}{|c|}{ 0.5 } & \multicolumn{5}{|c|}{ 0.6 } \\ \hline
$M_{R}'$ [GeV] $>$ & \multicolumn{5}{|c|}{ 700 } & \multicolumn{5}{|c|}{ 900 }  \\ \hline
\end{tabular}
\caption{\label{tab:Razor_SR} Overview of the selection criteria for the two signal regions used by the 0LRaz analysis. The 0LRaz\_SR$_{\rm tight}$ targets high masses of the heavy produced sparticle, and the 0LRaz\_SR$_{\rm loose}$ targets small mass splittings between the heavy produced sparticle and the LSP. Details of the construction of Razor variables $M_{R}^{'}$ and $R$ can be found in appendix \ref{app:razorvars}.}
\end{table}

\subsection{Final states with high-$\pt$ jets, missing transverse momentum and at least one electron or muon} \label{lepton_states}

Three types of searches addressing decays of squarks and gluinos in events containing electrons or muons, jets and missing transverse momentum are summarized here: searches with at least one isolated lepton, which have been extended with an additional signal region with high jet multiplicity, 
a search with two same-flavour opposite-sign leptons inconsistent with $Z$ boson decay (off-Z search),  
and searches in final states with a same-sign lepton pair or at least three leptons. 

The 1-lepton (soft+hard)/2-leptons + jets + \met (1L(S,H), 2L(S), 2LRaz) searches \cite{1lepPaper} require the presence of at least one isolated lepton (electron or muon) in the decay chains of strongly produced squarks or gluinos.  
Different categories of events are defined in order to cover a broad parameter space: first the events are separated 
by different requirements on the transverse momentum of the leptons, either using an electron or muon with $\pt > $25~\GeV\ in the hard lepton selection, or an electron (muon) with $\pt >$ 7~(6)~\GeV\  in the soft lepton selection. 
Each of these selections is further subdivided into a single-lepton and a dilepton  search channel. 
The soft and hard lepton channels are designed to be complementary, and are more sensitive to supersymmetric spectra with small or large mass splittings, respectively, while the different lepton multiplicities cover different production
and decay modes. To enhance the sensitivity to gluino or squark production, high and low
jet multiplicity signal regions, respectively, are defined.  
The single-lepton channels (1L(S,H)) use a statistically independent set of events, compared to the 0L search, allowing the statistical combination of the two searches in the models for which it is relevant. 
The hard dilepton channel (2LRaz) targets gluino and first- and second-generation squark production, as well as mUED searches. This channel uses a Razor variable set and is not designed to search for signal events in which a real $Z$ boson is present. 
In all search channels except the soft dimuon channel (2L(S)), two separate selections are performed for each jet multiplicity: one single-bin signal region optimized for discovery reach, which is also used to place limits on the visible cross-section, and one signal region which is binned in an appropriate variable in order to exploit the expected shape of the distribution of signal events when placing model-dependent limits.

An additional signal region with one hard lepton (electron or muon), high jet multiplicity and \met, referred to as 1L(H)\_7-jet in table~\ref{tab:signal_regions}, is considered in this paper. The selection is based on looser missing transverse momentum selection criteria compared to the value used in ref.~\cite{1lepPaper} together with the requirement for high jet multiplicity, which is suggested in refs.~\cite{Lisanti:2011tm,Evans:2013jna} in the search for natural SUSY. 
The 1L(H)\_7-jet signal region selection follows the concepts of the 1L(H) analysis \cite{1lepPaper}, only modifying the criteria for the signal, validation and control regions to take into account a selection of events with at least seven jets in the final state. Due to these changes, a re-evaluation of the systematic uncertainties, in particular of the theoretical uncertainties on the background, is also performed.  The selection criteria are summarized in table \ref{tab:onelepSRCRVR}.

\begin{sidewaystable}[htbp]
\centering
\small
\begin{tabular}{| l |c | c | c | c | c |}
\hline 
 & {\bf 1L(H)\_7-jet} & {\bf 1L(H)\_WR\_7-jet} & {\bf 1L(H)\_TR\_7-jet} & {\bf 1L(H)\_VR\_7-jet \met} & {\bf 1L(H)\_VR\_7-jet \mt} \\
\hline \hline
$N_{\rm{lep}}$ & \multicolumn{5}{c|}{ == 1} \\ \hline
$p_{\rm{T}}^{\rm{\ell_1}}$ [\GeV]  & \multicolumn{5}{c|}{$>$ 25 (20)} \\ \hline
$p_{\rm{T}}^{\rm{\ell_2}}$ [\GeV] & \multicolumn{5}{c|}{$<$ 10} \\ 
\hline
$N_{\rm{jet}}$  & \multicolumn{5}{c|}{$\geq$ 7} \\ \hline
$p_{\rm{T}}^{\rm{jet}}$ [\GeV] & \multicolumn{5}{c|}{$>$ 80, 25, 25, 25, 25, 25, 25} \\ \hline
N$_{b-\rm{tag}}$ & $-$ & == 0 & $\geq$ 1& $-$ & $-$ \\
\hline
\met\ [\GeV] & $>$ 180  & $\in$[100, 180] & $\in$[100, 180]  & $\in$[180, 500]  & $\in$[100, 180] \\ \hline
\mt\ [\GeV] & $>$ 120 & $\in$[40, 80] & $\in$[40, 120] & $\in$[60, 120] &  $\in$[120, 320] \\ \hline
$\meff^{\rm incl}$ [\GeV] & \multicolumn{5}{c|}{$>$ 750} \\
\hline
\end{tabular}
\caption[Overview of the multi-jet selection]{Overview of the
  selection criteria for the 1L(H) + 7 jets + \met analysis, 
  for the signal region (1L(H)\_7-jet), $W$+jets 
  and semileptonic \ttbar\  control regions (1L(H)\_WR\_7-jet and 1L(H)\_TR\_7-jet respectively) and two validation regions used to cross-check the background estimates (1L(H)\_VR\_7-jet \met and 1L(H)\_VR\_7-jet \mt). The \pt\ selections for leptons are given for electrons (muons).}
\label{tab:onelepSRCRVR}
\end{sidewaystable}

The 2-leptons off-Z (2L-offZ) search \cite{dilepton-edgePaper} targets events where the final state same-flavour opposite-sign leptons originate from the decay $\ninotwo \to \ell^+ \ell^- \ninoone$, where \ninotwo is produced in the decays of squarks and gluinos, e.g. $\squark \to q\ninotwo$ and $\gluino \to qq\ninotwo$. Compared to the decay $Z \to  \ell^+ \ell^-$, which leads to a peak in the $m_{\ell\ell}$ distribution around the $Z$ boson mass, the decay $\ninotwo \to \ell^+ \ell^- \ninoone$ leads to a rising distribution in $m_{\ell\ell}$ that terminates at an endpoint (``edge") \cite{Hinchliffe:1996iu} because events with larger $m_{\ell\ell}$ values are kinematically forbidden. 
Four signal regions are defined by requirements on jet multiplicity, $b$-tagged jet multiplicity and \met. The selection criteria are optimized for the simplified models of pair-produced squarks or gluinos followed by their two-step decays with sleptons, described in section \ref{subsubsec:onestep}.  The signal regions with a $b$-jet veto provide the best sensitivity in the two-step simplified models considered here, since the signal $b$-jet content is lower than that of the dominant \ttbar background. 
Signal regions with a requirement of at least one $b$-tagged jet target other signal models not explicitly considered here, such as those with bottom squarks that are lighter than the other squark flavours. One signal region with similar requirements to those used by the CMS experiment in a comparable search \cite{Khachatryan:2015lwa} which reported an excess of events above the SM background with a significance of 2.6 standard deviations, is also used for comparison purposes. No evidence for an excess is observed in this region.  

Another leptonic search channel \cite{SS3LPaper} is used for an analysis of final states with multiple jets, and either two leptons of the same electric charge or at least three leptons (SS/3L).
The motivation for searches using these final states is that pair-produced gluinos have the same probability to decay to pairs of leptons with the same charge as with opposite charge. Squark production (directly in pairs or through $\gluino\gluino$ or $\gluino\squark$ production with subsequent $\gluino \to q \squark$ decay) can also lead to same-sign lepton or three-lepton signatures when the squarks decay in cascades involving top quarks, charginos, neutralinos or sleptons. 
Requiring a pair of leptons with the same electric charge largely suppresses the background coming from the SM processes, giving a very clean and powerful signature to search for new physics processes. It also allows the use of relatively loose kinematic requirements on $\met$, increasing the sensitivity to scenarios with small mass differences between SUSY particles or with R-parity violation. 
Five statistically independent signal regions are defined: two signal regions requiring same-sign leptons and $b$-jets  (optimized for gluino decays via top squarks), a complementary signal region requiring a $b$-jet veto (optimized for the gluino decays via first- and second-generation squarks), and two signal regions requiring three leptons (designed for scenarios characterized by multi-step decays). 

\subsection{Final states with high-$\pt$ jets, missing transverse momentum and at least one hadronically decaying tau lepton}\label{tau_states}
  
A search for squarks and gluinos in events with large missing transverse momentum, jets and at least one hadronically decaying tau lepton \cite{TauStrongPaper} is motivated by naturalness arguments \cite{Barbieri:1987fn,deCarlos:1993yy}, 
and by the assumption that light sleptons could play a role in the co-annihilation with neutralinos in the early universe \cite{Ellis:1999mm}. In particular, models with light tau sleptons are consistent with dark-matter searches \cite{Vasquez:2011}. 
Four distinct topologies are studied in order to optimize the tau + jets + \met (TAU) search for various models: one hadronically decaying tau (1$\tau$) or two or more hadronically decaying taus (2$\tau$) in the final state with no additional light leptons ($e/\mu$) and one
or more tau leptons with exactly one lepton (one electron ($\tau+e$) or muon ($\tau+\mu$)). 
The different topologies (1$\tau$, 2$\tau$ and $\tau+\ell$, where $\ell$ is electron or muon) have been optimized separately, and, where relevant, are statistically combined to increase the analysis sensitivity.  
The same signal regions are used for additional model interpretations, not presented in ref. \cite{TauStrongPaper}.  
These are simplified models of squark- or gluino-pair production where the squark or gluino undergoes a two-step cascade decay via sleptons, as shown in figures \ref{fig:feynman-twostep}(a) and \ref{fig:feynman-twostep}(c), where the sleptons are assumed to be exclusively staus, since the first two generations of sleptons and sneutrinos are kinematically decoupled. 

\subsection{Final states with many $b$-jets and missing transverse momentum}\label{bjet_states}

A search requiring at least three $b$-jets \cite{3bjetsPaper} is one of the most sensitive searches to various SUSY models favoured by naturalness arguments, where top or bottom quarks are produced in the gluino decay chains. 
The search is carried out in statistically independent zero- and one-lepton channels (0/1L3B) which are combined to maximize the sensitivity. 
Three sets of signal regions, two for the zero-lepton channel and one for the one-lepton channel,  
are defined to enhance the sensitivity to the various models considered. 
They are characterized by having relatively hard \met\ requirements and at least four, 
six or seven jets, amongst which at least three are required to be $b$-jets. Signal
regions with zero leptons and at least four jets target SUSY models with sbottoms  
in the decay chain, while the one-lepton and the zero-lepton signal regions with at least six or seven jets aim to probe
SUSY models where top-quark-enriched final states are expected.

\section{Systematic uncertainties}
\label{sec:sysuncert}

Systematic uncertainties on background estimates in all searches included in this paper 
arise from the use of  transfer factors which relate observations in the control regions to background expectations in the signal regions, and from the MC modelling of minor backgrounds. Since CRs are designed to be kinematically as close as possible to the SRs, many sources of systematic uncertainty largely cancel. In searches which include leptons in the final state, systematic uncertainties also impact the estimation of jets misidentified as leptons or of non-prompt leptons.  The full details of all sources of systematic uncertainty and their impact on background predictions for each search included in this paper can be found in the corresponding original publication. Only the dominant uncertainties on the background estimations, common to all searches, are mentioned here.

Since at least one high-$\pt$ jet and significant missing transverse momentum are present in all searches summarized in this paper, the primary common sources of systematic uncertainty for the SM backgrounds estimated with transfer factors derived from MC simulation are the JES and the jet energy resolution (JER). 
The theoretical modelling of background processes and the limited number of data events in the CRs and in the MC simulation are also typically important. 

The JES uncertainty is estimated from a combination of simulation, test beam data and in-situ
measurements ~\cite{Aad:2011he,Aad:2012vm}, and depends on the $\pt$ and $\eta$ of the jet. Additional contributions accounting for the jet-flavour composition, the calorimeter response to different jet flavours, pile-up and $b$-jet calibration uncertainties are also taken into account. Uncertainties on the JER are obtained with an in-situ measurement \cite{Aad:2012ag} of the jet transverse momentum balance in dijet events. Uncertainties in jet measurements
are propagated to the $\met$, and additional subdominant uncertainties on $\met$ arising from the contribution from energy deposits not associated with reconstructed objects are also included. In signal regions designed for searches based on large jet multiplicities these uncertainties can be as large as 30\% of the estimated background yield in the SRs. 

Searches requiring the presence of tau leptons in the final state are subject to additional systematic uncertainties from the tau energy scale \cite{ATLAS:TES2013} and the tau lepton identification  \cite{ATLAS:TauID2013}. The uncertainties from the jet and tau energy scales are the largest experimental uncertainties in these searches, being as large as 13\% and 8\% of the  estimated background yield respectively, and are treated as uncorrelated, since the calibration methods differ. 
 
In searches that require the presence of $b$-jets in the final state, the uncertainty associated with flavour-tagging efficiencies is evaluated by varying the $\pt$- and flavour-dependent correction factors applied to each jet in the simulation within a range that reflects the systematic uncertainty on the measured tagging efficiencies and mistag rates. This uncertainty varies between 10\% and 16\% in the different SRs requiring at least three $b$-jets in the final state. 

Uncertainties arising from the theoretical modelling of background processes are typically evaluated by comparing the estimates to those obtained with different MC generators. The uncertainty due to the factorization and renormalization scales is computed by varying these scales up and down by a factor of two with respect to the nominal setting.
Uncertainties from PDFs are computed following the PDF4LHC recommendations \cite{Botje:2011sn}. These uncertainties vary across the different searches and in some signal regions are the dominant source of systematic uncertainties. 

The same sources of experimental uncertainty apply to the signal acceptance. Several theoretical uncertainties on the acceptance for the various signal models are taken into account. These uncertainties are estimated using the \Madgraph5+\pythia6 samples by varying the following parameters up and down by a factor of two: the \Madgraph\ scale used to determine the event-by-event renormalization and factorization scale, the parameters used to determine the scales for initial- and final-state QCD radiation and the parameters used for jet matching.  
The uncertainty on the modelling of initial-state radiation plays an important role in simplified models with small mass differences in the decay cascade, and is as large as 20--30\% in such regions. 
For all models, except the mUED model, the NLO+NLL cross-section uncertainty is taken from an envelope of cross-section predictions using different PDF sets and factorization and renormalization scales, as described in ref. \cite{Kramer:2012bx}. The mUED model cross-sections are based on a calculation at LO in QCD, and the events are generated with a leading-order MC event generator.  No theoretical uncertainties on the acceptance are considered for this case.

The overall background uncertainties for the two new signal regions defined in the 0LRaz search are estimated to be 6\% in  the 0LRaz\_SR$_{\rm loose}$ signal region and 9\% in the  0LRaz\_SR$_{\rm tight}$ signal region. These uncertainties are dominated by the modelling of the $Z$+jets process and by the uncertainties on diboson production due to renormalization and factorization scales and PDF uncertainties for which a conservative uniform 50\% uncertainty is applied. 
In the additional signal region for the 1-lepton (hard) + jets + \met\ analysis, 1L(H)\_7-jet, the overall background uncertainty is estimated to be 35\%, and it is dominated by the modelling of the \ttbar process in the events with high jet multiplicity. 
The estimated background uncertainties in the two additional signal regions for the 0-lepton + 2--6 jets + \met\ analysis, 0L\_4jt+ and 0L\_5jt, are consistent with the uncertainties obtained for the two closest signal regions (4jt and 5j) from the original publication \cite{0-leptonPaper}.

\section{Results for the new signal regions}
\label{sec:results}

The number of events observed in the data and expected from SM processes are shown for all new signal regions in 
tables~\ref{tab:0L4-5jets_results},  \ref{tab:0LRazor_results} and
\ref{tab:1L7jets_results}. %
Table~\ref{tab:0L4-5jets_results} summarizes the results for the two additional signal regions for the 0L analysis. 
Table~\ref{tab:0LRazor_results} displays the equivalent results for the two signal regions of the new 0LRaz analysis, and those for the additional region of the 1L(H) analysis are shown in table \ref{tab:1L7jets_results}. 
All results are determined using the background-only fit. The pre-fit background expectations are also shown in the tables, for comparison purposes.
The prediction of the $W/Z$+jets background processes by the simulation prior to the fit is found to be overestimated in the phase space of interest and is consequently decreased by the fit.  This is consistent with the behaviour observed in previous publications probing a similar phase space \cite{0-leptonPaper}.  In all new signal regions presented in this paper the number of events observed is consistent with the post-fit SM expectations.   
The observed and expected upper limits at 95\% CL on the number of BSM events ($S_{\rm obs}^{95}$ and $S_{\rm exp}^{95}$),  together with the upper limits on the visible cross-section of BSM physics ($\langle\epsilon\sigma\rangle_{\rm obs}^{95}$) and the $p$-value for the background-only hypothesis, are also presented in the tables  \ref{tab:0L4-5jets_results}--\ref{tab:1L7jets_results}. The confidence levels are calculated with  
the $CL_{\rm S}$ prescription \cite{Read:2002hq}. For an observed number of events lower than expected, the $p$-value is truncated at 0.5.

\begin{table}[H]
\begin{center}
{
\begin{tabular}{| l | c | c |}
\hline
  Signal region         & {\bf  0L\_4jt+}  &   {\bf  0L\_5jt}   \\ \hline   
 \multicolumn{3}{|c|}{Expected background events before the fit}    \\ \hline
       \ttbar (+ $V$) + single top         & $0.37$   & $2.9$     \\
        $W$+jets          & $0.75$   & $4.5$    \\
        $Z/\gamma^*$+jets          & $2.1$   & $4.8$    \\
        Diboson         & $-$    & $0.32$       \\ \hline
 \multicolumn{3}{|c|}{Fitted background events  }   \\ \hline
       \ttbar (+ $V$) + single top         & $0.39 \pm 0.32$   & $3.0 \pm 1.8$     \\
        $W$+jets          & $0.55 \pm 0.33$   & $2.0 \pm 1.5$    \\
        $Z/\gamma^*$+jets          & $0.10^{+0.17}_{-0.10}$   & $1.7 \pm 0.9$    \\
        Diboson         & $-$    & $0.32 \pm 0.16$    \\
        Multi-jet         & $-$   & $0.58^{+0.73}_{-0.58}$   \\ \hline
Total background          & $1.04 \pm 0.43$    & $7.6 \pm 1.9$  \\ \hline
Observed events          & $0$   & $8$   \\ \hline \hline 
$\langle\epsilon{\rm \sigma}\rangle_{\rm obs}^{95}$[fb]  & $0.17$ & $0.40$  \\ 
$S_{\rm obs}^{95}$  & $3.4$  & $8.2$  \\ 
$S_{\rm exp}^{95}$ & $3.5^{+1.3}_{-0.5}$ & $7.5^{+3.1}_{-2.0}$  \\
$p(s=0)$ & $0.50$ & $0.35$  \\ 
\hline
\end{tabular}
}
\end{center}
\caption{ 
The background expectations before the fit and the background fit results for the new 0L signal regions. Negligible contributions are marked as `$-$'. 
The uncertainties shown combine the statistical uncertainties on the event samples with the systematic uncertainties. 
Also shown are the 95\% CL upper limits on the visible cross-section
($\langle\epsilon\sigma\rangle_{\rm obs}^{95}$) and on the number of signal events ($S_{\rm obs}^{95}$ ).
The expected upper limit on the number of signal events ($S_{\rm exp}^{95}$) 
is calculated from the expected number of background events after fit, with uncertainties indicating the $\pm 1\sigma$ deviations from the expectation. The $p$-value ($p(s = 0)$) is also presented in the table. 
}
\label{tab:0L4-5jets_results}
\end{table}

\begin{table}[H]
\begin{center}
{
\begin{tabular}{| l | c | c |}
\hline
  Signal region         & {\bf 0LRaz\_SR$_{\rm loose}$}  &   {\bf  0LRaz\_SR$_{\rm tight}$}   \\ \hline   
 \multicolumn{3}{|c|}{Expected background events before the fit}    \\ \hline
        \ttbar       & $138$   & $1.8$      \\
       Single top         & $23.9$   & $1.6$     \\       
       \ttbar + $V$        & $4.7$   & $0.2$     \\
        $W$+jets          & $794$   & $49$    \\
        $Z$+jets          & $762$   & $58$    \\
        Diboson         & $112$    & $10$       \\ \hline
 \multicolumn{3}{|c|}{Fitted background events  }   \\ \hline
        \ttbar      & $117 \pm 22$     & $1.7 \pm 0.5$   \\
       Single top         & $24.9 \pm 2.6$   & $1.8 \pm 0.3$     \\
       \ttbar + $V$         & $3.7 \pm 1.0$   & $0.20 \pm 0.07$     \\
        $W$+jets          & $454 \pm 40$   & $27.0 \pm 3.0$    \\
        $Z$+jets          & $618 \pm 76$   & $45 \pm 6$    \\
        Diboson         & $94 \pm 49$    & $10 \pm 5$    \\
        Multi-jet         & $14 \pm 13$   & $2.4 \pm 2.4$   \\ \hline
Total background          & $1326 \pm 84$    & $88 \pm 8$  \\ \hline
Observed events          & $1322$   & $74$   \\ \hline \hline
$\langle\epsilon{\rm \sigma}\rangle_{\rm obs}^{95}$[fb]  & $6.17$ & $0.83$  \\ 
$S_{\rm obs}^{95}$  & $125.3$  & $16.8$  \\ 
$S_{\rm exp}^{95}$ & $135.1^{+64.8}_{-42.2}$ & $24.3^{+9.9}_{-6.9}$  \\ 
$p(s=0)$ & $0.49$ & $0.50$  \\ 
\hline
\end{tabular}
}
\end{center}
\caption{ 
The background expectations before the fit and the background fit results for the 0LRaz analysis. Negligible contributions are marked as `$-$'. 
The uncertainties shown combine the statistical uncertainties on the event samples with the systematic uncertainties. 
Also shown are the 95\% CL upper limits on the visible cross-section
($\langle\epsilon\sigma\rangle_{\rm obs}^{95}$) and on the number of signal events ($S_{\rm obs}^{95}$ ).
The expected upper limit on the number of signal events ($S_{\rm exp}^{95}$) 
is calculated from the expected number of background events after fit, with uncertainties indicating the $\pm 1\sigma$ deviations from the expectation. The $p$-value ($p(s = 0)$) is also presented in the table. 
}
\label{tab:0LRazor_results}
\end{table}

\begin{table}[H]
\begin{center}
{
\begin{tabular}{| l | c | }
\hline
  Signal region         & {\bf 1L(H)\_7-jet}   \\ \hline   
 \multicolumn{2}{|c|}{Expected background events before the fit}    \\ \hline
        \ttbar       & $81$       \\
        Single top         & $3.4$      \\
       \ttbar + $V$         & $2.8$      \\
        $W$+jets          & $11$     \\
        $Z$+jets          & $0.59$     \\
        Diboson         & $2.7$         \\ \hline
 \multicolumn{2}{|c|}{Fitted background events  }   \\ \hline
        \ttbar      & $56 \pm 27$      \\
        Single top         & $3.4 \pm 2.2$          \\
       \ttbar + $V$         & $2.8 \pm 1.0$      \\        
        $W$+jets          & $8 \pm 4$     \\
        $Z$+jets          & $0.59 \pm 0.17$     \\
        Diboson         & $2.7 \pm 1.5$      \\
        Multi-jet         & $2.4 \pm 2.3$    \\ \hline
Total background          & $76 \pm 27$    \\ \hline
Observed events          & $68$    \\ \hline \hline
$\langle\epsilon{\rm \sigma}\rangle_{\rm obs}^{95}$[fb]  & 2.06 \\ 
$S_{\rm obs}^{95}$  & 41.9 \\ 
$S_{\rm exp}^{95}$ & $45^{+12}_{-10}$ \\ 
$p(s=0)$ & $0.5$  \\ 
\hline
\end{tabular}
}
\end{center}
\caption{ 
The background expectations before the fit and the background fit
results for the new 1L(H) signal region. An overview of the selection criteria for the signal, validation and control regions used in this analysis is given in table \ref{tab:onelepSRCRVR}.  
The uncertainties shown combine the statistical uncertainties on the event samples with the systematic uncertainties. 
Also shown are the 95\% CL upper limits on the visible cross-section
($\langle\epsilon\sigma\rangle_{\rm obs}^{95}$) and on the number of signal events ($S_{\rm obs}^{95}$ ).
The expected upper limit on the number of signal events ($S_{\rm exp}^{95}$) 
is calculated from the expected number of background events after fit, with uncertainties indicating the $\pm 1\sigma$ deviations from the expectation. The $p$-value ($p(s = 0)$) is also presented in the table. 
}
\label{tab:1L7jets_results}
\end{table}

\section{Combination strategy}
\label{sec:combination}

Statistical combinations of the analyses, as listed in table~\ref{tab:AnalysesvsModels}, are performed in order to increase the exclusion reach in several SUSY models in which at least two analyses designed to be statistically independent in their signal and control region definitions provide comparable sensitivities. 
The conditions are satisfied for a combination of the 1L(S,H) and 0L searches. These analyses search for squarks and gluinos in final states containing jets and missing transverse momentum, either with at least one isolated electron or muon (1L(S,H)), or applying an explicit veto on events containing electrons or muons (0L). They are 
statistically independent due to a veto on any electron or muon with $\pt>10~\GeV$ in the case of the 0L search, and requiring an electron or muon with $\pt(e/\mu) > 25~\GeV$ (hard single-lepton) or $\pt(e/\mu) > 7/6~\GeV$ (soft single-lepton) in the case of the 1L(S,H) search. 
It has been checked explicitly that the difference in lepton-\pT\ thresholds in the 0L  and soft single-lepton analyses does not result in events selected by both analyses. The control regions used to estimate contributions from $W$+jets and top backgrounds used by the 0L search have been slightly modified with respect to the original regions \cite{0-leptonPaper} such that events which are selected by the respective control regions in the 1L(S,H) analyses are vetoed. This modification ensures complete statistical independence of the three analyses, which can therefore be combined where relevant. 

The statistical combination is obtained from the individual likelihoods of the analyses involved.
In the case of the 0L analysis the likelihood for the signal region that provides the best expected $CL_{\rm S}$ value for the signal model considered, and its corresponding control regions, is chosen. The choice of this signal region can vary as a function of sparticle masses. For the 1L(S,H) analyses, all the available signal regions are statistically independent and hence  a single likelihood that describes all of them serves as input for the combination procedure.
Some of the systematic uncertainties can be correlated when building the combined likelihood. 
The correlated uncertainties in the combination procedure are the luminosity uncertainty, the uncertainty on the SUSY cross-sections, $b$-tagging uncertainties, and the jet energy resolution and \met-related uncertainties. Other systematic uncertainties, such as theoretical uncertainties, are not correlated, e.g.\ the uncertainties due to different Monte Carlo generators used in the analyses considered. The jet energy scale uncertainty, which is subdominant, is not correlated due to the use of different prescriptions in the analyses involved.  
The combination of the analyses was carefully validated by ensuring that the combined likelihood did not lead to artificial correlations between fit parameters or major changes in post-fit values of nuisance parameters with respect to the individual analysis fits discussed in refs. \cite{0-leptonPaper,1lepPaper}.

Figures \ref{fig:combresult_SS} and \ref{fig:combresult_GG} show the result of the combination of the 1L(S,H) and 0L analyses for both squark-pair and gluino-pair production for the one-step decays of squarks and gluinos described in section~\ref{subsubsec:onestep}. The limits obtained improve the results of the separate analyses, reaching higher $\ninoone$ mass and approximately $50\,\GeV$ higher squark or gluino mass for massless neutralinos. The combined limit also approaches the diagonal $m(\ninoone) = m(\squark,\gluino)$ closer than the individual analyses.

\begin{figure}[H]
 \centering
\includegraphics[width=0.65\textwidth]{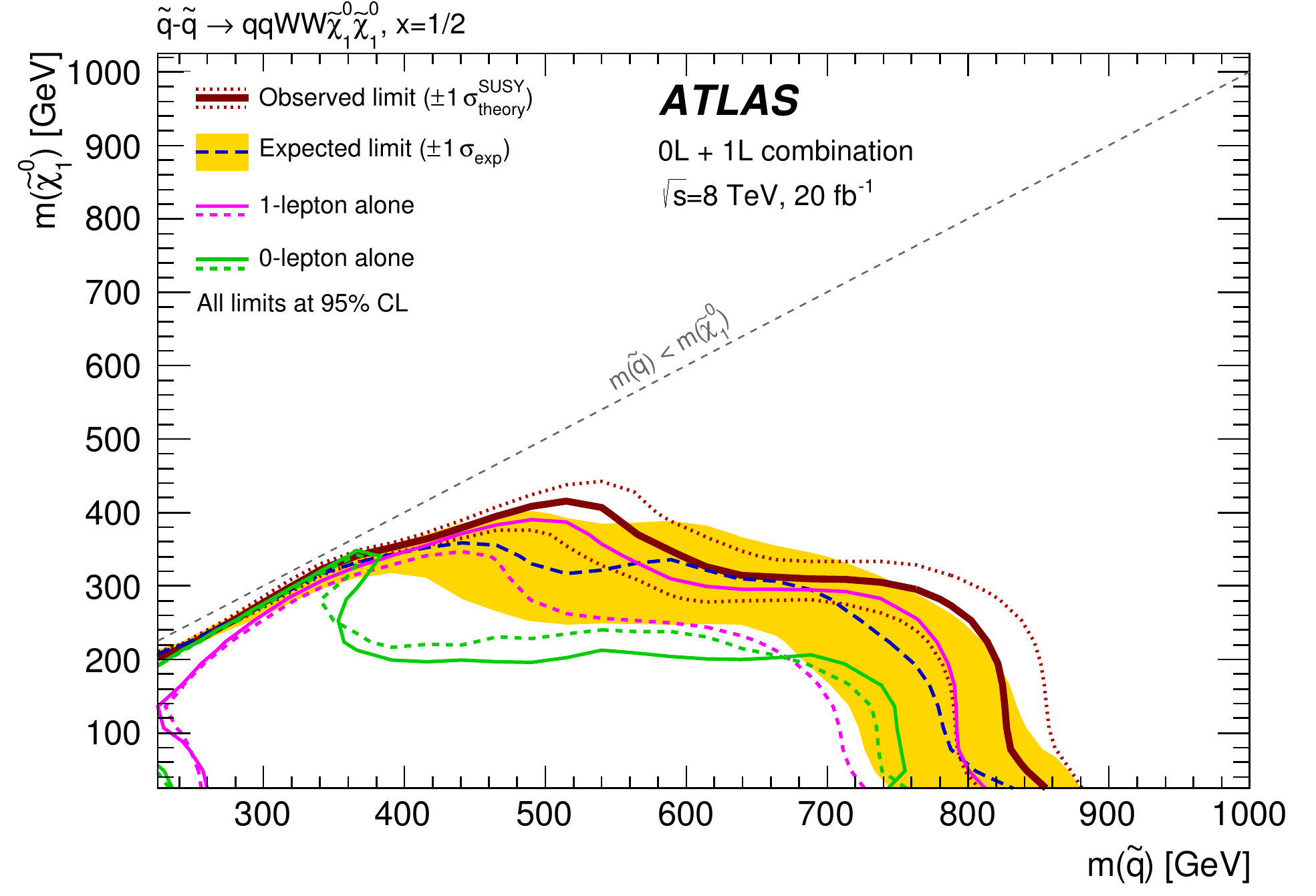}
 \caption{Observed and expected exclusion limits for simplified models of squark-pair production with one-step decays via the $\chinoonepm$ into a $W$ boson and the $\ninoone$.  %
The mass of the $\chinoonepm$ is chosen to be between the $m_{\squark}$ and $m_{\ninoone}$ and is determined by $x=(m_{\chinoonepm}-m_{\ninoone}) / (m_{\squark}-m_{\ninoone}) = 1/2$. 
Squark and neutralino masses in the area below the observed limit are excluded at 95\%~CL.
The yellow band includes all experimental uncertainties; the red dotted lines indicate the theory uncertainty on the cross-section. The individual limits from the 0L and the 1L(S,H) analyses are overlaid in green and magenta, respectively.}
 \label{fig:combresult_SS}
\end{figure}

\begin{figure}[H]
 \centering
\includegraphics[width=0.65\textwidth]{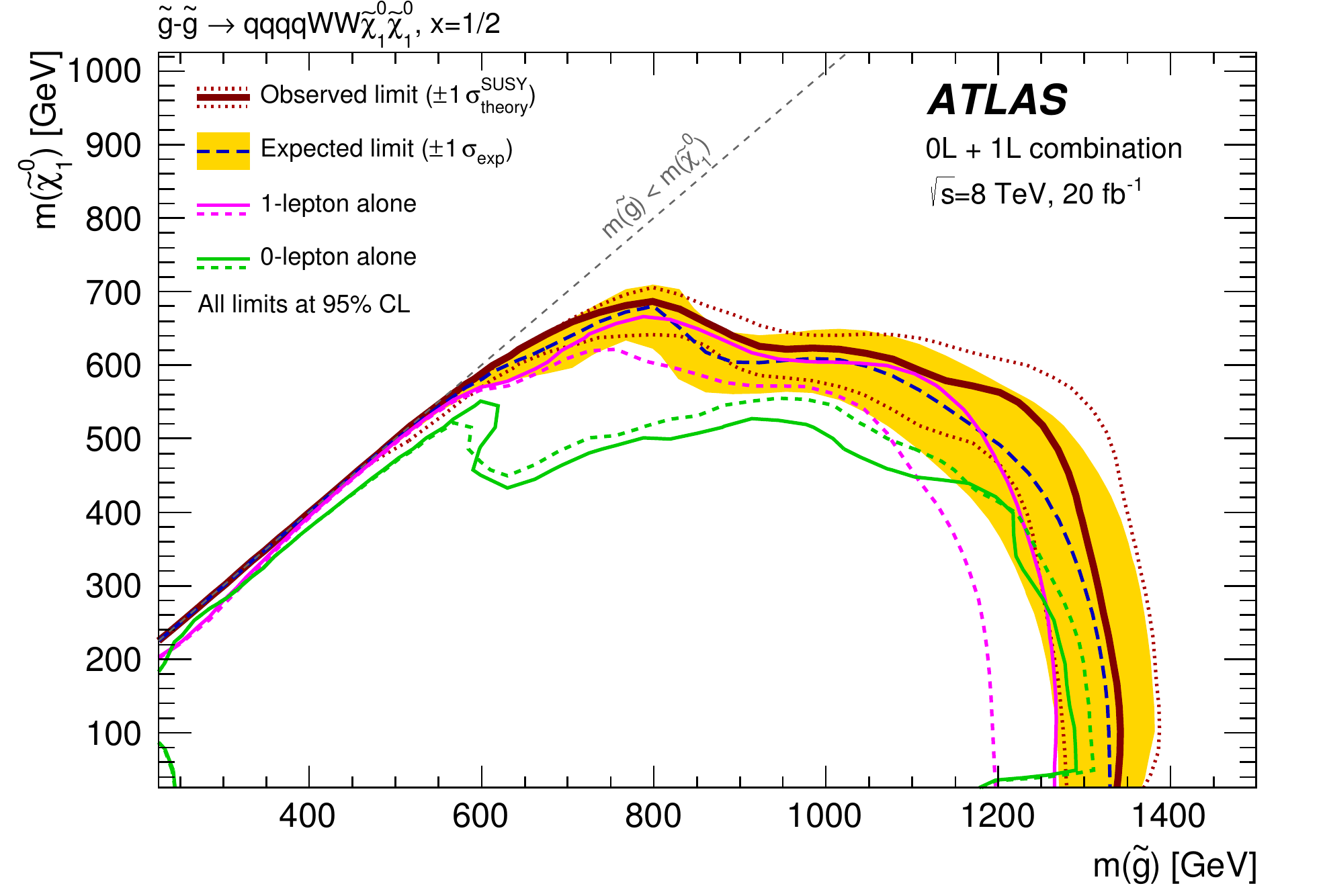}
 \caption{Observed and expected exclusion limits for simplified models of gluino-pair production with decays via the $\chinoonepm$ into a $W$ boson and the $\ninoone$. %
 The mass of the $\chinoonepm$ is chosen to be between the $m_{\gluino}$ and $m_{\ninoone}$ and is determined by $x=(m_{\chinoonepm}-m_{\ninoone}) / (m_{\gluino}-m_{\ninoone}) = 1/2$. 
Gluino and neutralino masses in the area below the observed limit are excluded at 95\%~CL.
 The yellow band includes all experimental uncertainties; the red dotted lines indicate the theory uncertainty on the cross-section. The individual limits from the 0L and the 1L(S,H) analyses are overlaid in green and magenta, respectively.}
 \label{fig:combresult_GG}
\end{figure}

\newpage
\section{Limits in SUSY models}
\label{sec:limits}

This section summarizes the exclusion limits placed in the various phenomenological and simplified models described in section~\ref{sec:susysignals} (there is a one-to-one correspondence between the subsections of section~\ref{sec:susysignals} and this section). The analyses and corresponding signal regions are referred to by their acronyms defined in table~\ref{tab:signal_regions}. An overview of all searches used to probe the phenomenological models described in section \ref{subsec:phenomodels} and the simplified models described in  section \ref{subsec:simplified} is given in table \ref{tab:AnalysesvsModels}. 
A limit obtained from the statistical combination of 1L(S,H) and 0L analyses is presented for models for which both analyses have comparable sensitivity and is used as a single contribution to the final combined limit. The final combined observed and expected 95\%~CL exclusion limits %
are obtained from the signal regions belonging to the contributing analyses that provide the best expected $CL_{\rm S}$ value. Expected limits from the individual analyses which contribute to the final combined limits are also presented for comparison. 
The $\pm1 \sigma^{\rm SUSY}_{\rm theory}$ lines around the observed limits in the figures %
 are obtained by changing the signal cross-section by one standard deviation
($\pm1\sigma$), as described in section~\ref{sec:sysuncert}. 
All mass limits on supersymmetric particles quoted later in this section are 
derived from the $-1 \sigma^{\rm SUSY}_{\rm theory}$  line.

\subsection{Limits in phenomenological models}
\label{sec:susymodels}

This section summarizes the exclusion limits placed on the phenomenological models described in section~\ref{subsec:phenomodels}. 

\subsubsection{A phenomenological MSSM model}

The measurements are interpreted in a phenomenological MSSM model, which possesses three parameters: $m_{\tilde{q}_{\rm L}}$, $M_1$ and $M_2$,  where $M_1$ and $M_2$ are the masses associated with the bino and wino fields. For the exclusion limits in figure~\ref{fig:limit-sqsq_pmssm} 
either $M_1$ is fixed to 60~\GeV\ and $M_2$ is varied independently, or $M_1$ is varied and $M_2$ is set to 
$ M_2=(M_1+m_{\tilde{q}_{\rm L}})/2$. The figures show limits in the ($m_{\tilde{q}}, m_{\chinoonepm,\ninotwo}$) and ($m_{\tilde{q}}, m_{\ninoone}$) planes, for various gluino masses, as obtained from the 0L analysis 
with the additional 0L\_4jt+ and 0L\_5jt signal regions optimized specifically for this model.
 As expected, for the relatively light gluino mass of 1600 GeV, a large range of squark masses (up to 1500~\GeV) and $\chinoonepm$/$\ninotwo$ masses (up to  
1150~\GeV) can be excluded. The exclusion reach decreases with increasing gluino mass.
\begin{figure}[htbp]
\centering
\subfigure[]{\includegraphics[width=0.48\textwidth]{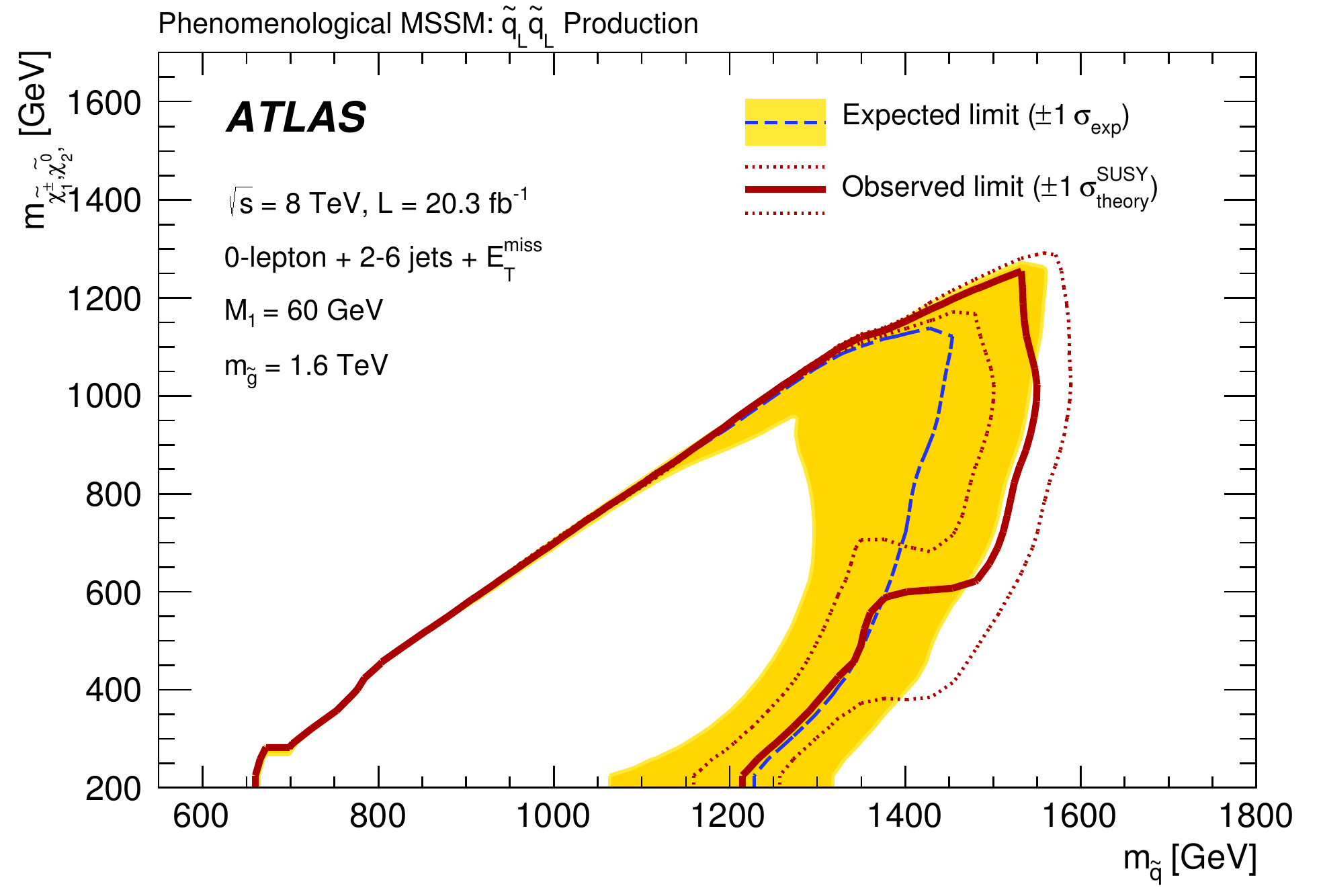}}
\subfigure[]{\includegraphics[width=0.48\textwidth]{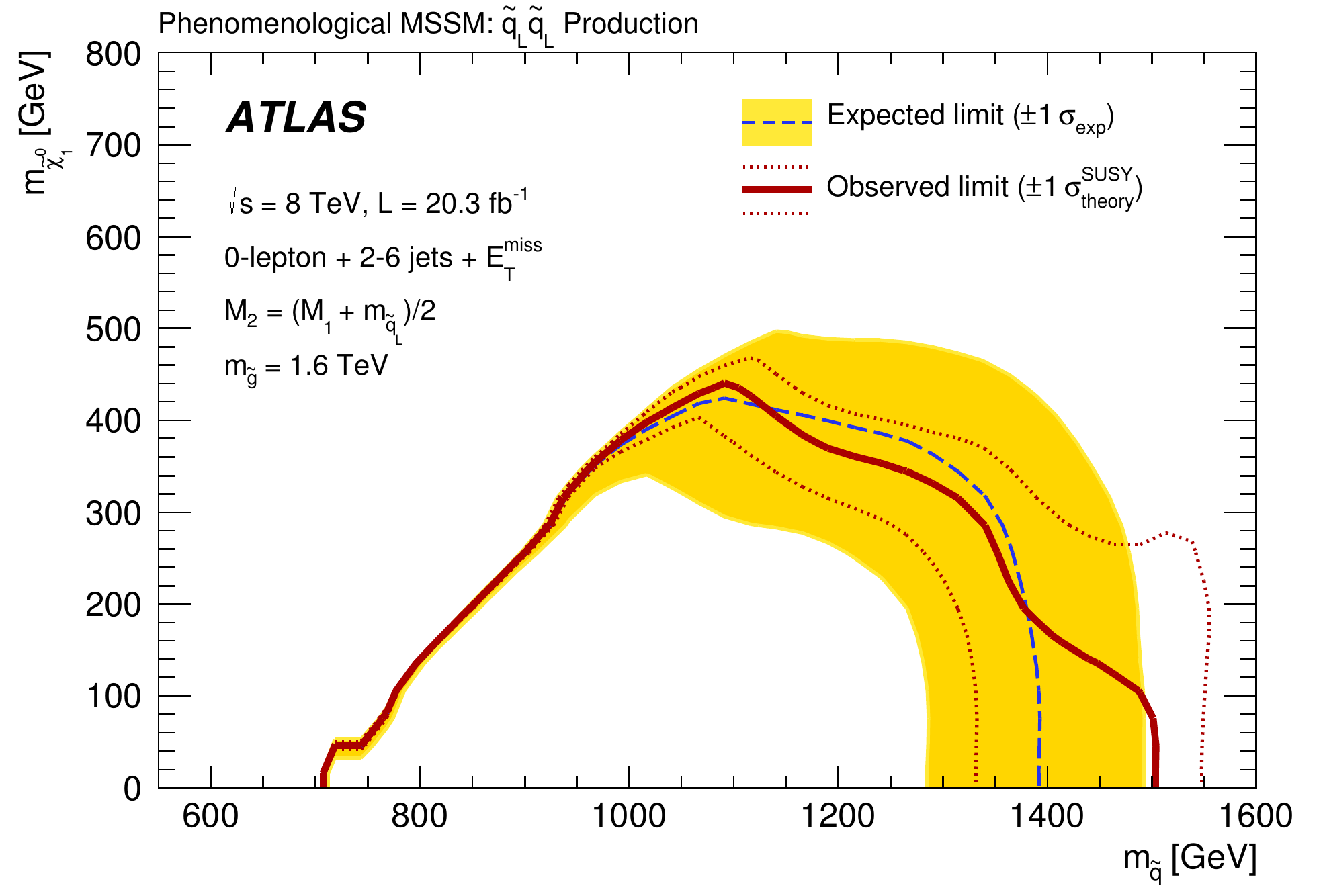}}
\subfigure[]{\includegraphics[width=0.48\textwidth]{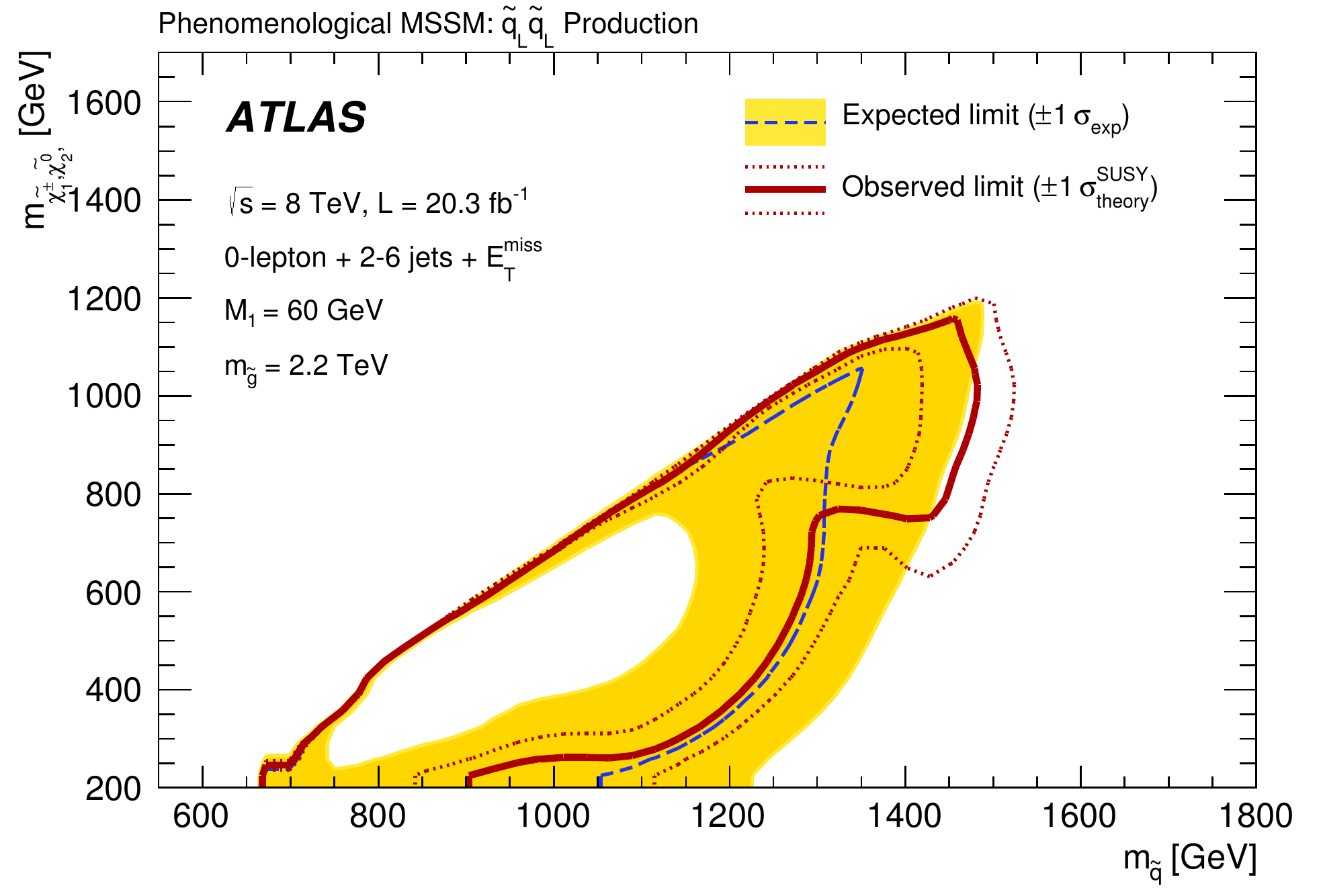}}
\subfigure[]{\includegraphics[width=0.48\textwidth]{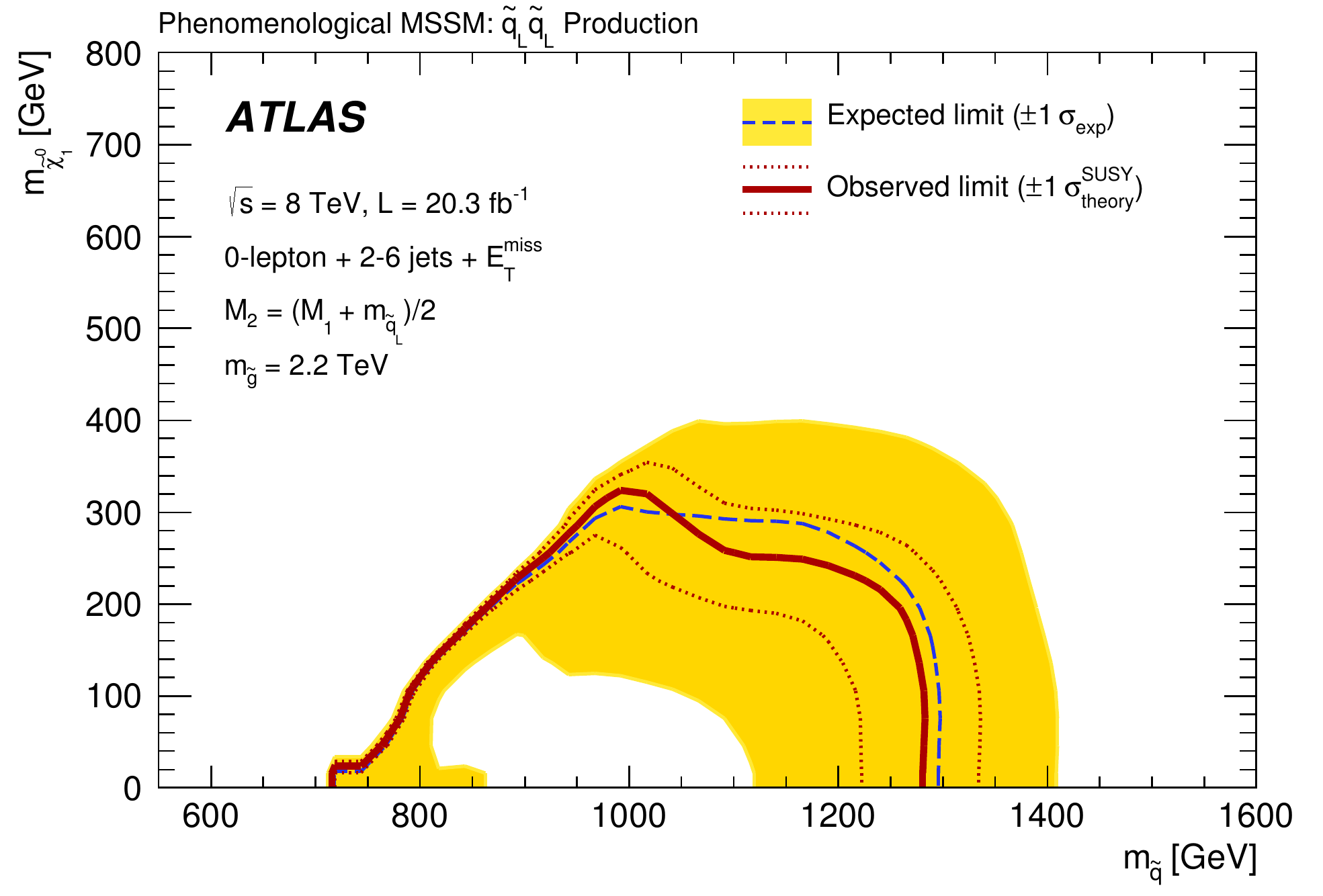}}
\subfigure[]{\includegraphics[width=0.48\textwidth]{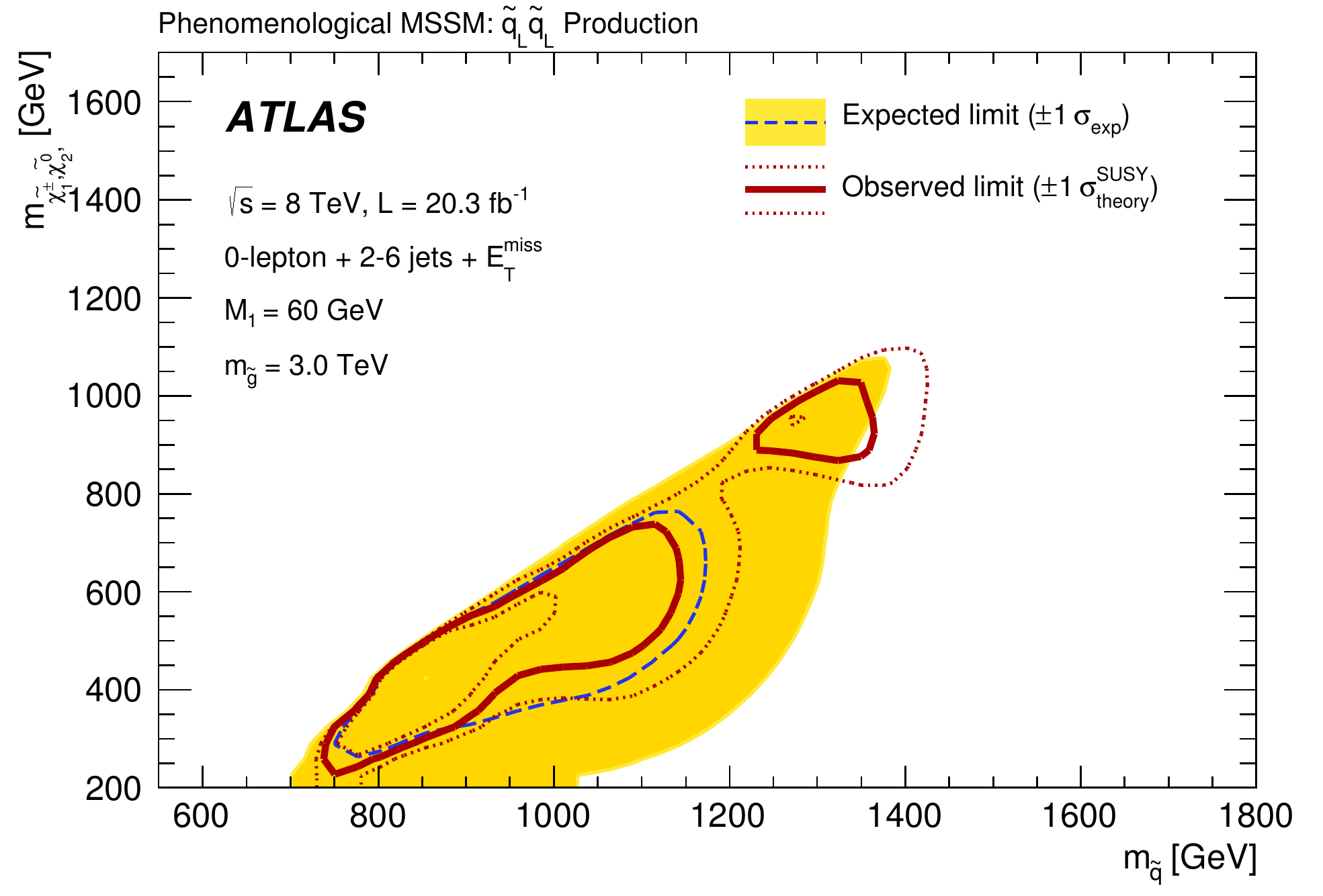}}
\subfigure[]{\includegraphics[width=0.48\textwidth]{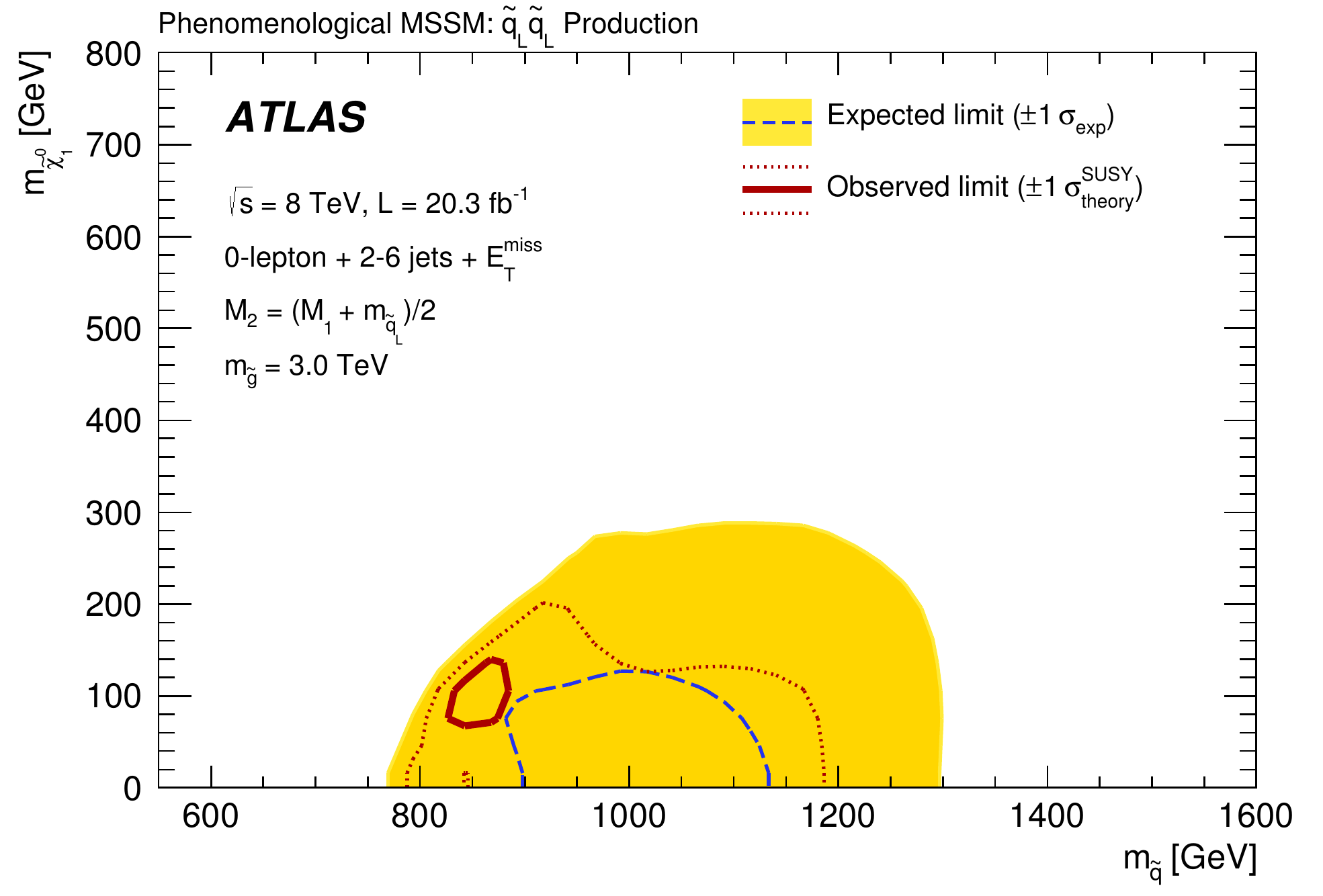}}
\caption{95\%~CL exclusion limits in the pMSSM considered in the search for left-handed squarks, with mass parameters $M_1$ and $M_2$, which are associated with the bino and wino masses, respectively. 
The limits are obtained from the 0L analysis with the additional 0L\_4jt+ and 0L\_5jt signal regions optimized specifically for this model. 
The parameter set considered here has either (a, c, e) $M_1 =60$~\GeV\  and $M_2$  varying or (b, d, f) $M_2=(M_1+m_{\tilde{q}_{\rm L}})/2$ and $M_1$ varying. For each $M_1$, $M_2$ combination three gluino masses are considered, $m_{\tilde{g}} =1600, 2200, 3000$~GeV. Gluino pair production is not included. 
 The solid red line and the dashed blue line show respectively the combined observed and expected 95\%~CL exclusion limits.
} \label{fig:limit-sqsq_pmssm}
\end{figure}

\subsubsection{Minimal Supergravity/Constrained MSSM and bilinear R-parity-violation models}

The exclusion limits in the ($m_{0}$, $m_{1/2}$) mSUGRA/CMSSM plane with $\tan\beta$ = 30, $A_0 = -2m_0$  and $\mu >$ 0
are shown in figure~\ref{fig:limit-msugra}. %
In the parameter space region with $m_0$ values smaller than about 1800~\GeV\ the best sensitivity is obtained with the (0+1)-lepton combination, which slightly improves the individual limit obtained by the 0L\_3jt signal region from the 0L search. 
For high $m_0$ values, final states with four top quarks dominate, and consequently the best sensitivity is provided by the  0/1L3B search. This search excludes gluino masses smaller than 1280~\GeV.

\begin{figure}[htbp]
\centering
\includegraphics[width=0.7\textwidth]{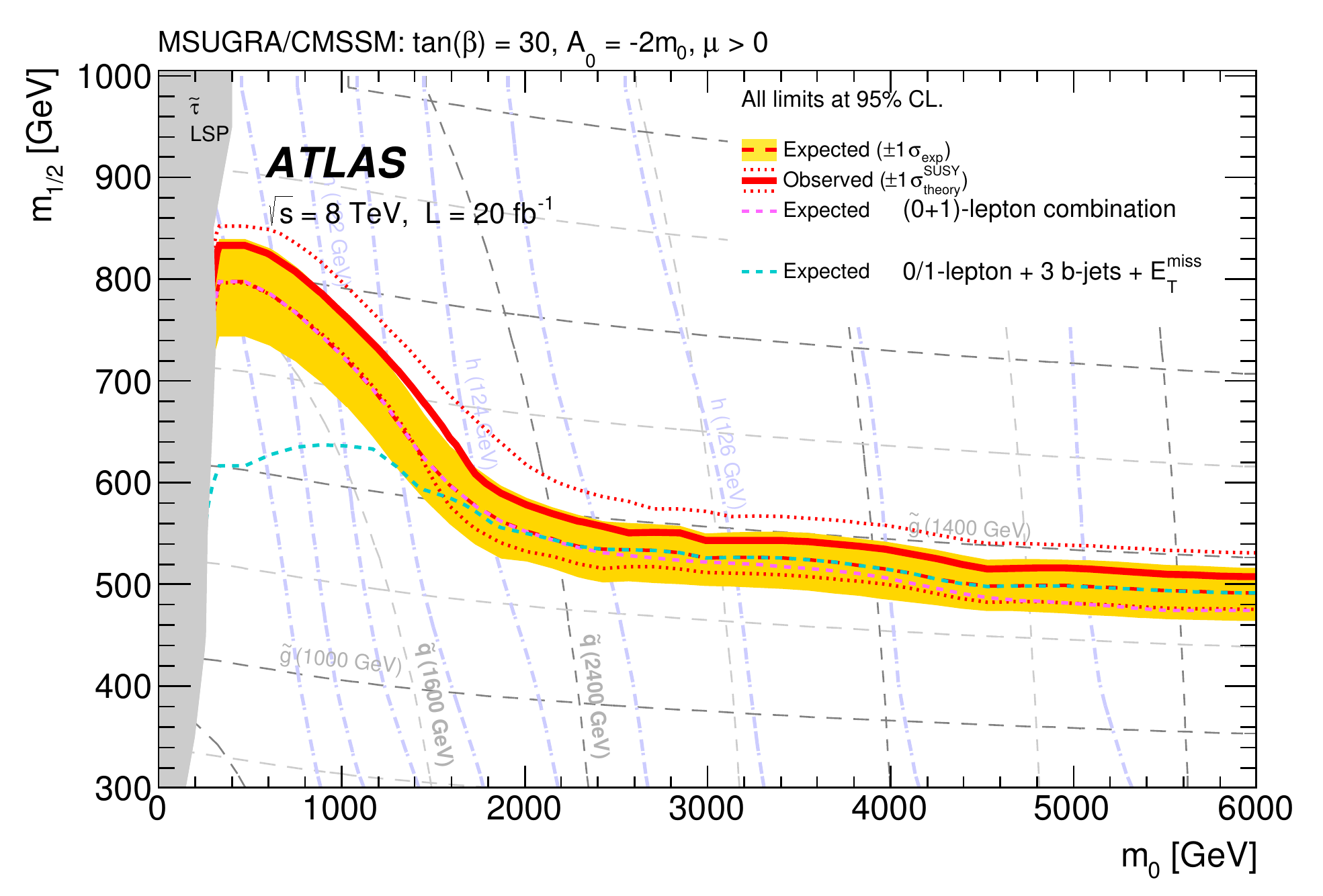}
\caption{Exclusion limits in the ($m_{0}$, $m_{1/2}$) plane for the mSUGRA/CMSSM model. %
 The solid red line and the dashed red line show respectively the combined observed and expected 95\%~CL exclusion limits. %
Expected limits from the individual analyses which contribute to the final combined limits are also shown for comparison.}
\label{fig:limit-msugra}
\end{figure}

The exclusion limits for the RPV model, which uses the same parameters as the mSUGRA/CMSSM but allows for bilinear R-parity-violating terms in the superpotential resulting in an unstable LSP, are shown in the ($m_{0}$, $m_{1/2}$) plane in figure~\ref{fig:limit-msugra-brpv}.
 The best sensitivity is provided by the TAU and the SS/3L searches. 
 For $m_0$ values smaller than approximately 750~\GeV\ the sensitivity is dominated by the TAU search which excludes $m_{1/2}$ values up to 680~\GeV\  using the combination of all final states considered in the search. 
At high $m_0$ values the best sensitivity is provided by the SS/3L\_SR3b signal region from the SS/3L search, which excludes values of $m_{1/2}$ between 200~\GeV\ and 490~\GeV.  For $m_0$ values below 2200~\GeV, signal models with $m_{1/2} < 200~\GeV$ are not considered because the lepton acceptance is significantly reduced due to the increased LSP lifetime in that region.

\begin{figure}[htbp]
\centering
\includegraphics[width=0.7\textwidth]{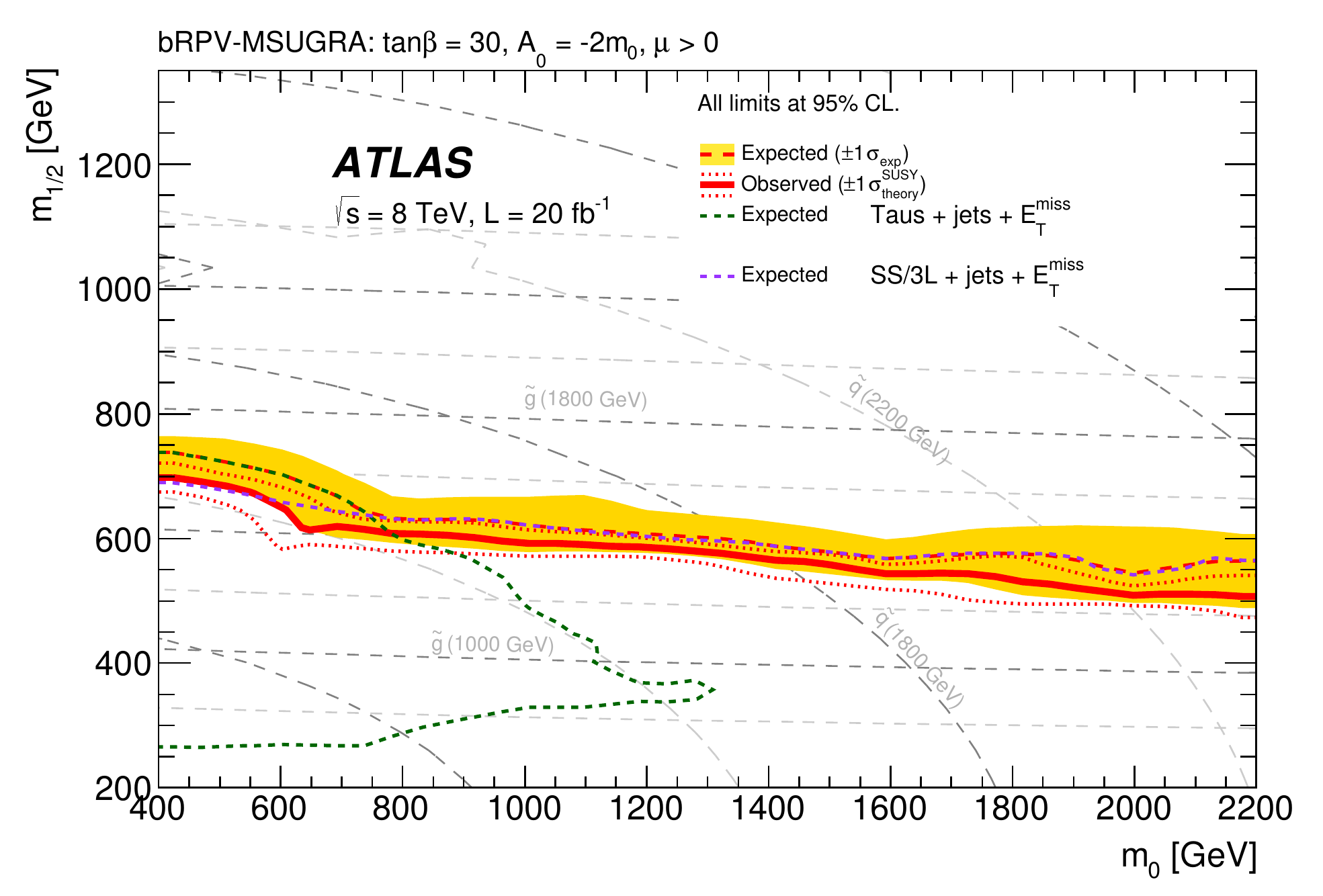}
\caption{Exclusion limits in the ($m_{0}$, $m_{1/2}$) plane for the bRPV model. The solid red line and the dashed red line show respectively the combined observed and expected 95\%~CL exclusion limits. %
Expected limits from the individual analyses which contribute to the final combined limits are also shown for comparison.} 
\label{fig:limit-msugra-brpv}
\end{figure}

\subsubsection{Minimal gauge-mediated supersymmetry breaking model}

The observed and expected limits for the mGMSB scenario are shown in figure~\ref{fig:limit-gmsb}, in the plane defined by the SUSY breaking scale $\Lambda$ and the $\tan\beta$ value. The region of small $\Lambda$ and large $\tan\beta$ just above the exclusion limit is excluded theoretically since it leads to tachyonic states. 
The SS/3L search provides the best sensitivity for this model and excludes 
values of $\Lambda$ up to about 75~\TeV. %

\begin{figure}[htbp]
\centering
\includegraphics[width=0.7\textwidth]{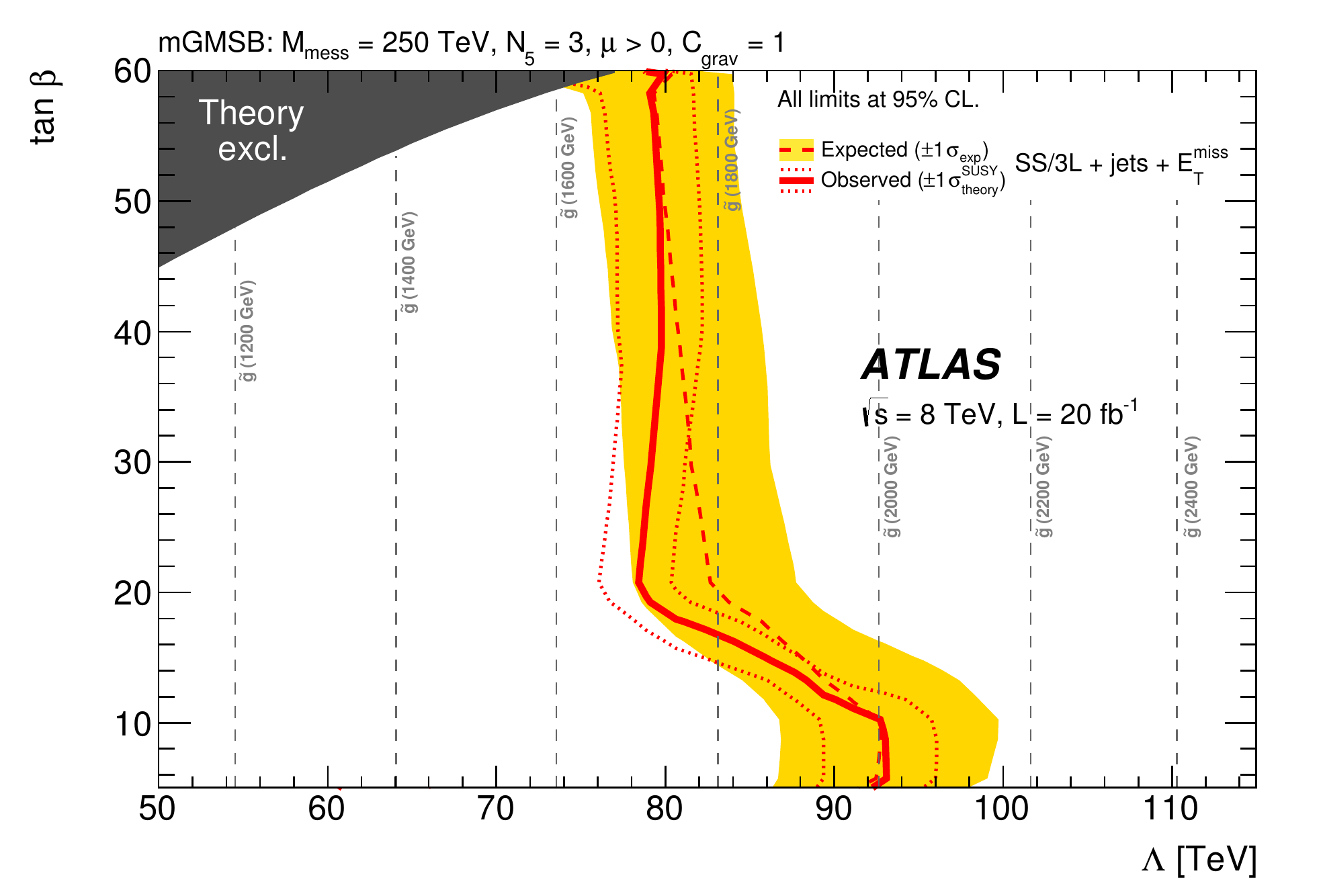}
\caption{Exclusion limits in the ($\Lambda$, $\tan\beta$) plane for the mGMSB model. The solid red line and the dashed red line show respectively the combined observed and expected 95\%~CL exclusion limits. %
The region of small $\Lambda$ and large $\tan\beta$ just above the exclusion limit is excluded theoretically since it leads to tachyonic states. } 
\label{fig:limit-gmsb}
\end{figure}

\subsubsection{Natural gauge mediation model}
The limits obtained for the nGM scenario %
are shown in figure~\ref{fig:limit-ngm} in the  ($m_{\stau}$, $m_{\gluino}$) plane. 
The best limits are obtained by the 1L(S,H) and TAU searches, resulting in an exclusion of gluino masses below approximately 1100~\GeV\ independent of the $\stau$ mass.

\begin{figure}[htbp]
\centering
\includegraphics[width=0.7\textwidth]{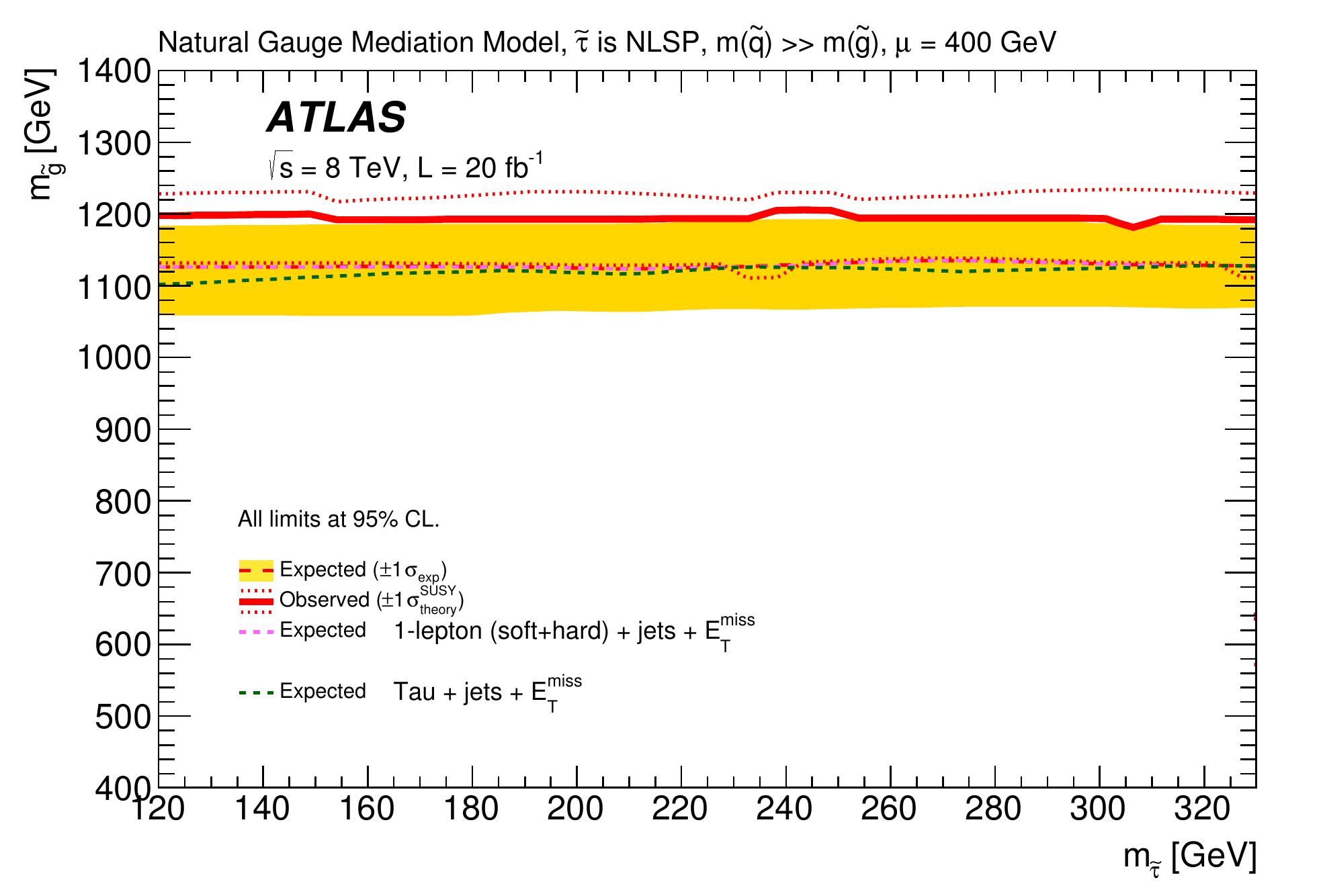}
\caption{Exclusion limits in the ($m_{\stau}$, $m_{\gluino}$) plane for the nGM model. The solid red line and the dashed red line show respectively the combined observed and expected 95\%~CL exclusion limits. %
Expected limits from the individual analyses which contribute to the final combined limits are also shown for comparison.} 
\label{fig:limit-ngm}
\end{figure}

\subsubsection{Non-universal Higgs mass model with gaugino mediation}

The exclusion limits in the context of a NUHMG model %
with parameters $m_0$ = 0, $\tan\beta$ = 10, $\mu >$ 0 and $m_{H_2}^{2}$ = 0 are shown in the ($m_{H_1}^{2}$, $m_{1/2}$) plane in figure~\ref{fig:limitnuhmg}. %
They are provided by the (0+1)-lepton combination.
A band in the ($m_{H_1}^{2}$, $m_{1/2}$) plane
can be excluded, extending up to the ranges $2000\times 10^3\GeV^2<m^2_{H_1} < 5400\times 10^3\GeV^2$ and 
$450\GeV< m_{1/2}<620\GeV$.

\begin{figure}[htbp]
\centering
\includegraphics[width=0.7\textwidth]{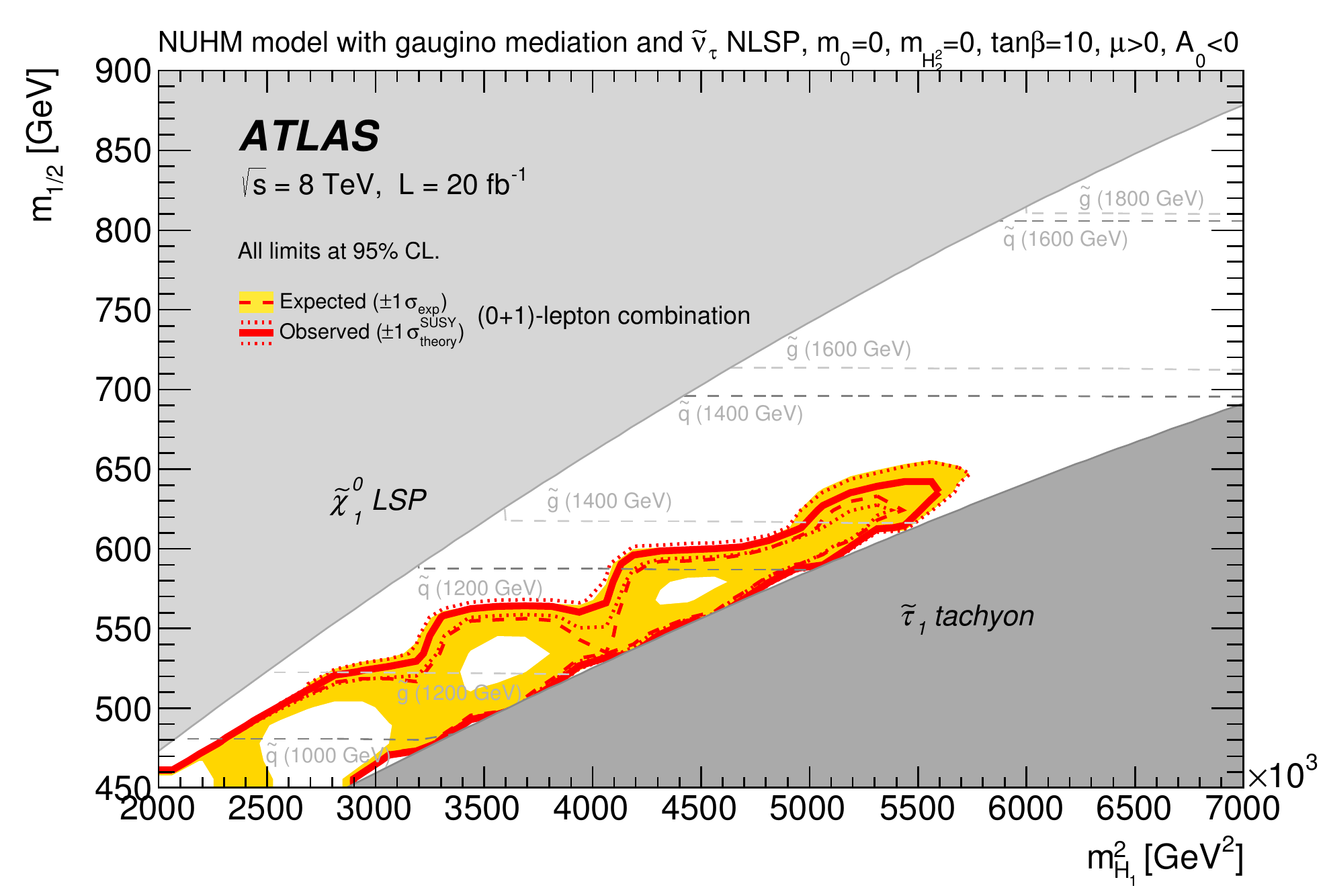}
\caption{Exclusion limits in the ($m_{H_1}^{2}$, $m_{1/2}$) plane for the NUHMG model. The solid red line and the dashed red line show respectively the combined observed and expected 95\%~CL exclusion limits. %
} 
\label{fig:limitnuhmg}
\end{figure}

\subsubsection{Minimal Universal Extra Dimension model}

Finally, the limits obtained for the mUED scenario %
are shown in figure~\ref{fig:limit-mued} in the ($1/R_{\rm c}$, $\Lambda R_{\rm c}$) plane. The 2L(S), 2LRaz and SS/3L searches provide competitive sensitivities for this model in which the mass spectrum is naturally degenerate and the decay chain of the Kaluza--Klein (KK) quark excitation to the lightest KK particle, the KK photon, gives a signature very similar to the supersymmetric decay chain of a squark via cascades involving top quarks, charginos, neutralinos or sleptons, which can subsequently produce many leptons. 
In the region where the cut-off scale times radius ($\Lambda R_{\rm c}$) is smaller than 13, the combined exclusion is dominated by the 2L(S) and 2LRaz searches which are combined based on the best expected $CL_{\rm S}$ value, while in the remaining regions of the parameter space the final exclusion is dominated by the SS/3L\_SR3Lhigh and SS/3L\_SR0b signal regions from the SS/3L search. Values of $1/R_{\rm c}$ below 850~\GeV\ are excluded. %

\begin{figure}[htbp]
\centering
\includegraphics[width=0.7\textwidth]{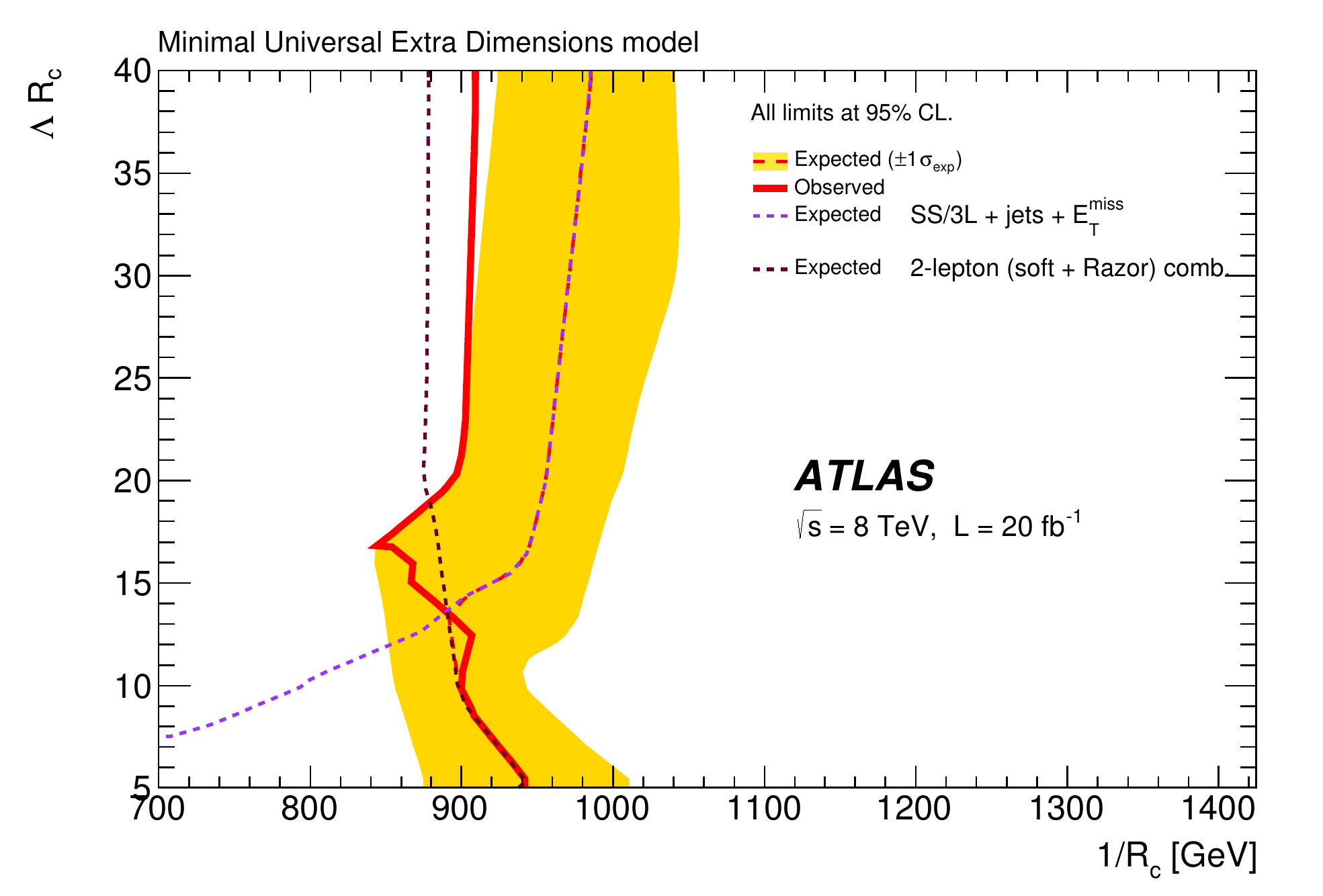}
\caption{Exclusion limits in the ($1/R_{\rm c}$, $\Lambda R_{\rm c}$) plane for the mUED model. The solid red line and the dashed red line show respectively the combined observed and expected 95\%~CL exclusion limits. %
Expected limits from the individual analyses which contribute to the final combined limits are also shown for comparison.} 
\label{fig:limit-mued}
\end{figure}

\subsection{Limits in Simplified Models}
\label{subsec:simplifiedlimits}

\subsubsection{Direct decays of squarks and gluinos}

This section summarizes the exclusion limits in simplified models with direct decays of gluinos and squarks of the first and second generation described in section~\ref{subsubsec:direct}. 
Here and in sections \ref{subsec:resultonestep} and \ref{subsec:resulttwostep}, unless otherwise stated, the eight light-flavoured squarks are always assumed to be mass-degenerate.

Figure~\ref{fig:limit-sqsq_direct} shows the exclusion limits in simplified models with squark-pair production and subsequent direct squark decays to a quark and the lightest neutralino.  
The expected limits from the three most sensitive searches (0L, MONOJ and 0LRaz) are presented individually along with the combined expected and observed exclusion limits.
The 0L and 0LRaz analyses yield in general higher expected mass limits, but the MONOJ search provides the best sensitivity close to the diagonal line, in the region of parameter space where the mass difference between the squark and the lightest neutralino is small. 
From the observed limits, neutralino masses below about 280~\GeV\ can be excluded for squark masses up to 
800~\GeV, and for a neutralino mass of 100~\GeV\ squark masses are excluded below 850~\GeV. In a scenario with only one light-flavour squark  produced, which affects only the cross-section but not the kinematics of the events, a lower limit on the squark mass of 440 \GeV\ is obtained with the 0L search \cite{0-leptonPaper}. 

\begin{figure}[htbp]
\centering
\includegraphics[width=0.7\textwidth]{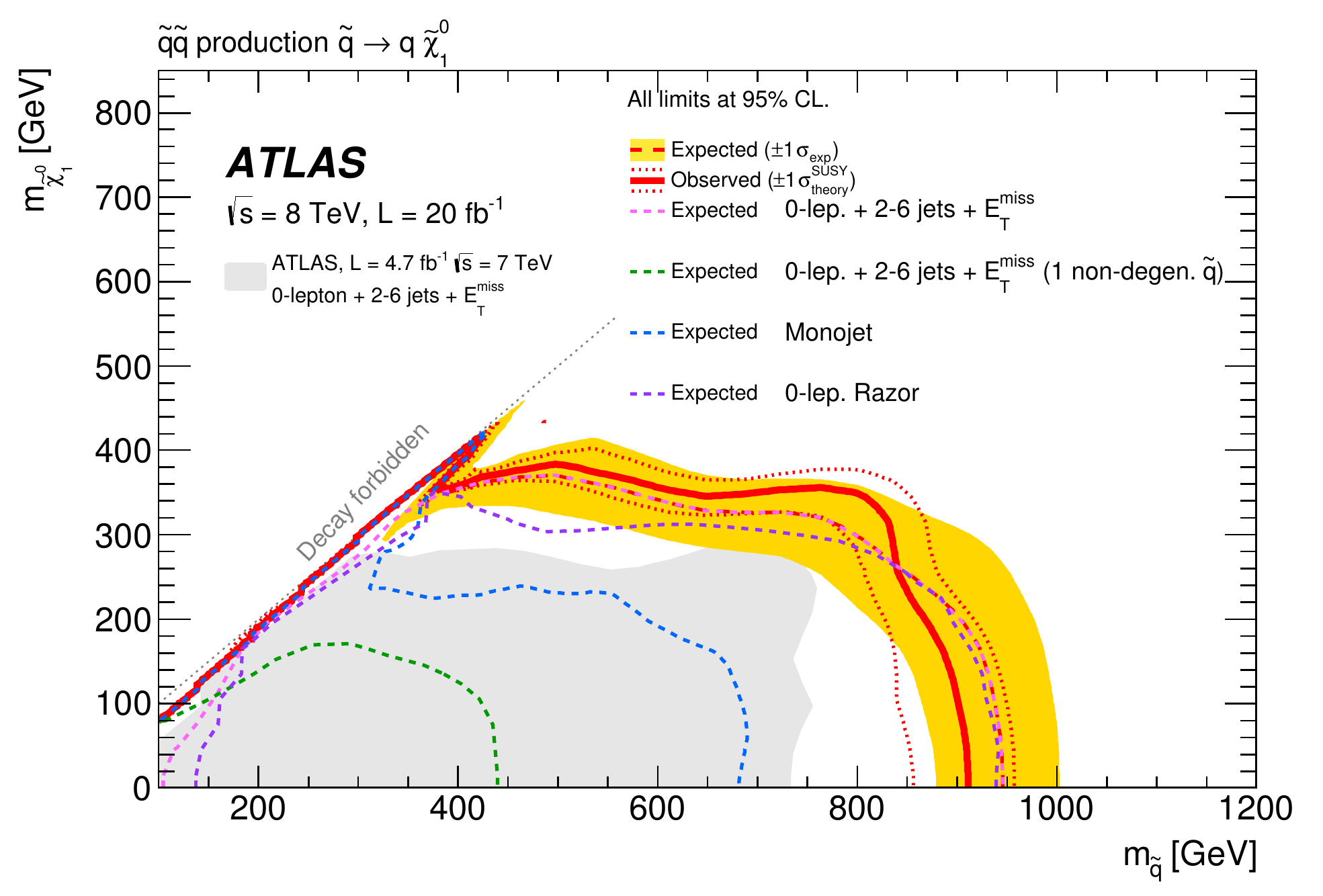}
\caption{Exclusion limits in the mass plane of the lightest neutralino and first- and second-generation squarks  assuming squark-pair production and direct decays $\squark \to q \ninoone$. The solid red line and the dashed red line show respectively the combined observed and expected 95\%~CL exclusion limits. %
Expected limits from the individual analyses which contribute to the final combined limits are also shown for comparison. A previous result from ATLAS ~\cite{Aad:2012fqa} using 7 \TeV\ proton--proton collisions is represented by the shaded (grey) area. A limit obtained with the 0L search  \cite{0-leptonPaper} in a scenario with only one light-flavour squark produced is also presented for completeness.} 
\label{fig:limit-sqsq_direct}
\end{figure}

\begin{figure}[htbp]
\centering
\includegraphics[width=0.7\textwidth]{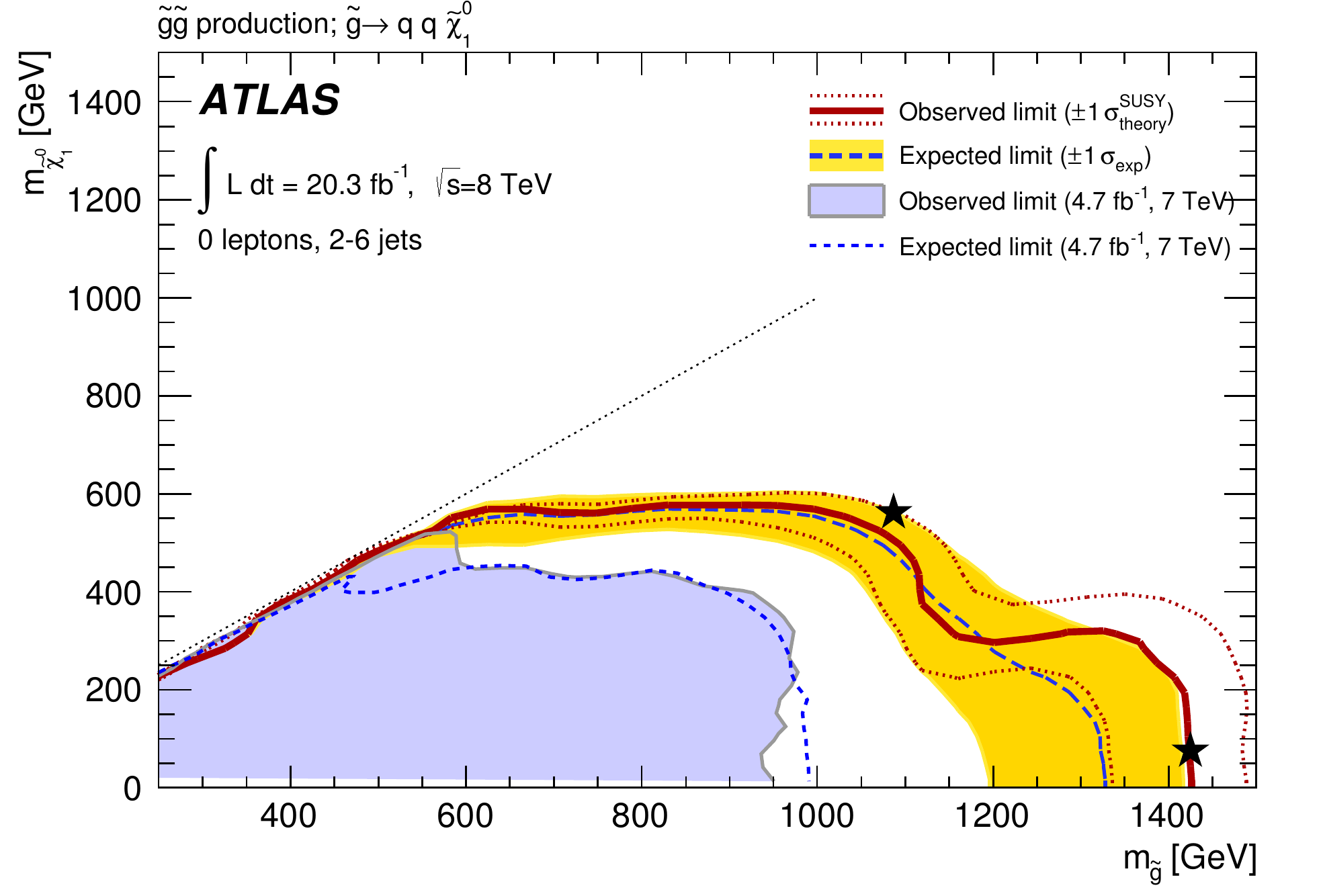}
\caption{Exclusion limits in the mass plane of the lightest neutralino and the gluino for gluino-pair production with direct decay $\tilde{g}\rightarrow qq\ninoone$ taken from  the 0L search \cite{0-leptonPaper}. The solid red line and the dashed blue line show respectively the observed and expected 95\%~CL exclusion limits.  The star symbols indicates two benchmark models which are investigated in more detail in the publication. A previous result from ATLAS ~\cite{Aad:2012fqa} using 7 \TeV\ proton--proton collisions is represented by the shaded (light blue) area. 
} \label{fig:limit-glgl_direct}
\end{figure}

Another example of a direct decay is shown in figure~\ref{fig:limit-glgl_direct}, taken from ref.~\cite{0-leptonPaper}, where gluino-pair production with the subsequent decay 
$\tilde{g}\rightarrow qq\ninoone$ is considered. %
Due to the higher production cross-sections compared to the squark-pair production, higher mass limits can be obtained. For gluino masses up to about 1000~\GeV, neutralino masses can be excluded below about 500~\GeV\ or close to the kinematic limit near the diagonal. For small neutralino masses the observed limit is as large as 1330~\GeV.

\begin{figure}[htbp]
\centering
\subfigure[]{\includegraphics[width=0.7\textwidth]{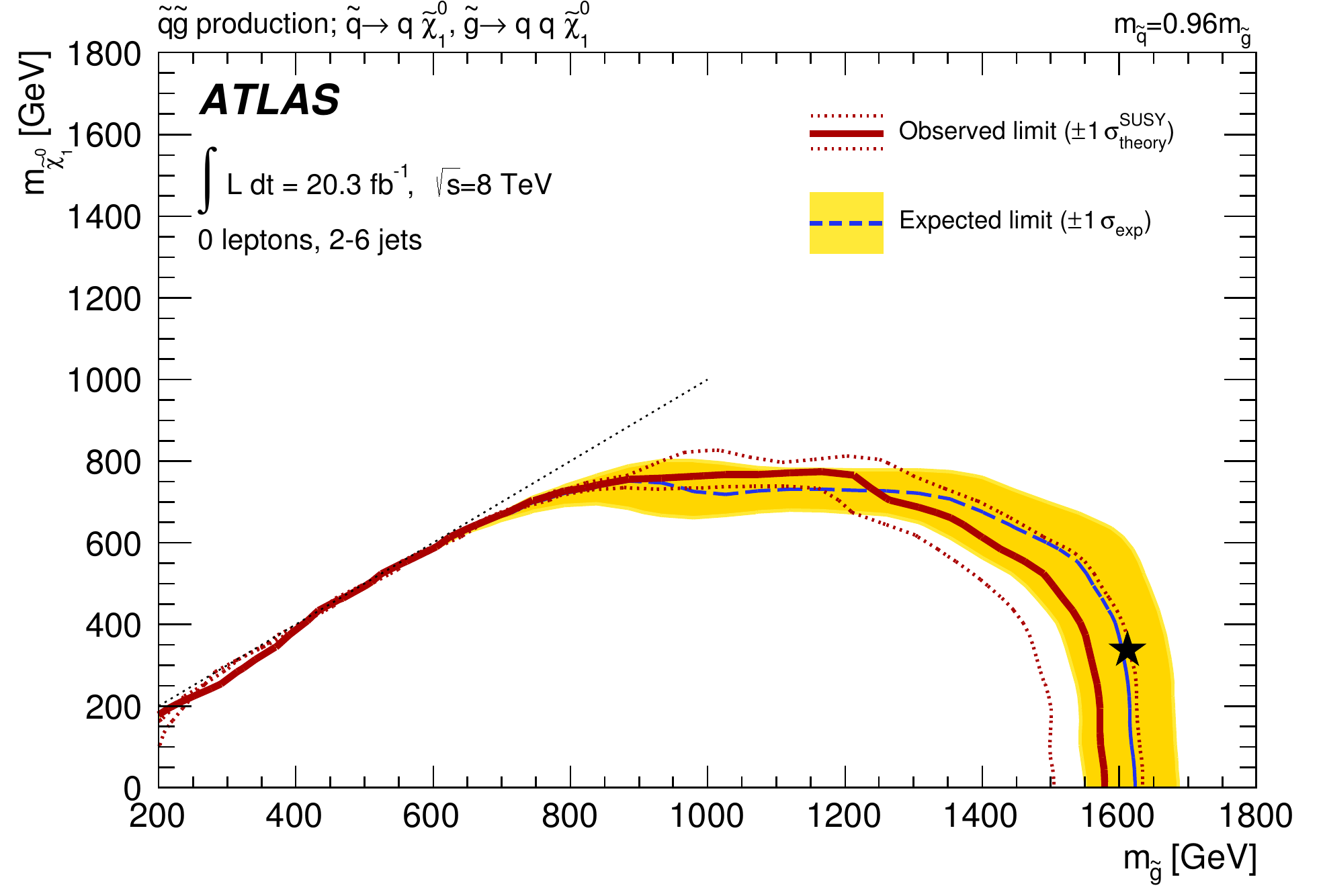}}\\
\subfigure[]{\includegraphics[width=0.7\textwidth]{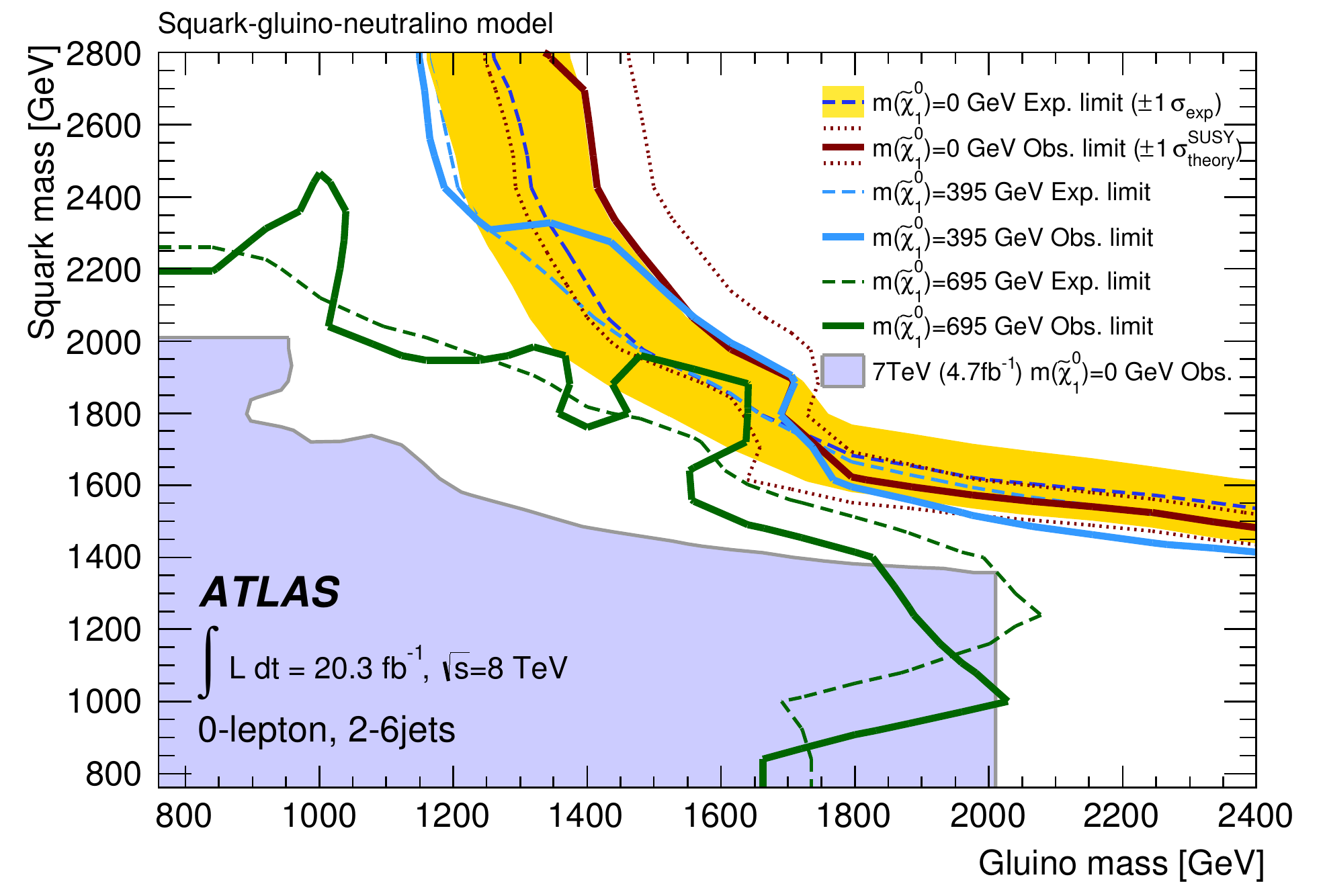}}
\caption{Exclusion limits on the production of a first- or second-generation squark and a gluino with direct decays of both particles, taken from the 0L search \cite{0-leptonPaper}. The solid red line and the dashed blue line show respectively the observed and expected 95\%~CL exclusion limits. The squark mass is fixed at
$0.96~m_{\tilde{g}}$ in (a) and allowed to vary freely in (b) with three assumptions on the neutralino mass of 0, 
395~\GeV\ or 695~\GeV.   %
 The black star indicates a benchmark model as discussed in ref. \cite{0-leptonPaper}.   
Previous results for a massless neutralino from ATLAS at 7~\TeV\   \cite{Aad:2012fqa} are represented by the shaded (light blue) area. The 7~\TeV\ results are valid for squark or gluino masses below 2000~\GeV, the mass range studied for that analysis.
} \label{fig:limit-sqgl_direct}
\end{figure}

A simplified model of $\tilde{q} \tilde{g}$ strong production with the direct decays of %
squarks $\tilde{q}\rightarrow q\ninoone$ and gluinos $\tilde{g}\rightarrow qq\ninoone$ is considered in figure~\ref{fig:limit-sqgl_direct}, taken from ref.~\cite{0-leptonPaper}, for the 0L analysis. 
The squark mass is fixed at $0.96 m_{\tilde{g}}$ in figure~\ref{fig:limit-sqgl_direct}(a), and gluinos can decay via on-shell squarks as $\gluino \to \squark q \to q q \ninoone $.  The exclusion limit for the neutralino mass is very close to the kinematic limit near the diagonal line and reaches 700~\GeV\ for gluino masses up to 1200~\GeV. For a massless neutralino, gluino masses below 1500~\GeV\ are excluded. 

Figure~\ref{fig:limit-sqgl_direct}(b) expresses the mass limits in the ($m_{\tilde{g}}, m_{\tilde{q}}$) plane in the model with combined production of squark pairs, gluino pairs, and of squark--gluino pairs, for different assumptions on the neutralino mass: $m_{\ninoone}=$ 0~\GeV, 395~\GeV\ or 695~\GeV. Depending on the mass hierarchy, the $\gluino \to \squark q$ and $\squark \to \gluino q $ one-step decays are taken into account. 
The masses of all other supersymmetric particles are set outside the kinematic reach.   A lower limit of 1650~\GeV\ for equal squark and gluino mass is found for the scenario with a massless $\ninoone$.

Figure~\ref{fig:limit-gluino_gluon} shows the cross-section times branching ratio limits for gluino-pair production with  direct gluino decays to a gluon and the lightest neutralino based on the 0L search. For a massless neutralino (figure~\ref{fig:limit-gluino_gluon}(a)), gluino masses below 1250~\GeV\ can be excluded.  The result can also be used to obtain lower mass limits on $\ninoone$, e.g. 550~\GeV\ for a gluino mass of 850~\GeV\ (figure~\ref{fig:limit-gluino_gluon}(b)). 
The cross-section exclusion for the $\gluino \rightarrow g \ninoone$ model is very similar to that for the $\squark \rightarrow q \ninoone$ as would be expected if there is not much difference between quark- and gluon-initiated jets.

\begin{figure}[htbp]
\centering
\subfigure[]{\includegraphics[width=0.7\textwidth]{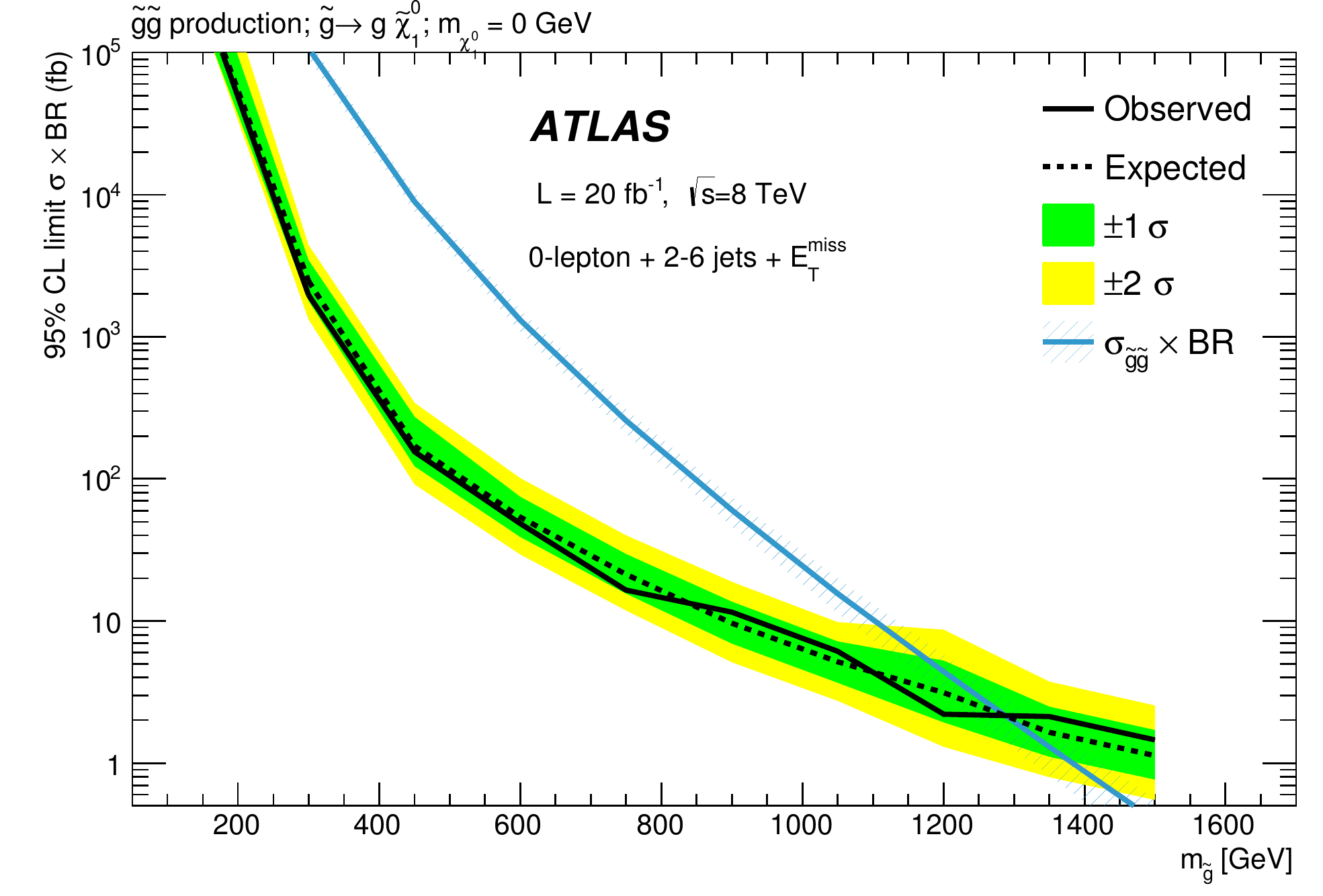}}\\
\subfigure[]{\includegraphics[width=0.7\textwidth]{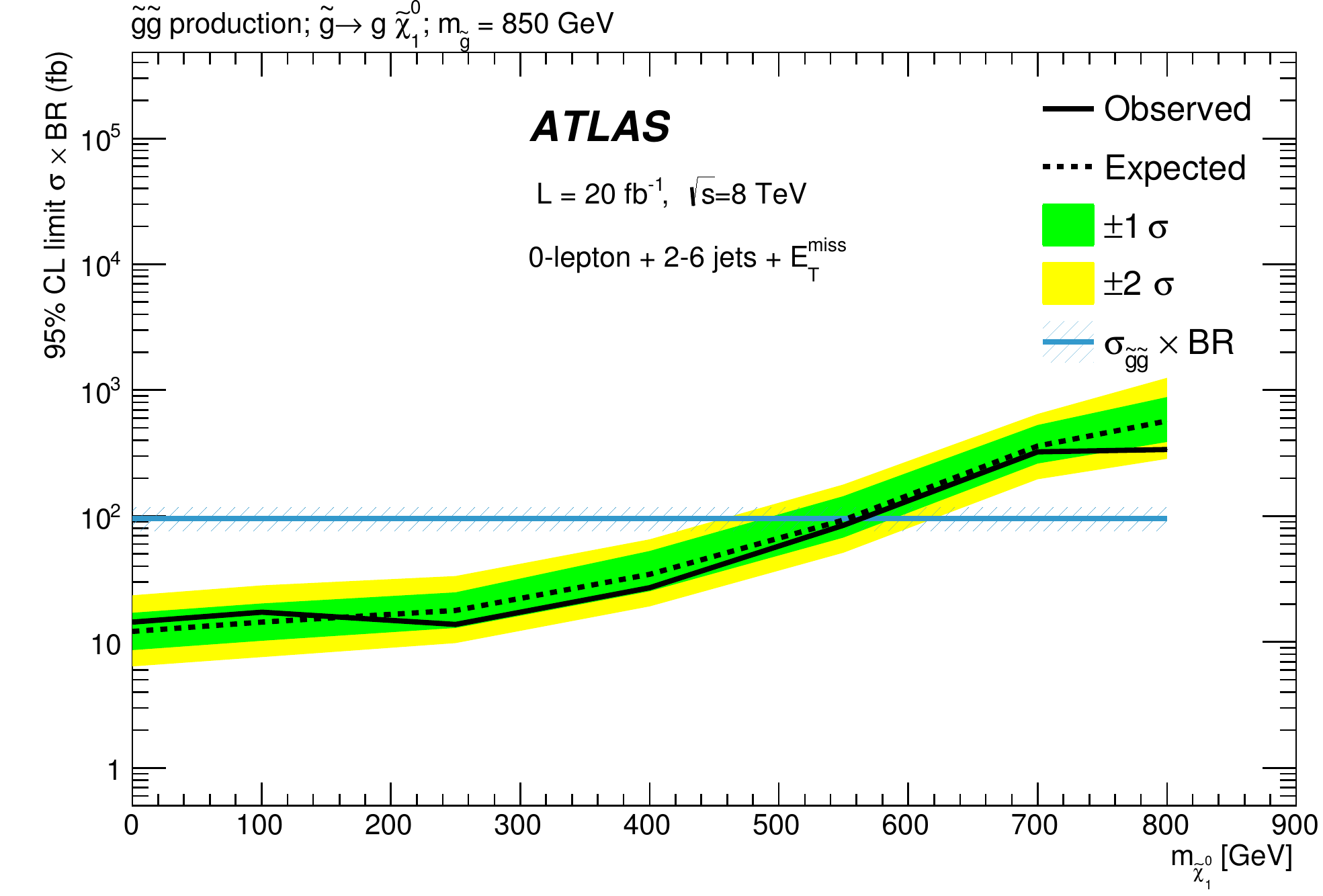}}
\caption{Limits at 95\%~CL on the production cross-section times branching ratio for gluino-pair production with direct decay of gluino to gluon and the lightest neutralino for (a) a massless neutralino as a function of the gluino mass, and (b) as a function of the $\ninoone$ mass for  a fixed gluino mass of $m_{\tilde{g}}= 850$~GeV. %
The solid black line shows the observed limit and the dashed line the expected limit. The solid medium dark (blue) line indicates the theoretical cross-section times branching ratio. The hatched (blue) bands around the theoretical $\sigma \cdot$BR curves denote the scale and PDF uncertainties.
} \label{fig:limit-gluino_gluon}
\end{figure}

\subsubsection{One-step decays of squarks and gluinos} 
\label{subsec:resultonestep}

This section presents the limits in simplified models with one-step decays of squarks and gluinos described in section \ref{subsubsec:onestep}.

Figure~\ref{fig:limit-sqsq_onestep} shows the exclusion limits for squark-pair production where the squark decays via an intermediate chargino (one step) to a quark, $W$ boson and neutralino.  For the model presented in figure~\ref{fig:limit-sqsq_onestep}(a) the chargino mass is fixed at $m_{\chinoonepm}=(m_{\tilde{q}}+m_{\ninoone})/2$ and the result is shown in the ($m_{\squark}, m_{\ninoone}$) plane. The best sensitivity is obtained by the (0+1)-lepton combination.
Neutralino masses up to 370~\GeV\ are excluded.
For a neutralino mass of 100~\GeV, squark masses are excluded below 790~\GeV.  Figure~\ref{fig:limit-sqsq_onestep}(b) shows the exclusion limits in the ($m_{\squark}, x$) plane, where $x$ is defined as $x=\Delta m(\chinoonepm, \ninoone)/ \Delta m ({\tilde{q}}, \ninoone)$, in models in which the neutralino mass is fixed at 60~GeV. Squark masses are excluded up to 830~\GeV\ for the most favourable $x$ values. The 1L(S,H)  search yields stronger limits than the 0L analysis for most of the parameter space of both types of models. 

\begin{figure}[htbp]
\centering
\subfigure[]{\includegraphics[width=0.7\textwidth]{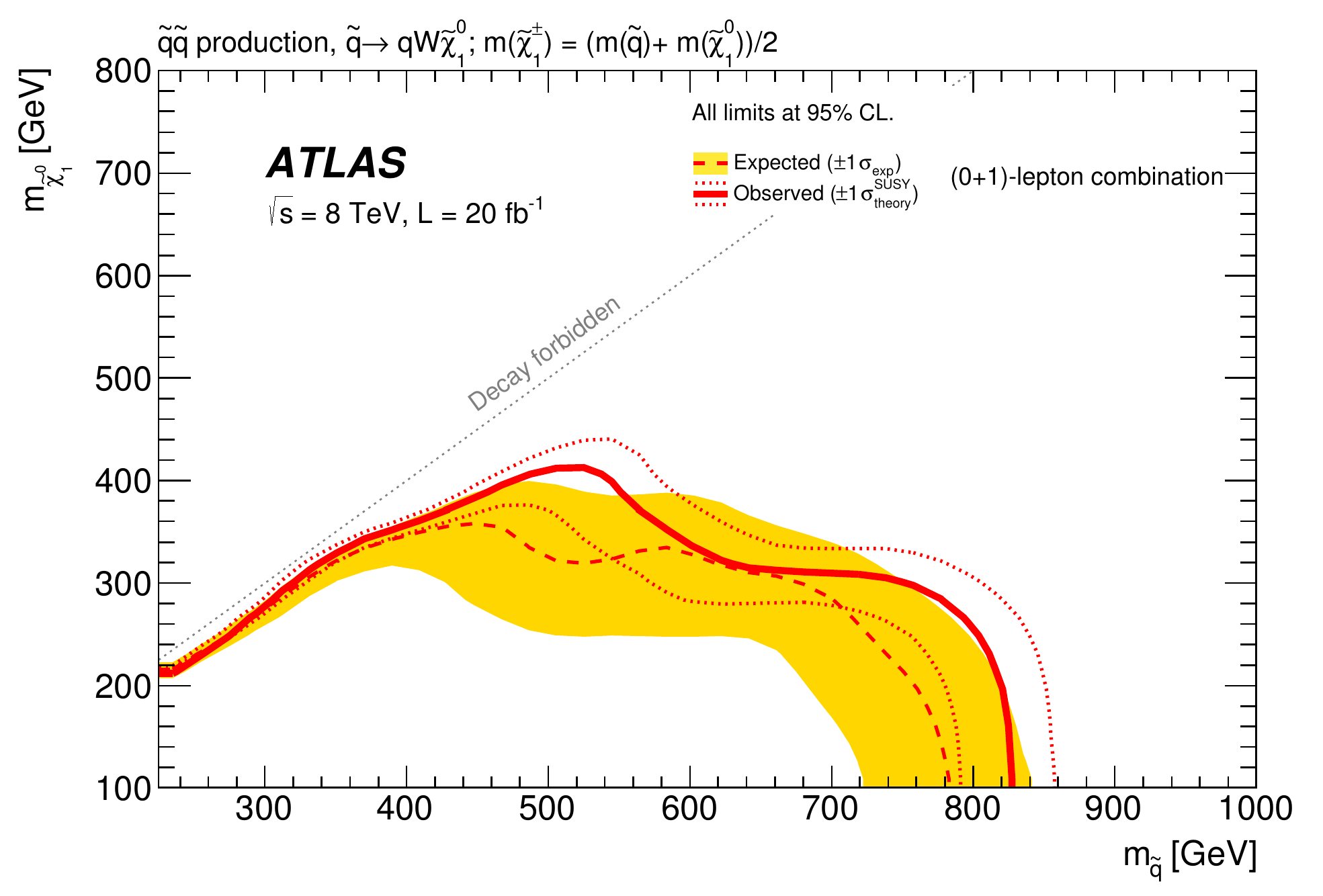}}\\
\subfigure[]{\includegraphics[width=0.7\textwidth]{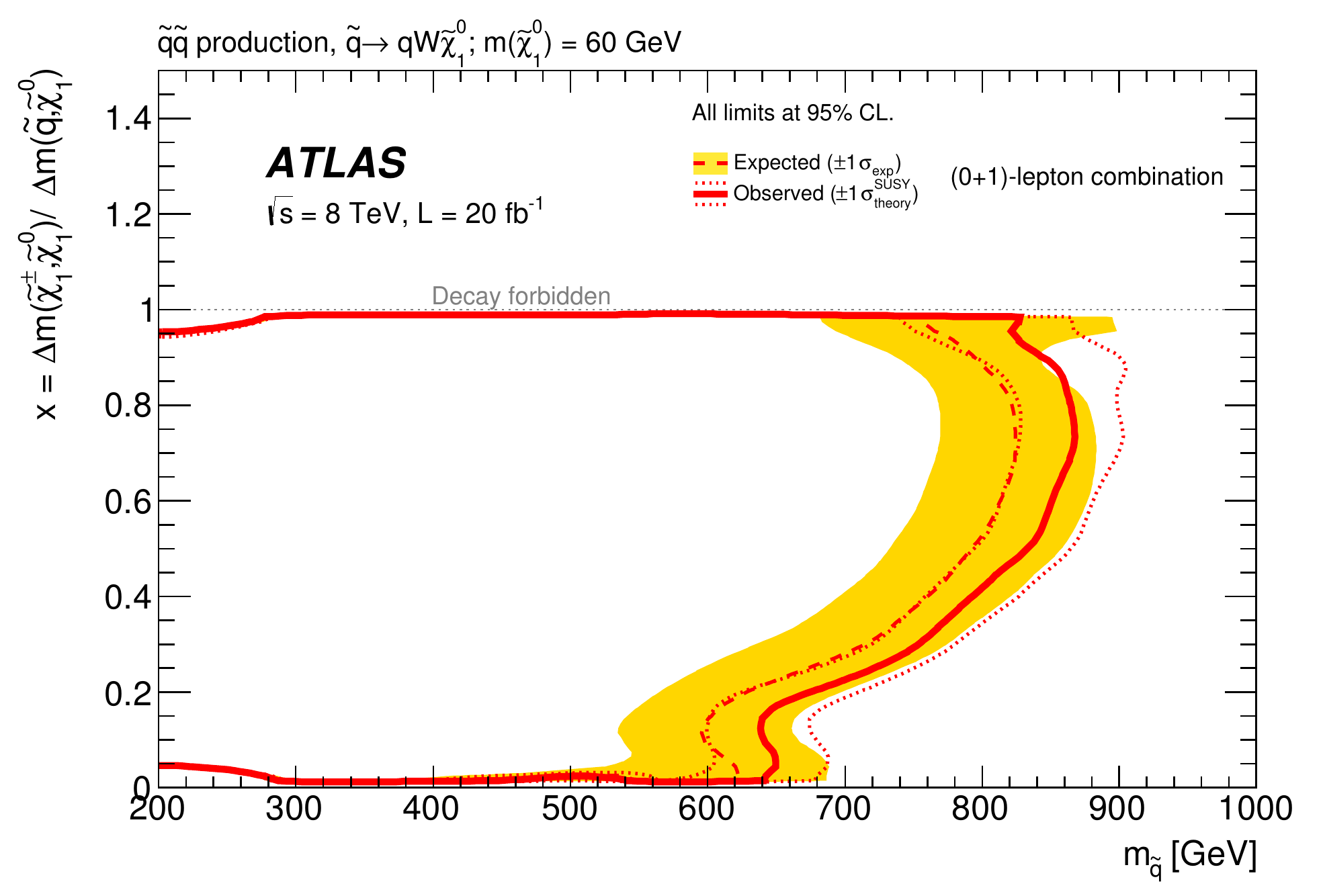}}
\caption{Exclusion limits for squark-pair production with a one-step decay via an intermediate chargino into $qW\ninoone$. 
Figure (a)  shows the limits in the ($m_{\squark}, m_{\ninoone}$) plane for a chargino mass fixed
 at $m_{\chinoonepm}=(m_{\tilde{q}}+m_{\ninoone})/2$.  Alternatively (b), the neutralino mass is fixed at 60~GeV and exclusion limits are given for $x=\Delta m(\chinoonepm, \ninoone)/ \Delta m ({\tilde{q}}, \ninoone)$ as function of the squark mass. The solid red line and the dashed red line show respectively the combined observed and expected 95\%~CL exclusion limits.} %
\label{fig:limit-sqsq_onestep}
\end{figure}

The results of the searches for gluino-pair production with a one-step decay via an intermediate chargino into $qqW\ninoone$ are shown in 
figure~\ref{fig:limit-glgl_onestep}. 
Figure~\ref{fig:limit-glgl_onestep}(a) shows the limit for a chargino mass fixed at $m_{\chinoonepm}=(m_{\tilde{g}}+m_{\ninoone})/2$, where the (0+1)-lepton combination provides the best sensitivity.
For a neutralino mass of 100~GeV, gluino masses below 1270~GeV are excluded.  Neutralino masses are excluded below 480~\GeV\ for gluino masses up to 1200~GeV.
Fixing the neutralino mass at 60~GeV (figure~\ref{fig:limit-glgl_onestep}(b)), one obtains limits on the variable $x=\Delta m(\chinoonepm, \ninoone)/ \Delta m ({\tilde{g}}, \ninoone)$. Nearly the whole range $0<x<1$ is excluded for gluino masses below 1100~GeV. 

\begin{figure}[htbp]
\centering
\subfigure[]{\includegraphics[width=0.7\textwidth]{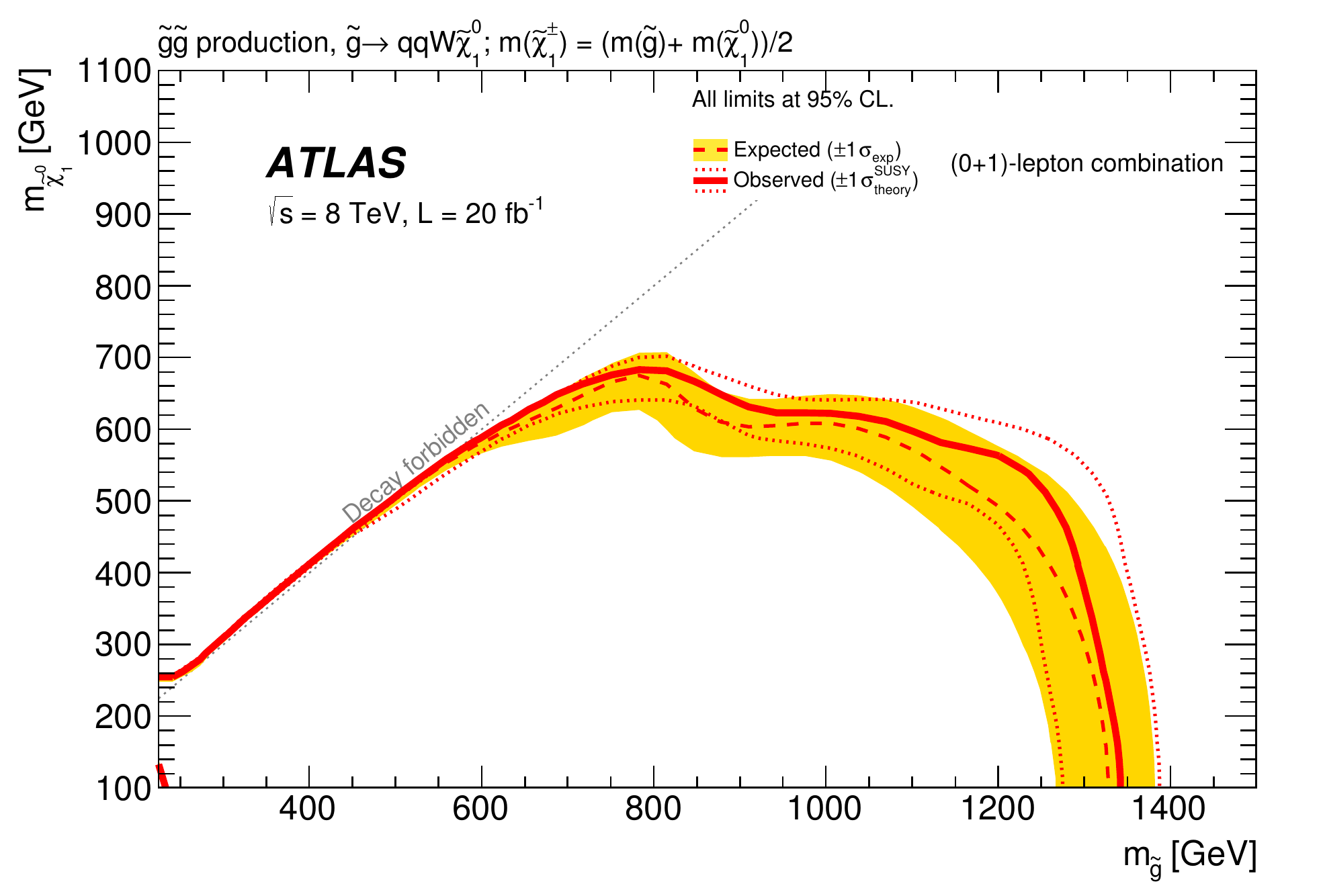}}\\
\subfigure[]{\includegraphics[width=0.7\textwidth]{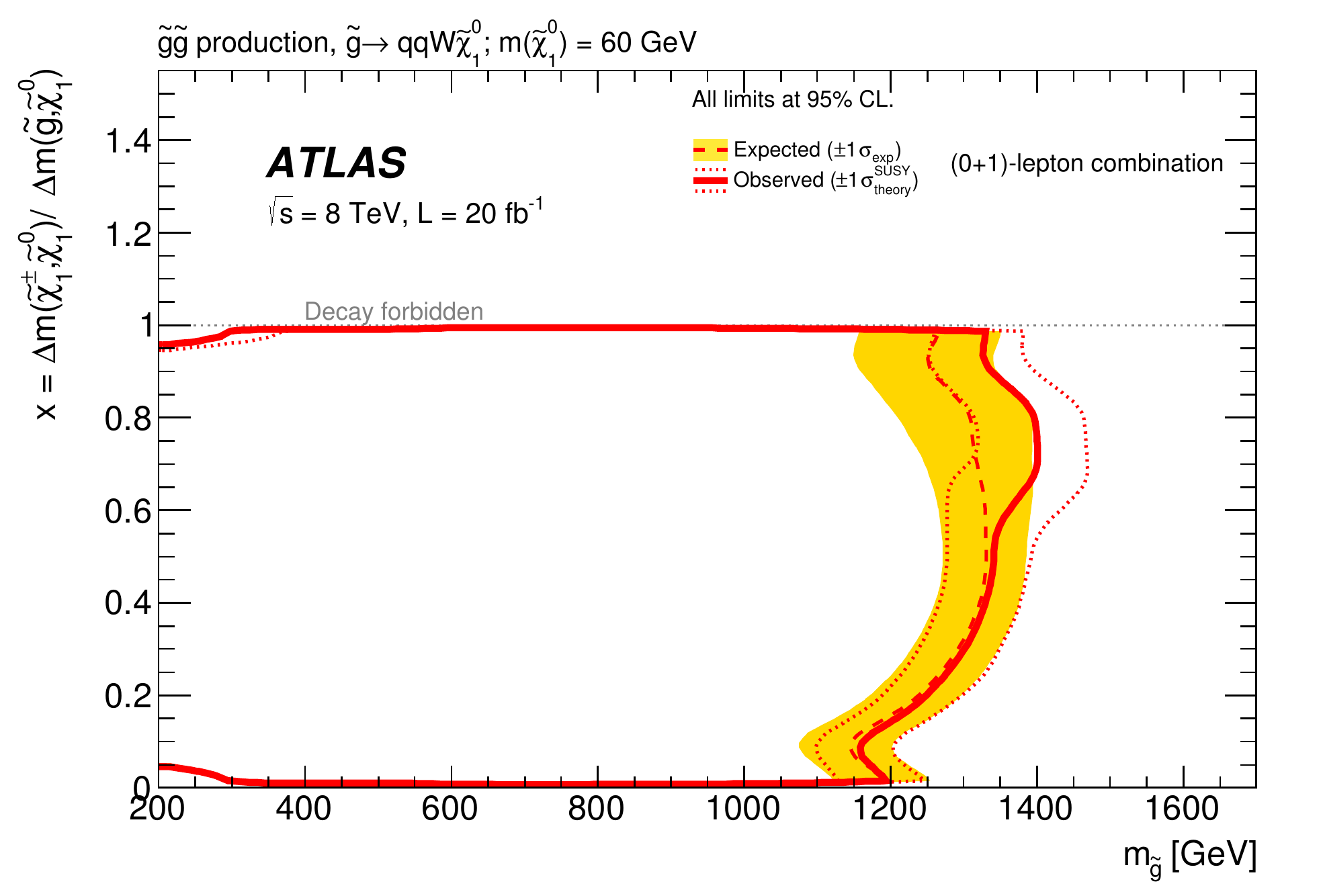}}
\caption{Exclusion limits for gluino-pair production with a one-step decay via an intermediate chargino into $qqW\ninoone$. 
Figure (a)  shows the limits in the ($m_{\gluino}, m_{\ninoone}$) plane for a chargino mass fixed
 at $m_{\chinoonepm}=(m_{\tilde{g}}+m_{\ninoone})/2$.  Alternatively (b), the neutralino mass is fixed at 60~GeV and exclusion limits are given for $x=\Delta m(\chinoonepm, \ninoone)/ \Delta m ({\tilde{g}}, \ninoone)$ as function of the gluino mass. The solid red line and the dashed red line show respectively the combined observed and expected 95\%~CL exclusion limits.} %
 \label{fig:limit-glgl_onestep}
\end{figure}

\subsubsection{Two-step decays of squarks and gluinos} 
\label{subsec:resulttwostep}

This section presents the limits in simplified models with two-step decays of squarks and gluinos described in section \ref{subsubsec:twostep}. 

Exclusion limits for squark-pair production with a subsequent two-step squark decay via a chargino and neutralino to $qWZ\ninoone$ are shown in figure~\ref{fig:limit-sqsq_twostepWWZZ}. Results are obtained with two searches, the 0L  and the SS/3L searches. The 0L  search is mainly sensitive in the low-mass region of 240~GeV $< m_{\tilde{q}}<$ 300~GeV, whereas the SS/3L  search is most sensitive for squark masses between  450 and 650~GeV, where $\ninoone$ masses below 250~GeV are excluded.

\begin{figure}[htbp]
\centering
\includegraphics[width=0.7\textwidth]{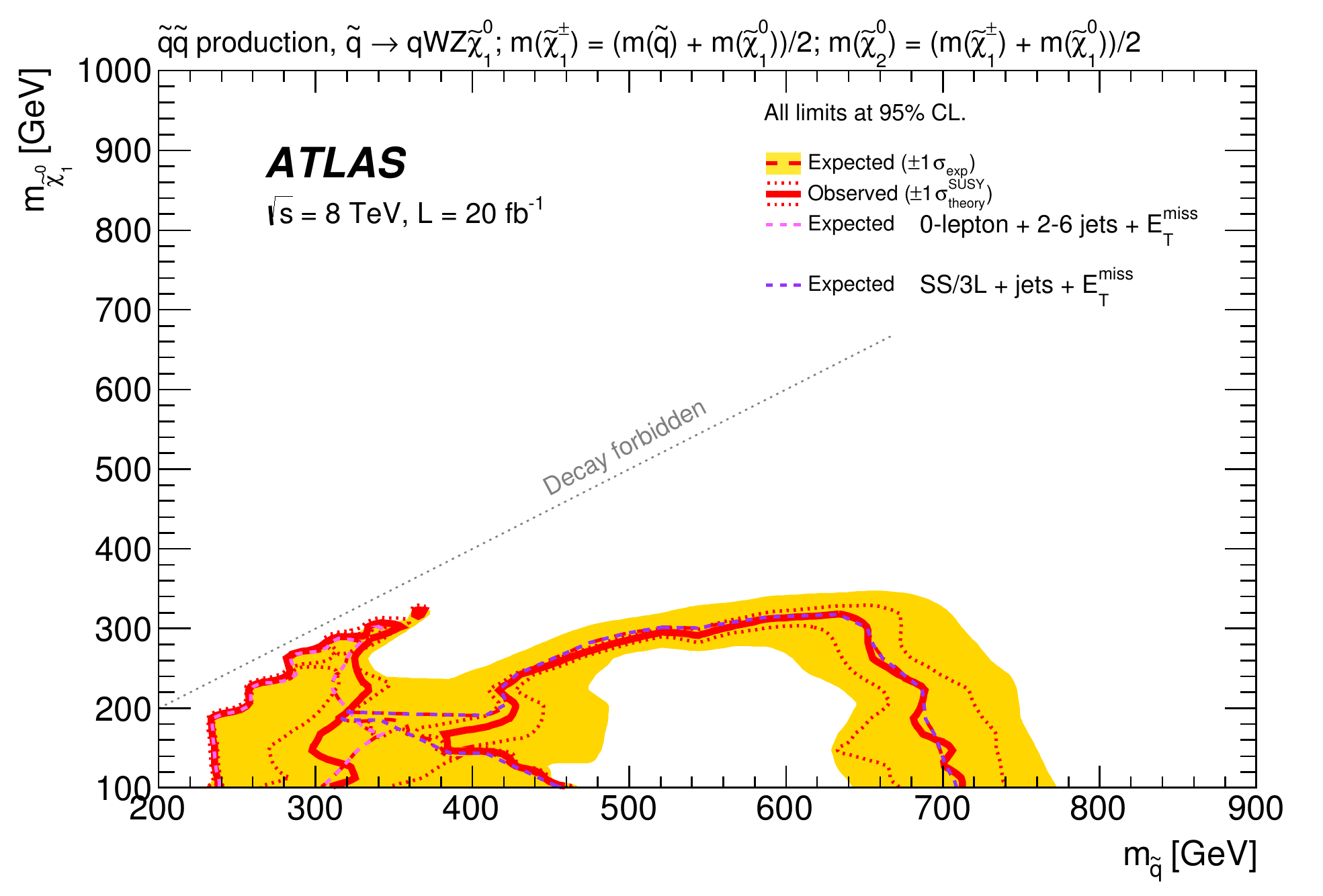}
\caption{Exclusion limits in the ($m_{\squark},  m_{\ninoone}$) plane for a simplified model of first- and second-generation squark-pair production with two-step decay into $qqWWZZ\ninoone\ninoone$ and missing transverse momentum. The solid red line and the dashed red line show respectively the combined observed and expected 95\%~CL exclusion limits. %
Expected limits from the individual analyses which contribute to the final combined limits are also shown for comparison.
}
\label{fig:limit-sqsq_twostepWWZZ}
\end{figure}

Exclusion limits in a simplified model of gluino-pair production with a subsequent two-step gluino decay via a chargino and neutralino to $qqWZ\ninoone$ are shown in figure~\ref{fig:limit-glgl_twostepWWZZ}. The results are obtained with the  (0+1)-lepton combination, %
MULTJ,  and the SS/3L  searches. 
The (0+1)-lepton combination provides the highest $\ninoone$ mass limits at low gluino masses. For the intermediate range around $m_{\tilde{g}}\approx 900$~GeV the SS/3L  search is most sensitive, while for high gluino masses the best limits are obtained by the  MULTJ analysis.
For gluino masses below 500~GeV, $\ninoone$ masses are excluded up to the kinematic limit indicated by the diagonal line, and in the range 500~GeV $< m_{\tilde{g}}<$ 1000~GeV lower limits on $\ninoone$ masses are set around 400~GeV. For $m_{\ninoone}= 100$~GeV, gluino masses are excluded below 1150~GeV.

\begin{figure}[htbp]
\centering
\includegraphics[width=0.7\textwidth]{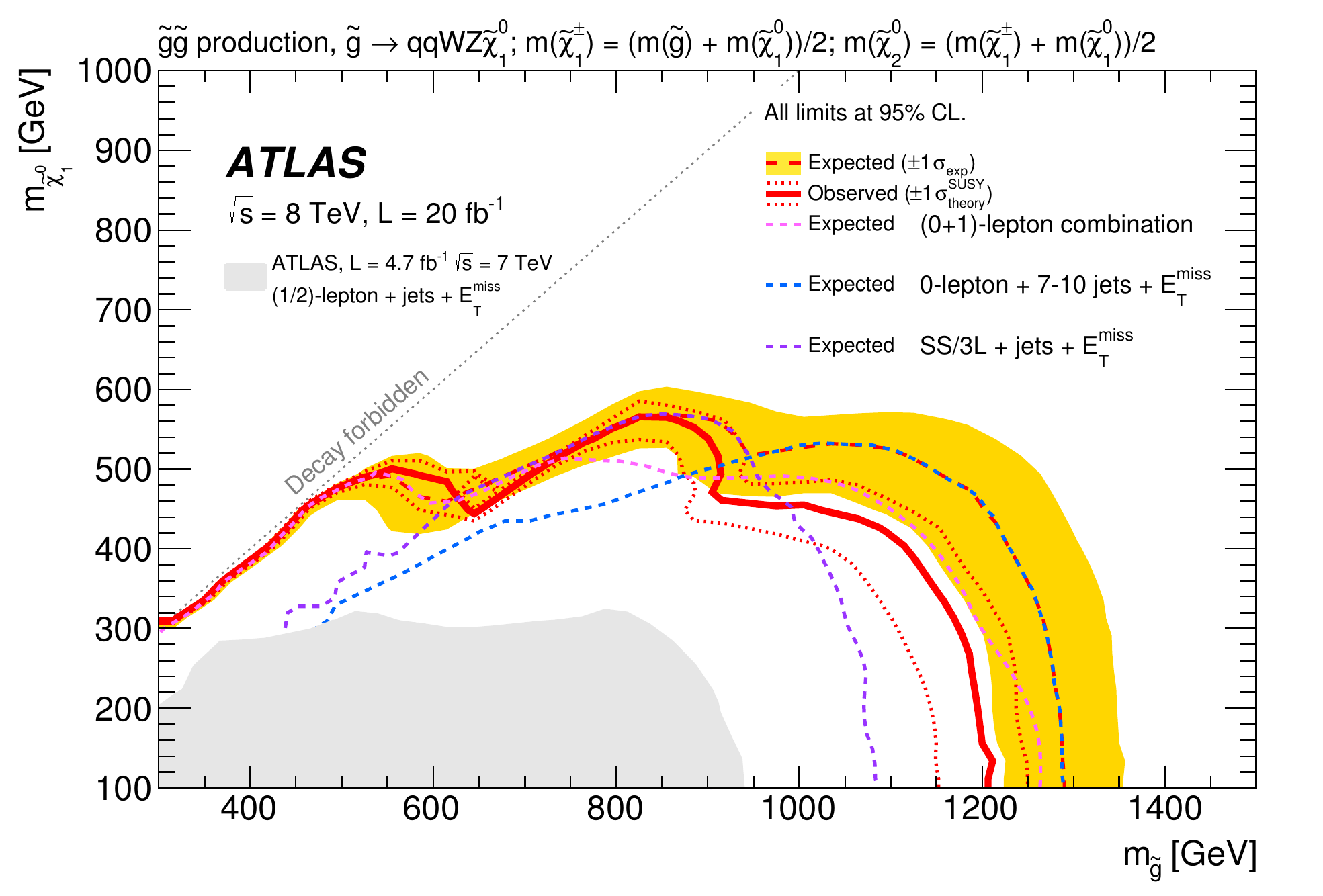}
\caption{Exclusion limits in the ($m_{\gluino},  m_{\ninoone}$) plane for a simplified model of gluino-pair production with two-step decay into $qqq'q'WWZZ\ninoone\ninoone$ and missing transverse momentum. The solid red line and the dashed red line show respectively the combined observed and expected 95\%~CL exclusion limits. %
Expected limits from the individual analyses which contribute to the final combined limits are also shown for comparison. A previous result from ATLAS ~\cite{Aad:2012ms} using 7 \TeV\ proton--proton collisions is represented by the shaded (grey) area.}
\label{fig:limit-glgl_twostepWWZZ}
\end{figure}

Another example of a simplified model with squark-pair production is considered in 
figure~\ref{fig:limit-sqsq_twostepSlepton}, where squarks decay through a two-step process via a chargino or neutralino and a slepton into final states with jets, leptons and missing transverse momentum. 
Figure~\ref{fig:limit-sqsq_twostepSlepton} shows the exclusion limits in the 
($m_{\tilde{q}}, m_{\ninoone}$) plane, for which the best results are obtained by the 2LRaz and the SS/3L searches.  Masses for the lightest neutralino can be excluded nearly up to the kinematic limit (diagonal line) for squark masses below 630~GeV. For $\ninoone$ masses below 100~GeV, squark masses can be excluded below 820~GeV. 

\begin{figure}[htbp]
\centering
\includegraphics[width=0.7\textwidth]{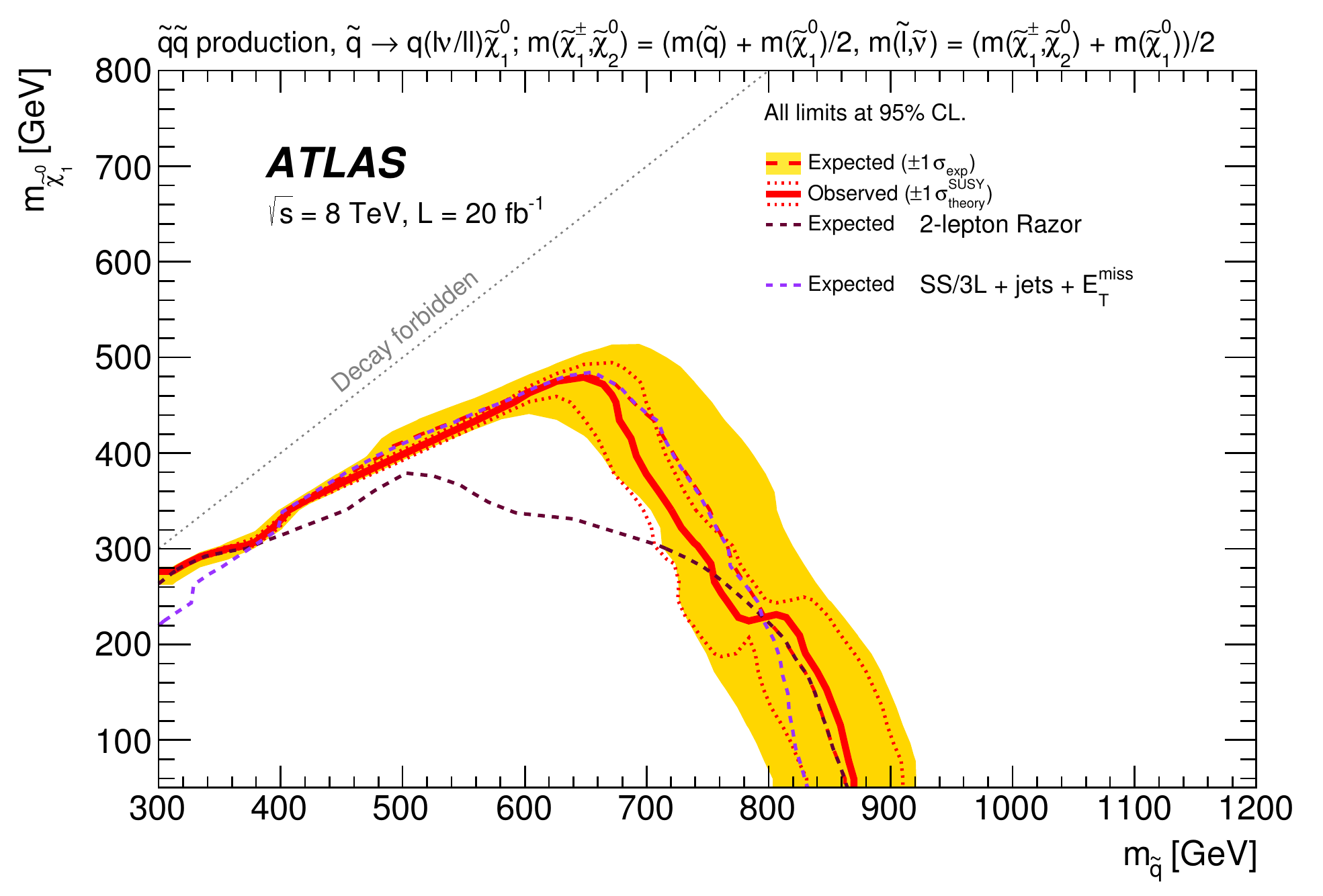}
\caption{Exclusion limits in the ($m_{\squark},  m_{\ninoone}$) plane for a simplified model of squark-pair production with two-step decay into jets, leptons and missing transverse momentum via sleptons. The solid red line and the dashed red line show respectively the combined observed and expected 95\%~CL exclusion limits.
Expected limits from the individual analyses which contribute to the final combined limits are also shown for comparison.}
\label{fig:limit-sqsq_twostepSlepton}
\end{figure}

Similarly,  a simplified model with gluino-pair production is considered in 
figure~\ref{fig:limit-glgl_twostepSlepton}, where  gluinos decay through a two-step process via a chargino or neutralino and sleptons into final states with jets, leptons and missing transverse momentum. 
Figure~\ref{fig:limit-glgl_twostepSlepton} shows the exclusion limits in the 
($m_{\gluino}, m_{\ninoone}$) plane.
The combined 1L(S,H)+2LRaz searches based on the best expected $CL_{\rm S}$ value and the SS/3L search provide the best sensitivities for this model. 
Masses for the lightest neutralino can be excluded nearly up to the kinematic limit (diagonal line) for gluino masses below 600~GeV. For $\ninoone$ masses below 100~GeV, gluino masses can be excluded below 1320~GeV. 

\begin{figure}[htbp]
\centering
\includegraphics[width=0.7\textwidth]{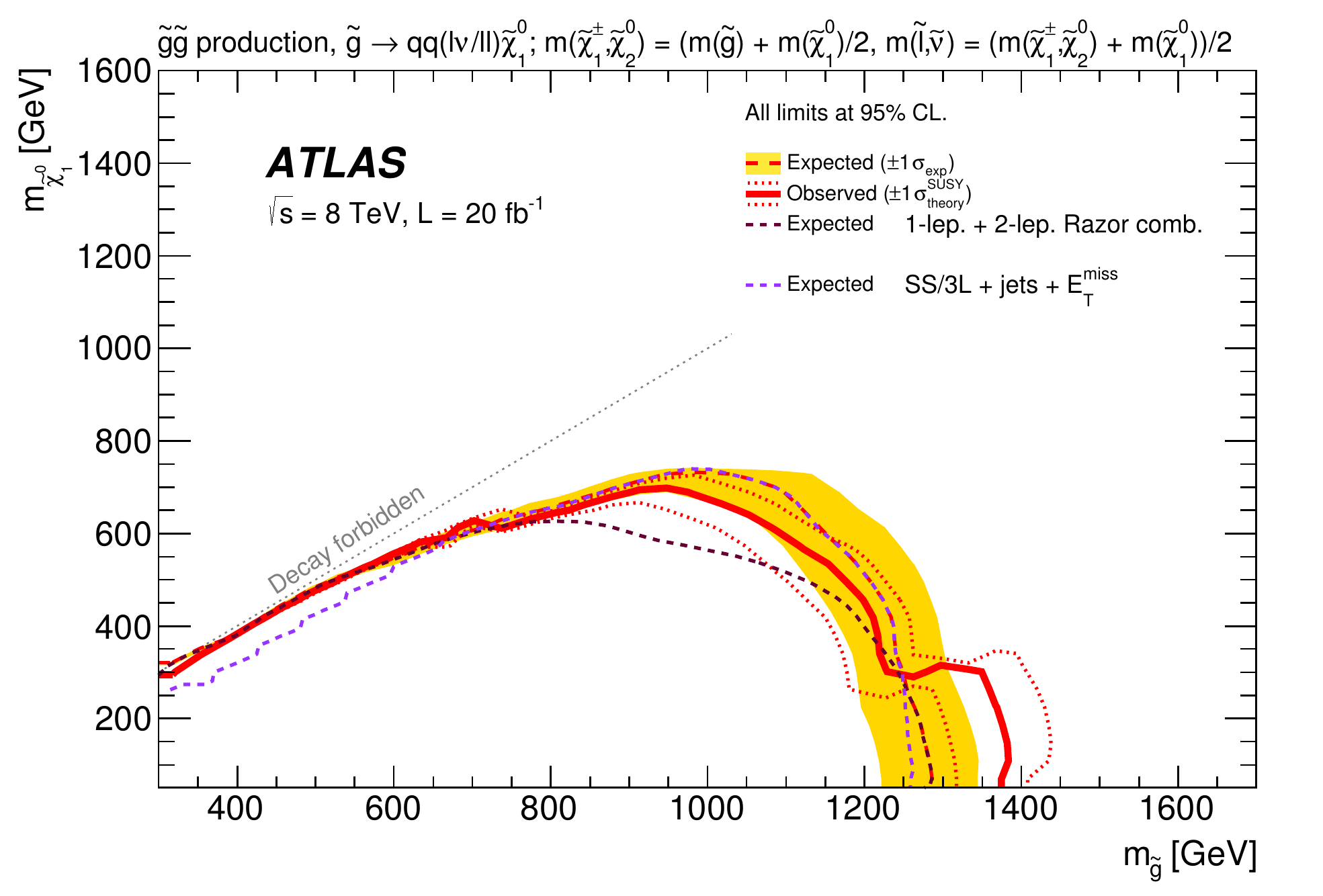}
\caption{Exclusion limits in the ($m_{\gluino},  m_{\ninoone}$) plane for a simplified model of gluino-pair production with two-step decay into jets, leptons and missing transverse momentum via sleptons. The solid red line and the dashed red line show respectively the combined observed and expected 95\%~CL exclusion limits. %
Expected limits from the individual analyses which contribute to the final combined limits are also shown for comparison.}
\label{fig:limit-glgl_twostepSlepton}
\end{figure}

A further example of a simplified model of %
squark-pair production and decay through a two-step process is shown in figure~\ref{fig:limit-sqsq_twostepTau}, where squarks
decay via charginos or neutralinos and staus. The exclusion limits obtained by the TAU search are indicated in the
($m_{\tilde{q}}, m_{\ninoone}$) plane. For light $\ninoone$ masses  around 50~GeV, squark masses below 850~GeV are excluded; and for light squark masses of 300~GeV, neutralino masses below 170~GeV are excluded.

\begin{figure}[htbp]
\centering
\includegraphics[width=0.7\textwidth]{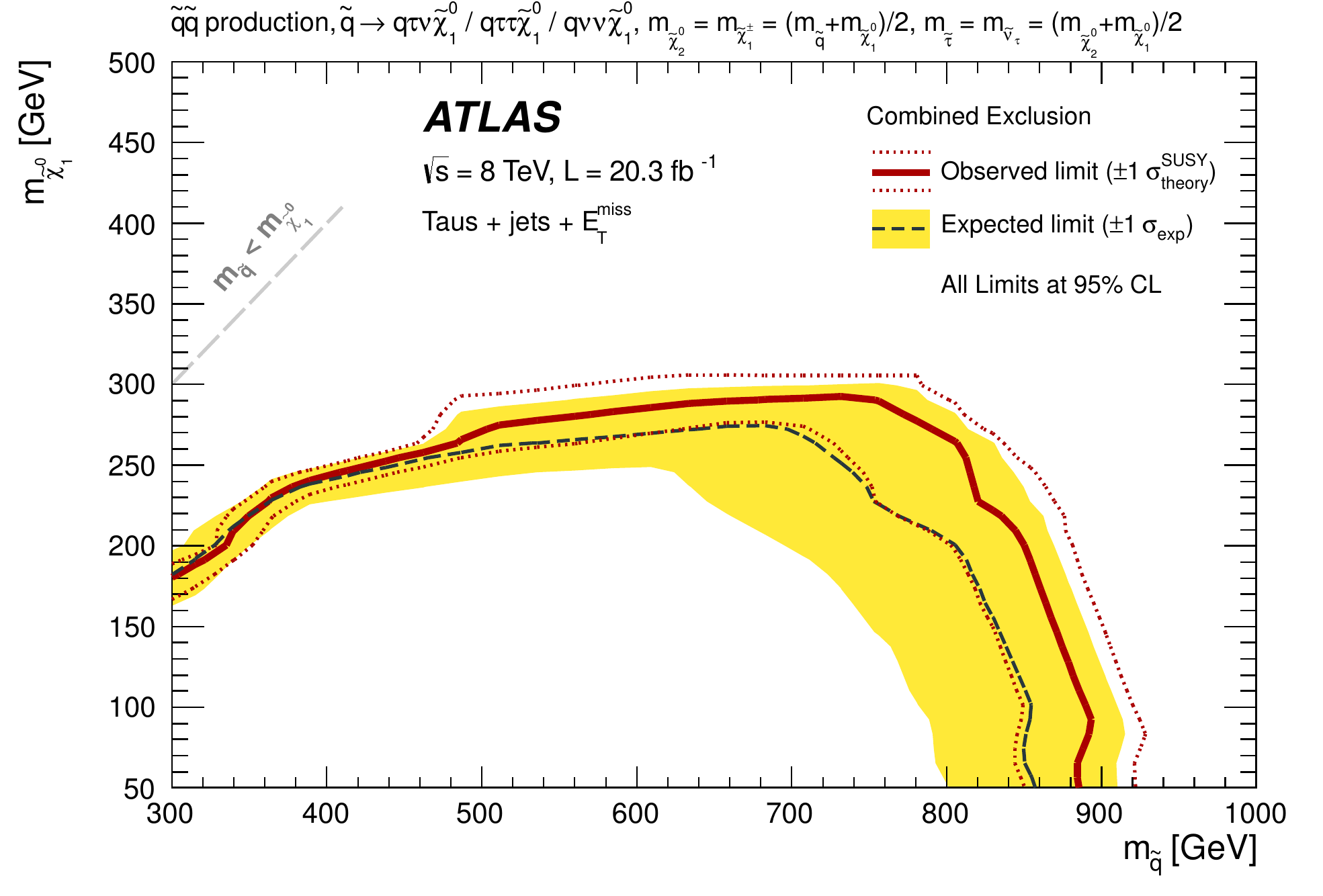}
\caption{95\%~CL exclusion limits in the ($m_{\squark},  m_{\ninoone}$) plane for a simplified model of squark-pair production with two-step decay via staus. 
The solid red line and the dashed black line show respectively the observed and expected 95\%~CL exclusion limits. %
}
 \label{fig:limit-sqsq_twostepTau}
\end{figure}

A simplified model of gluino-pair production and decay through a two-step process is shown in figure~\ref{fig:limit-glgl_twostepTau}, where gluinos
decay via charginos or neutralinos and staus. The exclusion limits obtained by the TAU search are indicated in the
($m_{\tilde{g}}, m_{\ninoone}$) plane. For light $\ninoone$ masses  around 100~GeV, gluino masses below 1220~GeV are excluded; and for light gluino masses of 400~GeV, neutralino masses below 280~GeV are excluded.

\begin{figure}[htbp]
\centering
\includegraphics[width=0.7\textwidth]{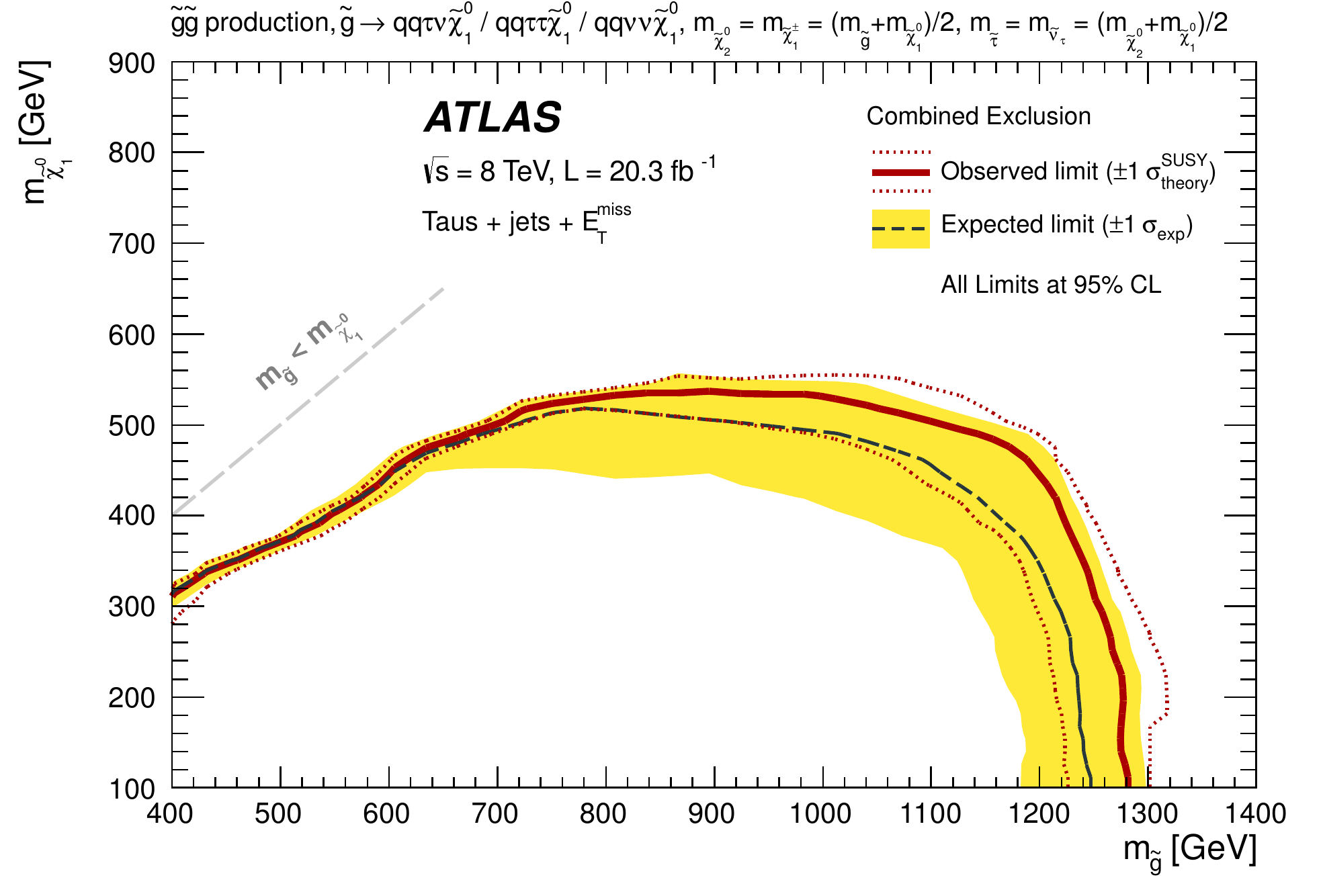}
\caption{95\%~CL exclusion limits in the ($m_{\gluino},  m_{\ninoone}$) plane for a simplified model of gluino-pair production with two-step decay via staus. 
The solid red line and the dashed black line show respectively the observed and expected 95\%~CL exclusion limits.
}
 \label{fig:limit-glgl_twostepTau}
\end{figure}

\subsubsection{Gluino decays via third-generation squarks}

This section summarizes the exclusion limits placed in the various simplified models with gluino decays via third-generation squarks described in section \ref{subsubsec:gttgbb}. 

The combined expected and observed exclusion limits for the gluino--off-shell--stop models  are given in the ($m_{\gluino}, m_{\ninoone}$) plane in figure \ref{fig:limit-Gtt-offshell}, where a 100\% branching ratio for the decay $\gluino \to t\bar{t}^{(*)}\ninoone$ via an off-shell stop is assumed.
In the regions where $m_{\gluino} < 2m_t + m_{\ninoone} $,  the three-body decays ($\gluino \to t\bar{t}\ninoone$) are replaced by the more complex multi-body decays proceeding via off-shell top quarks and $W$ bosons, as discussed in appendix \ref{AppGttExt}. 
The best sensitivity for this model is provided by the 0/1L3B and  the SS/3L searches.
 In the regions of parameter space where the mass difference between the gluino and the lightest neutralino is small, the most sensitive search is the SS/3L, and the sensitivity is dominated by the SS/3L\_SR3b signal region. In the regions with a large mass splitting between the gluino and the neutralino, where hard jets and large \met\ are expected, the sensitivity is dominated by the 0/1L3B\_SR-0l-7j signal regions from the  0/1L3B search. For these models, gluino masses below about 1310~\GeV\ are excluded for $m_{\ninoone} < $ 400~\GeV.

\begin{figure}[htbp]
\centering
\includegraphics[width=0.7\textwidth]{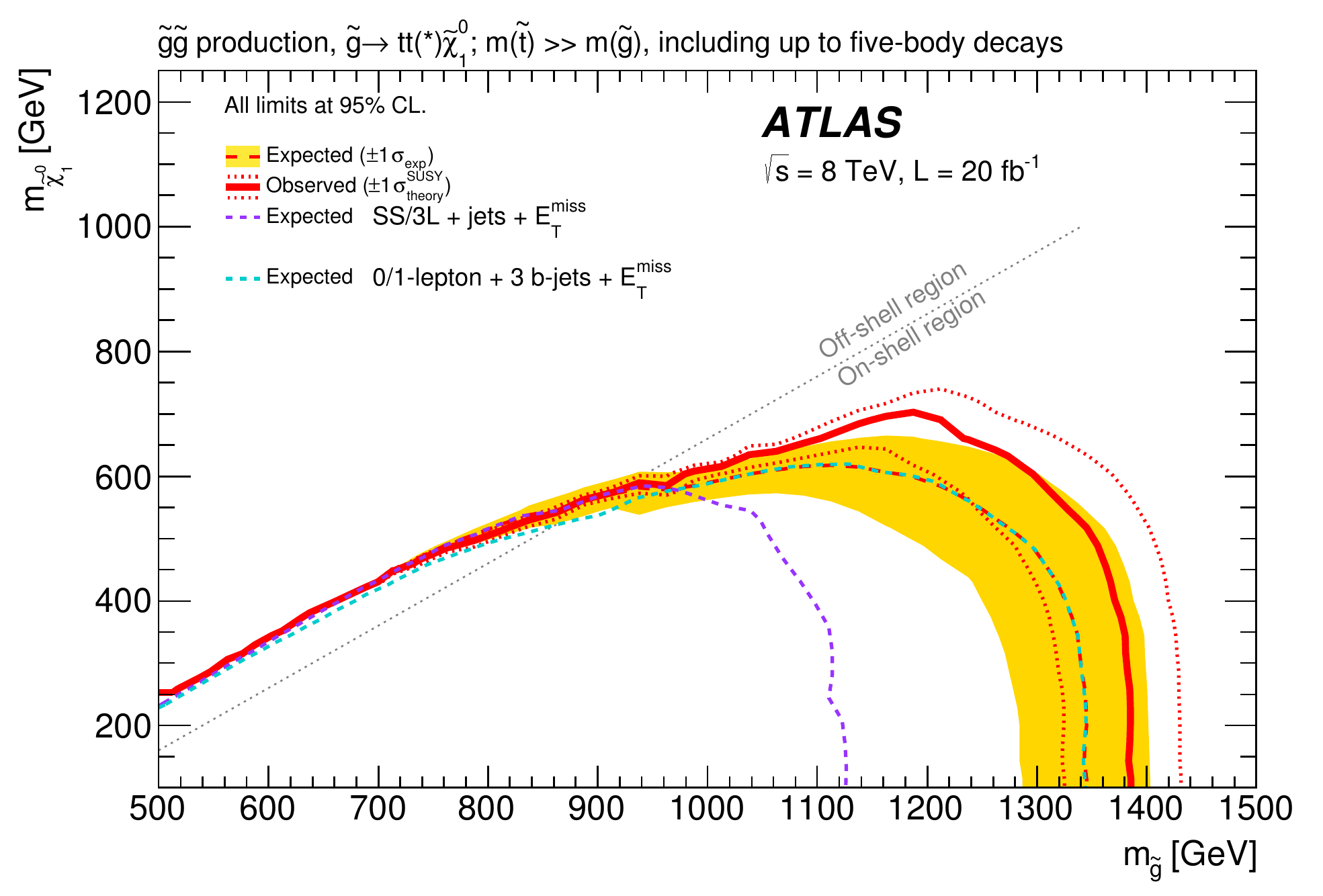}
\caption{Exclusion limits in the ($m_{\gluino}, m_{\ninoone}$) plane for the gluino--off-shell--stop simplified models in which the pair-produced gluinos decay via an off-shell stop, as $\gluino \to t\bar{t}\ninoone$.
In the region below the grey dashed line labelled ``On-shell region'', $m_{\gluino} > 2m_t + m_{\ninoone} $ and thus gluinos decay to two real top quarks. In the ``Off-shell region'', $m_{\gluino} < 2m_t + m_{\ninoone} $ and the decays of the gluino involve an off-shell top quark. %
 Only four-body  ($\gluino \to t W b \ninoone$) and five-body ($\gluino \to W b W b \ninoone$) decays are considered because for higher multiplicities the gluinos do not decay promptly. 
 The solid red line and the dashed red line show respectively the combined observed and expected 95\% CL exclusion limits. %
Expected limits from the individual analyses which contribute to the final combined limits are also shown for comparison.} 
 \label{fig:limit-Gtt-offshell}
\end{figure}

The exclusion limits for the gluino--stop simplified models are given in the ($m_{\gluino}, m_{\stopone}$) plane in figure \ref{fig:limit-Gtt-onshell}. 
The $\stopone$ is assumed to be the lightest squark while all other squarks are heavier than the gluino, and $m^{}_{\gluino}>m^{}_{\stopone}+m_t$ such that the branching ratio is 100\%\ for $\gluino \to \stopone t$ decays, and the top squark decays as  $\stopone \to t \ninoone$. 
The 0/1L3B search provides the best sensitivity in these models, excluding gluino masses below 1220~\GeV\ for stop masses up to 1000~\GeV.  
Limits for the same class of simplified models, but assuming the $\stopone \to b \chinoonepm$ decay of the top squark, are also given in the ($m_{\gluino}, m_{\stopone}$) plane, and summarized in figure~\ref{fig:limit-Gtt-bChi}. The mass of the lightest neutralino in these models is set to 60~\GeV\, and the mass of the chargino is assumed to be twice the mass of the neutralino. The chargino decays into a neutralino and a virtual $W$ boson. The strongest limits are provided by the  0/1L3B search.
Compared to the models where the top squark decays via $\stopone \to t \ninoone$, presented in figure~\ref{fig:limit-Gtt-onshell}, the sensitivity in these models is lower 
for most of the parameter space where soft \met and jets are expected from the chargino decay $\chinoonepm \rightarrow W^{*} \ninoone$. 
Gluino masses below 1180~\GeV\ are excluded for stop masses up to 1000~\GeV\ in these models. 

\begin{figure}[htbp]
\centering
\includegraphics[width=0.7\textwidth]{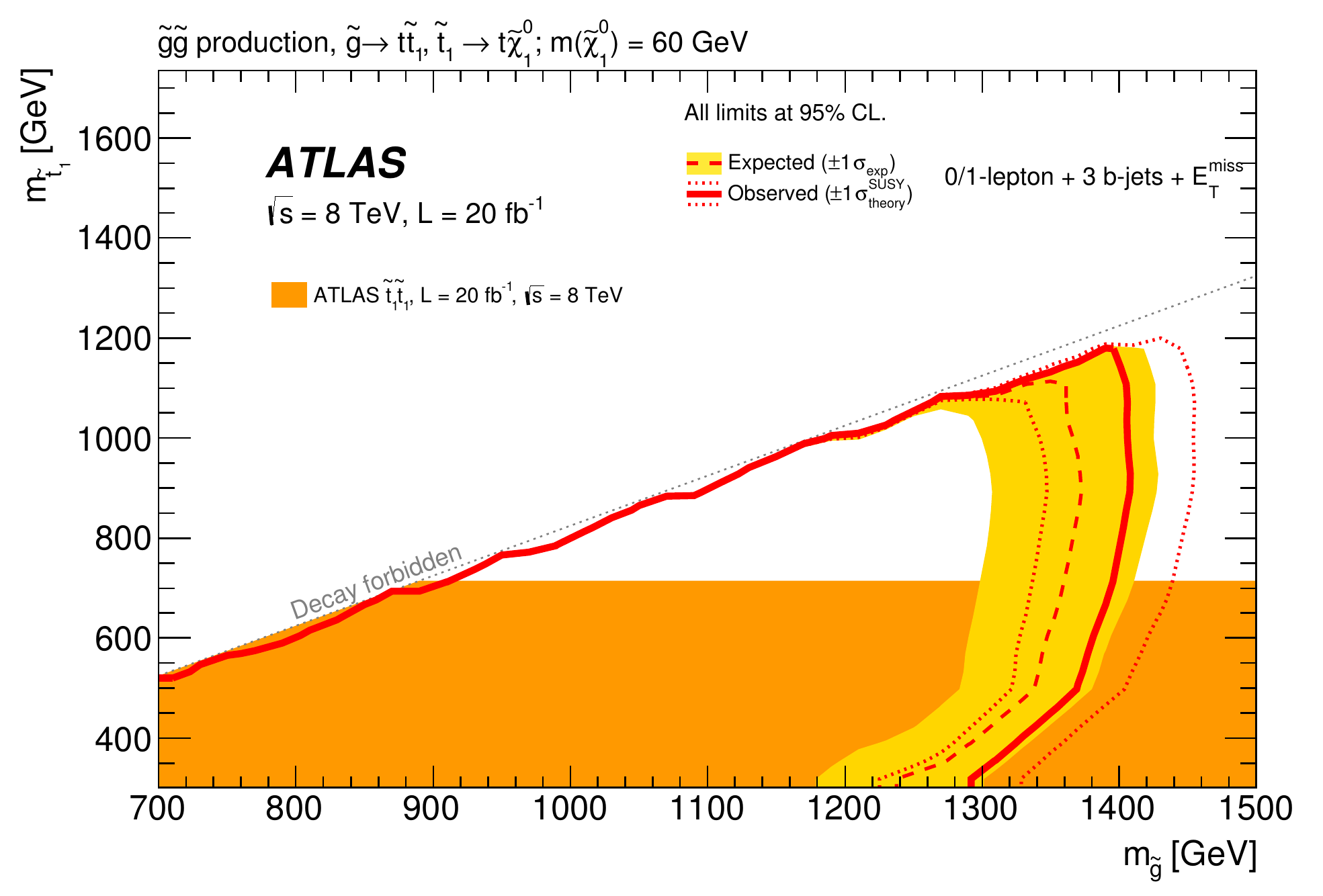}
\caption{Exclusion limits in the ($m_{\gluino}, m_{\stopone}$) plane for the gluino--stop simplified models in which the top squarks are produced in the decay of pair-produced gluinos and decay via $\stopone \to t \ninoone$. The neutralino mass is set to 60~\gev. The solid red line and the dashed red line show respectively the combined observed and expected 95\%~CL exclusion limits. %
Also shown for reference is the limit from the ATLAS search for direct stop-pair production \cite{3rdGen-summarypaper}.
} 
\label{fig:limit-Gtt-onshell}
\end{figure}

\begin{figure}[htbp]
\centering
\includegraphics[width=0.7\textwidth]{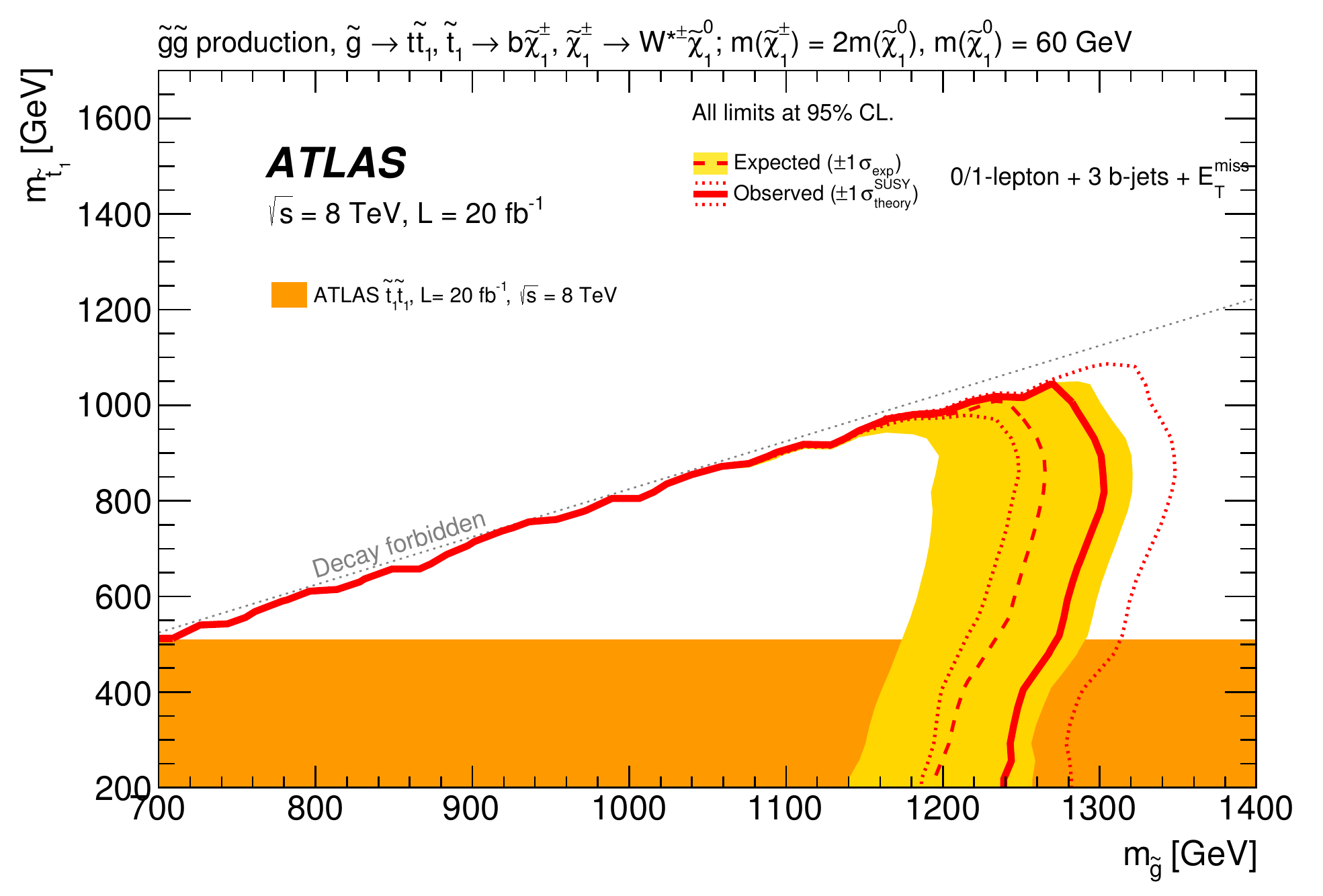}
\caption{Exclusion limits in the ($m_{\gluino}, m_{\stopone}$) plane for the gluino--stop simplified models in which the top squarks are produced in the decay of pair-produced gluinos and decay via  $\stopone \to b \chinoonepm$.  The neutralino mass is set to 60~\gev and the mass of the chargino is assumed to be twice the mass of the neutralino.  The solid red line and the dashed red line show respectively the combined observed and expected 95\%~CL exclusion limits. %
Also shown for reference is the limit from the ATLAS search for direct stop-pair production \cite{3rdGen-summarypaper}.
} 
 \label{fig:limit-Gtt-bChi}
\end{figure}

Another possible decay of the top squark, $\stopone \to c \ninoone$, is considered within the same class of simplified models, 
with the mass difference between the $\stopone$ and the lightest neutralino fixed to 20~\GeV. 
The (0+1)-lepton combination provides  the best sensitivity
 in these models. The 1L(S,H) search is complementary to the 0L search in that the expected limit for the single-lepton search is able to cover higher top squark masses at intermediate gluino masses (e.g. 80~\GeV\ higher at $m_{\gluino}$ = 900~\GeV). The resulting exclusion limit  is presented in the ($m_{\gluino}, m_{\stopone}$) plane in figure \ref{fig:limit-Gtt-charm}, and reaches gluino masses up to 1260~\GeV. 

\begin{figure}[htbp]
\centering
\includegraphics[width=0.7\textwidth]{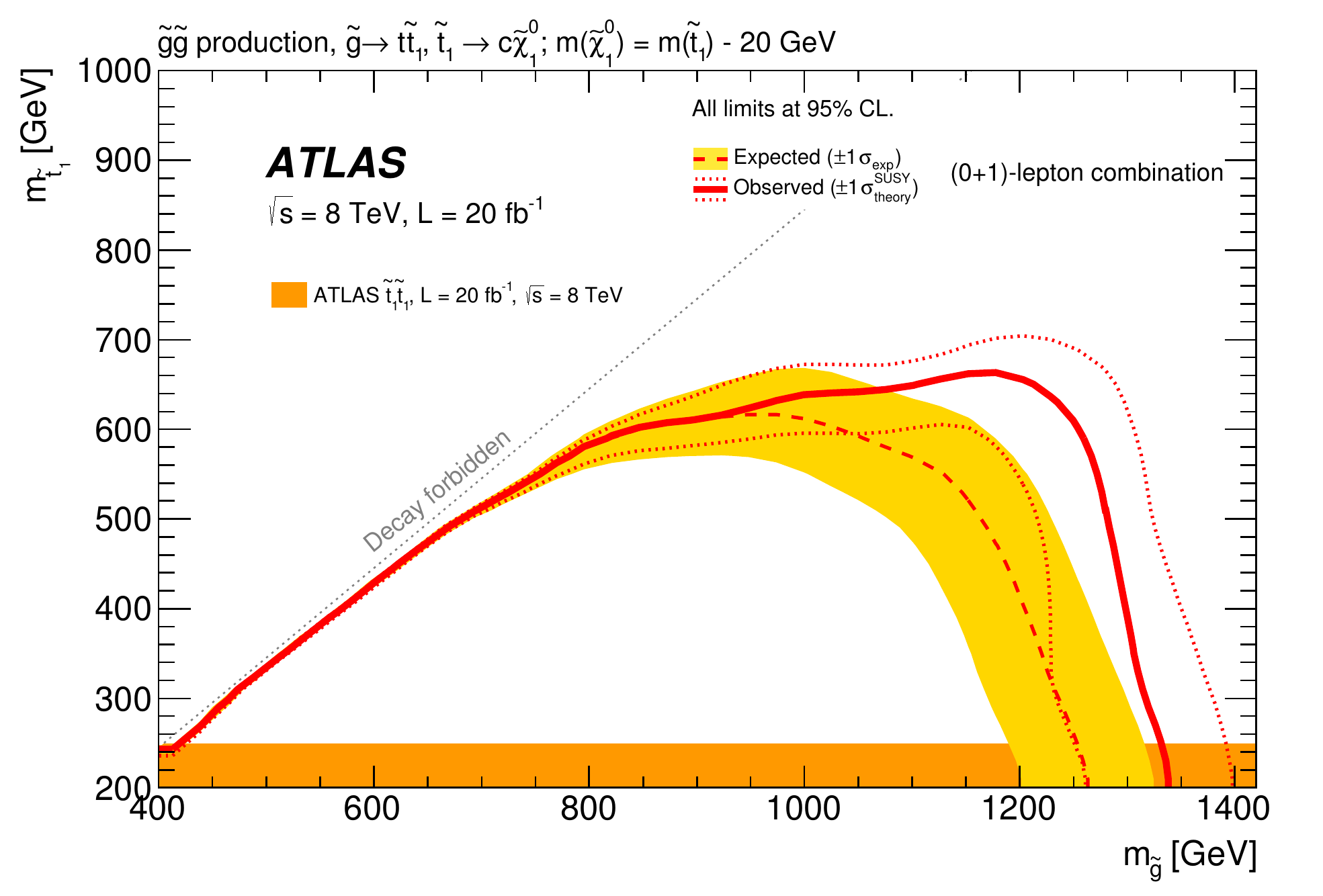}
\caption{Exclusion limits in the ($m_{\gluino}, m_{\stopone}$) plane for the gluino--stop simplified models in which the top squarks are produced in the decay of pair-produced gluinos and decay via  $\stopone \to c + \ninoone$. The mass difference between the $\stopone$ and the lightest neutralino is fixed to 20~\GeV.  The solid red line and the dashed red line show respectively the combined observed and expected 95\%~CL exclusion limits. %
Also shown for reference is the limit from the ATLAS search for direct stop-pair production \cite{3rdGen-summarypaper}.}
\label{fig:limit-Gtt-charm}
\end{figure}

A simplified model is also considered, in which the top squark decay, $\stopone \rightarrow s b$, involves R-parity and baryon number violation. %
The result is presented in the ($m_{\gluino}, m_{\stopone}$) plane in figure \ref{fig:limit-Gtt-RPV}, where the best limit is obtained by the MULTJ search. Gluino masses below 880~\GeV\ are excluded for top squark masses ranging from 400~\GeV\ to 1000~\GeV. 

\begin{figure}[htbp]
\centering
\includegraphics[width=0.7\textwidth]{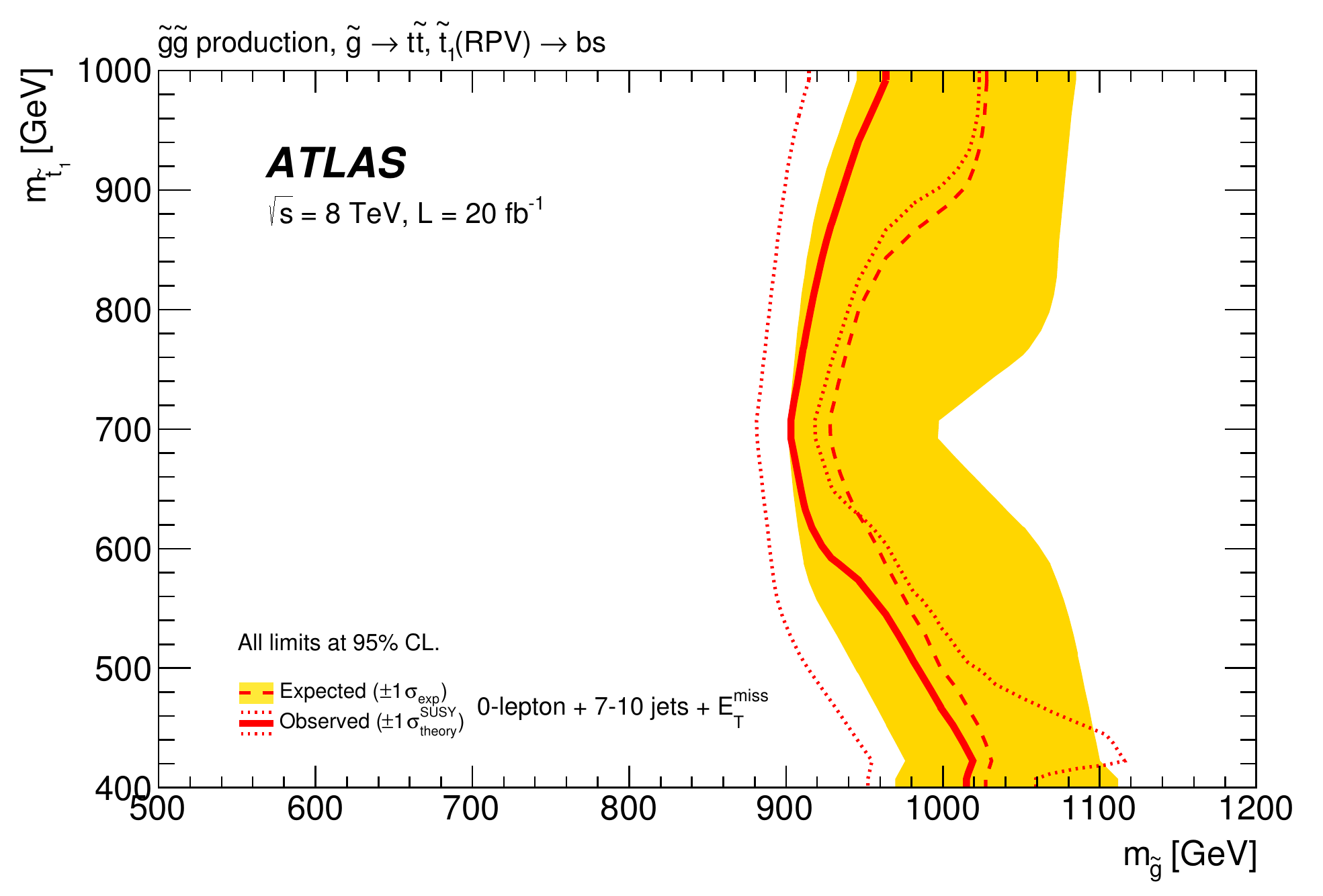}
\caption{Exclusion limits in the ($m_{\gluino}, m_{\stopone}$) plane for the gluino--stop simplified models in which the top squarks are produced in the decay of pair-produced gluinos and decay via R-parity and baryon number violation  $\stopone \rightarrow s b$. The solid red line and the dashed red line show respectively the combined observed and expected 95\%~CL exclusion limits. %
 Expected limits from the individual analyses which contribute to the final combined limits are also shown for comparison. }
\label{fig:limit-Gtt-RPV}
\end{figure}

The sensitivity in the gluino--sbottom simplified models in which the branching ratio for $\gluino \to \sbottomone b$ decays is 100\% and the bottom squarks are assumed to decay exclusively via $\sbottomone \to b\ninoone$ is provided only by the 0/1L3B search \cite{3bjetsPaper} and the result is presented for completeness in figure~\ref{fig:limit-Gbbonshell}. 
The search excludes gluino masses below 1200~\GeV\ for sbottom masses up to about 1100~\GeV. 

\begin{figure}[htbp]
\centering
\includegraphics[width=0.7\textwidth]{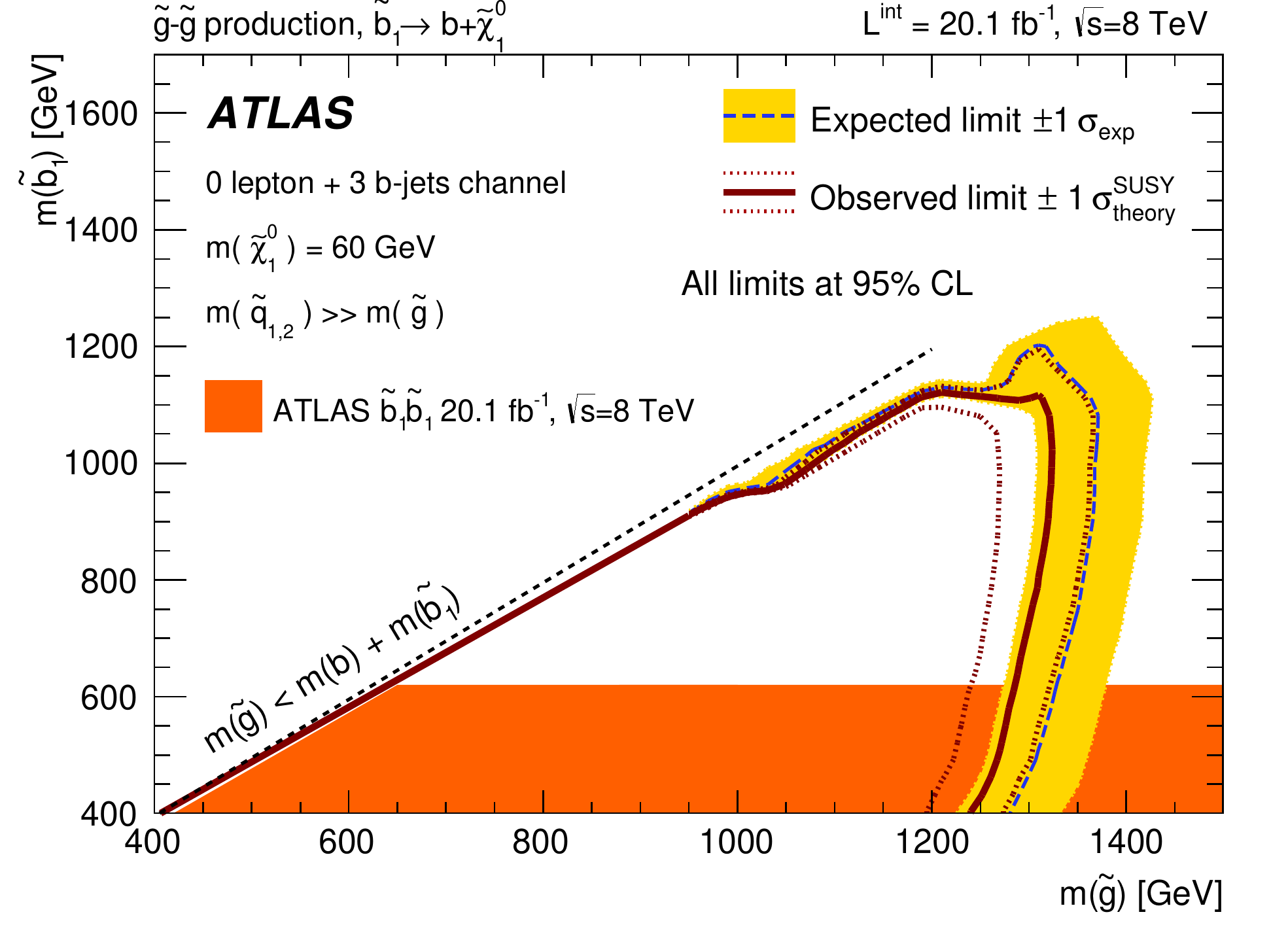}
\caption{Exclusion limits in the ($m_{\gluino}, m_{\sbottomone}$) plane for the gluino--sbottom simplified models, taken from ref.~\cite{3bjetsPaper}. Also shown for reference is the limit from the ATLAS search for direct sbottom-pair production \cite{3rdGen-summarypaper}.} 
\label{fig:limit-Gbbonshell}
\end{figure}

The limits for the gluino--off-shell--sbottom simplified models, which assume 
100\%\ branching ratio for the 
gluino three-body decay $\gluino \to b\bar{b}\ninoone$ via an off-shell sbottom,  are given in the ($m_{\gluino}, m_{\ninoone}$) plane in 
figure~\ref{fig:limit-Gbb}. The best sensitivities are provided by two searches, the 
0/1L3B and the 0L search.
The former is the most sensitive search in regions of the parameter space with a large mass splitting between the gluino and the lightest neutralino, and the latter in regions with a small mass splitting where softer jets and smaller \met are expected in the final state. In these models, gluino masses below 1250~\GeV\ are excluded for $m_{\ninoone} <$ 400~\GeV\ while neutralino masses below 600~\GeV\ are excluded in the gluino mass range between 700 and 1200~\GeV.

\begin{figure}[H]
\centering
\includegraphics[width=0.7\textwidth]{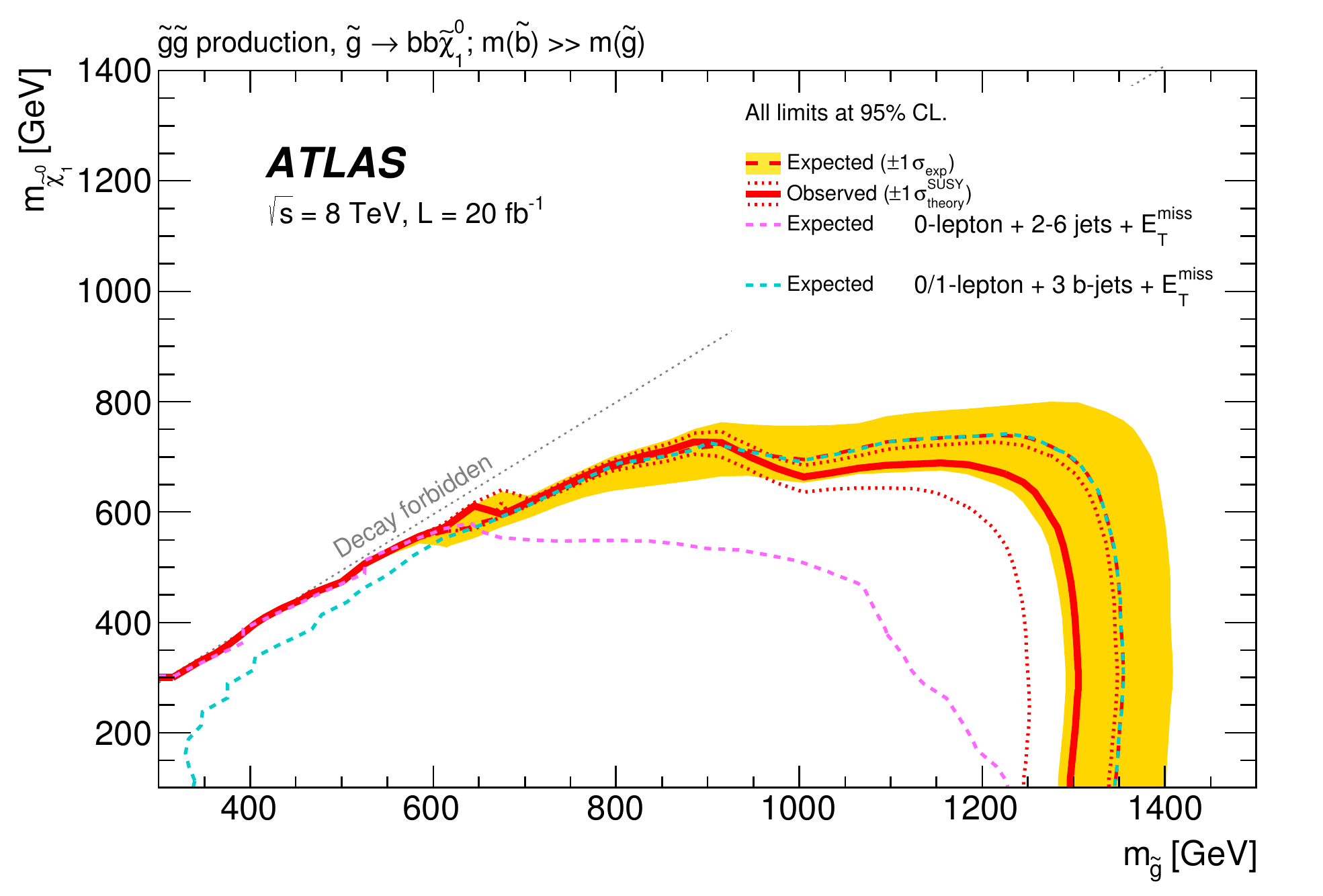}
\caption{Exclusion limits in the ($m_{\gluino}, m_{\ninoone}$) plane for the gluino--off-shell--sbottom simplified models in which the pair-produced gluinos decay via an off-shell sbottom as $\gluino \to b\bar{b}\ninoone$. The solid red line and the dashed red line show respectively the combined observed and expected 95\%~CL exclusion limits. %
Expected limits from the individual analyses which contribute to the final combined limits are also shown for comparison. } 
\label{fig:limit-Gbb}
\end{figure}

The sensitivity in the gluino--off-shell--stop/sbottom simplified models in which gluinos decay via virtual stops or sbottoms 
is provided only by the 0/1L3B search \cite{3bjetsPaper}. Here
the mass difference between the particles is set such that the gluino decays result in an effectively three-body final state ($bt\ninoone$). 
The exclusion limit is presented in figure~\ref{fig:limit-Gtb} for completeness. 
For neutralino masses of 500~\GeV, gluino masses are excluded between 750 and 1250~\GeV.

\begin{figure}[H]
\centering
\includegraphics[width=0.7\textwidth]{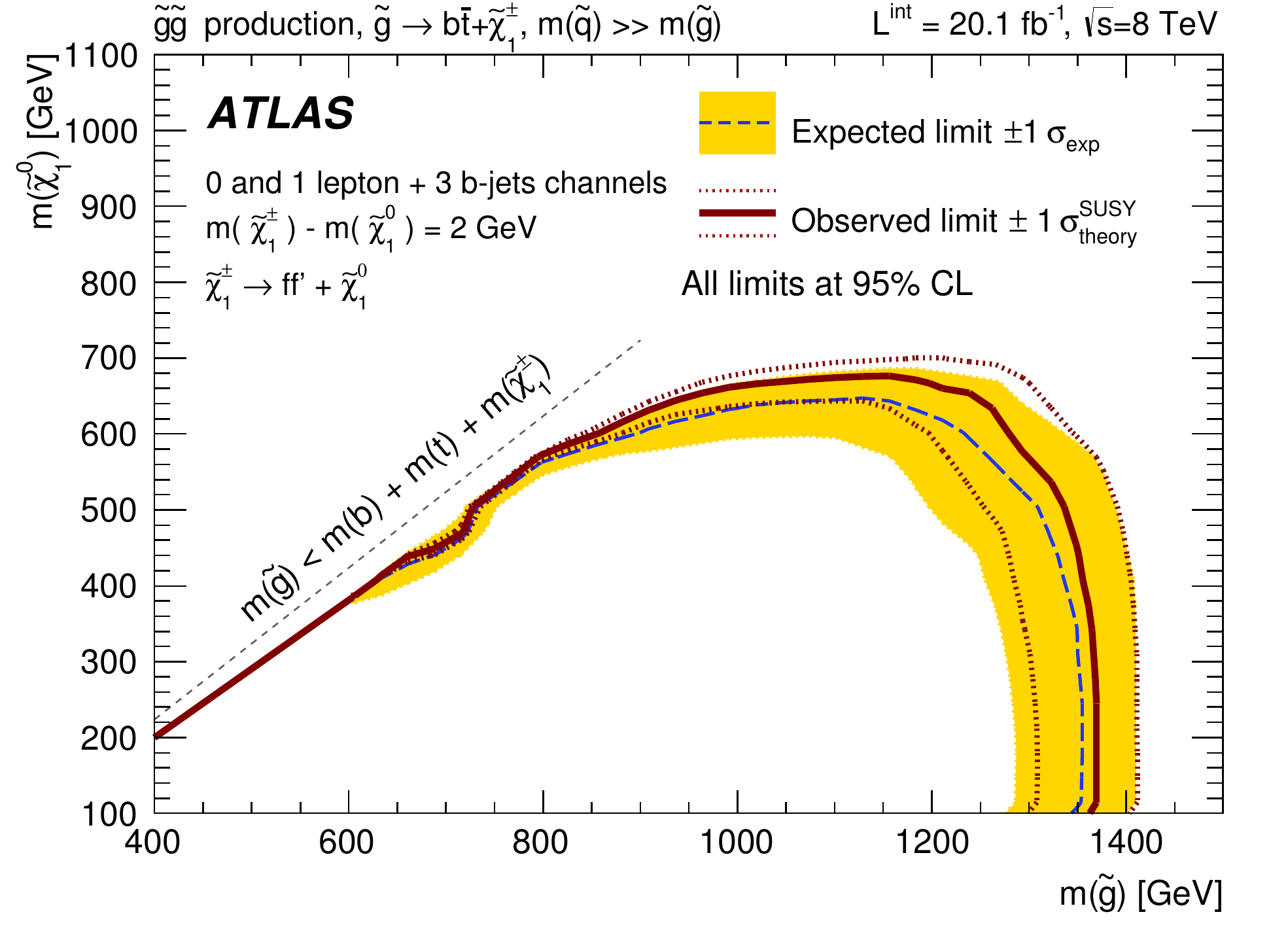}
\caption{Exclusion limits in the ($m_{\gluino}, m_{\ninoone}$) plane for the gluino--off-shell--stop/sbottom simplified models, taken from ref.~\cite{3bjetsPaper}. } 
\label{fig:limit-Gtb}
\end{figure}

\section{Conclusions}
\label{sec:conclusion}

A search for squarks and gluinos in inclusive final states containing high-$\pt$ jets and missing
transverse momentum, with or without leptons or $b$-jets, is presented. The data were recorded in
2012 by the ATLAS experiment with
$\sqrt{s} = 8$~TeV proton--proton collisions at the Large Hadron
Collider, with a total integrated luminosity up to 20.3~fb$^{-1}$. 
Earlier ATLAS searches have been extended and combined with new search techniques,
thus improving the sensitivity for supersymmetric models.
Good agreement is found with the predictions from SM processes. 
The data are therefore used to set exclusion limits for a variety of simplified and phenomenological SUSY models. 

Limits in simplified models with gluinos and squarks of the first and
second generations are derived for direct and one- or two-step decays of squarks and gluinos, and gluino decays via third-generation squarks. %
In all the considered simplified models that assume R-parity conservation, the limit on the gluino mass exceeds 1150 GeV at 95\% CL, for an LSP mass smaller than 100 GeV.
Additional limits are set in a phenomenological MSSM model used in the search for left-handed squarks, a minimal Supergravity/Constrained MSSM model, R-parity-violation scenarios, a minimal gauge-mediated supersymmetry breaking model, a natural gauge mediation model, a non-universal Higgs mass model with gaugino mediation and a minimal model of universal extra dimensions.
These limits are either new or extend the region of parameter space excluded by previous searches with the ATLAS detector. 

\newpage

\section*{Acknowledgements}

We thank CERN for the very successful operation of the LHC, as well as the
support staff from our institutions without whom ATLAS could not be
operated efficiently.

We acknowledge the support of ANPCyT, Argentina; YerPhI, Armenia; ARC,
Australia; BMWFW and FWF, Austria; ANAS, Azerbaijan; SSTC, Belarus; CNPq and FAPESP,
Brazil; NSERC, NRC and CFI, Canada; CERN; CONICYT, Chile; CAS, MOST and NSFC,
China; COLCIENCIAS, Colombia; MSMT CR, MPO CR and VSC CR, Czech Republic;
DNRF, DNSRC and Lundbeck Foundation, Denmark; EPLANET, ERC and NSRF, European Union;
IN2P3-CNRS, CEA-DSM/IRFU, France; GNSF, Georgia; BMBF, DFG, HGF, MPG and AvH
Foundation, Germany; GSRT and NSRF, Greece; RGC, Hong Kong SAR, China; ISF, MINERVA, GIF, I-CORE and Benoziyo Center, Israel; INFN, Italy; MEXT and JSPS, Japan; CNRST, Morocco; FOM and NWO, Netherlands; BRF and RCN, Norway; MNiSW and NCN, Poland; GRICES and FCT, Portugal; MNE/IFA, Romania; MES of Russia and NRC KI, Russian Federation; JINR; MSTD,
Serbia; MSSR, Slovakia; ARRS and MIZ\v{S}, Slovenia; DST/NRF, South Africa;
MINECO, Spain; SRC and Wallenberg Foundation, Sweden; SER, SNSF and Cantons of
Bern and Geneva, Switzerland; NSC, Taiwan; TAEK, Turkey; STFC, the Royal
Society and Leverhulme Trust, United Kingdom; DOE and NSF, United States of
America.

The crucial computing support from all WLCG partners is acknowledged
gratefully, in particular from CERN and the ATLAS Tier-1 facilities at
TRIUMF (Canada), NDGF (Denmark, Norway, Sweden), CC-IN2P3 (France),
KIT/GridKA (Germany), INFN-CNAF (Italy), NL-T1 (Netherlands), PIC (Spain),
ASGC (Taiwan), RAL (UK) and BNL (USA) and in the Tier-2 facilities
worldwide.

\newpage

\appendix
\section{Extension of the $\gluino \to \ttbar \ninoone$ simplified model to include decays with off-shell top quarks}
\label{AppGttExt}

In this appendix, further details are provided about the extension 
of the gluino-mediated off-shell stop model $\gluino\to t\bar t\ninoone$ to the region $m_t+m_W+m_b\le m_{\tilde g}-m_{\ninoone}\le 2 m_t$, 
where three-body decays are replaced by more complex multi-body decays proceeding via off-shell top quarks and $W$ bosons. 
This region is delimited by the kinematic boundaries corresponding to three- and four-body gluino decays. 
In principle, the extension could have been performed up to mass gaps as small as $m_{\gluino}-m_{\ninoone}\ge 2 m_b$,  
but it was found that for a $100\%$ branching ratio hypothesis for the $\gluino\to t\bar t\ninoone$ mode, 
mass gaps smaller than the four-body kinematic bound quickly lead to large gluino lifetimes, 
resulting in displaced gluino decays. %
This is verified even in the scenario leading to the smallest gluino lifetime ($m_{\tilde t_1}=m_{\gluino}$, $\tilde t_1=\tilde t_R$). 
Since the results of the various searches reported in this paper are all based on prompt objects, 
the expected sensitivity to these scenarios with small mass gaps is very small, 
and they are addressed by dedicated searches~\cite{metastableGluinos}. 
Therefore, the probed parameter space is limited to  $m_{\tilde g}-m_{\ninoone}\ge m_t+m_W+m_b$. 

Despite the restriction of the model parameter space to regions 
where four-body gluino decays $\gluino\to t Wb\ninoone$ are always kinematically allowed, 
more complex decays (mainly five-body) can occur concurrently with a significant branching ratio, 
and even become dominant when the mass gap approaches its lower bound. 
Consideration of these alternative decay modes also sometimes leads to large differences in kinematic distributions of the decay products. 
For example, the $b$-quark in the decay $\gluino\to tWb\ninoone$ becomes too soft for experimental detection 
when $m_{\gluino}\approx m_{t}+m_{W}+m_{b}+m_{\ninoone}$, 
whereas it can still get sizeable momentum in the alternative decay $\gluino\to WWbb\ninoone$. 
Since the most sensitive searches for this model rely on selections with at least three $b$-jets, 
one can see that the acceptance of these selections would vanish at the kinematic bound if considering only four-body decays, 
although it is not the case thanks to the alternative decay modes. 
To summarize, quantitative studies showed that it is important to consider at least the five-body decay as well in this region of the parameter space. 

Finally, signal events were generated following a configuration defined and validated by comparing 
generator predictions for several observables in a region at low neutralino mass where only three-body decays contribute. 
The reference was provided by \herwig++ (the generator used for the $m_{\gluino}-m_{\ninoone}>2m_t$ region of this scenario), 
characterized notably by the use of a matrix element amplitude for the gluino decay, and the preservation of spin correlation between the decay products. 
It was first observed that narrow-width approximations for the gluino decay compared poorly to the reference, 
hence imposing the need for decay amplitudes computed from matrix elements for the four- and five-body decays, 
and as a consequence the choice of the \Madgraph{} generator. 
However, the computing requirements to obtain the amplitude associated with such a $2 \to 10\,(+1)$ hard process (with an extra parton) are too demanding. 
An approximation is used instead, consisting in the separate generation of pair-produced gluinos, 
and gluino decays into fermions $\gluino\to f\bar f'f''\bar f''' bb\ninoone$. 
The two stages are then combined by boosting the gluino decay products according to the gluino's directions and momenta 
defined by the hard process, while preserving the gluino spin orientations. 
The use of the inclusive seven-body gluino decay, instead of the minimally required five-body decay, 
came at no additional computing cost and allowed, in particular, 
proper propagation of the various spin correlations along the gluino decay chain. 
This setup provided agreement with the reference at the level of $5\%$, 
for the shapes of various generator-level kinematic distributions 
(a few of which are presented in figure~\ref{fig:gtt5body_mcval}),
as well as for fiducial acceptances of typical event selections used as signal regions in the relevant SUSY searches.

\begin{figure} [H]
\centering
\subfigure[Distribution of the number of prompt leptons (electrons and muons), $\pt>20$~\GeV, $|\eta|<2.5$]{\includegraphics[width=0.49\textwidth]{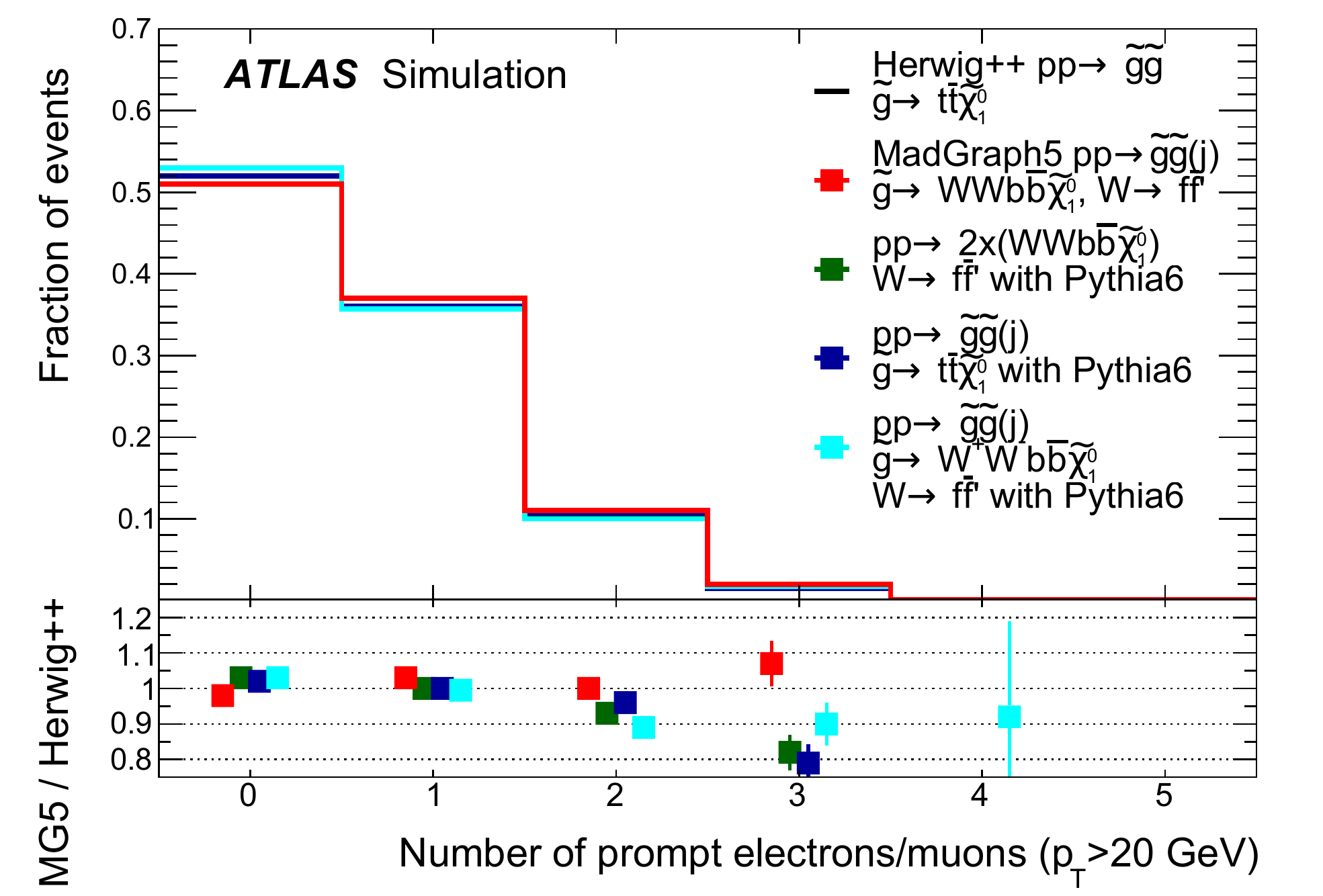}}
\subfigure[Cumulative distribution of the effective mass built from selected leptons, jets and non-interacting particles]{\includegraphics[width=0.49\textwidth]{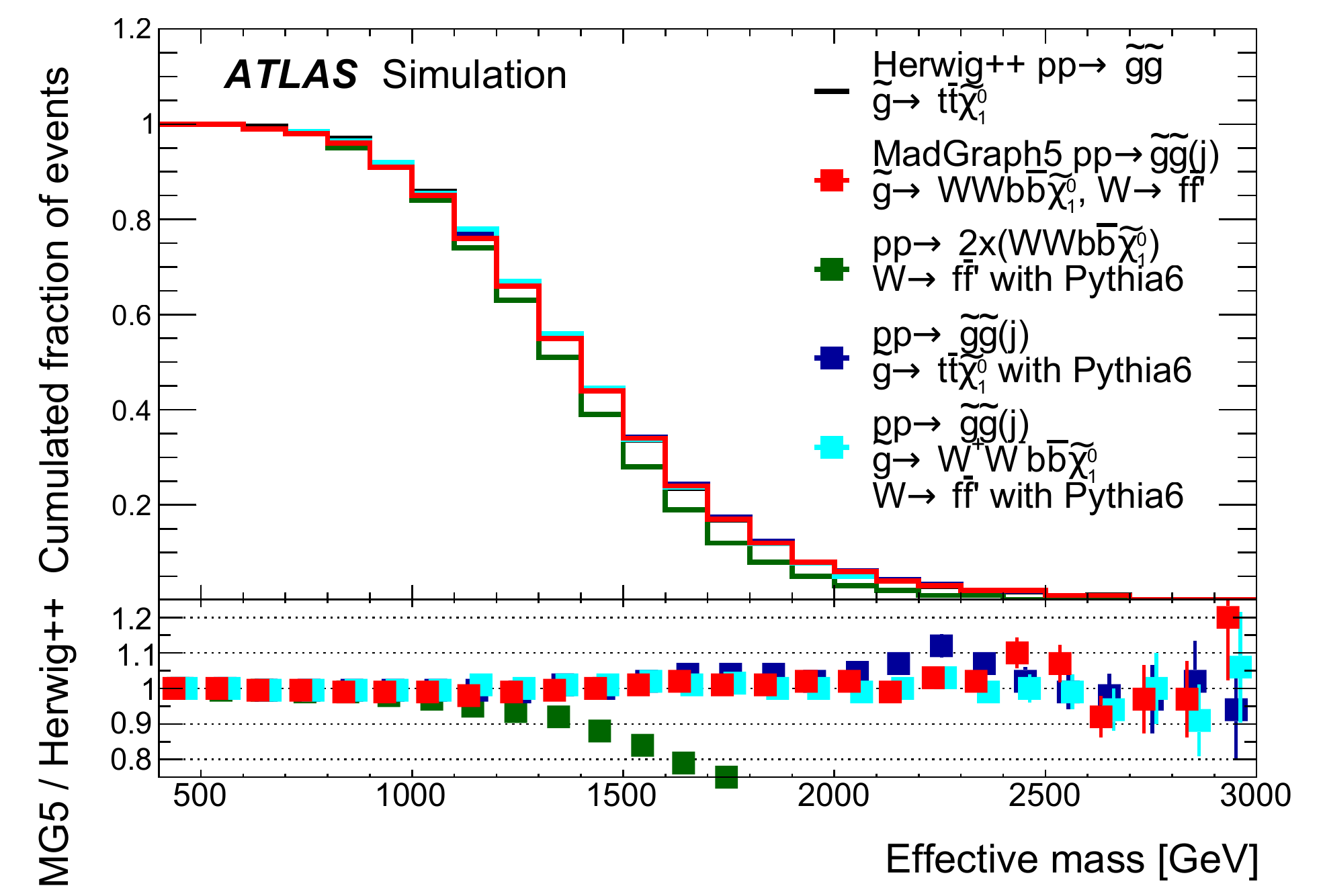}}
\subfigure[Cumulative distribution of the number of jets (anti-$k_t$ algorithm, $R$ = 0.4), $\pt>40$~\GeV, $|\eta|<2.8$]{\includegraphics[width=0.49\textwidth]{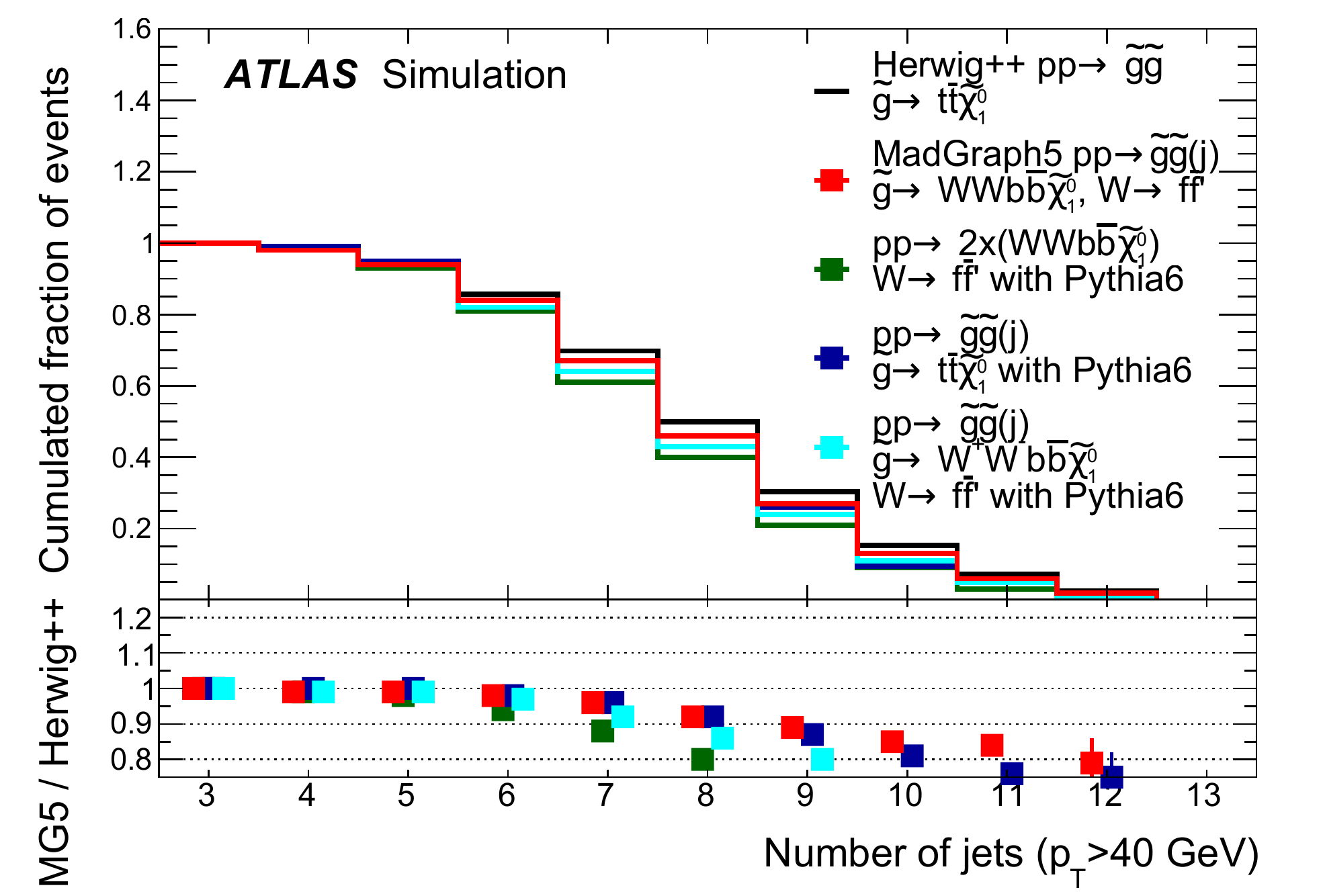}}
\subfigure[Cumulative distribution of the number of $b$-quarks, $\pt>50$~\GeV, $|\eta|<2.5$]{\includegraphics[width=0.49\textwidth]{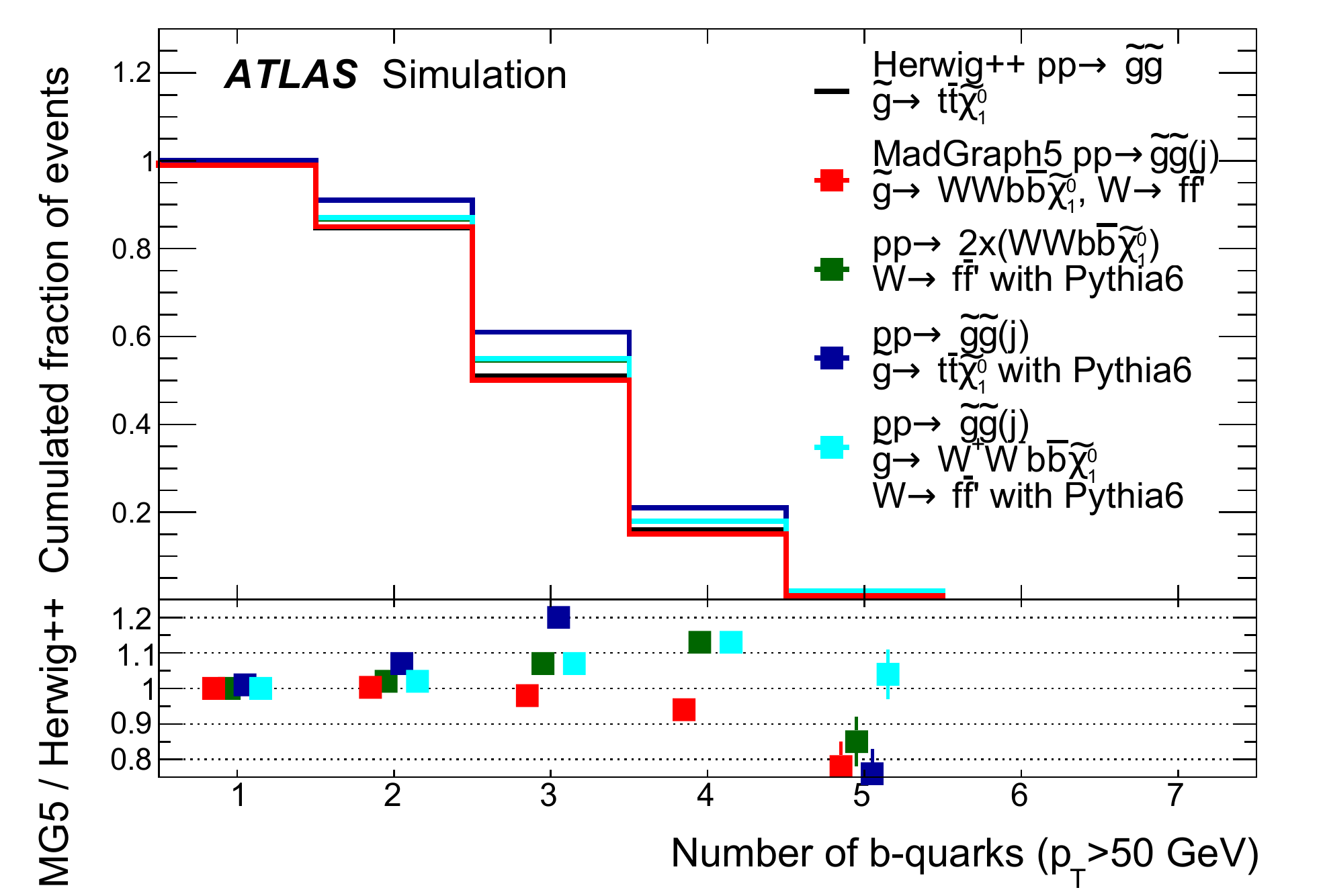}}
\caption{Validation of the {\sc Madgraph5+Pythia6} setup used to generate five-body gluino decays $\gluino\to W^{+}W^{-}b\bar{b}\ninoone$:  
kinematic distributions obtained with the nominal configuration (red markers)
are compared to the reference \herwig++ sample (black), 
for a signal scenario featuring only three-body gluino decays $\gluino\to t\bar{t}\ninoone$ ($m_{\gluino}=1$~\TeV, $m_{\ninoone}=100$~\GeV). 
A few alternative (and simpler) generator configurations are also shown (other coloured markers),
but they fail to reach a satisfactory level of agreement with the reference,
for the typical signal region requirements (cf. SR3b in table~\ref{tab:SS3LSRdefs}).
The distributions are built from the outgoing particles provided by the generators. 
\label{fig:gtt5body_mcval}}
\end{figure}
\newpage
\section{Summary of selection criteria}
\label{AppSRdefs}

Tables \ref{tab:MonojetSRdefs}--\ref{tab:3bjetsSRdefs} summarize the selection criteria for signal regions listed in table \ref{tab:signal_regions} which have been defined in previous ATLAS publications \cite{MonojetPaper,0-leptonPaper,multijetsPaper,1lepPaper,dilepton-edgePaper,SS3LPaper,TauStrongPaper,3bjetsPaper}.

\begin{table}[H]
  \footnotesize
  \begin{center}\renewcommand\arraystretch{1.4}
    \begin{tabular}{|c| c | c | c |}
      \hline
      \multirow{2}{*}{Requirement}      &\multicolumn{3}{|c|}{Signal region} \\
  \cline{2-4}
  & {\bf ~~M1~~} & {\bf ~~M2~~} & {\bf M3} \\ \hline \hline
Jets & \multicolumn{3}{|c|}{At most three jets with }\\ 
preselection & \multicolumn{3}{|c|}{$\pt > 30$~GeV and $|\eta|<2.8$}\\ \hline
$\pt^{\rm jet_1}$ [GeV] $>$ & 280  & 340  & 450 \\ \hline
\met [GeV] $>$  &  220  &  340  & 450 \\ \hline
$\Delta\phi(\textrm{jet},\bm{E}\mathrm{^{miss}_T})$ $>$ &\multicolumn{3}{|c|}{0.4} \\ \hline
\end{tabular}
\caption{\label{tab:MonojetSRdefs} Selection criteria used to define the three signal regions in the search with at least one high-$\pt$ jet and large missing transverse momentum (Monojet) \cite{MonojetPaper}. The azimuthal separation $\Delta\phi(\textrm{jet},\bm{E}\mathrm{^{miss}_T})$ is calculated between the missing transverse momentum direction and each of the selected jets. }
  \end{center}
\end{table}

\begin{table}[H]
  \footnotesize
  \begin{center}\renewcommand\arraystretch{1.4}
    \begin{tabular}{|l|c |c|c| c|c| c|}
      \hline
      \multirow{2}{*}{Requirement}      &\multicolumn{6}{|c|}{Signal region} \\
  \cline{2-7}
  & {\bf 2jl} & {\bf 2jm} & {\bf 2jt} & {\bf 2jW} & {\bf 3j} & {\bf 4jW}\\
 \hline  \hline
\met [GeV] $>$&\multicolumn{6}{|c|}{ 160 }\\ \hline
$\pt^{\rm jet_1}$ [GeV] $>$&\multicolumn{6}{|c|}{ 130 }\\ \hline
$\pt^{\rm jet_2}$ [GeV] $>$&\multicolumn{6}{|c|}{ 60 }\\ \hline
$\pt^{\rm jet_3}$ [GeV] $>$&\multicolumn{4}{|c|}{--} &60 &40 \\ \hline 
$\pt^{\rm jet_4}$ [GeV] $>$&\multicolumn{5}{|c|}{--} &40 \\ \hline 
$\Delta\phi(\textrm{jet}_{1,2,(3)},\bm{E}\mathrm{^{miss}_T})_\mathrm{min}$ $>$ &\multicolumn{6}{|c|}{0.4} \\ \hline
$\Delta\phi(\textrm{jet}_{i>3},\bm{E}\mathrm{^{miss}_T})_\mathrm{min}$ $>$ &\multicolumn{5}{|c|}{--} &0.2 \\ \hline
$W$ candidates &\multicolumn{3}{|c|}{--} &2$(W\to j)$ &--  &$(W\to j)\,+\,(W\to jj)$ \\ \hline
$\met/\sqrt{H_{\rm T}}$ [GeV$^{1/2}$] $>$ &8 &\multicolumn{2}{|c|}{15} &\multicolumn{3}{|c|}{--}  \\ \hline
$\met/\meff^{N_{\rm{j}}}$ $>$ &\multicolumn{3}{|c|}{--} & 0.25 & 0.3 & 0.35 \\ \hline
$ \meff^{\rm incl}$ [GeV] $>$ & 800 &1200 &1600 &1800 &2200 &1100 \\ \hline
\end{tabular}
    \begin{tabular}{|l|c |c|c|c|c|c|c|c|c|}
      \hline
      \multirow{2}{*}{Requirement}      &\multicolumn{9}{|c|}{Signal region} \\
  \cline{2-10}
   & {\bf 4jl-} & {\bf 4jl} & {\bf 4jm} & {\bf 4jt} & {\bf 5j} & {\bf 6jl} & {\bf 6jm} & {\bf 6jt} & {\bf 6jt+} \\
 \hline  \hline
\met [GeV] $>$&\multicolumn{9}{|c|}{ 160 }\\ \hline
$\pt^{\rm jet_1}$ [GeV] $>$&\multicolumn{9}{|c|}{ 130 }\\ \hline
$\pt^{\rm jet_2}$ [GeV] $>$&\multicolumn{9}{|c|}{ 60 }\\ \hline
$\pt^{\rm jet_3}$ [GeV] $>$ &\multicolumn{9}{|c|}{60}  \\ \hline
$\pt^{\rm jet_4}$ [GeV] $>$ &\multicolumn{9}{|c|}{60}  \\ \hline
$\pt^{\rm jet_5}$ [GeV] $>$&\multicolumn{4}{|c|}{--} &\multicolumn{5}{|c|}{60} \\ \hline
$\pt^{\rm jet_6}$ [GeV] $>$&\multicolumn{5}{|c|}{--} &\multicolumn{4}{|c|}{60}  \\ \hline
$\Delta\phi(\textrm{jet}_{1,2,(3)},\bm{E}\mathrm{^{miss}_T})_\mathrm{min}$ $>$ &\multicolumn{9}{|c|}{0.4} \\ \hline
$\Delta\phi(\textrm{jet}_{i>3},\bm{E}\mathrm{^{miss}_T})_\mathrm{min}$ $>$ &\multicolumn{9}{|c|}{0.2}\\ \hline
$\met/\sqrt{H_{\rm T}}$ [GeV$^{1/2}$] $>$  &\multicolumn{2}{|c|}{10} &\multicolumn{7}{|c|}{--} \\ \hline
$\met/\meff^{N_{\rm{j}}}$ $>$ &\multicolumn{2}{|c|}{--} &0.4 &0.25 &\multicolumn{3}{|c|}{0.2} &0.25 &0.15 \\ \hline
$ \meff^{\rm incl}$ [GeV] $>$ &700 &1000 &1300 & 2200 &1200 &900 &1200 &1500 &1700\\ \hline
\end{tabular}
\caption{\label{tab:0LeptonSRdefs} Selection criteria used to define the fifteen signal regions in the search with at least two to at least six jets, significant $\met$ and the absence of isolated electrons or muons (0-lepton + 2--6 jets + $\met$) \cite{0-leptonPaper}. Each signal region is labelled with 
the inclusive jet multiplicity considered (`2j', `3j', etc.) together with the degree of background rejection.
The latter is denoted by labels `l-' (`very loose'), `l' (`loose'), `m' (`medium'), `t' (`tight') and `t+' (`very tight'). The $\met/\meff^{N_{\rm{j}}}$ requirement in any $N_{\rm j}$-jet channel uses a value of $\meff$ constructed from only the leading $N_{\rm j}$ jets ($\meff^{N_{\rm{j}}}$).  The final $\meff^{\rm incl}$ selection, which is used to define the signal regions, includes all jets with $\pt>40~\GeV$. The variable $H_{\rm T}$ is defined as the scalar sum of the transverse momenta of all jets with $\pt >$ 40 \GeV. The azimuthal separation $\Delta\phi(\textrm{jet},\bm{E}\mathrm{^{miss}_T})_\mathrm{min}$ is defined to be the smallest of the azimuthal separations between $\bm{E}\mathrm{^{miss}_T}$ and the reconstructed jets. 
In SR 2jW and SR 4jW a requirement 60 GeV $<m(W_{\rm cand})<$ 100 GeV is placed on the masses of candidate resolved or unresolved hadronically decaying $W$ bosons. Candidate $W$ bosons are reconstructed from single high-mass jets (unresolved candidates; $W \to j$ in the table) or from pairs of jets (resolved candidates; $W \to jj$ in the table.}
  \end{center}
\end{table}

\begin{table}[H]
\footnotesize
\renewcommand\arraystretch{1.4}
\begin{center}
\begin{tabular}{| c | c | c | c | c | c | c | c | c | c | c | c | c | c |}
      \hline
  \multirow{2}{*}{Requirement} & \multicolumn{13}{|c|}{Signal regions in multi-jet + flavour stream} \\ 
  \cline{2-14}  
 & \multicolumn{3}{|c|}{~\BSR{8j50}~} & \multicolumn{3}{|c|}{ ~\BSR{9j50}~} &
            { ~\BSR{10j50}~} &
              \multicolumn{3}{|c|}{ ~\BSR{7j80}~} & \multicolumn{3}{|c|}{ ~\BSR{8j80}~} \\ \hline \hline
 $|\eta|^{\rm jet}$ $<$ &
  \multicolumn{13}{|c|}{$2.0$} \\ \hline
 $\pt^{\rm jet}$ [GeV] $>$&
  \multicolumn{7}{|c|}{$50$} & 
  \multicolumn{6}{|c|}{$80$} \\ \hline
$N_{\rm jet}$ &
  \multicolumn{3}{|c|}{$=8$} &   \multicolumn{3}{|c|}{$=9$} & $\geq10$ & 
  \multicolumn{3}{|c|}{$=7$} &   \multicolumn{3}{|c|}{$\geq8$}  \\ \hline
$N_{b\mathrm{-jet}}$ & \multirow{2}{*}{0} & \multirow{2}{*}{1}  & \multirow{2}{*}{$\geq 2$} &
                             \multirow{2}{*}{0} & \multirow{2}{*}{1} & \multirow{2}{*}{$\geq 2$} & \multirow{2}{*}{---} & 
                             \multirow{2}{*}{0} & \multirow{2}{*}{1} & \multirow{2}{*}{$\geq 2$} &
                             \multirow{2}{*}{0} & \multirow{2}{*}{1} & \multirow{2}{*}{$\geq 2$} \\ 
$(\pt> 40\,\GeV, |\eta|<2.5)$ & & & &
                              & &  &  & 
                              & &  &
                              & &  \\ \hline
$\met/\sqrt{H_{\rm T}}$ [GeV$^{1/2}$] $>$&  \multicolumn{13}{|c|}{$4 $} \\
\hline
\end{tabular}

\begin{tabular}{| c |  c | c | c |}
      \hline
   \multirow{2}{*}{Requirement} & \multicolumn{3}{|c|}{Signal regions in multi-jet + \MJ{} stream} \\ 
 \cline{2-4}  
 &    { ~\BSR{8j50}~} & { ~\BSR{9j50}~} & { ~\BSR{10j50}~}\\ \hline \hline
 $|\eta| ^{\rm jet}$ $<$ &
  \multicolumn{3}{|c|}{$2.8$} \\ \hline
 $\pt^{\rm jet}$ [GeV] $>$&
  \multicolumn{3}{|c|}{$50$} \\ \hline
$N_{\rm jet}$  &
  $\geq8$ & $\geq 9$ & $\geq10$ \\ \hline
\MJ{} [GeV]  & \multicolumn{3}{|c|}{$>340$ and $>420$ for each case}  \\ \hline
$\met/\sqrt{H_{\rm T}}$ [GeV$^{1/2}$] $>$&  \multicolumn{3}{|c|}{$4 $}\\
\hline
\end{tabular}
\caption{ \label{tab:MultijetsSRdefs} Selection criteria used to define the nineteen signal regions in the search with at least seven to at least ten jets, significant $\met$ and the absence of isolated electrons or muons (0-lepton + 7--10 jets + $\met$) \cite{multijetsPaper}. 
The four-momenta of the $R$=0.4 jets satisfying $\pT>20\,$GeV and $|\eta|<2.8$ are used as inputs
to a second iteration of the anti-$k_t$ jet algorithm, this time using the larger distance parameter $R$=1.0.
The resulting larger objects are denoted as composite jets.
The selection variable \MJ{} is then defined to be the sum 
of the masses of the composite jets: 
$\MJ \equiv \sum_{j} m_{j}^{R=1.0}$, where the sum is over the composite jets that satisfy 
$p_{\rm T}^{R=1.0} > 100 \GeV$ and  $|\eta^{R=1.0}|<1.5$. 
The variable  $H_{\rm T}$ is defined as the scalar sum of $\pt$ of all jets with $\pt > $ 40 \GeV\ and $|\eta| < $2.8.}
\end{center}
\end{table}

\begin{table}[H]
\begin{center}
\footnotesize
\renewcommand\arraystretch{1.4}
\vspace*{-0.03\textheight}\begin{tabular}{|l|c|c|c|c|}
\hline
      \multirow{3}{*}{Requirement}      &\multicolumn{4}{|c|}{Signal region} \\
  \cline{2-5}
 & \multicolumn{3}{c|}{\bf Single-bin (binned) soft single-lepton} & {\bf Soft dimuon} \\
   \cline{2-5}
 & {\bf 3-jet} & {\bf 5-jet} & {\bf 3-jet inclusive} & {\bf 2-jet} \\\hline \hline
$N_{\ell}$ & \multicolumn{3}{c|}{1 electron or muon} & 2 muons\\\hline
\ptl [\GeV]   & \multicolumn{3}{c|}{ [7,25] for electron, [6,25] for muon} & [6,25]\\\hline
Lepton veto & \multicolumn{4}{c|}{ No additional electron or muon with \pt$>$ 7 \GeV~or 6 \GeV, respectively}  \\\hline
$m_{\mu\mu}$ [\GeV] & $-$ & $-$ & $-$ & [15,60] \\\hline
$N_{\mathrm{jet}}$  & [3,4] & $\geq$ 5 & $\geq$ 3 & $\geq 2$ \\\hline
\pt$^{\mathrm{jet}}$[\GeV] $>$ & 180, 25, 25 & 180, 25, 25, 25, 25 & 130, 100, 25 & 80, 25 \\\hline
$N_{b\mathrm{-jet}}$ &  $-$ &$-$ & 0 & 0\\\hline
\hline
\met\ [\GeV]  $>$ & 400 & 300 & \multicolumn{2}{c|}{ 180} \\\hline
\mt\ [\GeV]  $>$  & \multicolumn{2}{c|}{100 } &120 & $40$ \\\hline
\met/$m_{\mathrm{eff}}^{\mathrm{incl}}$ $>$ & \multicolumn{2}{|c|}{$0.3$ $(0.1)$} & $0.1$  & $ 0.3$ \\\hline
$\Delta R_{\mathrm{min}}(\mathrm{jet},\ell)$ $>$  & $1.0$ & $-$ & $-$ & $1.0$ ($\mathrm{2^{nd}}$ muon)\\\hline
Binned variable & \multicolumn{3}{c|}{(\met/$m_{\mathrm{eff}}^{\mathrm{incl}}$ in 4 bins)} & $-$ \\\hline
Bin width & \multicolumn{3}{c|}{(0.1, $4^{\mathrm{th}}$ is inclusive)} & $-$ \\\hline
\end{tabular}

\begin{tabular}{|l|c|c|c|}
\hline
      \multirow{3}{*}{Requirement}      &\multicolumn{3}{|c|}{Signal region} \\
  \cline{2-4}
 & \multicolumn{3}{c|}{\bf Single-bin (binned) hard single-lepton} \\
   \cline{2-4}
 & {\bf 3-jet} & {\bf 5-jet} & {\bf 6-jet} \\\hline \hline
$N_{\ell}$ & \multicolumn{3}{c|}{1 electron or muon} \\\hline
\ptl [\GeV]  $>$  & \multicolumn{3}{c|}{ 25 }\\\hline
Lepton veto & \multicolumn{3}{c|}{ \pt$^{\mathrm{2^{nd} lepton}}<$ 10 \GeV} \\\hline
$N_{\mathrm{jet}}$  & $\geq$ 3 & $\geq$ 5 & $\geq$ 6 \\\hline
\pt$^{\mathrm{jet}}$[\GeV] $>$ & 80, 80, 30 & 80, 50, 40, 40, 40 &  80, 50, 40, 40, 40, 40\\\hline
Jet veto & (\pt$^{\mathrm{5^{th} jet}}< 40$ \GeV) & (\pt$^{\mathrm{6^{th} jet}}< 40$ \GeV) & $-$ \\\hline
\met\ [\GeV] $>$  & 500 (300) & 300&  350 (250) \\\hline
\mt\ [\GeV] $>$   & 150  & 200 (150) & 150 \\\hline
\met/$m_{\mathrm{eff}}^{\mathrm{excl}}$ $>$ &  0.3 & $-$ & $-$  \\\hline
$m_{\mathrm{eff}}^{\mathrm{incl}}$ [\GeV] $>$ & \multicolumn{2}{c|}{ 1400 (800)} & 600 \\\hline
Binned variable & \multicolumn{2}{c|}{($m_{\mathrm{eff}}^{\mathrm{incl}}$ in 4 bins)} & (\met~in 3 bins) \\\hline  
Bin width & \multicolumn{2}{c|}{(200 \GeV, $4^{\mathrm{th}}$ is inclusive)} & (100 \GeV, $3^{\mathrm{rd}}$ is inclusive) \\\hline      
\end{tabular}
\caption{\label{tab:1LeptonSRdefs} Selection criteria used to define the signal regions in the search requiring at least one isolated lepton (1-lepton (soft+hard) + jets + \met) and in the search requiring two soft muons (2-leptons + jets + \met) \cite{1lepPaper}. 
For each jet multiplicity in the single-lepton channel, two sets of requirements are defined: one single-bin signal region optimized for discovery reach, which is also used to place limits on the visible cross-section, and one signal region which is binned in an appropriate variable in order to exploit the expected shape of the distribution of signal events when placing model-dependent limits. 
The requirements of the binned signal region are shown in parentheses when they differ from those of the single-bin signal region. 
The transverse mass ($m_{\rm{T}}$) of the lepton ($\ell$)
and ${\boldsymbol E}_{\mathrm{T}}^\mathrm{miss}$ is defined as
    $m_{\mathrm{T}} =
\sqrt{2 p_{\mathrm{T}}^{\ell} \met
  (1-\cos[\Delta\phi(\vec{\ell},{\boldsymbol E}_{\mathrm{T}}^\mathrm{miss})])}$. 
The inclusive effective mass ($m_{\rm{eff}}^{\rm{inc}}$) is computed as the scalar sum of
the \pt~of the lepton(s), the jets and \met:
   $ m_{\mathrm{eff}}^{\mathrm{inc}} = 
    \sum_{i=1}^{{N}_{\ell}}p_{\mathrm{T},i}^{\ell} +  \sum_{j=1}^{{N}_\mathrm{jet}}
    p_{\mathrm{T},j} + \met$, 
where the index $i$ identifies all the signal leptons and the index $j$ all the signal jets in the event. 
The exclusive effective mass ($m_{\rm{eff}}^{\rm{excl}}$) is defined in a similar way to $m_{\rm{eff}}^{\rm{inc}}$, with
the exception that only the three leading signal jets are considered. The minimum angular separation $\Delta R_{\mathrm{min}}$ calculated between the signal lepton $\ell$ and all preselected jets is used to reduce the background coming from misidentified or non-prompt leptons in the soft-lepton signal region with three jets and in the soft dimuon signal region. In the latter case, the subleading signal muon 
is used to compute $\Delta R_{\mathrm{min}}$.
}
\end{center}
\end{table}

\begin{table}[H]
\begin{center}
\footnotesize
\renewcommand\arraystretch{1.4}
\begin{tabular}{|l|c|c|c|c|c|}
\hline
      \multirow{4}{*}{Requirement}      &\multicolumn{4}{|c|}{Signal region} \\
  \cline{2-5}
 & \multicolumn{4}{c|}{\bf Single-bin (binned) hard dilepton} \\   \cline{2-5}
 & \multicolumn{2}{c|}{{\bf Low-multiplicity ($\le2$-jet) }} & \multicolumn{2}{c|}{{\bf 3-jet}}  \\   \cline{2-5}
 & {\bf $ee/\mu\mu$} & {\bf $e\mu$} & {\bf $ee/\mu\mu$} & {\bf $e\mu$}\\   \hline \hline
$N_{\ell}$ & \multicolumn{4}{c|}{$2$, $2$ of opposite sign or $\ge 2$} \\\hline
\ptl [\GeV]  $>$  & \multicolumn{4}{c|}{ 14,10 } \\\hline
$N_{\ell\ell}$ with 81$<m_{\ell\ell}<$101 \GeV & 0 & $-$ & 0 & $-$ \\\hline
$N_{\mathrm{jet}}$  & \multicolumn{2}{c|}{$\leq$ 2} & \multicolumn{2}{c|}{$\geq$ 3} \\\hline
\pt$^{\mathrm{jet}}$[\GeV] $>$ & \multicolumn{2}{c|}{50,50} & \multicolumn{2}{c|}{ 50, 50, 50} \\\hline
$N_{b\mathrm{-jet}}$ &\multicolumn{4}{c|}{0} \\\hline
$R$   & \multicolumn{2}{c|}{$>$0.5} & \multicolumn{2}{c|}{$>$0.35} \\\hline
$M_{R}^{'}$ [\GeV] $>$ & \multicolumn{2}{c|}{$600$ ($400$ in 8 bins)} & \multicolumn{2}{c|}{$800$ ($800$ in 5 bins)} \\\hline
$M_{R}^{'}$ bin width [\GeV] & \multicolumn{4}{c|}{(100, the last is inclusive )} \\\hline
\end{tabular}
\caption{\label{tab:dileptonSRdefs} Selection criteria used to define the signal regions in the search requiring two hard leptons (2-leptons + jets + \met) \cite{1lepPaper}. 
For each jet multiplicity two sets of requirements are defined: one single-bin signal region optimized for discovery reach, which is also used to place limits on the visible cross-section, and one signal region which is binned in an appropriate variable in order to exploit the expected shape of the distribution of signal events when placing model-dependent limits. 
The requirements of the binned signal region are shown in parentheses when they differ from those of the single-bin signal region. 
Details of the construction of Razor variables $M_{R}^{'}$ and $R$ can be found in the appendix \ref{app:razorvars}. In this case, mega-jets are constructed using the final-state jets and leptons. 
}
\end{center}
\end{table}

\begin{table}[H]
\begin{center}
\footnotesize
\renewcommand\arraystretch{1.4}
\begin{tabular}{|l|c|c|c|c|c|}
\hline
      \multirow{2}{*}{Requirement}      &\multicolumn{5}{|c|}{Signal region} \\
  \cline{2-6}
 & {\bf SR-2j-bveto} & {\bf SR-2j-btag}  & {\bf SR-4j-bveto}  & {\bf SR-4j-btag} & {\bf SR-loose}  \\ \hline \hline
$N_{\rm jet}$ $\geq$ &  $\geq$ 2 & $\geq$ 2  & $\geq$ 4  & $\geq$ 4  & (2, $\geq$ 3)  \\ \hline
$N_{b\mathrm{-jet}}$  & = 0  & $\geq$ 1  & = 0  & $\geq$ 1  &  -- \\ \hline
$\met$ [GeV] $>$ & 200  & 200  & 200  & 200  & (150, 100)  \\ \hline
$m_{\rm \ell \ell}$ [GeV] $\slashed{\in}$ & [80, 110]  & [80, 110]  & [80, 110]  & [80, 110]  & [80, 110]  \\ \hline

\end{tabular}
\caption{\label{tab:offZSRdefs} Selection criteria used to define the signal regions in the search requiring two same-flavour opposite-sign electrons or muons (2-leptons off-Z) \cite{dilepton-edgePaper}.  If more than two leptons are present, the two with the largest values of \pt\ are selected. The leading lepton in the event must have $\pt >$ 25 \GeV\ and the subleading lepton is required to have $\pt >$ 20 \GeV. These two leptons are used to define the dilepron invariant mass, $m_{\rm \ell \ell}$. In addition, one SR with the same requirements as those used in the CMS search  \cite{Khachatryan:2015lwa}, which reported an excess of events above the SM background with a significance of 2.6 standard deviations, is defined (SR-loose) for comparison purposes.  
}
\end{center}
\end{table}

\begin{table}[H]
\footnotesize
\renewcommand\arraystretch{1.4}
\begin{center}

\begin{tabular}{|l|c|c|c|c|c|}
\hline
      \multirow{2}{*}{Requirement}      &\multicolumn{5}{|c|}{Signal region} \\
  \cline{2-6}
 & {\bf SR3b} & {\bf SR0b}  & {\bf SR1b}  & {\bf SR3Llow} & {\bf SR3Lhigh}  \\ \hline \hline
 Leptons & SS or 3L & SS & SS & 3L & 3L \\ \hline 
$N_{b\mathrm{-jet}}$  & $\geq$ 3 & =0 & $\geq$ 1 & - & - \\ \hline
$N_{\rm jet}$ $\geq$ &  5 & 3 & 3 & 4 & 4 \\ \hline
$\met$ [GeV] & &  $>$ 150 & $>$ 150 & 50 $<$ \met $<$ 150 & $>$ 150  \\ \hline
$\mt$ [GeV] $>$ & - & 100 & - & - & - \\ \hline
Veto  & - & - & SR3b & $Z$ boson, SR3b & SR3b\\ \hline
$\meff$ [GeV] $>$ & 350 & 400 & 700 & 400 & 400 \\ \hline
\end{tabular}
\caption{\label{tab:SS3LSRdefs} Selection criteria used to define the signal regions in the search with multiple jets, and either two leptons of the same electric charge (same-sign leptons) or at least three leptons (SS/3L + jets + \met)  \cite{SS3LPaper}. 
The effective mass (\meff) is computed from all selected leptons and selected jets in event, as 
$m_{\mathrm{eff}}=\sum_{i=1}^{{N}_{\ell}}p_{\mathrm{T},i}^{\ell} +  \sum_{j=1}^{{N}_\mathrm{jet}} p_{\mathrm{T},j} + \met$. 
The transverse mass ($\mt$) is computed from the highest-\pt\ lepton ($\ell_1$) and ${\boldsymbol E}_{\mathrm{T}}^\mathrm{miss}$ 
as $\mt= \sqrt{2 \pt^{\ell_1} \met(1-\cos[\Delta\phi(\vec{\ell_1}, {\boldsymbol E}_{\mathrm{T}}^\mathrm{miss})])}$. 
}
\end{center}
\end{table}

\begin{table}[H]
\footnotesize
\renewcommand\arraystretch{1.4}
\begin{center}

 \begin{tabular}{| l|c|c |}
  \hline
        \multirow{2}{*}{Requirement}      &\multicolumn{2}{|c|}{Signal region} \\
  \cline{2-3}
&{\bf \onetau Loose SR} &{\bf \onetau Tight SR}\\ \hline \hline
Taus		&\multicolumn{2}{|c|}{$N_\tau^\text{medium}=1$}\\
         &\multicolumn{2}{|c|}{$\pT>30$\,GeV}\\  \hline
$ \Delta\phi(\text{jet}_{1,2},\bm{E}\mathrm{^{miss}_T}) >$	&\multicolumn{2}{|c|}{ 0.4 }\\ \hline
$ \Delta \phi(\tau,\bm{E}\mathrm{^{miss}_T}) >$ & \multicolumn{2}{|c|}{0.2 }\\ \hline
$\mt^{\tau}$ [GeV] $>$  & \multicolumn{2}{|c|}{140} \\ \hline
$\met$ [GeV] $>$ & 200 & 300 \\ \hline
$\HT$ [GeV] $>$ & 800 & 1000 \\ \hline
\end{tabular}

 \begin{tabular}{| l|c|c|c|c |}
  \hline
       \multirow{2}{*}{Requirement}      &\multicolumn{4}{|c|}{Signal region} \\
  \cline{2-5} 
			&{\bf \twotau Inclusive SR} &{\bf \twotau GMSB SR} &{\bf \twotau nGM SR}&{\bf \twotau bRPV SR}\\ \hline \hline
Taus		&\multicolumn{4}{|c|}{$N_\tau^\text{loose}\geq2$}\\
         &\multicolumn{4}{|c|}{$\pT>20$\,GeV} \\ \hline
$\Delta\phi(\text{jet}_{1,2},\bm{E}\mathrm{^{miss}_T})\geq $	&\multicolumn{4}{|c|}{0.3}\\ \hline
$m_{\rm T}^{\rm \tau_1} + m_{\rm T}^{\rm \tau_2}$ [GeV] $\geq$ & 150 & 250 & 250 & 150 \\ \hline
$H_{\rm T}^{\rm 2j}$ [GeV] $>$ & 1000 & 1000 & 600 & 1000 \\ \hline
 $N_{\mathrm{jet}}\geq $ & - & \multicolumn{3}{|c|}{4}\\ \hline
 \end{tabular}

 \begin{tabular}{| l|c|c|c|c |}
  \hline
        \multirow{2}{*}{Requirement}      &\multicolumn{4}{|c|}{Signal region} \\
  \cline{2-5}
  &{\bf \tauleps GMSB SR} &{\bf \tauleps nGM SR} & {\bf \tauleps bRPV SR} &{\bf \tauleps mSUGRA SR} \\ \hline \hline
  Taus &\multicolumn{4}{|c|}{$N_\tau^\text{loose}\geq1$}\\[-1.0ex]
  & \multicolumn{4}{|c|}{$\pT>20$\,GeV} \\
  \hline
$N_\ell = $ 	&\multicolumn{4}{|c|}{1}\\ \hline
$m_\text{T}^\ell$ [GeV] $>$ 	&\multicolumn{4}{|c|}{ 100 }\\ \hline
$\meff$ [GeV] $>$ & 1700 & - &1300 & - \\ \hline
$\met$ [GeV] $>$ & - & 350 & - & 300 \\ \hline
$N_{\mathrm{jet}}\geq $ & - & 3 & 4 & 3 \\ \hline
 \end{tabular}
\caption{ Selection criteria used to define the signal regions in the search requiring large missing transverse momentum, jets and at least one hadronically decaying tau lepton (taus + jets + \met) \cite{TauStrongPaper}.
The transverse mass $\mt^{\tau}$ is formed from the ${\boldsymbol E}_{\mathrm{T}}^\mathrm{miss}$ and the \pt ~of  the tau lepton in the \onetau channel as: 
  $\mt^{\tau}=\sqrt{2\pt^{\tau}\met(1-\cos(\Delta\phi(\tau,{\boldsymbol E}_{\mathrm{T}}^\mathrm{miss})))}$.
In addition, the variable $m_{\rm T}^{\rm \tau_1} + m_{\rm T}^{\rm \tau_2}$ is used as a discriminating variable in the \twotau channel. 
The transverse mass $m_\text{T}^{\ell}$ is similarly formed from the ${\boldsymbol E}_{\mathrm{T}}^\mathrm{miss}$ and the \pt ~of the light leptons. 
The variable $\HT$ is defined as the scalar sum of the transverse momenta of the tau, light
 lepton and jets ($\pt^{\rm jet}>$ 30 \GeV): $\HT=\sum_{i=1}^{{N}_{\ell}}\pt^\ell+\sum_{j=1}^{{N}_\mathrm{\tau}}\pt^\tau + \sum_{k=1}^{{N}_\mathrm{jet}}\pt^\mathrm{jet}$. 
 The variable $H_{\rm T}^{\rm 2j}$ is defined as the scalar sum of the transverse momenta of the tau and light lepton candidates, 
and the two jets with the largest transverse momenta in the event: 
$H_{\rm T}^{\rm 2j}=\sum_{i=1}^{{N}_{\ell}}\pt^\ell+\sum_{j=1}^{{N}_\mathrm{\tau}}\pt^\tau + \sum_{k=1,2} \pt^{\mathrm{jet}_k}$.
The effective mass (\meff) is defined as $\meff=H_{\rm T}^{\rm 2j}+\met$.}
\label{tab:TauStrongSRdefs} 
 \end{center}
\end{table}

\begin{table}[H]
\footnotesize
\renewcommand\arraystretch{1.4}
\begin{center}

\begin{tabular}{|l|c|c|c|c|c|c|}
\hline
      \multirow{2}{*}{Requirement}      &\multicolumn{6}{|c|}{Signal region} \\
  \cline{2-7}
 & {\bf SR-0$\ell$-4j-A} & {\bf SR-0$\ell$-4j-B}  & {\bf SR-0$\ell$-4j-C* }  & {\bf SR-0$\ell$-7j-A } & {\bf SR-0$\ell$-7j-B} & {\bf SR-0$\ell$-7j-C} \\ \hline \hline
Baseline 0-lepton selection &  \multicolumn{6}{|c|}{ lepton veto, $p_{\mathrm{T}}^{\rm jet_1} > 90$ \GeV, \met $>$ 150 \GeV}\\ \hline 
 $N$ jets (\pt\ [\GeV]) $\geq$ & 4 (50) & 4 (50) & 4 (30) & 7 (30) & 7 (30) & 7 (30)  \\ \hline
 \met\ [\GeV] $>$ & 250 & 350 & 400 & 200 & 350 & 250 \\ \hline
 $\meff^{\rm incl}$ [\GeV] $>$ & -  & - & - & 1000 & 1000 & 1500  \\ \hline
 $\meff^{\rm 4j}$ [\GeV] $>$ &1300 & 1100 & 1100 & - & - & -  \\ \hline
  \met/$\sqrt{H_{\rm T}^{\rm 4j}}$  [$\sqrt{\GeV}$] $>$   & - & - & 16 & - & - & - \\ \hline 
 \end{tabular}
 
 \begin{tabular}{|l|c|c|c|}
\hline
      \multirow{2}{*}{Requirement}      &\multicolumn{3}{|c|}{Signal region} \\
  \cline{2-4}
 & {\bf SR-1$\ell$-6j-A} & {\bf SR-1$\ell$-6j-B}  & {\bf SR-1$\ell$-6j-C}   \\ \hline \hline
Baseline 1-lepton selection &  \multicolumn{3}{|c|}{$\geqslant$ 1 signal lepton ($e$,$\mu$),  $p_{\mathrm{T}}^{\rm jet_1} > 90$ \GeV, \met $>$ 150 \GeV}\\ \hline 
 $N$ jets (\pt\ [\GeV]) $\geq$ & 6 (30) & 6 (30) & 6 (30)   \\ \hline
 \met\ [\GeV] $>$ & 175 & 225 & 275  \\ \hline
 \mt [\GeV] $>$ &140 & 140 & 160  \\ \hline
 $\meff^{\rm incl}$ [\GeV] $>$ & 700  & 800 & 900   \\ \hline  
\end{tabular}
\caption{Selection criteria used to define the signal regions in the search that requires at least three jets tagged as $b$-jets, no or at least one lepton, jets and large missing transverse momentum (0/1-lepton + 3$b$-jets + \met) \cite{3bjetsPaper}. 
The jet \pt\ threshold requirements are also applied to $b$-jets.  
The notation SR-0$\ell$-4j-C* means that the leading jet is required to fail the $b$-tagging requirements, in order to target the region close to the kinematic boundary in the gluino--sbottom simplified models. In the 0-lepton selection, the inclusive effective mass $\meff^{\rm incl}$ is defined  as the scalar sum of the \met\ and the $\pt$ 
of all jets with $\pt >$ 30 \GeV. In the 1-lepton selection the  $\meff^{\rm incl}$ is defined as for the 0-lepton selection with the addition of the \pt\ of all selected leptons with $\pt > $ 20 \GeV. The exclusive effective mass ($m_{\mathrm{eff}}^{\mathrm{4j}}$) is defined as the scalar sum of the \met\ and the $\pt$ 
of the four leading jets. The transverse mass (\mt) is computed from the leading lepton and the missing transverse
momentum. The variable $H_{\rm T}^{\rm 4j}$  is defined as the scalar sum of the transverse momenta of the four leading jets. }
 \label{tab:3bjetsSRdefs}
 \end{center}
 \end{table}

\newpage
\section{0-lepton Razor analysis details}
\label{AppRazor}

Many kinematical variables have been used to search for SUSY at hadron colliders, typically making use of the expected heavy mass scale of the SUSY particles produced, and the missing transverse momentum originating from the LSP. The analysis described here searches for squarks in final states with high-$\pt$ jets, missing transverse momentum and no electrons or muons, using the Razor variable set \cite{RazorVariables}. These variables provide longitudinal and transverse information about each event, contribute to the rejection of the background from the multi-jet processes that dominate hadronic collisions, and can be used as an approximation of the mass scale of the produced particles. 

The basic analysis approach relies on the definition of statistically independent regions in the Razor variable phase space, $R$ and $M_R'$, explained in details in the remainder of this section, that are rich in SUSY-like events, and other regions, each dominated by one SM background component.
Following the strategy explained in section \ref{sec:strategy}, the Monte Carlo expectations are normalized to the data in each background control region, and those normalization factors are then transferred to the MC prediction in the SUSY signal regions. 
The baseline object selection and event cleaning, as well as the choice of MC generators for SM background processes and the approach for calculating systematic uncertainties exactly follow those of the 0-lepton + 2--6 jets + $\met$ (0L) search \cite{0-leptonPaper}, and are not discussed here.

\subsection{The Razor variables} \label{app:razorvars}

The Razor variable set is designed to group together visible final-state particles associated with heavy produced sparticles, and in doing so contains information about the mass scale of the directly produced sparticles.
The final-state jets are grouped into two hemispheres called ``mega-jets'', where 
all visible objects from one side of the di-sparticle decay are collected together to create 
a single four-vector, representing the decay products of a single sparticle.
The mega-jet construction involves iterating over all possible combinations of the four-vectors of the visible reconstructed objects,
with the favoured combination being that which minimizes the sum of the squared masses of the mega-jet four-vectors.
Using this mega-jet configuration, with some simplifying assumptions (e.g. symmetric sparticle production), 
the rest frame of the sparticles (the so-called ``$R$-frame'' described in ref.~\cite{RazorVariables}) can be reconstructed, 
and a characteristic mass $M_R'$ can be defined in this frame:

\begin{equation}
	M_{R}' = \sqrt{ ( j_{1,E} + j_{2,E} )^2 - ( j_{1,{\rm L}} + j_{2,{\rm L}} )^2 } ,
\end{equation}

\noindent where $j_{i,{\rm L}}$ denotes the longitudinal momentum, and $j_{i,E}$ the energy in the $R$-frame, of the mega-jet $i$. 

To help reduce the SM backgrounds, a second variable,  $M^R_{\rm T}$, is defined that includes information about the transverse quantities, including the total missing transverse momentum and its angular distance to the two mega-jets.  %
In the di-sparticle decay there are two mega-jets, each with associated \met\ from the escaping LSPs. 
Assigning half of the missing transverse momentum per event to each of the LSPs, $M^R_{\rm T}$ is defined as

\begin{equation}
	M_{\rm T}^R = \sqrt{\frac{ {E}_{\mathrm{T}}^\mathrm{miss}( j_{1,{\rm T}} + j_{2,{\rm T}}) -  {\bm E}_{\mathrm{T}}^\mathrm{miss} \cdot ( {\bm j}_{1,{\rm T}} + {\bm j}_{2,{\rm T}} ) }{2}} ,
\end{equation}

\noindent where ${\bm {j}_{i,{\rm T}}}$ denotes the transverse momentum of the mega-jet $i$. 
The variable $M_{\rm T}^R$ is designed such that for small values of \met, $M_{\rm T}^R$ is also small. If the multi-jet event were perfectly reconstructed then \met $= 0$ and $M_{\rm T}^R$ would also be zero, while in the case where a jet is mis-calibrated, the fake ${\bm E}_{\mathrm{T}}^\mathrm{miss}$ tends to align with one of the mega-jets (that are back-to-back), also creating small values of $M_{\rm T}^R$. For SUSY-like events where the mega-jets tend not to be back-to-back, and their vector sum is opposite to the ${\bm E}_{\mathrm{T}}^\mathrm{miss}$, the quantity $M_{\rm T}^R$ is large. The kinematic endpoint of $M_{\rm T}^R$ is the mass difference between the heavy and the light sparticles. 

Finally, the razor variable is defined as:
\begin{equation}
	R = \frac{M_{\rm T}^R}{M_{R}'} .
\end{equation}

For SUSY-like events, when the mass splitting between the heavier and light sparticles is large, $M_R'$ peaks near the mass of the heavier sparticle and $M_{\rm T}^R$ has a kinematical endpoint at the mass of the heavier sparticle. For SM processes, $R$ tends to have a low value, while it tends to have a broad distribution centred around 0.5 for SUSY-like events. Thus $R$ can be used as a discriminant between signal and background.

\subsection{Signal regions}\label{subsec:SROverview}

The SUSY models targeted by this search are expected to have final states characterized by the presence of
jets, missing transverse momentum, and no leptons. 
In the simplest case of squark-pair production with direct decays to quarks and neutralinos, there are at least two jets visible in the detector, so the baseline inclusive signal regions require at least two jets. This is also the minimum number of visible objects in the final state necessary to construct the Razor variables.
Figure~\ref{fig:RVsMRTwoPtsSquarkDirect} shows the values of $R$ and  $M_R'$ for two simplified model signal points, one with $m_{\squark}=450$ \GeV\ and $m_{\rm LSP}=400$ \GeV, where the $\Delta m =50$ \GeV\ is small (referred to as small-$\Delta m_{\rm signal}$)
and the other with $m_{\squark}=850$ \GeV\ and $m_{\mathrm{LSP}} =100$ GeV, where the $\Delta m =750$ \GeV\ is large (referred to as large-$\Delta m_{\rm signal}$), after requiring no leptons and at least two jets in the final state. 

\begin{figure}[H]                                                                     
\centering
\subfigure[]{\includegraphics[width=0.45\textwidth]{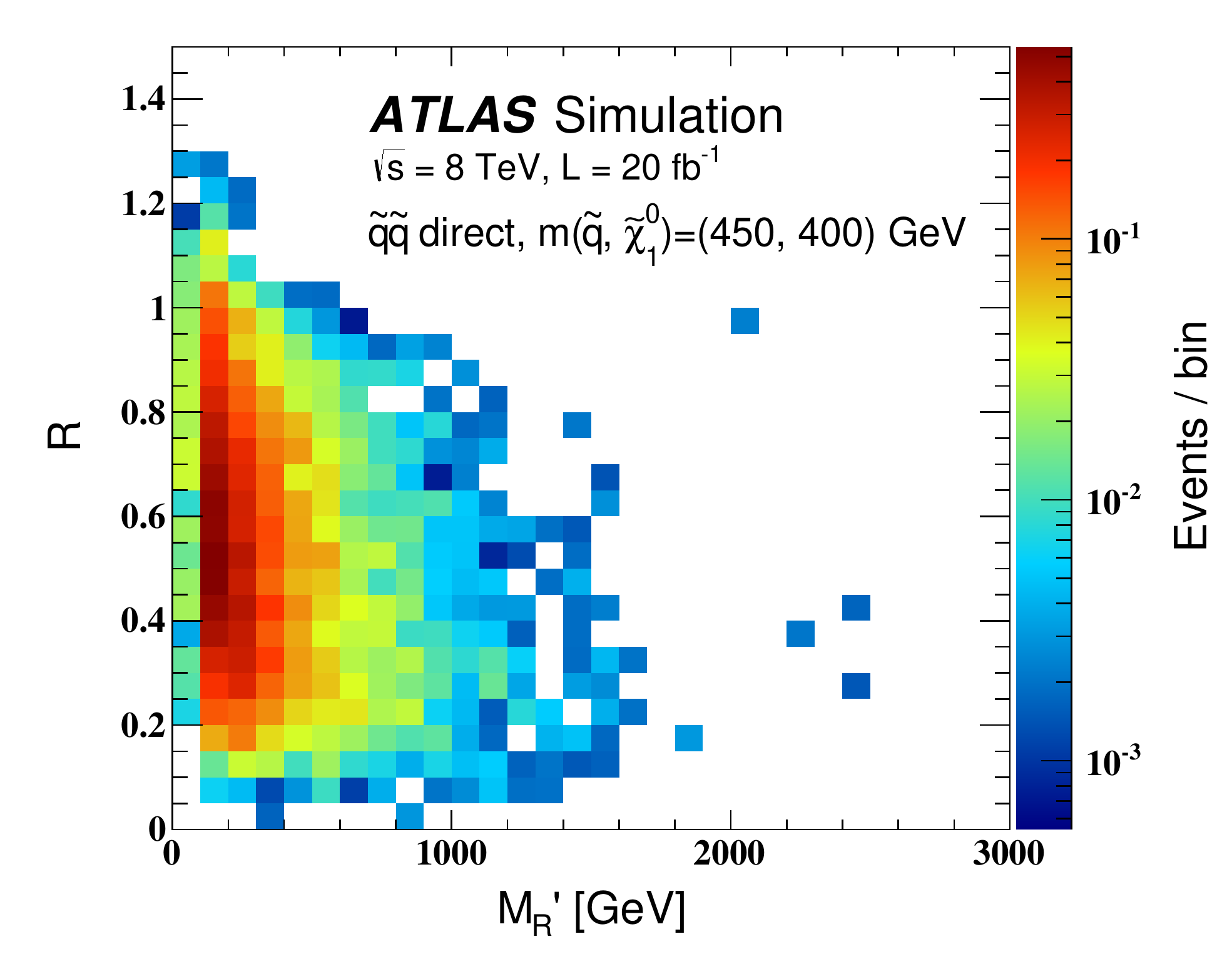}}
\subfigure[]{\includegraphics[width=0.45\textwidth]{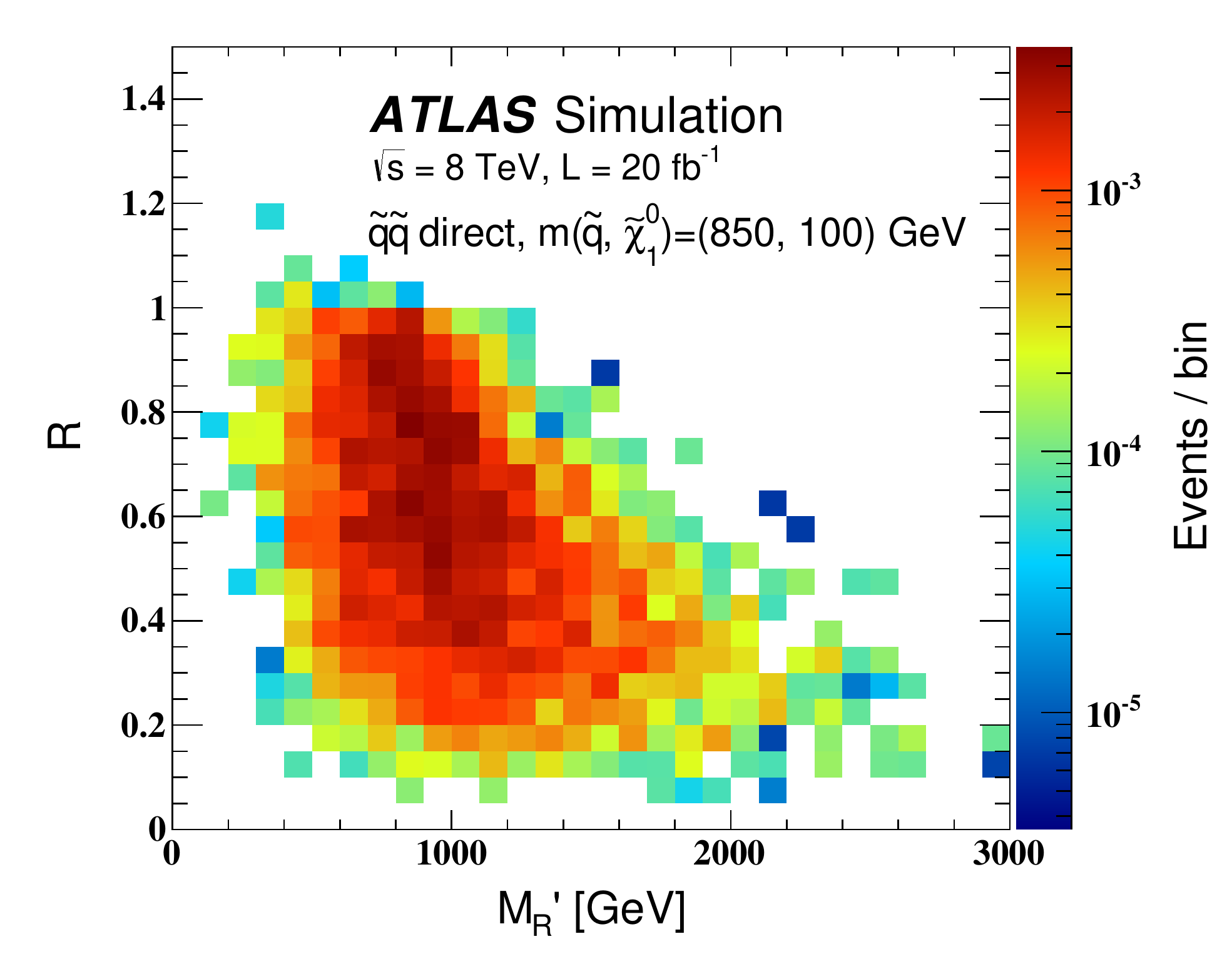}}
\caption{Distributions of the Razor variable $R$ versus the characteristic mass in the $R$-frame $M_R'$ for two points of the simplified model with squark-pair production assuming the direct decay of squarks, after requiring no leptons and at least two jets. Figure (a) shows the case where $m_{\mathrm{\squark}} = 450$~GeV and $m_{\mathrm{LSP}} = 400$~GeV, and (b) where $m_{\mathrm{\squark}} = 850$~GeV and $m_{\mathrm{LSP}} = 100$~GeV. \label{fig:RVsMRTwoPtsSquarkDirect}}
\end{figure}
\begin{figure}[H]                                                                     
\centering
\subfigure[]{\includegraphics[width=0.45\textwidth]{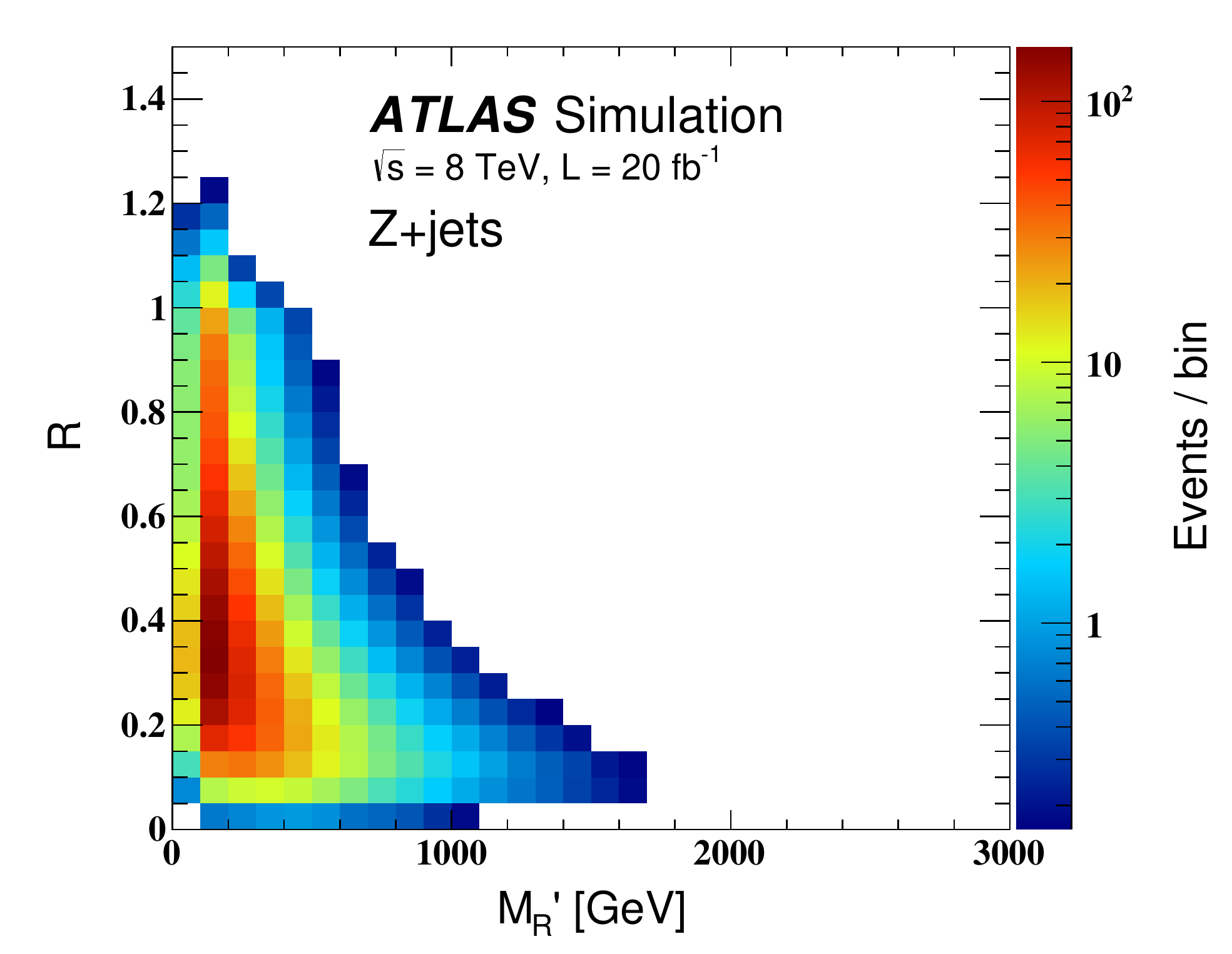}}
\subfigure[]{\includegraphics[width=0.45\textwidth]{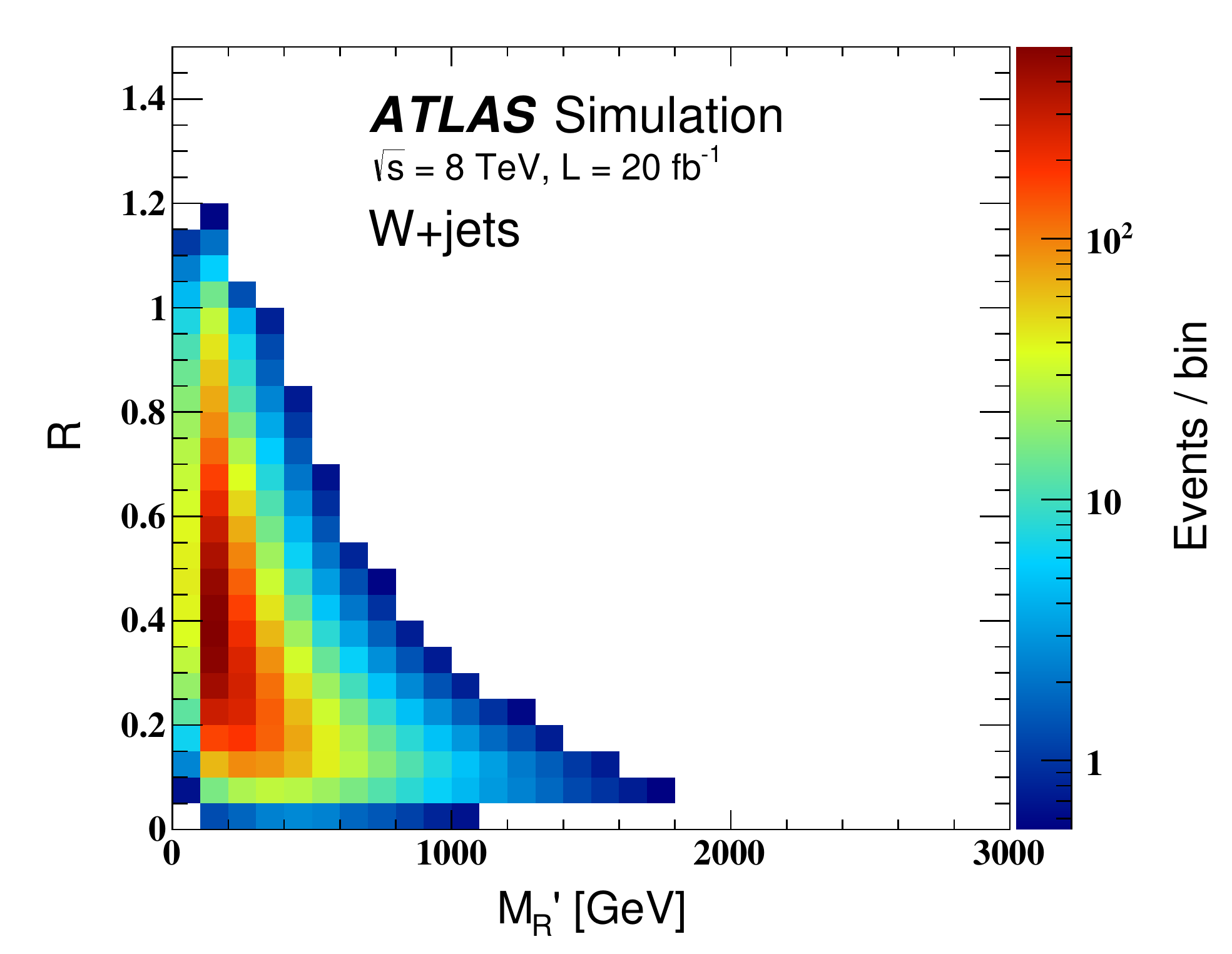}}
\subfigure[]{\includegraphics[width=0.45\textwidth]{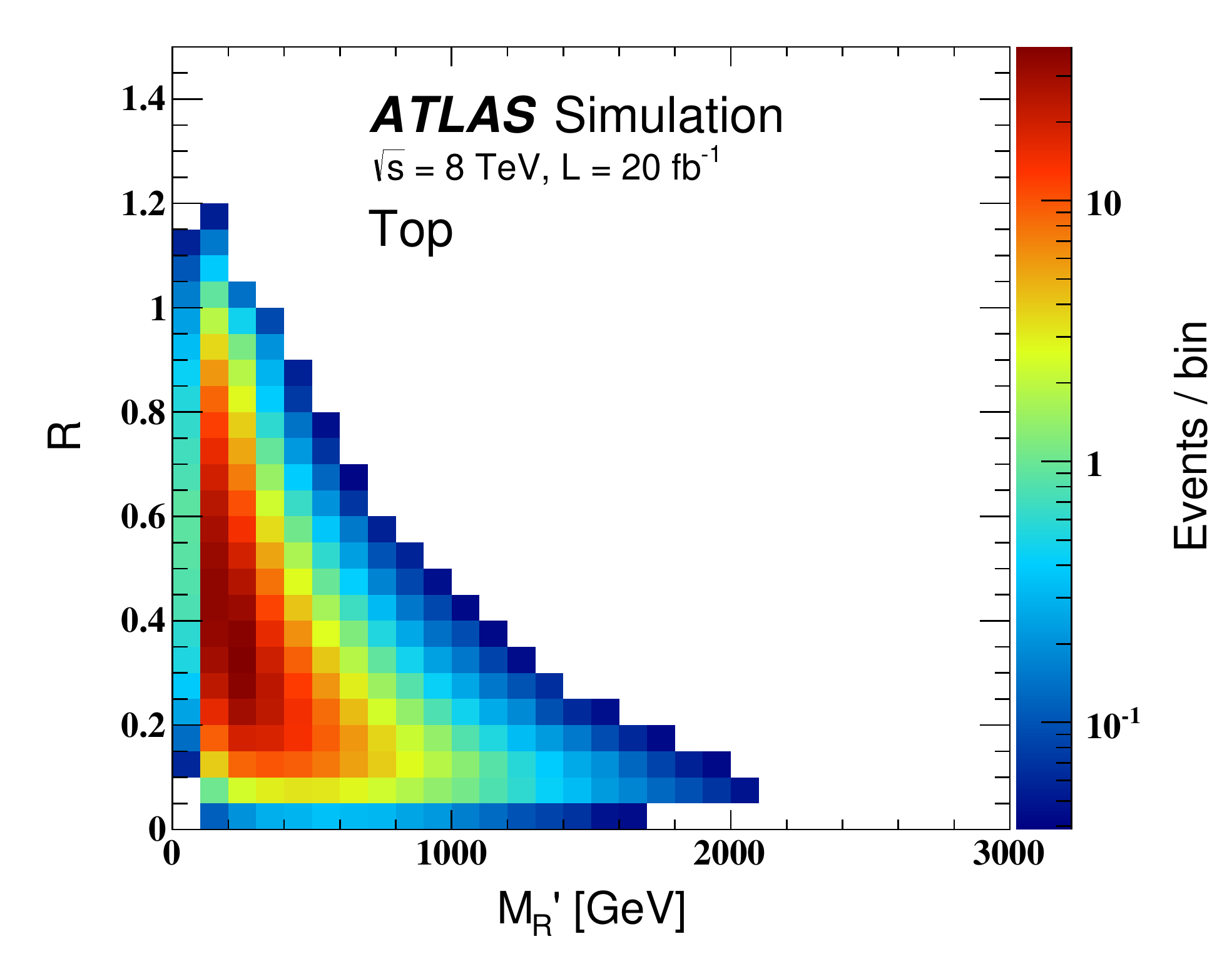}}
\caption{Distributions of the Razor variable $R$ versus the characteristic mass in the $R$-frame $M_R'$ for the dominant Standard Model backgrounds: (a) $Z$+jets, (b) $W$+jets and (c) $\ttbar$, after requiring no-leptons and at least two jets in the final state. \label{fig:RVsMRBaselineBackgrounds}}
\end{figure}

Since the variable $M_R'$ is related to the mass difference between the squark and the neutralino, going from the small-$\Delta m_{\rm signal}$ to the large-$\Delta m_{\rm signal}$, the events tend to populate higher $M_R'$ regions.
Extending this to all points of the model with squark-pair production followed by the direct decay of squarks, the average value of $M_R'$ is approximately constant for a fixed mass splitting between the LSP and the squark, and increases with increasing $\Delta m$, while the average $R$-value tends to be around 0.5.

To select events for this search, the combination of two \met\ triggers, which 
are fully efficient in events having offline reconstructed \met$ > 160$ \GeV\ is used.  Two signal regions which target different regions of the ($m_{\squark}, m_{\ninoone}$) plane are defined:  SR$_{\rm loose}$ and SR$_{\rm tight}$. 
Signal region SR$_{\rm loose}$ has a lower requirement on $R$ and targets regions of the ($m_{\squark}, m_{\ninoone}$) plane with small mass splitting, which typically have softer visible objects. Signal region SR$_{\rm tight}$ was chosen to target high squark masses which typically contain harder visible objects. An overview of the selection criteria for these two signal regions is given in table \ref{tab:Razor_SR}. 

\subsection{Control and validation regions for SM background processes}

The dominant SM background processes which contribute to the event counts in the signal regions are: $Z$+jets, $W$+jets, top quark pairs, and multiple jets. For each of these processes a dedicated control region is defined. 
The production of boson ($W/Z$) pairs in which at least one boson decays to charged leptons and/or neutrinos (referred to as `dibosons' below), the single-top production, and the $\ttbar+W/Z$ boson production are small components of the total background and are estimated with MC simulated data. 

Figure~\ref{fig:RVsMRBaselineBackgrounds} shows the values of $R$ and $M_R'$ for the major SM backgrounds in this search. As previously discussed, the SM backgrounds tend to occupy regions with lower values of $M_R'$ and $R$, which is taken into account while defining control regions for the main background processes. A summary of the selection criteria used to define the control and validation regions in this search is shown in table~\ref{tab:Razor_CRVR}. %

\begin{sidewaystable}[htbp]
\small
\centering
\begin{tabular}{|c|c|c|c|c|} 
\hline
\multicolumn{5}{|c|}{Preselection}\\
\hline 
Exactly two opposite-sign,                           & Exactly one     & Exactly one                   & No leptons & No leptons\\ 
same-flavour  electrons                       &  electron or muon    &   electron or muon    &    &\\
or muons                &         &              &  &\\ \hline
$-$ &  No $b$-jets   & At least one $b$-jet    & $-$ & $-$ \\ \hline
66$<m(\ell\ell)<$116~GeV & $-$ & $-$  & $\pt^{\rm jet_{1,2}}>150$~GeV  & $\pt^{\rm jet_{1,2}}>200$~GeV \\ \hline
$E_{\rm T}^{\rm miss'} = \met + \pt(\ell\ell)$ & Treat lepton as a jet  & Treat lepton as a jet  & $-$ & $-$ \\ \hline
\hline
{\bf 0LRaz\_CRZ} & {\bf 0LRaz\_CRW} & {\bf 0LRaz\_CRT} & {\bf 0LRaz\_CRQ$_{\rm loose}$} & {\bf 0LRaz\_CRQ$_{\rm tight}$}\\ \hline \hline 
$-$ & $-$ & $-$ & $\Delta\phi(\pt^{\rm jet_{2}}, \met)<0.2$   & $\Delta\phi(\pt^{\rm jet_{2}}, \met)<0.2$  \\ \hline
$0.3<R<0.55$                     &   $0.3<R<0.55$                     & $0.3<R<0.55$                     &  $0.35<R<0.45$   &  $0.5<R<0.55$ \\ \hline
$M_{R}'>800$ GeV & $M_{R}'>800$ GeV & $M_{R}'>800$ GeV & $M_{R}'>1000$ GeV & $M_{R}'>1000$ GeV\\ 
\hline
\hline
{\bf 0LRaz\_VRZ} & {\bf 0LRaz\_VRW} & {\bf 0LRaz\_VRT}  & {\bf 0LRaz\_VRQ$_{\rm loose}$} & {\bf 0LRaz\_VRQ$_{\rm tight}$}\\
\hline \hline
$-$ & $-$ & $-$  & $\Delta\phi(\pt^{\rm jet_{2}}, \met)<0.4$   & $\Delta\phi(\pt^{\rm jet_{2}}, \met)<1.4$  \\ \hline
$0.55<R<1.0$                   & $0.55<R<1.0$                     & $0.55<R<1.0$                    & $0.45<R<0.5$ & $0.55<R<0.6$\\ \hline
$400<M_{R}'<1000$ GeV & $400<M_{R}'<1000$ GeV & $400<M_{R}'<1000$ GeV & $M_{R}'>900$ GeV & $M_{R}'>900$ GeV \\
\hline
\end{tabular}
\caption{\label{tab:Razor_CRVR} Overview of the selection criteria for the 
$Z$+jets, $W$+jets, semileptonic \ttbar\ and multi-jet control regions (0LRaz\_CRZ, 0LRaz\_CRW, 0LRaz\_CRT and 0LRaz\_CRQ$_{\rm loose/tight}$ respectively) and the corresponding validation regions (0LRaz\_VRZ, 0LRaz\_VRW, 0LRaz\_VRT, 0LRaz\_VRQ$_{\rm loose/tight}$) 
used by the 0-lepton Razor analysis. Details of the construction of Razor variables $M_{R}^{'}$ and $R$ can be found in the appendix \ref{app:razorvars}.}
\end{sidewaystable}

The largest potential background for a search with no leptons is expected to originate from the QCD-induced multi-jet event. 
However, the Razor variables were constructed to be able to distinguish this background from a SUSY-like signal, which minimizes the contribution of the multi-jet events after the SR event selection has been applied. 
To estimate the contribution of multi-jet background events in the final signal region selection, a data-driven technique \cite{Aad:2012fqa}, which applies a resolution function to well-measured multi-jet events in order to estimate the impact of jet energy mis-measurement and heavy-flavour semileptonic decays on \met{} and other variables, is used. Two dedicated control regions, CRQ$_{\rm loose}$ and CRQ$_{\rm tight}$, which use different selection criteria on $R$ and $\Delta\phi(p_{T}^{j2}, \met)$, correspond to the loose and tight signal regions respectively, and select samples of events with similar kinematics to the SR but enriched in multi-jet background events. 

Since the QCD multi-jet background is significantly reduced by use of the Razor variables, the largest remaining backgrounds come from the production of $W/Z$ bosons with additional jets and semileptonic $t\bar{t}$ decays, where the leptons can be mis-reconstructed as jets, non-prompt or be outside the lepton identification criteria. For each of these backgrounds a control region, rich in the respective process, is defined. The trigger requirements for these lepton-rich control regions follow those used by the corresponding control regions for the the 0L search \cite{0-leptonPaper}. The Razor variables are used to preselect a region which is dominated by the particular process. Following this preselection, to control the $t\bar{t}$ background, the control region CRT requires at least one jet tagged as a $b$-jet and exactly one electron or muon, while the $W$+jets control region, CRW, applies a veto on the presence of $b$-jets and requires exactly one electron or muon. In both cases the lepton is treated as a jet in the reconstruction of the Razor variables. This treatment of leptons as jets is motivated by the observation that  $\sim$75$\%$ of W($\rightarrow$ $\ell\nu$)+jets and semileptonic $t\bar{t}$ events appearing in the SRs possess leptons which have fake jets, either through the misidentification of electrons or by the production of tau leptons decaying hadronically (identification of hadronic tau decays is not used in this analysis).
The $Z$+jets control region, CRZ, is required to have exactly two opposite-sign same-flavour leptons. The $Z$+jets contribution to the signal region largely originates from $Z$ decays to neutrinos. To mimic the behaviour of $Z \rightarrow \nu\nu$ in the control region, the two leptons have been treated as invisible and are used to re-calculate ${\bm E_{\rm{T}}^{\rm{miss}}}$ as ${\bm E_{\rm{T}}^{\rm{miss'}} = \bm E_{\rm{T}}^{\rm{miss}} + \bm p_{\rm T}(\ell \ell)}$ with the invariant mass of the leptons falling within a $Z$-mass window, $66 < m(\ell \ell) < 116$~GeV. The Razor variables are also calculated with this methodology where the leptons are treated as invisible objects.

To validate the normalization parameters extracted in the background control regions, validation regions VRZ, VRW, VRT, VRQ$_{\rm loose}$ and VRQ$_{\rm tight}$ for $Z$+jets, $W$+jets, $t\bar{t}$ and multi-jet backgrounds respectively, are defined (table \ref{tab:Razor_CRVR}). These validation regions are statistically independent from the signal and control regions previously defined, and are expected to have minimal contribution from any signal, if present.

Following the definition of the control and validation regions, the backgrounds from $Z$+jets, $W$+jets, $\ttbar$ and multi-jet processes are estimated by using the background-only fit, as described in section \ref{sec:strategy}. Figures \ref{fig:MRinCRs} and \ref{fig:MRinVRs} show the $M_R'$ distributions for these control and validation regions after the fit. Good agreement is seen between the fitted and observed yields in all regions. 

\begin{figure}[htbp]
\centering
\subfigure[]{\includegraphics[width=0.4\textwidth]{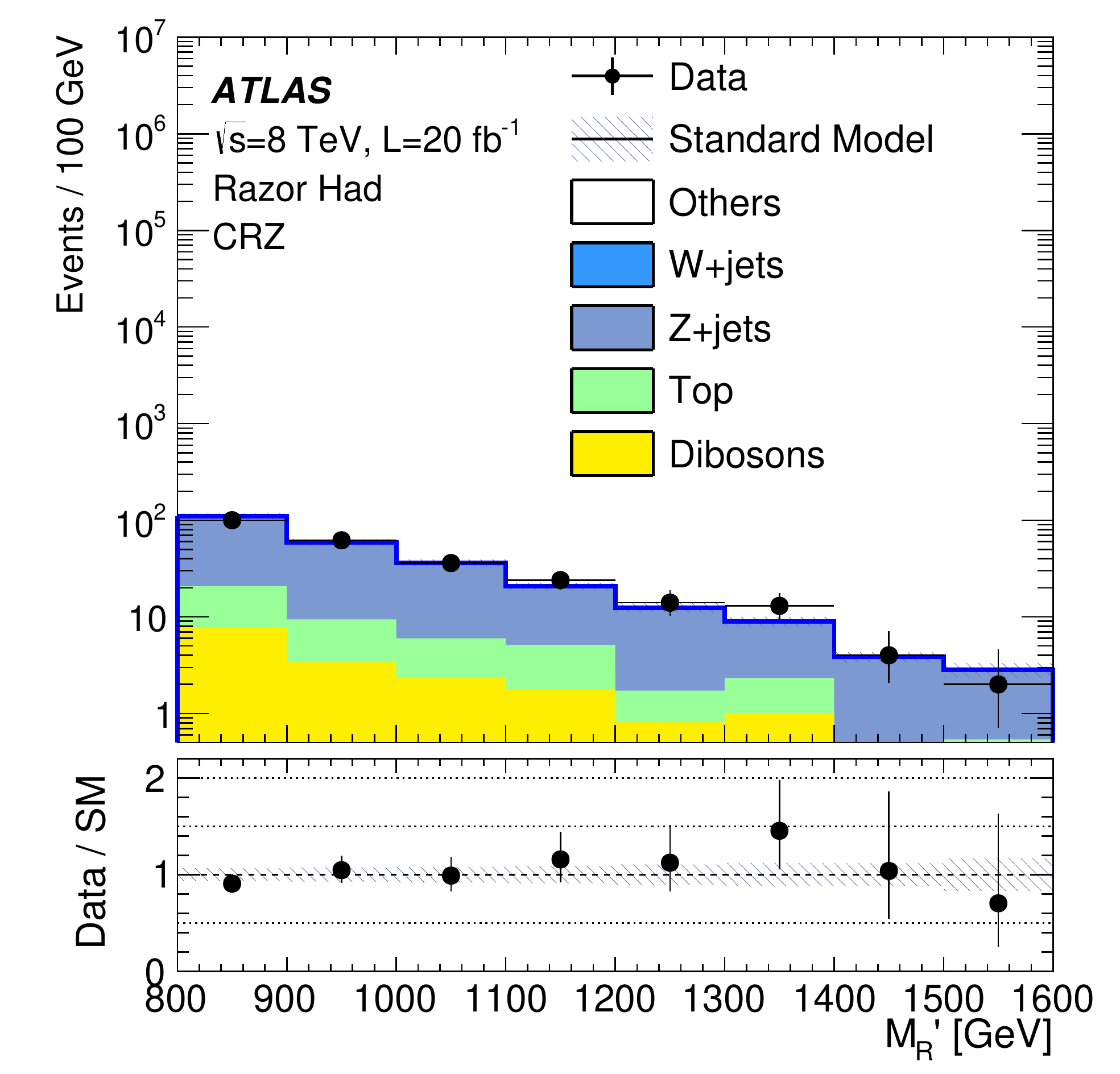}}
\subfigure[]{\includegraphics[width=0.4\textwidth]{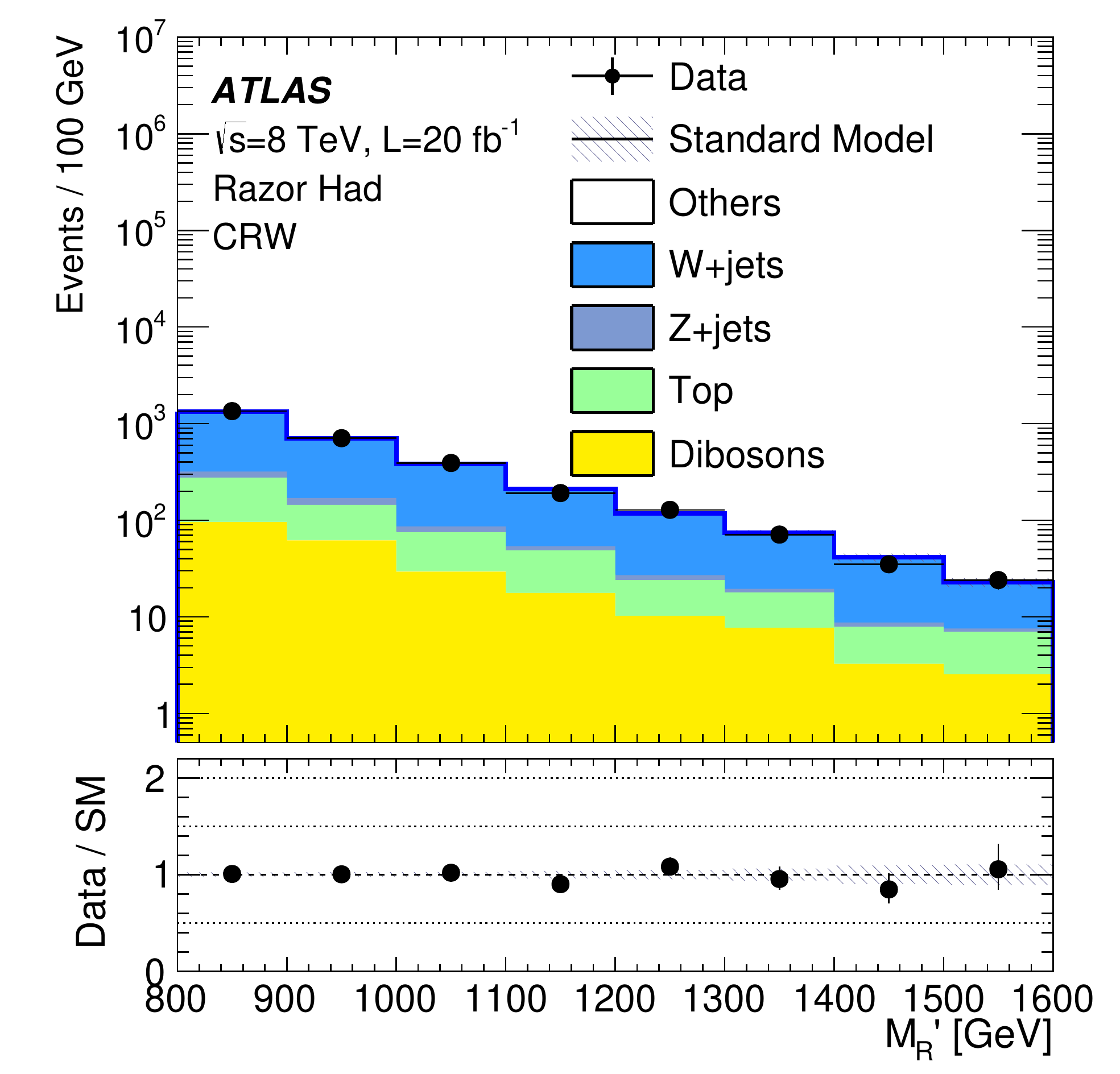}}
\subfigure[]{\includegraphics[width=0.4\textwidth]{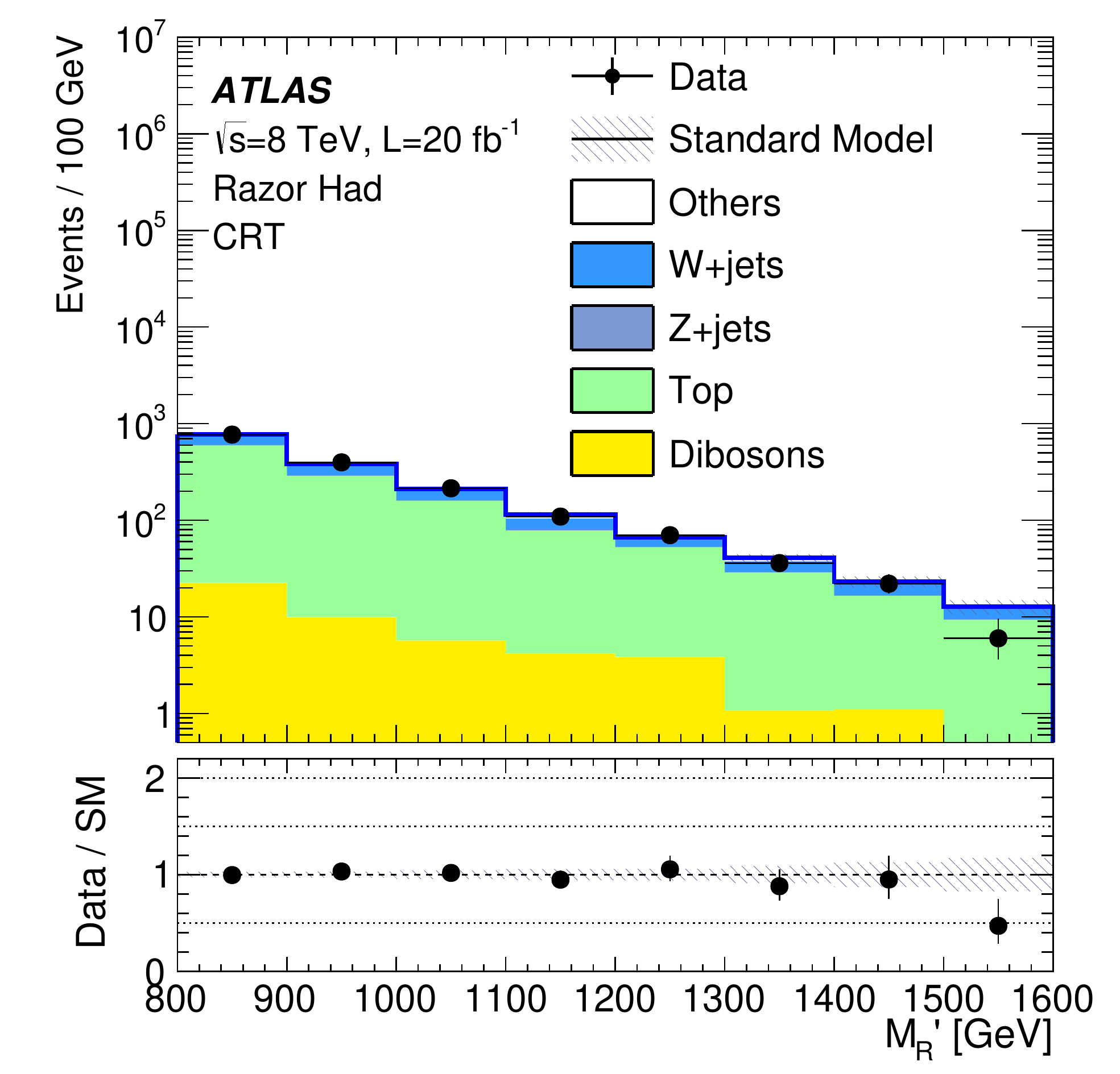}}
\subfigure[]{\includegraphics[width=0.4\textwidth]{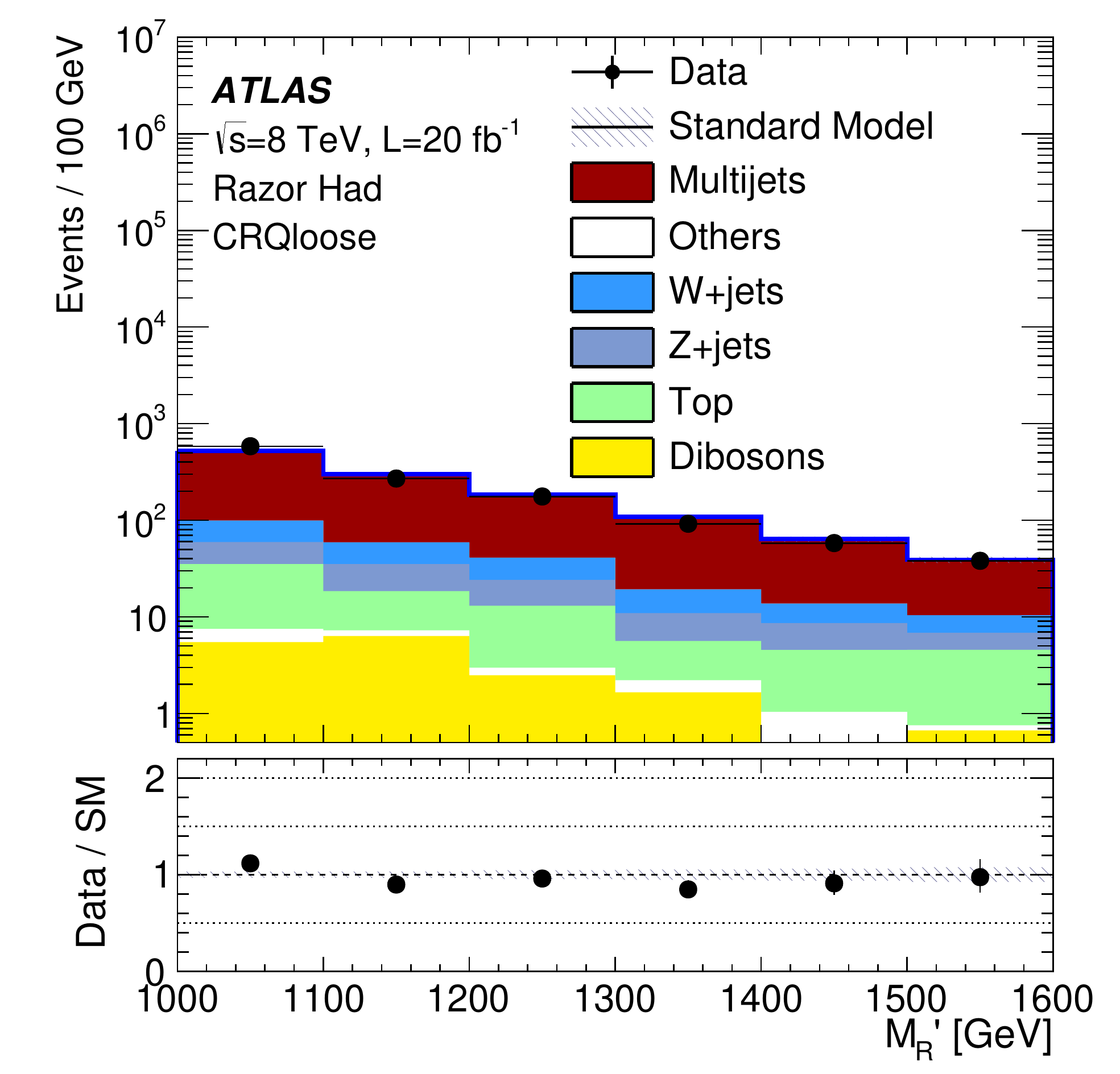}}
\subfigure[]{\includegraphics[width=0.4\textwidth]{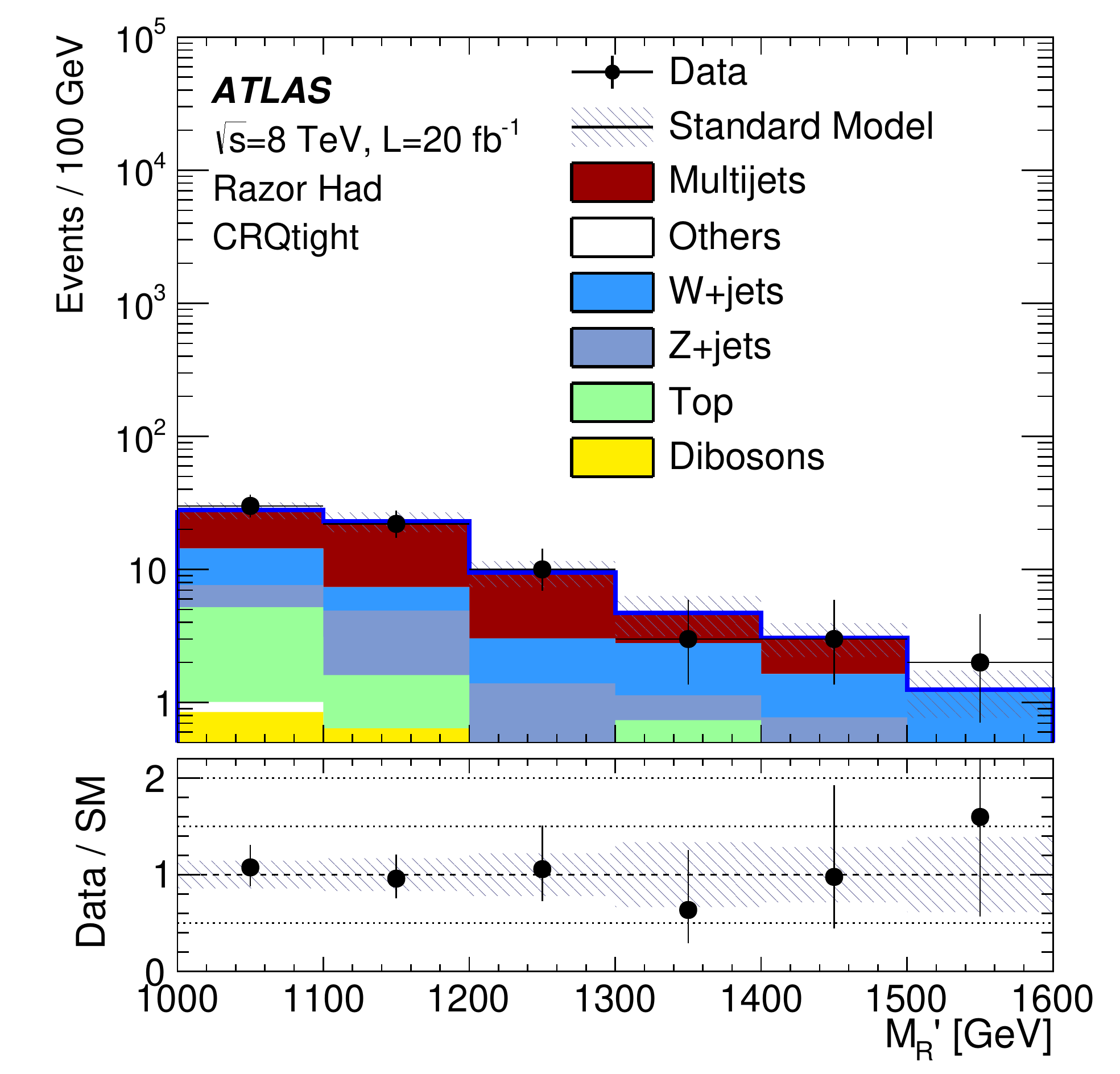}}
\caption{Observed $M_{R}'$ distributions in control regions for (a) $Z$+jets, (b) $W$+jets, (c) $\ttbar$ and multi-jet backgrounds for (d) loose and (e) tight selection. The ``Top" label includes all top-quark-related
backgrounds ($\ttbar$, single top and $\ttbar+V$), while the ``Others" includes the contributions of the jets misidentified as leptons or of non-prompt leptons, and the $\gamma$+jets background which is estimated with MC simulated data.   
All distributions are after the background-only fit has been performed. 
} \label{fig:MRinCRs}
\end{figure}

\begin{figure}[htbp]
\centering
\subfigure[]{\includegraphics[width=0.4\textwidth]{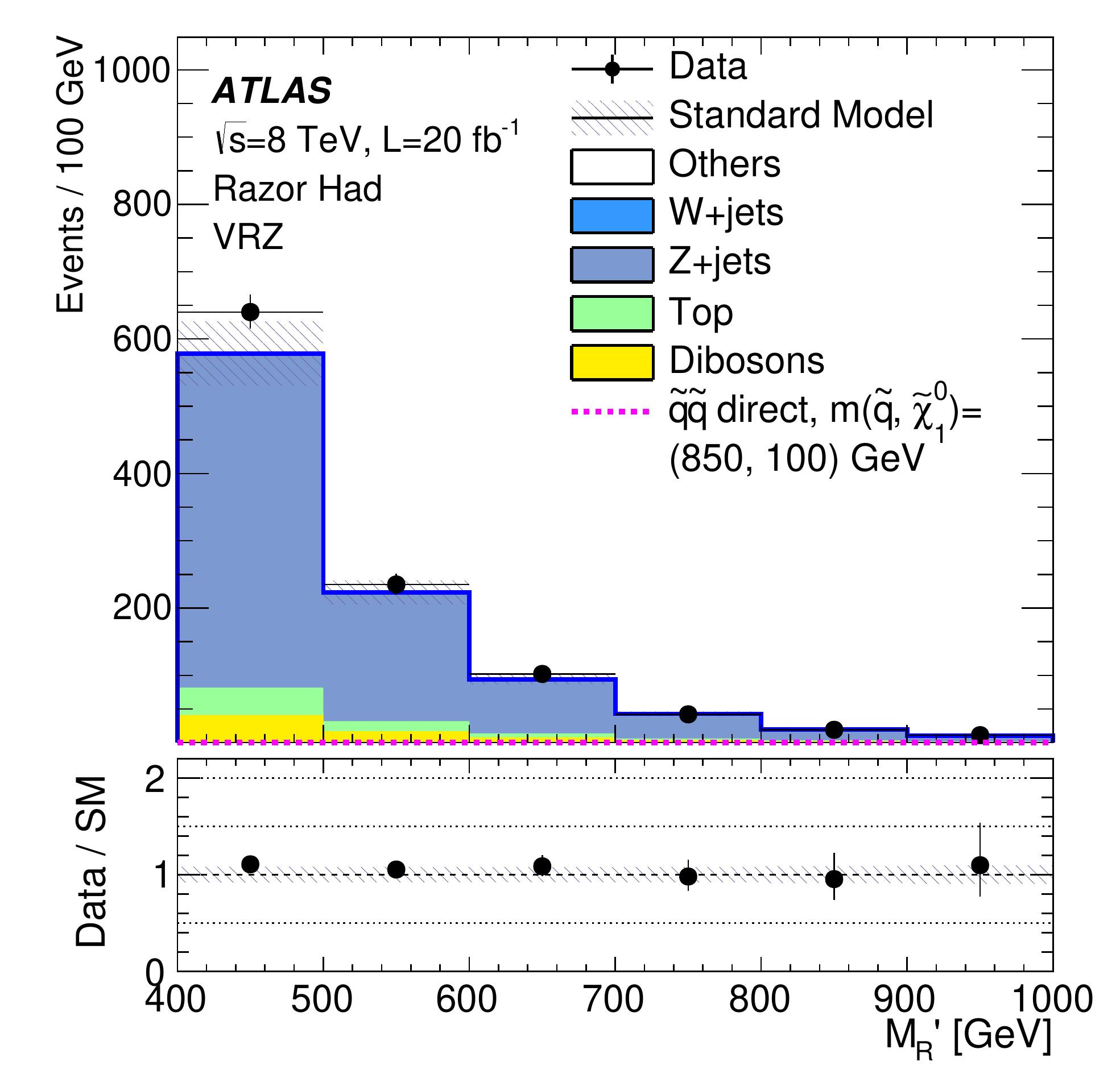}}
\subfigure[]{\includegraphics[width=0.4\textwidth]{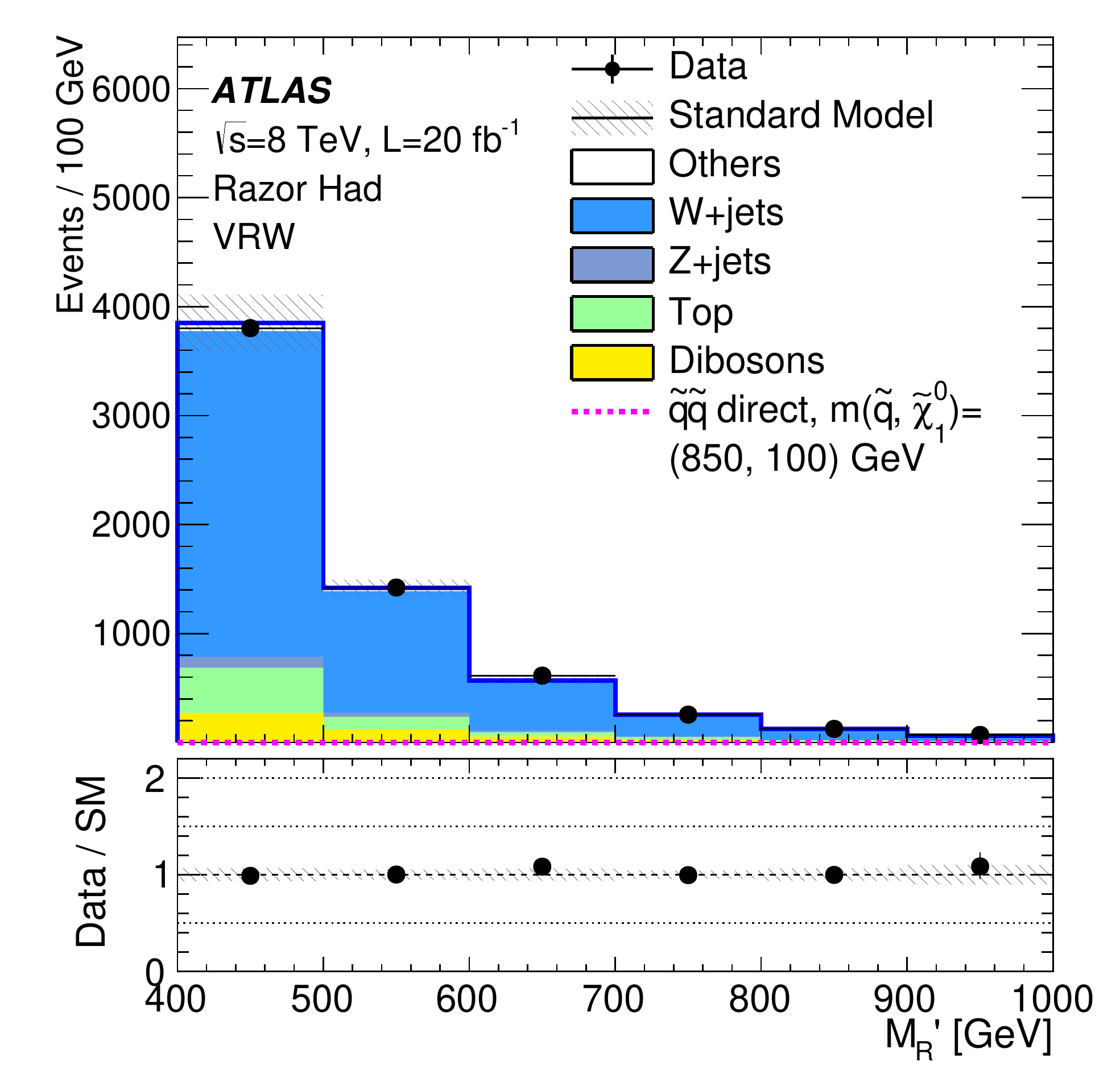}}
\subfigure[]{\includegraphics[width=0.4\textwidth]{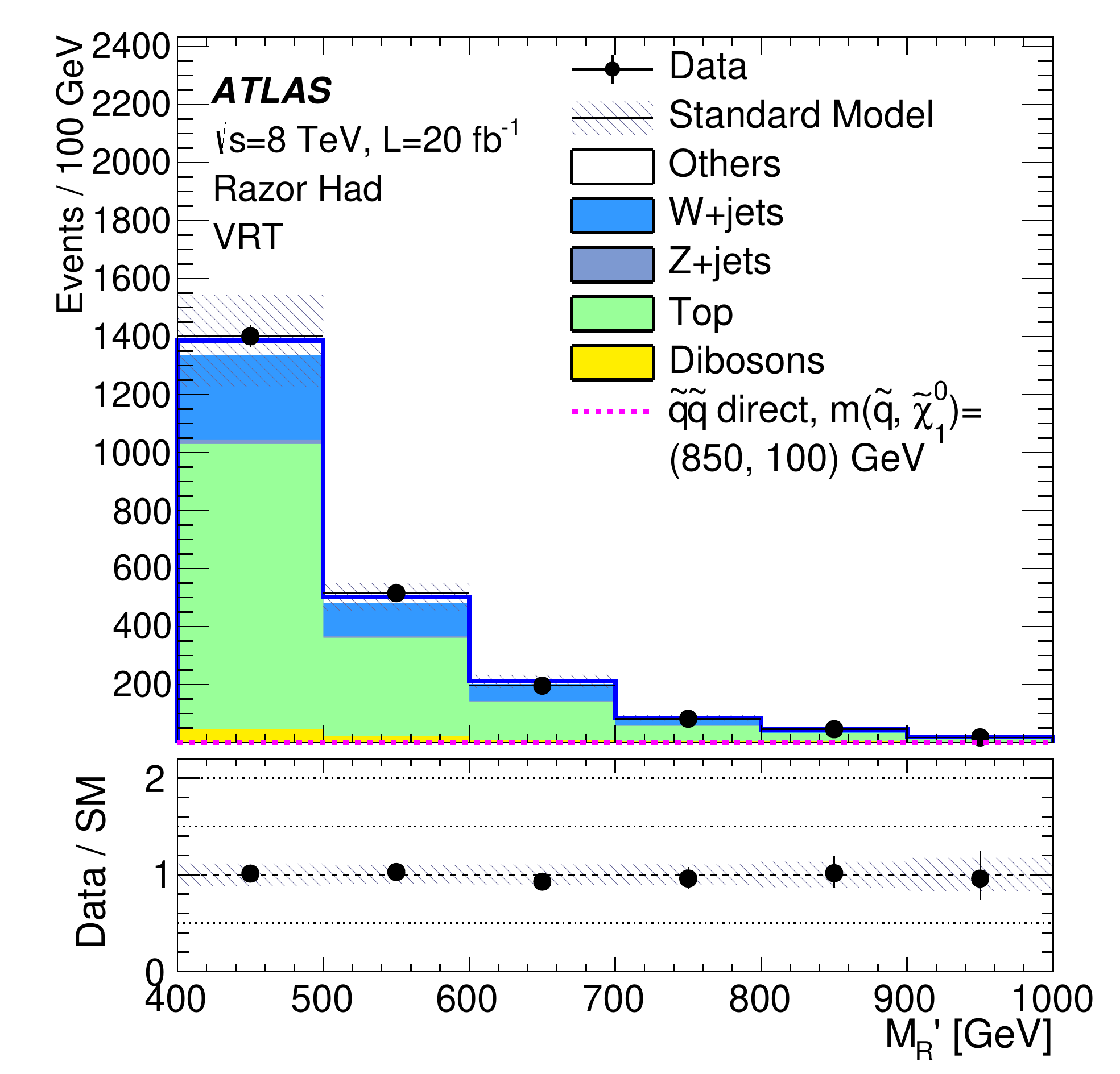}}
\subfigure[]{\includegraphics[width=0.4\textwidth]{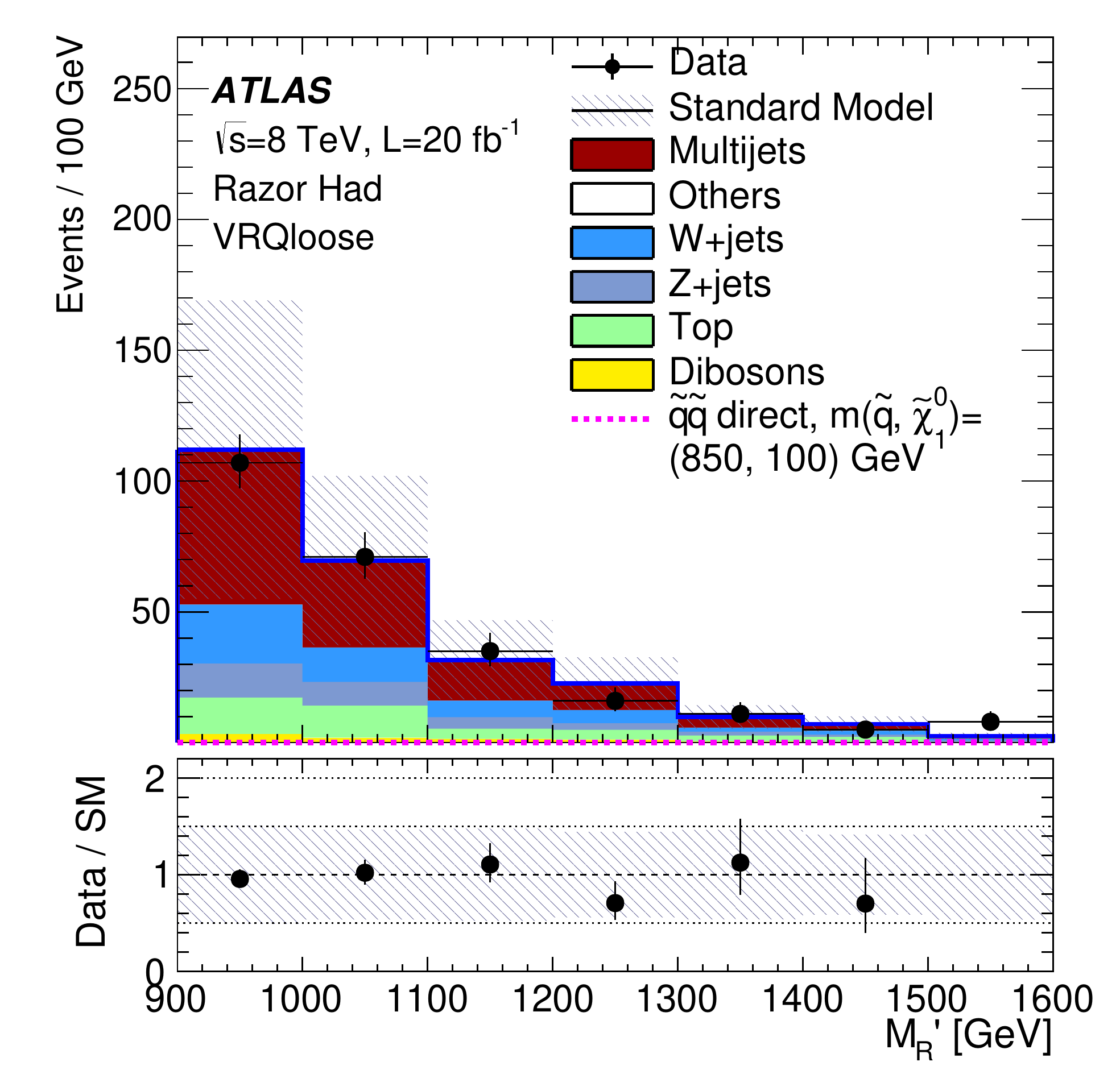}}
\subfigure[]{\includegraphics[width=0.4\textwidth]{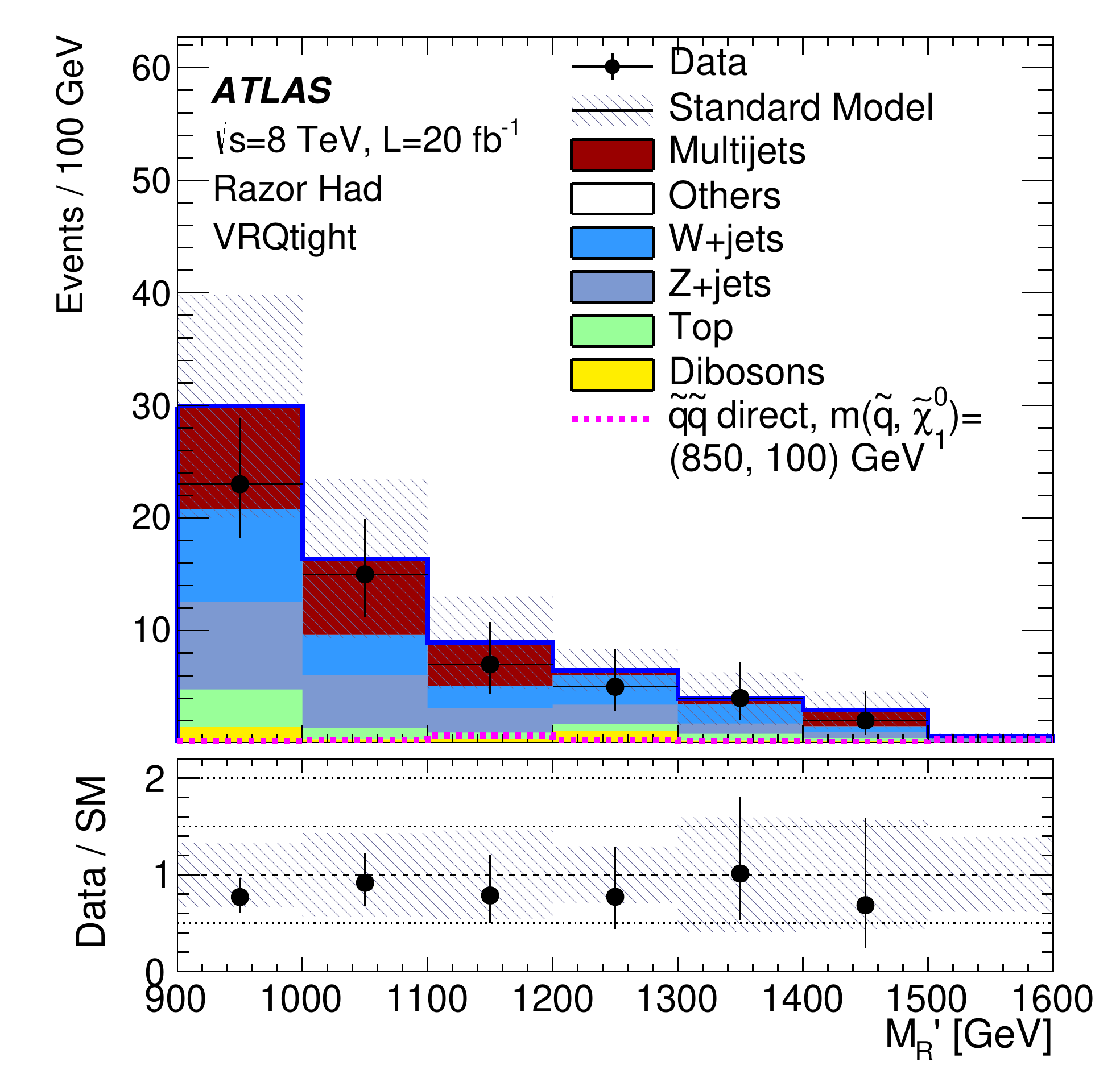}}
\caption{Observed $M_{R}'$ distributions in validation regions for (a) $Z$+jets, (b) $W$+jets, (c) $\ttbar$ and multi-jet backgrounds for (d) loose and (e) tight selection. The ``Top" label includes all top-quark-related
backgrounds ($\ttbar$, single top and $\ttbar+V$), while the ``Others" includes the contributions of the jets misidentified as leptons or of non-prompt leptons, and the $\gamma$+jets background which is estimated with MC simulated data. 
All distributions are after the background-only fit has been performed. 
} \label{fig:MRinVRs}
\end{figure}

The Razor variable distributions, $M_{R}'$ and $R$, for the SM backgrounds and a simplified model point with 
large-$\Delta m_{\rm signal}$ are shown in figure \ref{fig:RazorVarsinSRs} for SR$_{\rm loose}$ and SR$_{\rm tight}$.

\begin{figure}[htbp]
\centering
\subfigure[]{\includegraphics[width=0.4\textwidth]{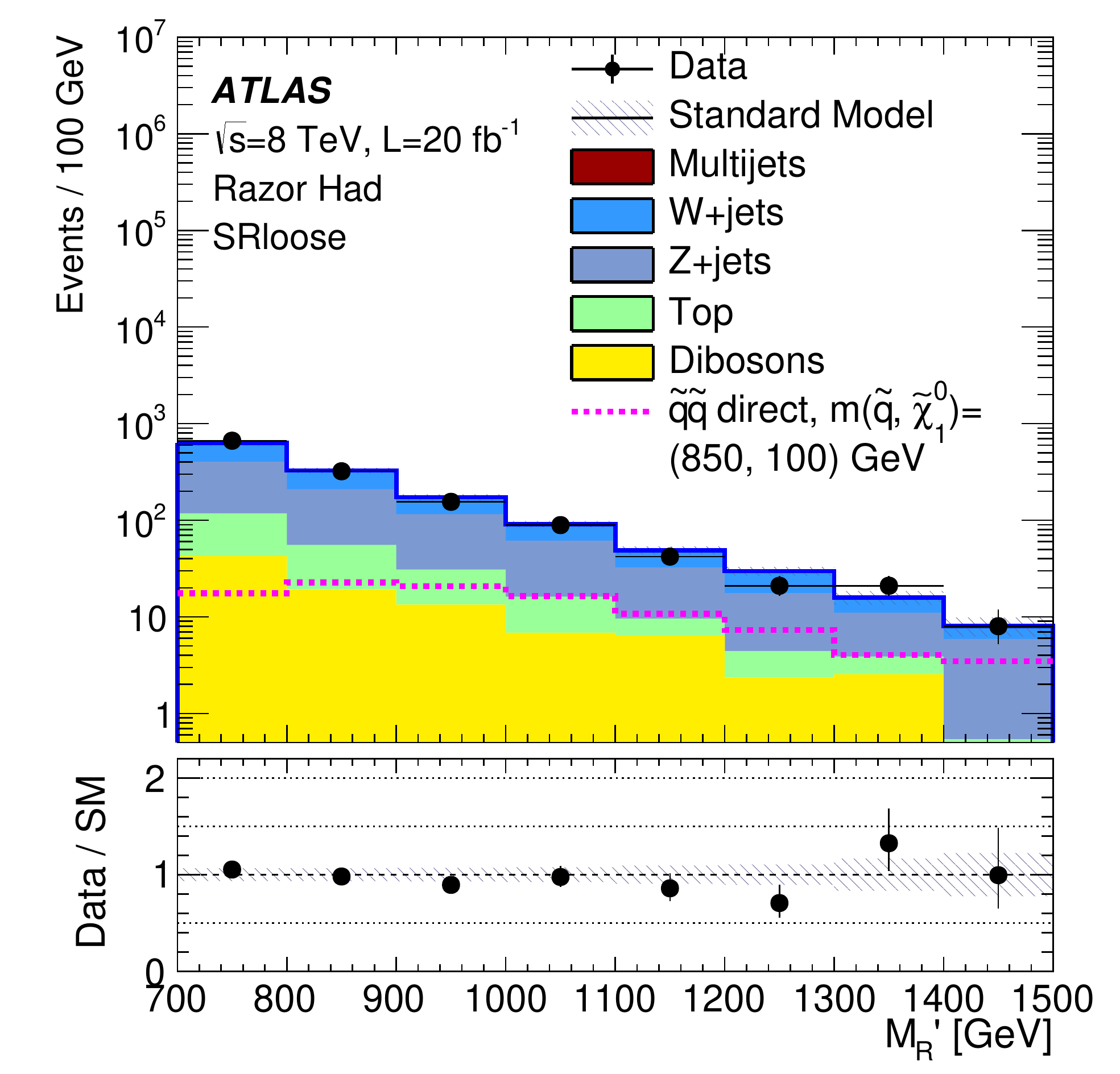}}
\subfigure[]{\includegraphics[width=0.4\textwidth]{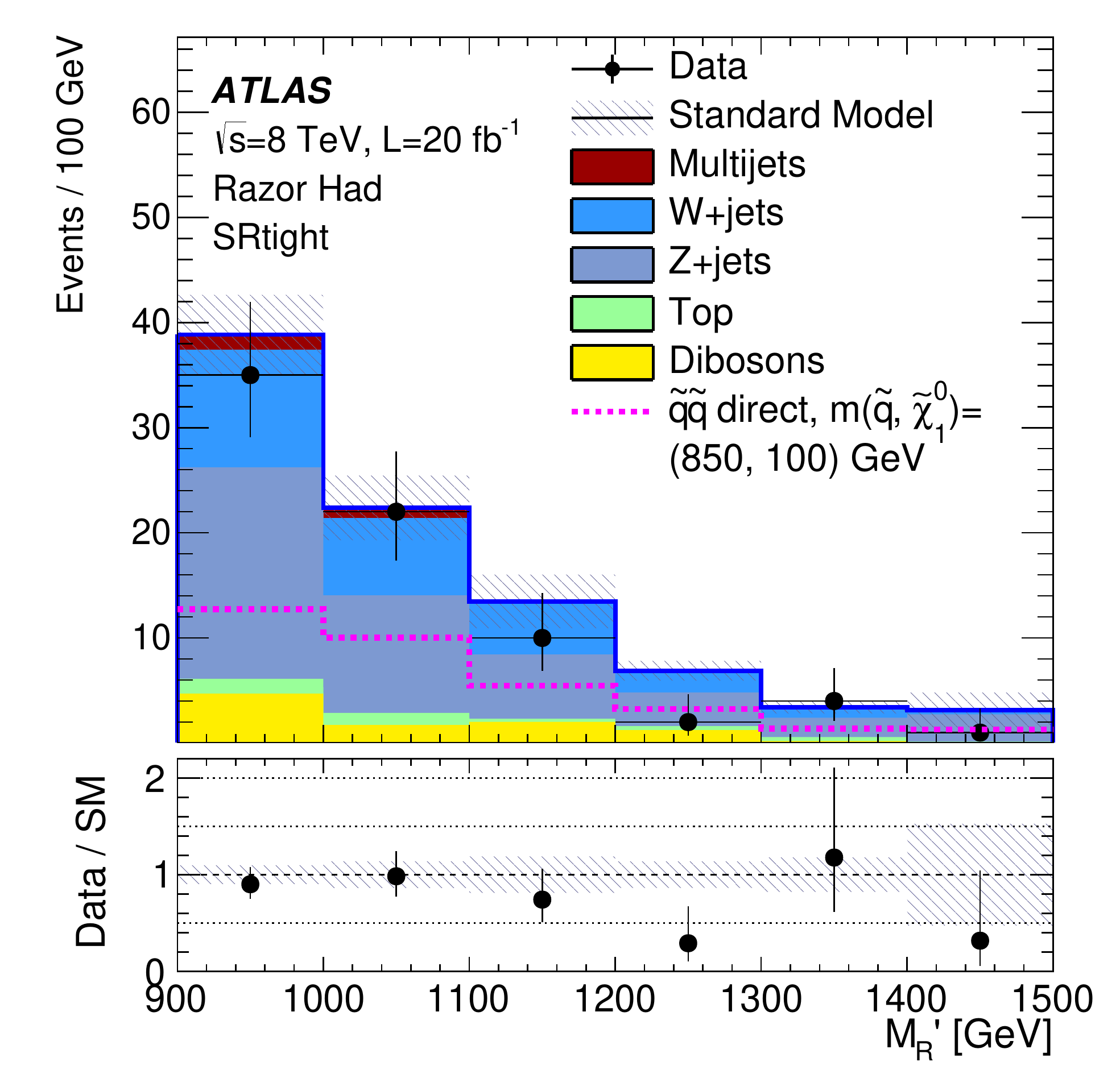}}
\subfigure[]{\includegraphics[width=0.4\textwidth]{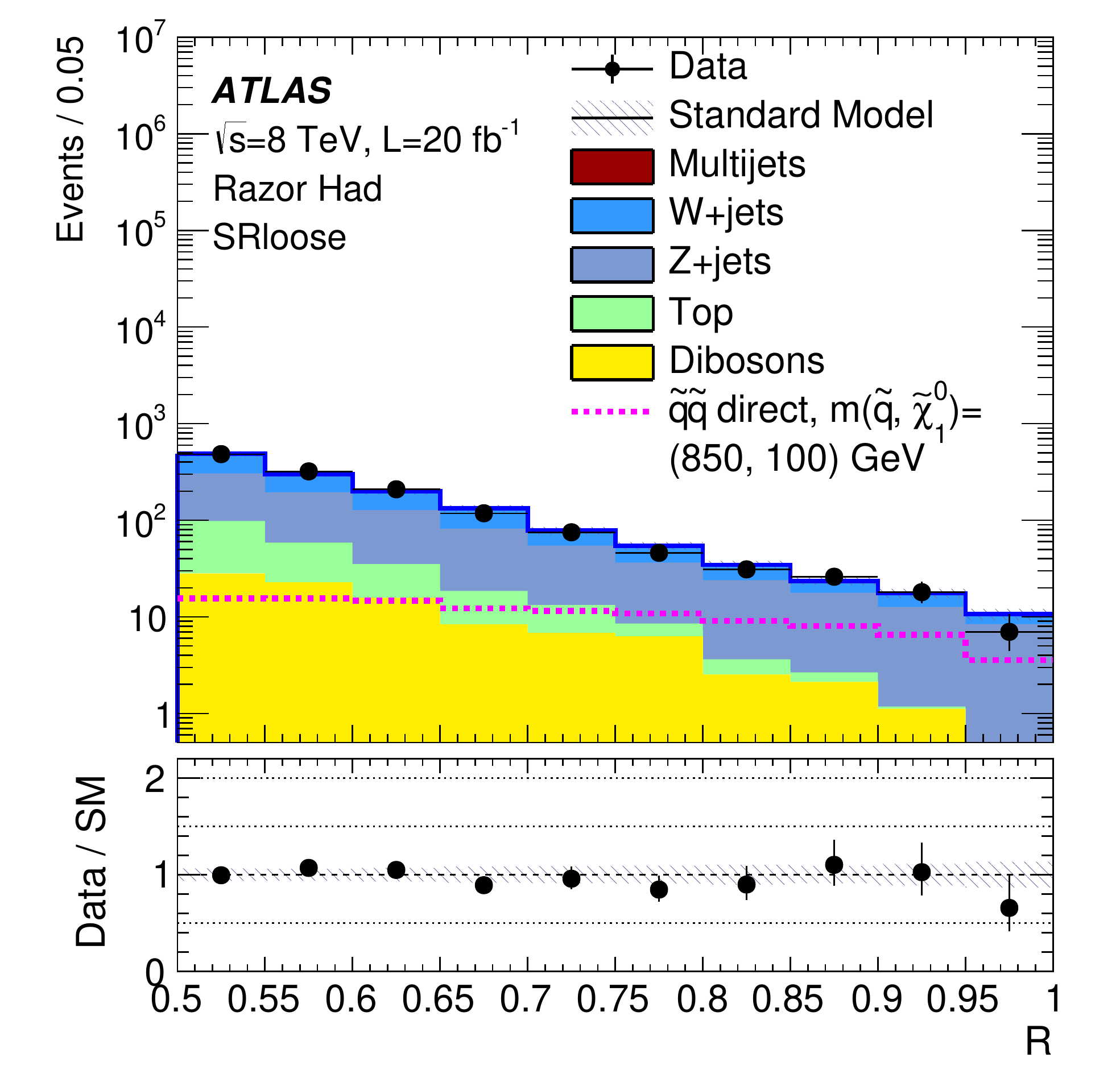}}
\subfigure[]{\includegraphics[width=0.4\textwidth]{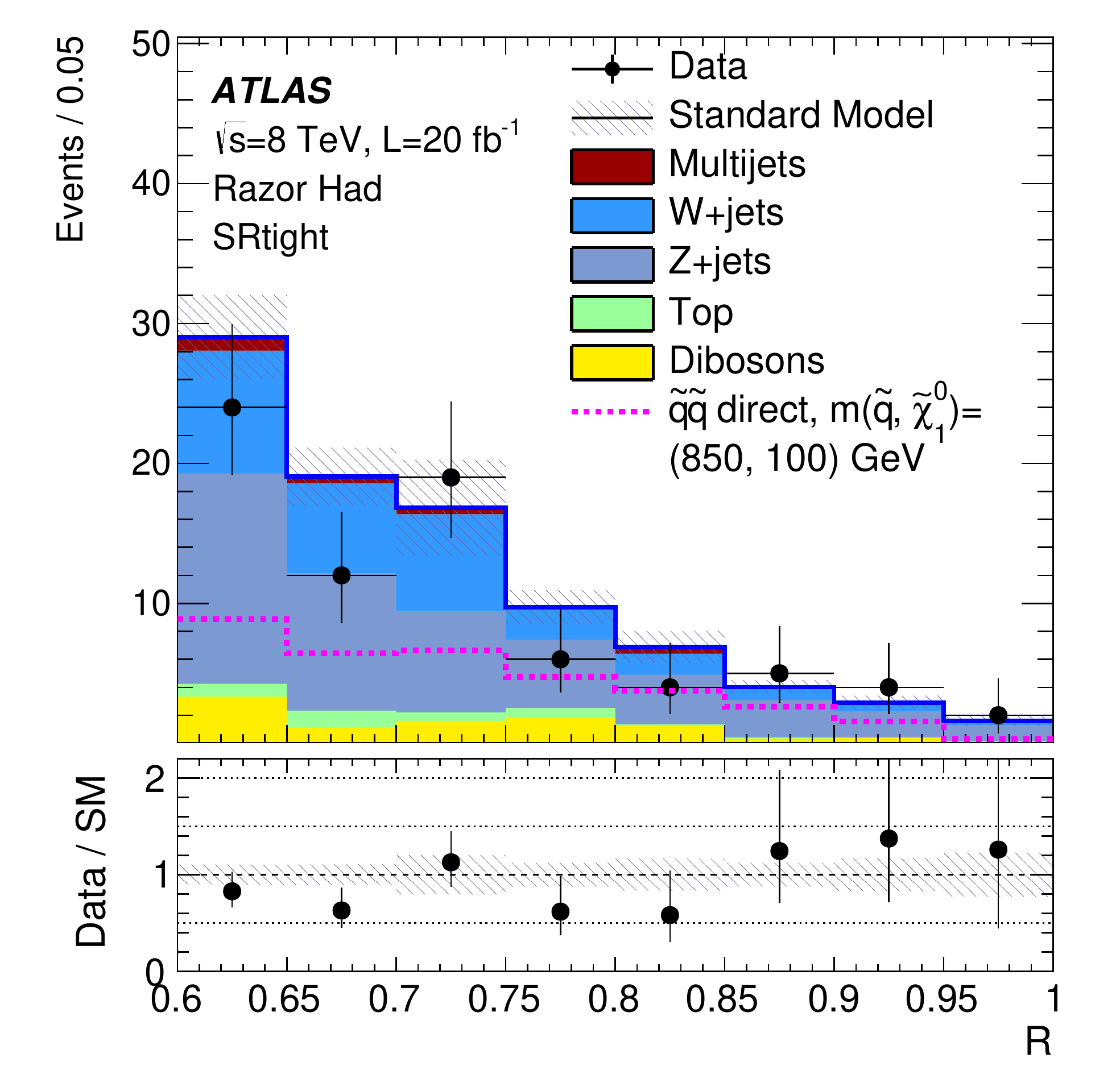}}
\caption{Observed (a, c) $M_{R}'$ and (b, d) $R$ distributions in (a, b) loose and (c, d) tight signal region selections listed in table \ref{tab:Razor_SR}. The ``Top" label includes all top-quark-related
backgrounds ($\ttbar$, single top and $\ttbar+V$).   
All distributions are after the background-only fit has been performed. 
} \label{fig:RazorVarsinSRs}
\end{figure}

\newpage
\printbibliography

\newpage 
\begin{flushleft}
{\Large The ATLAS Collaboration}

\bigskip

G.~Aad$^{\rm 85}$,
B.~Abbott$^{\rm 113}$,
J.~Abdallah$^{\rm 151}$,
O.~Abdinov$^{\rm 11}$,
R.~Aben$^{\rm 107}$,
M.~Abolins$^{\rm 90}$,
O.S.~AbouZeid$^{\rm 158}$,
H.~Abramowicz$^{\rm 153}$,
H.~Abreu$^{\rm 152}$,
R.~Abreu$^{\rm 116}$
% ...
\bigskip
\\
% ...
$^{11}$ Institute of Physics, Azerbaijan Academy of Sciences, Baku, Azerbaijan\\
% ...
$^{85}$ CPPM, Aix-Marseille Universit{\'e} and CNRS/IN2P3, Marseille, France\\
% ...
$^{90}$ Department of Physics and Astronomy, Michigan State University, East Lansing MI, United States of America\\
% ...
$^{107}$ Nikhef National Institute for Subatomic Physics and University of Amsterdam, Amsterdam, Netherlands\\
% ...
$^{113}$ Homer L. Dodge Department of Physics and Astronomy, University of Oklahoma, Norman OK, United States of America\\
% ...
$^{116}$ Center for High Energy Physics, University of Oregon, Eugene OR, United States of America\\
% ...
$^{151}$ Institute of Physics, Academia Sinica, Taipei, Taiwan\\
$^{152}$ Department of Physics, Technion: Israel Institute of Technology, Haifa, Israel\\
$^{153}$ Raymond and Beverly Sackler School of Physics and Astronomy, Tel Aviv University, Tel Aviv, Israel\\
% ...
$^{158}$ Department of Physics, University of Toronto, Toronto ON, Canada\\
% ...
\end{flushleft}


\end{document}